\pdfoutput=1
\documentclass[11pt,letterpaper]{article}
\usepackage[letterpaper,margin=1in]{geometry}
\usepackage{microtype}
\usepackage{amsmath,amssymb,amsthm,mathtools}
\usepackage{booktabs,array,tabularx}
\usepackage[inline]{enumitem}
\usepackage{xcolor}
\usepackage{tikz}
\usepackage{pgfplots}
\pgfplotsset{compat=1.18}
\usetikzlibrary{calc,decorations.pathreplacing,arrows.meta,patterns}
\usepackage{subcaption}
\usepackage{algorithm}
\usepackage{algpseudocode}
\usepackage[colorlinks=true,linkcolor=blue!50!black,citecolor=green!35!black,urlcolor=blue!50!black]{hyperref}
\usepackage[capitalise,noabbrev,nameinlink]{cleveref}

\makeatletter\providecommand{\theHALG@line}{}\renewcommand{\theHALG@line}{\thealgorithm.\arabic{ALG@line}}\makeatother
\hypersetup{hypertexnames=false}

\setlist{itemsep=1pt,topsep=3pt}

\theoremstyle{plain}
\newtheorem{theorem}{Theorem}[section]
\makeatletter
\@ifundefined{@amsthm@saved@ynthm}%
  {\RequirePackage{aliascnt}%
   \newcommand*{\sharedtheorem}[2]{\newaliascnt{#1}{theorem}\newtheorem{#1}[#1]{#2}\aliascntresetthe{#1}}}
  {\newcommand*{\sharedtheorem}[2]{\newtheorem{#1}[theorem]{#2}}}
\makeatother
\sharedtheorem{lemma}{Lemma}
\sharedtheorem{proposition}{Proposition}
\sharedtheorem{corollary}{Corollary}
\theoremstyle{definition}
\sharedtheorem{definition}{Definition}
\numberwithin{equation}{section}
\usepackage{thm-restate}
\makeatletter
\newcommand{\thissection}{this section}
\newcommand{\restate}[2]{\begingroup
  \def\thmhead##1##2##3{\hyperref[#2]{\thmname{##1}\thmnumber{\@ifnotempty{##1}{ }\@upn{##2}}}\thmnote{ {\the\thm@notefont(##3)}}}%
  #1*\endgroup}
\makeatother

\newcommand{\R}{\mathbb{R}}
\newcommand{\Q}{\mathbb{Q}}
\newcommand{\Z}{\mathbb{Z}}
\newcommand{\Oh}{\mathcal{O}}
\DeclareMathOperator*{\argmax}{arg\,max}
\DeclareMathOperator{\dist}{dist}
\DeclareMathOperator{\med}{med}
\DeclareMathOperator{\clamp}{clamp}
\DeclareMathOperator{\cut}{cut}
\DeclareMathOperator{\relint}{relint}
\DeclareMathOperator{\aff}{aff}

\newcommand{\Pp}[1]{P_{(+)}(#1)}
\newcommand{\Bpm}[1]{B_{(\pm)}(#1)}

\newcommand{\xs}{x^{\star}}
\newcommand{\as}{\alpha^{\star}}
\newcommand{\UA}{U_{\text{A}}}
\newcommand{\UP}{U_{\text{P}}}
\newcommand{\rbar}{\bar r}
\newcommand{\FO}{\mathrm{FO}}
\newcommand{\binv}{\beta^{-1}}

\newcommand{\Ccut}{\mathcal{C}_{\mathrm{cut}}}
\newcommand{\Fin}{\mathfrak{F}}
\newcommand{\muin}{\mu^{\mathrm{in}}}
\newcommand{\aknob}{\alpha_{\circ}}
\newcommand{\enc}[1]{\langle #1\rangle}
\newcommand{\SSq}{S_{2}}

\newcommand{\iverson}[1]{[#1]}

\begin{document}

\pagenumbering{roman}
\begin{titlepage}
\begin{center}
{\LARGE\bfseries The Complexity of Continuous Contract Design \newline over Polymatroids\par}
\vspace{1.5em}
{\large
\begin{tabular}{c@{\hspace{3em}}c@{\hspace{3em}}c}
Max Klimm & Paul Lippke & Philipp Warode\\[2pt]
\small TU Berlin & \small TU Berlin & \small HU Berlin\\
\small klimm@math.tu-berlin.de & \small lippke@math.tu-berlin.de & \small philipp.warode@hu-berlin.de
\end{tabular}\par}
\end{center}
\vspace{2em}
\begin{abstract}
We study optimal contracts when an agent allocates continuous effort subject to polymatroid capacity constraints. Effort produces a linear expected reward and incurs a cost. With binary outcomes and limited liability, the principal chooses a single reward share. For linear costs, an optimal contract can be computed in polynomial time. For quadratic costs, the agent has a unique best response computable by polynomially many submodular minimizations, yet optimizing the principal's share is NP-hard even for explicitly represented cut-based rank functions. With $n$ activities, the best-response path can have $2^{n-\Oh(\log n)}$ breakpoints even when each reward coefficient and each rank value is an integer representable in $\Oh(n\log n)$ bits. The construction also answers a question of Allman, Lo, and McCormick on polynomial bit length for source-sink monotone two-parameter minimum cuts.

Despite this complexity, an optimal contract for quadratic costs always exists with a share of at most $1/2$. Consequently, $\Oh(1/\varepsilon)$ best-response computations suffice for a $(1-\varepsilon)$-approximation. For rational quadratic costs on explicitly described rational polytopes containing zero, this gives a fully polynomial-time approximation scheme. More generally, a share of at most $1/p$ suffices for continuous convex costs homogeneous of degree $p>1$, on any compact convex action set containing zero. For general convex costs with an exact value and bounded-subgradient oracle, we obtain an oracle approximation scheme using approximate best responses and a supplied lower bound on positive welfare.
\end{abstract}

\begingroup
\setcounter{tocdepth}{1}
\makeatletter
\let\originaltocsection\l@section
\renewcommand*{\l@section}[2]{\begingroup\let\addvspace\@gobble\originaltocsection{#1}{#2}\endgroup}
\makeatother
\tableofcontents
\endgroup
\end{titlepage}
\pagenumbering{arabic}

\section{Introduction}\label{sec:intro}

Contract design studies how a principal can incentivize an agent to undertake costly effort. An employer hiring a worker or a firm commissioning a project cannot directly observe the agent's effort, but can commit to payments based on observable outcomes. The contract must balance the incentives it provides with the payment it requires. Classical multitask models study how incentives affect effort across activities~\cite{HM87,HM91}. A central analytical tool is the \emph{first-order approach}, which replaces the agent's global optimality requirement by a first-order condition~\cite{Hol79}. Establishing when this replacement is valid is itself an important part of the theory~\cite{Mir99,Rog85}.

Recent years have seen substantial progress in the algorithmic study of contract design. A \emph{linear contract}, which pays the agent a fixed share of the realized reward, admits tight parameterized approximation guarantees relative to an optimal contract~\cite{DRT19}. For combinatorial action spaces, efficient algorithms compute optimal contracts under gross-substitutes success probabilities~\cite{Dut+21}, and approximation schemes extend to general monotone success probabilities~\cite{DEFK25}. These results reveal how the structure of the agent's choices affects the computational complexity of providing incentives. The surveys of D\"utting, Feldman, and Talgam-Cohen~\cite{DFTC24} and of Feldman~\cite{Fel25} give broader overviews.

Much of this algorithmic progress concerns \emph{finite} action spaces, including exponentially large families of subsets. In many applications, however, an agent chooses not only which activities to undertake, but also how much time or effort to devote to each. These choices range over a \emph{continuum} and are coupled by shared resource limits. Although continuous effort is central to classical economic models, the algorithmic results above do not directly resolve the complexity of computing contracts in this setting. With finitely many actions, the best response changes only when the agent switches between actions, giving a finite set of candidate shares. With continuous effort, the response can vary throughout an interval, so this reduction no longer applies. Discretizing effort also requires controlling how the approximation changes the agent's incentives and the principal's utility. This leaves a computational question: \emph{Which structures of continuous effort costs and feasible allocations permit efficient exact or approximate contract design?}

This paper develops an algorithmic theory of contract design for continuous effort allocations under polymatroid constraints. Polymatroids capture several natural forms of resource sharing. A contractor may allocate working time among activities grouped into projects, with budgets for individual projects, departments, and the contract as a whole. When these groups are nested or disjoint, their effort limits define a laminar polymatroid~\cite{FG86}.
Polymatroid constraints also arise when activities share machines, but not every machine can serve every activity. Suppose work can be split among eligible machines. The total workload of any set $S$ of activities is bounded by the combined capacity $f(S)$ of the machines that can serve at least one activity in $S$. This weighted coverage function is submodular, and the max-flow min-cut theorem shows that these bounds suffice: a nonnegative workload vector $x$ is feasible whenever $\sum_{e\in S}x_e\le f(S)$ for every~$S$~\cite[(44.13)--(44.15)]{Sch03}. The model thus captures service providers sharing compatible equipment among clients. Similarly, in a capacitated network with a single source, feasible delivery volumes form a polymatroid whose rank on a set of destinations is the maximum flow to that set~\cite{Gro91}. Contracts influence both total effort and its allocation across these shared resources.

In our model, the agent chooses an effort vector~$x$ in the independence polyhedron $\Pp{f}$ of a polymatroid on a ground set~$E$ of $n$ activities, incurs a cost $c(x)$, and generates expected reward $r(x)=m^\top x$ for the principal. The outcome is binary, success or failure, with success probability proportional to $r(x)$. In this setting, it suffices to use a linear contract, as for finitely many actions~\cite{DFTC24}. The principal chooses a share $\alpha\in[0,1]$ to maximize $\UP(\alpha)=(1-\alpha)\,r(\xs(\alpha))$, where $\xs(\alpha)$ is the effort allocation maximizing the agent's utility $\alpha m^\top x-c(x)$ over $\Pp{f}$.
The results depend on the agent's cost function.

\begin{description}
\item[Linear costs $c(x) = \sum_e c_e x_e$.] Edmonds' greedy algorithm~\cite{Edm03} gives a best response that is a step function with $\Oh(n^2)$ candidate shares. Checking all of them computes an optimal contract with $\Oh(n^3)$ rank-value queries (\cref{thm:linear}). With integral capacities and positive linear costs, our model admits an exact expansion into a gross-substitutes combinatorial contract~\cite{Dut+21}, with one discrete action per unit of each activity's capacity (\cref{sec:A-linear}). Our algorithm works directly on the polymatroid and avoids this pseudo-polynomial expansion.

\item[Quadratic costs $c(x) = \sum_e \beta_e x_e^2/2$.] For $\beta_e>0$, the agent has a unique best-response allocation~$\xs(\alpha)$ at each share $\alpha$. Water-filling computes this allocation with $\Oh(n^2)$ submodular minimizations (\cref{thm:waterfilling}), yet optimizing the share is NP-hard even for explicitly represented cut-based rank functions (\cref{thm:nphard}). The response is continuous and piecewise affine. If $B$ counts its \emph{breakpoints} in $(0,1)$, where the affine formula changes, exact optimization takes polynomially many operations in $n$ and $B$ (\cref{thm:pieces}). This gives polynomial-time algorithms for partition polymatroids and laminar polymatroids of logarithmic height (\cref{rem:subclasses}). However, $B$ can be $2^{n-\Oh(\log n)}$ even when each reward coefficient and each rank value is an integer representable in $\Oh(n\log n)$ bits (\cref{thm:polybits}). Our construction also answers the question of Allman, Lo, and McCormick~\cite{ALM22} whether exponentially many two-parameter minimum cuts can occur with polynomial-bit data.

Nevertheless, a share of at most $1/2$ suffices (\cref{thm:prule}). This insight gives rise to an oracle FPTAS using $\Oh(1/\varepsilon)$ best-response computations at shares independent of the instance, and a bit-complexity FPTAS for explicitly described rational polytopes containing zero (\cref{cor:fptasquad}). Every fixed grid with this guarantee needs $\Omega(1/\varepsilon)$ shares (\cref{prop:sharesets}).

\item[Homogeneous convex costs $c(x)$ with $c(tx) = t^p c(x)$ for all $x \geq 0$ and $t > 0$.] For continuous convex costs homogeneous of degree $p>1$, a share of at most $1/p$ suffices. This bound is tight and remains valid on any compact convex action set containing zero (\cref{thm:prule}). Combining this share bound with an approximation scheme, an exact best-response oracle gives a $(1-\varepsilon)$-optimal contract with $\Oh(1+\varepsilon^{-1}\log(p/(p-1)))$ computations, independently of the instance.

\item[General convex costs.] Exact best responses may be transcendental, so we give an oracle approximation scheme using approximately optimal responses. Given an exact value and bounded-subgradient oracle for the costs and a supplied lower bound on positive welfare $W:=\max_{x\in\Pp{f}}\bigl(r(x)-c(x)\bigr)$, the scheme uses polynomially many submodular minimizations, oracle queries, and exact real-arithmetic operations (\cref{thm:convexfptas}). Its approximation guarantee holds for every best response to the returned share. The underlying scheme applies to any compact action set containing zero, with continuous nonnegative costs vanishing at zero, given approximate responses and their reward and cost values (\cref{prop:approxresponses}).
\end{description}

Taken together, our results bridge a gap between classical economic models of unconstrained continuous effort and algorithmic contract design with finite action sets. With unconstrained quadratic costs, raising the share simply scales the agent’s effort vector. Polymatroid capacity constraints can change which activities receive additional effort as the share rises, giving a continuous model a combinatorial structure. The principal still chooses just one number, and each share induces a unique, efficiently computable best response. Yet the response path can have exponentially many breakpoints, and finding an optimal contract is NP-hard.

We formalized the mathematical content of all results of this paper in Lean~4, including bounds on the numbers of arithmetic operations and oracle calls. Independent agents checked the formalizations against the statements in the paper. Some running-time bounds in the paper follow from the formalized results by a standard bit-complexity analysis. That final step lies outside the formalization because our framework does not model Turing machines. The proofs are available in a \href{https://github.com/philippwarode/contract-design-polymatroids-lean}{public repository} and archived on Zenodo~\cite{KLW26Lean}.

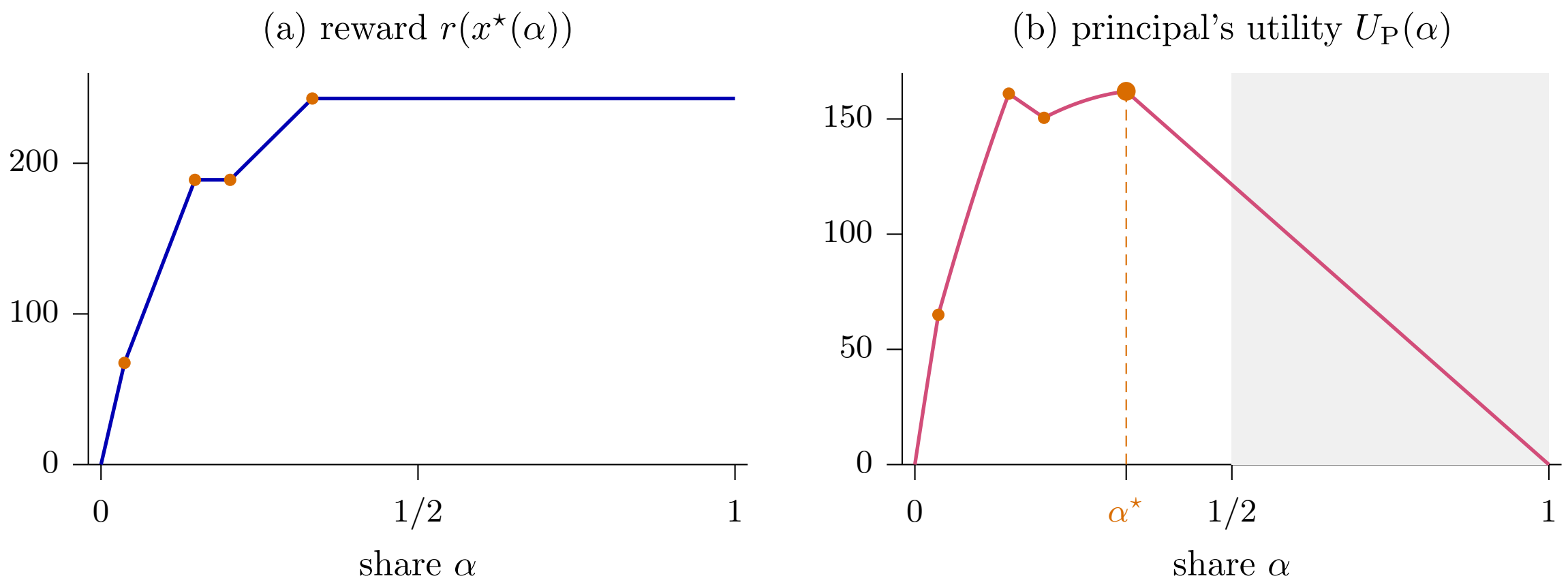
\begin{figure}[t]
\centering
\colorlet{figintroReward}{blue!70!black}%
\colorlet{figintroUtility}{purple!70}%
\colorlet{figintroMark}{orange!85!black}%
\pgfplotsset{figintro/.style={
  scale only axis, width=6.4cm, height=3.8cm,
  axis lines=left, axis line style={-}, xmin=0, xmax=1, ymin=0,
  enlarge x limits={abs=0.02}, enlarge y limits=false,
  xtick={0,0.5,1}, xticklabels={$0$,$1/2$,$1$},
  xlabel={share $\alpha$}, xlabel style={yshift=3pt},
  tick label style={font=\footnotesize}, label style={font=\small},
  title style={font=\small, yshift=-3pt},
  tick align=outside, tick style={black, thin}, typeset ticklabels with strut,
  clip=false,
}}
\begin{tikzpicture}
\begin{axis}[figintro, name=figintroA, ymax=260, ytick={0,100,200},
  title={(a) reward $r(\xs(\alpha))$}]
\addplot[figintroReward, line width=1pt, line join=round] coordinates {
    (0,0) (0.037037,67.5) (0.148148,189) (0.203704,189)
    (0.333333,243) (1,243)
};
\addplot[only marks, mark=*, mark size=1.5pt, figintroMark] coordinates {
    (0.037037,67.5) (0.148148,189) (0.203704,189) (0.333333,243)
};
\end{axis}
\begin{axis}[figintro, at={($(figintroA.south east)+(1.5cm,0)$)}, anchor=south west,
  ymax=170, ytick={0,50,100,150},
  xtick={0,0.333333,0.5,1}, xticklabels={$0$,$\color{figintroMark}\as$,$1/2$,$1$},
  title={(b) principal's utility $\UP(\alpha)$}]
\fill[black!6] (axis cs:0.5,0) rectangle (axis cs:1,170);
\addplot[figintroUtility, line width=1pt, line join=round] coordinates {
    (0,0) (0.002315,4.209) (0.00463,8.3984) (0.006944,12.5684)
    (0.009259,16.7188) (0.011574,20.8496) (0.013889,24.9609) (0.016204,29.0527)
    (0.018519,33.125) (0.020833,37.1777) (0.023148,41.2109) (0.025463,45.2246)
    (0.027778,49.2188) (0.030093,53.1934) (0.032407,57.1484) (0.034722,61.084)
    (0.037037,65) (0.041868,69.7353) (0.046699,74.4197) (0.05153,79.0529)
    (0.056361,83.6352) (0.061192,88.1664) (0.066023,92.6465) (0.070853,97.0756)
    (0.075684,101.4537) (0.080515,105.7807) (0.085346,110.0567) (0.090177,114.2817)
    (0.095008,118.4556) (0.099839,122.5784) (0.10467,126.6503) (0.109501,130.6711)
    (0.114332,134.6408) (0.119163,138.5595) (0.123994,142.4272) (0.128824,146.2439)
    (0.133655,150.0095) (0.138486,153.724) (0.143317,157.3875) (0.148148,161)
    (0.203704,150.5) (0.208689,151.2012) (0.213675,151.8817) (0.218661,152.5414)
    (0.223647,153.1805) (0.228632,153.7988) (0.233618,154.3964) (0.238604,154.9734)
    (0.24359,155.5296) (0.248575,156.0651) (0.253561,156.5799) (0.258547,157.074)
    (0.263533,157.5473) (0.268519,158) (0.273504,158.432) (0.27849,158.8432)
    (0.283476,159.2337) (0.288462,159.6036) (0.293447,159.9527) (0.298433,160.2811)
    (0.303419,160.5888) (0.308405,160.8757) (0.31339,161.142) (0.318376,161.3876)
    (0.323362,161.6124) (0.328348,161.8166) (0.333333,162) (1,0)
};
\addplot[only marks, mark=*, mark size=1.5pt, figintroMark] coordinates {
    (0.037037,65) (0.148148,161) (0.203704,150.5) (0.333333,162)
};
\draw[figintroMark, densely dashed, thin] (axis cs:0.333333,0) -- (axis cs:0.333333,162);
\addplot[only marks, mark=*, mark size=2.4pt, figintroMark] coordinates {(0.333333,162)};
\end{axis}
\end{tikzpicture}
\caption{An instance with two activities, ${\Pp{f}=\{x\in\R^2_{\ge0}:x_1\le1,\ x_1+x_2\le3\}}$, rewards $m=(27,81)$, and quadratic costs $\smash{c(x)=\frac{1}{2}x_1^2+3x_2^2}$. In both curves,  dots mark the breakpoints $1/27$, $4/27$, $11/54$, and $1/3$ of~$\xs$. (a)~The reward is continuous, piecewise affine, and nondecreasing. It is constant on $[4/27,11/54]$, where $x_1\le1$ and $x_1+x_2\le3$ are both tight, and rises again as the agent moves effort to the more rewarding activity~2, until $x_1=0$ at $1/3$. (b)~$\UP$ has two local maxima, $\UP(4/27)=161$ and $\UP(1/3)=162$, so the optimal contract is $\as=1/3$. By our share bound (\cref{thm:prule}), $\UP$ is decreasing on the shaded interval $[1/2,1]$.}\label{fig:intro}
\end{figure}

\paragraph{Techniques.} With finitely many actions, a linear contract changes the agent's choice only at \emph{critical values} of the share~\cite{DRT19,Dut+21}. For linear costs on a polymatroid, polynomially many jumps play this role. For quadratic costs, the response instead follows a piecewise affine path. The principal's utility is quadratic on each piece and can be optimized there directly, but there may be exponentially many pieces. Even a two-activity instance can have two local maxima (\cref{fig:intro}).

For quadratic costs, we describe the optimal effort allocation $\xs(\alpha)$ through its marginal utilities $\rho_e(\alpha)=\alpha m_e-\beta_e\xs_e(\alpha)$. At a water level $\theta>0$, the set of activities with marginal utility at least $\theta$ is the maximal minimizer of
$
  S\longmapsto f(S)-\alpha\sum_{e\in S}\frac{m_e}{\beta_e}
                       +\theta\sum_{e\in S}\frac1{\beta_e}$ (\cref{thm:levelsets}).
The two parameters are thus the contract share $\alpha$ and the water level $\theta$. For cut-based rank functions, $f(S)$ is a directed cut capacity plus a sum of weights of the elements in $S$. The two parameter terms simply change these weights, which can be represented by arcs from a source or to a sink. Minimizing the displayed function therefore becomes a minimum-cut problem with capacities depending on $\alpha$ and $\theta$. This cut formulation lets us apply a construction from the parametric minimum-cut literature. Allman, Lo, and McCormick~\cite{ALM22} give a two-parameter network in which every cut is uniquely optimal for some parameter values. We realize these cuts as marginal level sets. Since a single affine piece admits only $\Oh(n^2)$ distinct level sets, the best-response path must have exponentially many pieces.

Exponential response complexity alone does not establish hardness of choosing a contract. For that, we need to control the reward along the path so that optimizing the share encodes a decision problem. We construct \emph{combs}, instances whose induced reward $r(\xs(\alpha))$ stays below an affine function and touches it exactly at designated shares, called \emph{contacts}. These contacts encode subset sums. A \emph{translation extension} of the submodular function copies the gap between the line and reward into two windows of shares, doubling the contacts at each step (\cref{lem:translation,lem:dmt}). Combining two combs on disjoint activity sets adds their rewards and their lines, so the combined reward touches its line exactly at their common contacts. Suitable combs make the existence of a common contact equivalent to a Subset Sum instance having a solution. Finally, rescaling the construction and adding an auxiliary activity turn this contact test into a threshold test for the principal's utility (\cref{sec:hardness}).

For costs homogeneous of degree $p>1$, comparing optimal allocations along feasible segments gives $r(\xs(\alpha'))\le (\alpha'/\alpha)^{1/(p-1)}r(\xs(\alpha))$ for $0<\alpha<\alpha'$. It follows that $\UP$ is nonincreasing on $[1/p,1]$ (proof of~\cref{thm:prule}). We then adapt the geometric discretization of~\cite{Dut+21}. Rounding shares upward on a grid geometric in $1-\alpha$ loses little of the principal's retained fraction and cannot decrease the reward. For general convex costs, a shift to a slightly larger share converts an approximately optimal response into a reward guarantee for every exact response (\cref{lem:shift}); welfare provides the share bound.

\paragraph{Related work.} Our model follows the hidden-action principal--agent framework of Mirrlees~\cite{Mir99}, Holmstr\"om~\cite{Hol79}, and Grossman and Hart~\cite{GH83}. Linear contracts with convex effort costs are central to classical multitask models~\cite{HM87,HM91}; we study the computational effect of shared capacity constraints. The agent is risk-neutral, but \emph{limited liability} requires nonnegative payments in every outcome~\cite{DFTC24}. Thus the principal cannot charge a fixed fee in exchange for the entire realized reward, which would align the agent's incentives with welfare. It must instead trade off induced effort against the reward fraction it retains.

For combinatorial action spaces, D\"utting, Ezra, Feldman, and Kesselheim~\cite{Dut+21} compute optimal contracts for gross-substitutes success probabilities in polynomial time and prove that the problem is NP-hard for budget-additive success probabilities, for which already the agent's demand problem encodes Subset Sum. D\"utting, Feldman, and Gal-Tzur~\cite{DFGT24}, and concurrently Deo-Campo Vuong, Dughmi, Patel, and Prasad~\cite{DDPP24}, enumerate the critical values with a number of demand queries linear in their number, an instance of the Eisner--Severance technique~\cite{ES76}. \Cref{thm:pieces} is the continuous counterpart, with the breakpoints in the role of the critical values, and it needs a number of best-response computations linear in $B$ up to a factor $\Oh(n^2)$. Baldwin, D\"utting, Feldman, and Schlesinger~\cite{BDFS26} bound the number of critical values by $\Oh(n^2)$ for additive costs and a class of success probabilities that contains the gross-substitutes, the supermodular, and the ultra functions. \Cref{rem:subclasses} gives polynomial bounds on the number of breakpoints for partition polymatroids and laminar polymatroids of logarithmic height. D\"utting, Feldman, and Gal-Tzur~\cite[\S1.1]{DFGT24} ask whether optimizing a contract is hard for matching-based rewards, which admit an efficient demand oracle but can have superpolynomially many critical values. \Cref{thm:nphard} is a result of this kind in our continuous model, with an easy agent's problem.

D\"utting, Feldman, Gal-Tzur, and Rubinstein~\cite{DFGR26} show that finding an optimal contract for submodular success probabilities and additive costs needs exponentially many demand queries, using \emph{equal-revenue} instances with sparse demand, in which exponentially many contracts are optimal, as in the construction behind our best-response query lower bound (\cref{prop:querylb}).

For approximation, D\"utting, Ezra, Feldman, and Kesselheim~\cite{Dut+21} give a geometric-grid approximation scheme whose query bound depends on the numerical encoding length. Their later work~\cite{DEFK25} gives an FPTAS for general monotone success probabilities and additive costs, using polynomially many value and demand queries independently of the numerical encoding length. D\"utting, Feldman, Gal-Tzur, and Rubinstein~\cite{DFGR26} extend this guarantee to monotone subadditive costs using value and best-response queries. Our approximation results adapt this discretization to continuous action sets. The additional ingredients are a share bound derived from homogeneous costs and a guarantee that uses approximately optimal responses for general convex costs.

Separable concave maximization over polymatroids is a classical resource allocation problem~\cite{FG86,Gro91,Hoc94}, and for quadratic objectives Fujishige~\cite{Fuj80}, without linear terms, and, for structured constraints, Hochbaum and Hong~\cite{HH95} give algorithms. Our ray computations use the discrete Newton method~\cite{Din67,GGJ17}, whose number of iterations along nonnegative directions is bounded by Topkis's monotonicity theorem~\cite[Thms.~6.1--6.2]{Top78}.

For quadratic costs, the best response is the solution path of a parametric quadratic program over $\Pp{f}$ (\cref{lem:pwaffine}). Such paths can have superpolynomially many breakpoints for network problems~\cite{Car83,Mul99} and for linear and quadratic programs~\cite{Mur80,GJM12,MY12}, and \cref{thm:expbp,thm:polybits} show that this happens already over polymatroids. Consider a network in which increasing one parameter can only raise source-arc capacities and lower sink-arc capacities, with all other capacities fixed. At each parameter value, take the inclusionwise maximal source side of a minimum cut. These sets form a chain, so their number is linear in the number of vertices~\cite{ES76,GGT89}. With two parameters, exponentially many distinct cuts can each be uniquely optimal at some parameter value~\cite{ALM22}. Our construction realizes their source sides, with the source removed, as marginal level sets of the best response.

\section{Preliminaries}\label{sec:prelim}\label{sec:model}

\paragraph{Polymatroids and notation.}
Let $E$ be a finite nonempty ground set, let $n:=|E|$, and write $[n]:=\{1,\dots,n\}$. For $x\in\R^E$ and $S\subseteq E$, write $x(S):=\sum_{e\in S}x_e$. A set function $f\colon 2^E\to\R$ is \emph{submodular} if $f(X)+f(Y)\ge f(X\cup Y)+f(X\cap Y)$ for all $X,Y \in 2^E$. It is a \emph{rank function} if it is also \emph{normalized} (i.e., $f(\emptyset)=0$) and \emph{monotone} (i.e., $f(X)\le f(Y)$ for $X\subseteq Y$). Then $(E,f)$ is a \emph{polymatroid}, with independence polyhedron
\[
\Pp{f}:= \bigl\{x\in\R^E_{\ge0}:x(S)\le f(S)\text{ for all }S\subseteq E \bigr\}.
\]
For $x\in\Pp{f}$, a set $S\subseteq E$ is \emph{tight} if $x(S)=f(S)$. We access the rank function $f$ through a value oracle and write $\FO$ for the cost of one evaluation.

\paragraph{Contract model.}
There is one principal and one agent. The agent chooses an action $x\in\Pp{f}$, incurs a cost $c(x)$, and generates expected reward $\smash{r(x):=m^\top x}$, where $\smash{m\in\R^E_{\ge0}}$ and $\smash{c\colon\R^E_{\ge0}\to\R_{\ge0}}$ satisfies $c(0)=0$. There are two outcomes, failure with reward $0$ and success with a given reward~$\rbar>0$, where $r(x)\le\rbar$ for every $x\in\Pp{f}$. Success occurs with probability $r(x)/\rbar$, so the expected reward to the principal is $r(x)$.\footnote{If $r$ vanishes on $\Pp{f}$, every share is optimal. Otherwise, one convenient choice is $\rbar=\sum_em_ef(\{e\})$, since $x_e\le f(\{e\})$ for every feasible action. Any positive upper bound on the feasible expected rewards gives the same expected utilities below.}
A contract pays the agent a nonnegative amount in each outcome. For binary outcomes, it suffices to use a \emph{linear contract}, which pays $0$ for failure and $\alpha\rbar$ for success, for a \emph{share} $\alpha\in[0,1]$. This is well-known for contract models with a discrete action set \cite{Dut+21,DFTC24}. We give a formal proof of this fact for our continuous action sets in \Cref{lem:linear-contracts} in \Cref{app:proofs-prelim}. We identify a linear contract with its share $\alpha$. The agent's and the principal's expected utilities are
\[
\UA(x\mid\alpha):=\alpha\,r(x)-c(x),\qquad \UP(x\mid\alpha):=(1-\alpha)\,r(x).
\]
Unless stated otherwise, the agent chooses a best response $\xs(\alpha)\in\argmax_{x\in\Pp{f}}\UA(x\mid\alpha)$, breaking ties in favor of the principal~\cite[\S2]{DFTC24}. We assume that this maximum is attained and, when using this tie-breaking rule, that the reward attains its maximum over the set of best responses. Both hold when $c$ is continuous on the compact set $\Pp{f}$. Write $\UP(\alpha):=\UP(\xs(\alpha)\mid\alpha)$ and, when it exists, $\as\in\argmax_{\alpha\in[0,1]}\UP(\alpha)$ for an \emph{optimal contract}.

\paragraph{Warm-up: linear costs.}\label{sec:linear}
Let us assume that the agent's costs are linear, i.e., $c(x)=\sum_ec_ex_e$ with~$c_e\ge0$.
Then, $\UA(x\mid\alpha)=\rho(\alpha)^\top x$ with $\rho_e(\alpha):=\alpha \,m_e-c_e$, so the agent maximizes a linear function over $\Pp{f}$.  
Edmonds' greedy algorithm~\cite{Edm03} computes such an optimal solution, and the course of the algorithm reveals valuable insights on how optimal solutions behave as a function of $\alpha$. For fixed $\alpha$, the algorithm orders the activities by nonincreasing coefficient $\rho_e(\alpha)$. Writing this order as $e_1,\dots,e_n$ and $S_i:=\{e_1,\dots,e_i\}$ with $S_0:=\emptyset$, it assigns $x_{e_i}=f(S_i)-f(S_{i-1})$ to each activity with nonnegative coefficient and zero to the others. Thus each profitable activity receives the additional capacity available when it is added to the preceding activities. This maximizes the agent's linear objective.
Ordering the elements by nonincreasing $\rho_e(\alpha)$, breaking ties by nonincreasing~$m_e$ and then by a fixed enumeration, makes the greedy vector optimal for the principal among the agent's best responses. The response can change only at a share where some $\rho_e$ changes sign, $\alpha=c_e/m_e$ for $m_e>0$, or where two coefficients change order, $\alpha=(c_e-c_{e'})/(m_e-m_{e'})$ for $m_e>m_{e'}$. At a crossing, the tie rule puts the coefficient with larger slope $m_e$ first, as does the greedy order immediately to the right. Between crossings the response is constant, so the principal's utility is affine and nonincreasing. It therefore suffices to test share $0$ together with the $\Oh(n^2)$ \emph{candidate shares} in $[0,1]$ given by the two formulas above.
The proof is in \cref{sec:A-linear}.

\begin{restatable}{theorem}{rsLinear}\label{thm:linear}
For linear costs,
an optimal contract can be computed in time $\Oh(n^3)\cdot\FO+\Oh(n^3\log n)$.
\end{restatable}

\Cref{sec:A-linear} also shows that, for an integral rank function and positive costs, linear costs are a special case of the gross-substitutes contracts of~\cite{Dut+21}, after a pseudo-polynomial reduction.

\section{Exact computation for quadratic costs}\label{sec:bestresponse}

The linear-cost warm-up solves the principal's problem by describing the agent's response as the share varies. Specifically, the response is constant between finitely many candidate shares, and the principal optimizes over these intervals. We now pursue the same strategy for quadratic costs~$c(x)=\sum_e\beta_ex_e^2/2$ with curvatures $\beta_e>0$. The response can move continuously, but it still has a finite description by affine pieces, on each of which the principal's utility is quadratic. Thus, exact contract optimization reduces to finding these pieces and optimizing over them.

Quadratic costs are particularly suited to this approach because their marginal costs are linear in effort. Each response piece has a simple affine representation, and rational data yield rational breakpoints and an optimal contract of polynomial bit length. These properties are not guaranteed for general convex costs, where even a single best response or the share of an optimal contract can be transcendental (\cref{app:convex}). A simple representation of each piece, however, does not imply that the entire response curve has polynomial size.

We first compute the best response at a fixed share by water-filling (\cref{thm:waterfilling}) and characterize its marginal level sets by submodular minimization (\cref{thm:levelsets}). We then establish the piecewise-affine structure and bit-length bounds (\cref{lem:pwaffine,thm:bitlength}), and give an exact algorithm whose operation count is polynomial in the number of pieces (\cref{thm:pieces}). Finally, we determine when this approach is efficient. The number of pieces can be exponential even for polynomial-bit data (\cref{thm:polybits}), but is polynomial for partition polymatroids and laminar polymatroids of logarithmic height (\cref{rem:subclasses}). This separates efficient response computation from the difficulty of optimizing over the whole response curve, which motivates the hardness analysis in \cref{sec:hardness}.

To describe the agent's best response, note that at a given share $\alpha\ge0$, the agent chooses an effort allocation $x\in\Pp{f}$ to maximize $\smash{\alpha \,m^\top x-c(x)}$. This objective is strictly concave, so the best response $\xs(\alpha)$ is unique.
Write $t_e:=m_e/\beta_e$, $t(S):=\sum_{e\in S}t_e$, and $\binv(S):=\sum_{e\in S}1/\beta_e$, and also use $t$ and $\binv$ for the corresponding vectors. The agent's \emph{marginal utility} of effort on activity $e$ is the partial derivative of its objective with respect to $x_e$. If we ignore the polymatroid capacity constraints, the activities can be optimized independently by setting these derivatives to zero. Thus the marginal utilities and the effort allocation preferred in the absence of capacity constraints are, respectively,
\begin{equation}\label{eq:marg}
\rho_e(x):=\alpha \,m_e-\beta_ex_e,\qquad z_e(\alpha):=\alpha \,t_e.
\end{equation}
The vector $z(\alpha)$ is nonnegative but may violate the constraints defining $\Pp{f}$. If it is feasible, it equals $\xs(\alpha)$; otherwise, the agent must adjust its effort to respect the constraints. To characterize this adjustment, for $\theta>0$ let $A_\theta(x):=\{e\in E:\rho_e(x)\ge\theta\}$ be the set of activities whose marginal utility is at least $\theta$.
The proof is in \cref{sec:A-bestresponse}.

\begin{restatable}{proposition}{rsOptchar}\label{prop:optchar}
A point $x\in\Pp{f}$ is the best response to $\alpha$ if and only if $\rho(x)\ge0$ and every level set $A_\theta(x)$, $\theta>0$, is tight, i.e., $x(A_\theta(x))=f(A_\theta(x))$. In particular, $0\le\xs_e(\alpha)\le\alpha \,t_e$ for all $e$.
\end{restatable}

The characterization suggests a water-filling algorithm, a continuous form of the greedy method proved optimal over integral polymatroids by Federgruen and Groenevelt~\cite{FG86}. Starting from zero, it raises the efforts with the largest marginal utility until a capacity becomes tight, another activity joins, or the marginal reaches zero. The discrete Newton method~\cite{GGJ17} computes each step using submodular minimization; we use the algorithm of Iwata and Orlin~\cite{IO09}, which returns the maximal minimizer. For quadratic costs, a \emph{best-response computation} means one run of water-filling at a given share.
The proof is in \cref{sec:A-bestresponse}.

\begin{restatable}{theorem}{rsWaterfilling}\label{thm:waterfilling}
The best response $\xs(\alpha)$ can be computed in time $\Oh(n^7\log n)\cdot\FO+\Oh(n^8\log n)$.
\end{restatable}

The level sets of the best response have a second description that lets us track how the allocation changes with the share. For $\alpha\ge0$, $\theta>0$, write $\rho(\alpha):=\rho(\xs(\alpha))$ and $A_\theta(\alpha):=A_\theta(\xs(\alpha))$ for the marginals and the level sets at the best response, and let
$F_{\alpha,\theta}(S):=f(S)-\alpha\,t(S)+\theta\,\binv(S)$ for $S \subseteq E$.
For a fixed share, this is the classical minimizer characterization of separable convex optimization over a polymatroid~\cite[Thm.~9.5]{Fuj05}.
The proof is in \cref{sec:A-bestresponse}.

\begin{restatable}{theorem}{rsLevelsets}\label{thm:levelsets}
For all $\alpha\ge0$, $\theta>0$, the level set $A_\theta(\alpha)$ is the unique maximal minimizer of $F_{\alpha,\theta}$.
\end{restatable}

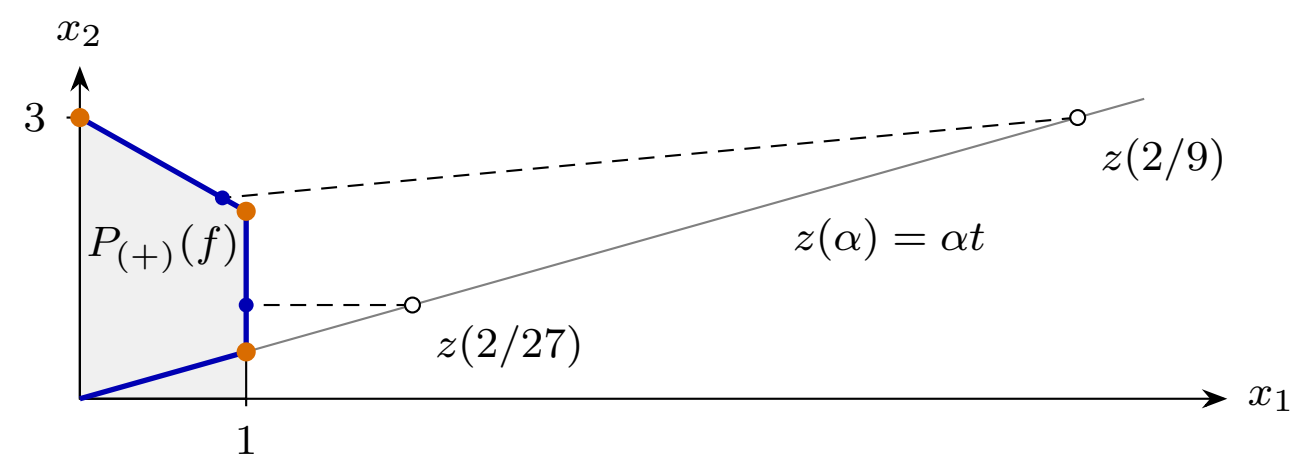
\begin{figure}[t]
\centering
\colorlet{figprojResponse}{blue!70!black}%
\colorlet{figprojMark}{orange!85!black}%
\begin{minipage}[c]{0.53\textwidth}
\centering
\begin{tikzpicture}[x=1.1cm, y=0.62cm, >=Stealth, font=\scriptsize]
\fill[black!6] (0,0) -- (1,0) -- (1,2) -- (0,3) -- cycle;
\draw[thin] (0,0) -- (1,0) -- (1,2) -- (0,3) -- cycle;
\node at (0.5,1.6) {$\Pp{f}$};
\draw[->] (0,0) -- (6.9,0) node[right] {$x_1$};
\draw[->] (0,0) -- (0,3.55) node[above] {$x_2$};
\draw (1,0) -- (1,-0.08) node[below] {$1$};
\draw (0,3) -- (-0.08,3) node[left] {$3$};
\draw[gray] (0,0) -- (6.4,3.2);
\node[anchor=north west, inner sep=1pt] at (4.25,2) {$z(\alpha)=\alpha t$};
\draw[densely dashed, thin] (2,1) -- (1,1);
\draw[densely dashed, thin] (6,3) -- (0.8571,2.1429);
\draw[figprojResponse, line width=1pt, line join=round] (0,0) -- (1,0.5) -- (1,2) -- (0,3);
\filldraw[fill=white, draw=black] (2,1) circle (1.4pt);
\filldraw[fill=white, draw=black] (6,3) circle (1.4pt);
\fill[figprojResponse] (1,1) circle (1.4pt);
\fill[figprojResponse] (0.8571,2.1429) circle (1.4pt);
\fill[figprojMark] (0,3) circle (1.8pt);
\fill[figprojMark] (1,0.5) circle (1.8pt);
\fill[figprojMark] (1,2) circle (1.8pt);
\node[anchor=north west, inner sep=1pt] at (2.1,0.85) {$z(2/27)$};
\node[anchor=north west, inner sep=1pt] at (6.1,2.85) {$z(2/9)$};
\end{tikzpicture}
\end{minipage}\hfill
\begin{minipage}[c]{0.45\textwidth}
\caption{The best response (thick) of the instance of \cref{fig:intro}, computed exactly. Up to $\alpha=1/27$ it is $z(\alpha)=\alpha t$, then the projection of $z(\alpha)$ onto $\Pp{f}$ (\cref{lem:pwaffine}), dashed for $\alpha=2/27$ and $\alpha=2/9$. The dots mark $\xs$ at the breakpoints $1/27$, $4/27$ and $11/54$ (both at the corner $(1,2)$), and $1/3$.}\label{fig:projection}
\end{minipage}
\end{figure}

We next turn from computing the response at one share to describing how it changes with the contract. Completing the square gives
\[
\begin{aligned}
  \UA(x\mid\alpha)
  &=\sum_{e\in E}\left(\alpha m_ex_e-\frac{\beta_e}{2}x_e^2\right) =\frac{\alpha^2}{2}\sum_{e\in E}\frac{m_e^2}{\beta_e}
    -\frac12\sum_{e\in E}\beta_e\left(x_e-\frac{\alpha m_e}{\beta_e}\right)^{\!2}.
\end{aligned}
\]
The first term is independent of $x$. Thus maximizing the agent's utility over $\Pp{f}$ is equivalent to minimizing the weighted squared distance $\sum_e\beta_e(x_e-\alpha t_e)^2$. In other words, $\xs(\alpha)$ is the projection of the preferred allocation $z(\alpha)=\alpha t$ onto $\Pp{f}$ in the inner product $\smash{u^\top\operatorname{diag}(\beta)\,v}$. \Cref{fig:projection} illustrates this projection for the instance of \cref{fig:intro}. As $\alpha$ varies, the projected point follows finitely many affine pieces, giving a finite description of the response curve.

A \emph{breakpoint} is a share $\alpha>0$ at which $\xs$ is not affine on any neighborhood, and $B$ denotes the number of breakpoints in $(0,1)$. A \emph{piece} is the closure of a connected component of the non-breakpoint shares in $(0,1)$.
The proof is in \cref{sec:A-structure}.

\begin{restatable}{lemma}{rsPwaffine}\label{lem:pwaffine}
The map $\alpha\mapsto\xs(\alpha)$ is continuous and piecewise affine on $[0,\infty)$ with finitely many breakpoints. Indeed, $\xs(\alpha)$ is the projection of $z(\alpha)$ onto $\Pp{f}$ in the inner product $u^\top\operatorname{diag}(\beta)\,v$.
\end{restatable}

We use this piecewise-affine structure to optimize the principal's utility separately on each piece. We first explain how to optimize a known piece, then establish the structural and encoding bounds needed to enumerate the pieces exactly.
On a piece $I$, write $r(\xs(\alpha))=r^0_I+r^1_I\alpha$. Then $\UP(\alpha)=(1-\alpha)(r^0_I+r^1_I\alpha)$ is a quadratic polynomial, maximized over $I$ at an endpoint of $I$ or, if $r^1_I>0$ and it lies in $I$, at the vertex $(r^1_I-r^0_I)/(2r^1_I)$. Thus, once the pieces and their affine responses are known, optimizing the principal's utility reduces to checking finitely many endpoints and quadratic maxima.

A second useful property concerns the reward along this curve. Increasing the share cannot decrease the reward induced by the agent's best response. This monotonicity is already established for combinatorial actions by D\"utting, Ezra, Feldman, and Kesselheim~\cite{Dut+21}. Their proof also applies to the continuous setting we consider here and is in \cref{sec:A-structure}.

\begin{restatable}{lemma}{rsMonreward}\label{lem:monreward}
For every cost function $c$, all shares $0\le\alpha<\alpha'$, and any maximizers $x$ of $\UA(\cdot\mid\alpha)$ and $x'$ of $\UA(\cdot\mid\alpha')$ over $\Pp{f}$, we have $r(x)\le r(x')$.
\end{restatable}

In the quadratic setting, \cref{lem:monreward} implies $r^1_I\ge0$ on every piece.
The finite piecewise-affine description leaves a numerical question. Even with rational input data, how many bits are needed to specify a breakpoint or an optimal contract? The next theorem shows that both admit short rational representations. Its bound depends on the dimension and the encoding length of the data, rather than on how many changes the response has already undergone.
The proof is in \cref{sec:C-bitlength}.

\begin{restatable}{theorem}{rsBitlength}\label{thm:bitlength}
Suppose that every $m_e$, every $\beta_e$, and every value $f(S)$ is a quotient of two integers of absolute value less than $2^\nu$, and $n\le\nu$. Then every breakpoint of $\xs$ and some optimal contract are rationals whose numerators and denominators have bit length $\Oh(n\nu)$.
\end{restatable}

The theorem controls the precision needed to represent the candidate contracts exactly and will also be used to establish membership in NP in \cref{sec:hardness}. It does not imply a polynomial-time algorithm, because the number of pieces can still be exponential. We next give an algorithm whose operation count is polynomial in the number of pieces, before showing that exponentially many pieces can occur even with polynomial-bit data.

We enumerate the response pieces by extending a computed best response along its affine piece and repeating until all of $[0,1]$ is covered. The following theorem bounds the total work polynomially in $n$ and the number $B+1$ of pieces; optimizing the principal's utility on these pieces then yields an optimal contract.
The proof is in \cref{sec:A-pieces}.

\begin{restatable}{theorem}{rsPieces}\label{thm:pieces}
All pieces of $\xs$ in $[0,1]$, their affine responses, and an optimal contract can be computed in time
  $\Oh(n^9(B+1)\log n)\cdot\FO+\Oh(n^{10}(B+1)\log n+n^4(B+1)^2)$.
\end{restatable}

The number of pieces can nevertheless be exponential. We use the two-parameter cut construction of Allman, Lo, and McCormick~\cite{ALM22}, in which every subset is a unique minimizer for some parameter values.
This formulation is derived in \cref{sec:A-expbp}.

\begin{restatable}[Allman, Lo, and McCormick~\cite{ALM22}]{theorem}{rsAlm}\label{thm:alm}
For every $n\ge1$ and $E=[n]$, there are $a,b\in\R^E_{>0}$ and a submodular $\tilde f\colon2^E\to\R_{\ge0}$ with $\tilde f(\emptyset)=0$ such that every $S\subseteq E$ is the unique minimizer of $\smash{\tilde f-\lambda_Sa-\theta_Sb}$ for some $(\lambda_S,\theta_S)\in(0,1)^2$.
\end{restatable}

Here $\tilde f$ is a directed cut function. Adding a suitable modular function makes it a polymatroid rank function; choosing rewards and quadratic costs to match $a$ and $b$ then turns its parametric minimizers into best-response level sets, after a change of parameters (\cref{lem:embed} in \cref{sec:A-expbp}). Thus every subset occurs as a level set, whereas a single affine piece admits only $\Oh(n^2)$ distinct level sets, forcing exponentially many pieces.
The proof is in \cref{sec:A-expbp}.

\begin{restatable}{theorem}{rsExpbp}\label{thm:expbp}
For every $n\ge1$, there is an instance on $n$ elements whose best response has at least $2^n/(1+n(n+1)/2)-1$ breakpoints in $(0,1)$, which is $2^{n-\Oh(\log n)}$.
\end{restatable}

The construction of Allman, Lo, and McCormick uses doubly exponential capacities; they ask whether polynomially many bits suffice~\cite[Cor.~5.1, \S5]{ALM22}. We answer this affirmatively by constructing a submodular function one element at a time. Each step preserves a parameter point at which each old subset is the unique minimizer and translates that point by a common vector to obtain a witness for the subset containing the new element. Thus the number of witnessed subsets doubles at every step, while their parameter points are sums of the translation vectors. Choosing these translations and the new coefficients carefully bounds the encoding length of each coefficient and each witness coordinate by $\Oh(n\log n)$. The resulting construction gives both the desired two-parameter cut networks and exponentially many best-response breakpoints, even with equal curvatures.
The proof is in \cref{sec:C-polybits}.

\begin{restatable}{theorem}{rsPolybits}\label{thm:polybits}
For every $n\ge1$, the following hold.
\begin{enumerate}[label=(\roman*)]
\item There is an instance on $n$ elements with $\beta=1$, in which each reward coefficient $m_e$ and each rank value $f(S)$ is an integer representable in $\Oh(n\log n)$ bits, whose rank function $f$ is evaluated in polynomial time, and whose best response has at least $2^n/(1+n(n+1)/2)-1$ breakpoints in $(0,1)$.
\item There is a network of type \cite{ALM22} with a source $s$, a sink, and $n$ further nodes. The arc from $s$ to node $j$ has capacity $a_j\lambda+b_j\mu$ for parameters $(\lambda,\mu)$ and positive integers $a_j,b_j$. Each coefficient and each fixed arc capacity is an integer representable in $\Oh(n\log n)$ bits. Every cut $\{s\}\cup S$ is the unique minimum cut at some point of $(0,1)^2$ whose coordinates each have $\Oh(n\log n)$ bits.
\end{enumerate}
\end{restatable}

For structured polymatroids, polynomial bounds on $B$ instead yield polynomial-time exact optimization via \cref{thm:pieces}. A polymatroid is \emph{laminar} if it has a description $\Pp{f}=\{x\ge0:x(L)\le f(L)\text{ for }L\in\mathcal L\}$ with nonempty constraint sets that are pairwise disjoint or nested. Its \emph{height} $h$ is the number of sets in a longest chain in $\mathcal L$; disjoint constraint sets give a \emph{partition polymatroid}.
The two bounds follow from \cref{prop:partition,thm:laminar}, respectively, in \cref{sec:A-subclasses}.

\begin{proposition}\label{rem:subclasses}
The number of breakpoints satisfies $B\le n$ for partition polymatroids and ${B\le n+2h(n+1)|\mathcal L|\,3^{h-1}}$ for laminar polymatroids.
\end{proposition}

\section{Hardness for quadratic costs}\label{sec:hardness}

\Cref{sec:bestresponse} showed that the best-response curve can have exponentially many breakpoints, even for instances of polynomial encoding length. We use the same recursive translation idea to derive a hardness reduction. This requires precise control of the reward curve, beyond producing many breakpoints, and the construction is technically involved. The central idea is to build two reward curves whose contacts with affine upper bounds encode subset sums; combining the curves makes a sufficiently profitable contract exist exactly when the Subset Sum instance has a solution.

The reduction uses rank functions of the form $f(S)=\cut_D(S)+\omega(S)$, where $\cut_D(S)$ is the total capacity of the arcs of a digraph $D$ on $E$ that leave $S$, and $\omega$ is modular. We write $\Ccut$ for this class of instances, with rational data given in binary; see also \cref{def:cutform} in the appendix for a formal definition.
The proof is in \cref{sec:D-npcomplete}.
\begin{restatable}{theorem}{rsNphard}\label{thm:nphard}
Given an instance of $\Ccut$ and $\vartheta \in \mathbb{Q}$, it is NP-complete to decide whether $\UP(\as)\ge\vartheta$.
\end{restatable}

For the proof, we reduce from Subset Sum.
We sketch the reduction below; \cref{app:hard} gives the construction and its analysis.

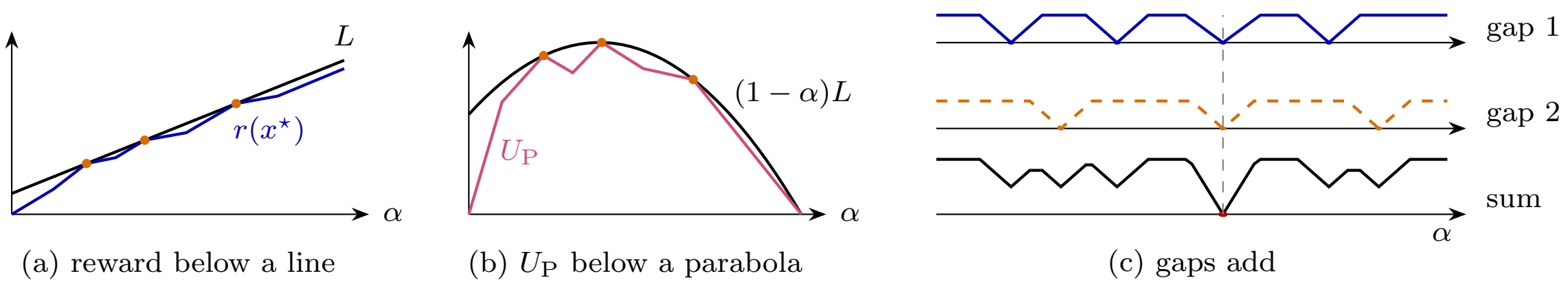
\begin{figure}[b]
\centering
\begin{tikzpicture}[>=Stealth]
\begin{scope}[scale=0.8]
\draw[->] (0,0)--(4.3,0) node[right,font=\scriptsize]{$\alpha$};
\draw[->] (0,0)--(0,2.2);
\draw[thick] (0,0.25)--(4,1.85) node[above,font=\scriptsize]{$L$};
\draw[thick,blue!70!black] (0,0)--(0.5,0.3)--(0.9,0.61)--(1.25,0.68)--(1.6,0.89)--(2.1,0.98)--(2.7,1.33)--(3.2,1.42)--(4,1.75);
\foreach \x/\y in {0.9/0.61,1.6/0.89,2.7/1.33} \fill[orange!85!black] (\x,\y) circle (1.6pt);
\node[blue!70!black,font=\scriptsize] at (3.1,1.0) {$r(\xs)$};
\node[font=\scriptsize] at (2,-0.6) {(a) reward below a line};
\end{scope}
\begin{scope}[xshift=4.4cm,scale=0.8]
\draw[->] (0,0)--(4.3,0) node[right,font=\scriptsize]{$\alpha$};
\draw[->] (0,0)--(0,2.2);
\draw[thick,domain=0:4,samples=40] plot (\x,{(1-\x/4)*(1.2+1.4*\x)});
\node[font=\scriptsize] at (3.9,1.45) {$(1-\alpha)L$};
\draw[thick,purple!70] (0,0)--(0.4,1.35)--(0.9,1.9065)--(1.25,1.7)--(1.6,2.064)--(2.1,1.75)--(2.7,1.6185)--(3.2,1.0)--(4,0);
\foreach \x in {0.9,1.6,2.7} \fill[orange!85!black] (\x,{(1-\x/4)*(1.2+1.4*\x)}) circle (1.6pt);
\node[purple!70,font=\scriptsize] at (0.62,0.75) {$\UP$};
\node[font=\scriptsize] at (2,-0.6) {(b) $\UP$ below a parabola};
\end{scope}
\begin{scope}[xshift=8.9cm,xscale=0.6,yscale=0.33]
\foreach \y in {5,2.5,0} \draw[->] (0,\y)--(8.5,\y);
\node[font=\scriptsize,below left] at (8.5,0) {$\alpha$};
\draw[thick,blue!70!black] (0,5.8)--(0.7,5.8)--(1.2,5)--(1.7,5.8)--(2.4,5.8)--(2.9,5)--(3.4,5.8)--(4,5.8)--(4.6,5)--(5.2,5.8)--(5.8,5.8)--(6.3,5)--(6.8,5.8)--(8.2,5.8);
\foreach \x in {1.2,2.9,4.6,6.3} \fill[blue!70!black] (\x,5) circle (1.6pt);
\draw[thick,dashed,orange!85!black] (0,3.3)--(1.5,3.3)--(2,2.5)--(2.5,3.3)--(4.1,3.3)--(4.6,2.5)--(5.1,3.3)--(6.6,3.3)--(7.1,2.5)--(7.6,3.3)--(8.2,3.3);
\foreach \x in {2,4.6,7.1} \fill[orange!85!black] (\x,2.5) circle (1.6pt);
\draw[thick] (0,1.6)--(0.7,1.6)--(1.2,0.8)--(1.5,1.28)--(1.7,1.28)--(2,0.8)--(2.4,1.44)--(2.5,1.44)--(2.9,0.8)--(3.4,1.6)--(4,1.6)--(4.1,1.4667)--(4.6,0)--(5.1,1.4667)--(5.2,1.6)--(5.8,1.6)--(6.3,0.8)--(6.6,1.28)--(6.8,1.28)--(7.1,0.8)--(7.6,1.6)--(8.2,1.6);
\fill[red!70!black] (4.6,0) circle (2.2pt);
\draw[gray,dashed] (4.6,0)--(4.6,5.8);
\node[font=\scriptsize,anchor=west] at (8.6,5.4) {gap 1};
\node[font=\scriptsize,anchor=west] at (8.6,2.9) {gap 2};
\node[font=\scriptsize,anchor=west] at (8.6,0.4) {sum};
\node[font=\scriptsize] at (4.1,-1.45) {(c) gaps add};
\end{scope}
\end{tikzpicture}
\caption{Basic ideas of hardness proof: (a) If the reward stays below a line $L$ and touches it at some shares (dots), then (b) $\UP$ stays below the parabola $(1-\alpha)L$ and touches it there. (c) In a direct sum of two combs, the gaps add, so the gap of the sum vanishes only at the common contacts (gray line).}\label{fig:parabola}\label{fig:combsum}
\end{figure}

We construct a reward curve $r(\xs(\alpha))$ that stays below an affine function $L$ and touches it exactly at a prescribed finite set of shares~$\Gamma$. Then on $[0,1)$ the utility $\UP(\alpha)=(1-\alpha)\,r(\xs(\alpha))$ stays below the parabola $(1-\alpha)L(\alpha)$ and touches it on~$\Gamma\cap[0,1)$ (\cref{fig:parabola}(a) and~(b)). Such an instance is a \emph{comb} with \emph{contacts}~$\Gamma$ and \emph{width}~$\delta$ if, in addition, the \emph{gap} $L(\alpha)-r(\xs(\alpha))$ grows fast enough with the distance from $\Gamma$, up to the width $\delta$ (\cref{def:comb}). In a direct sum of two instances on disjoint ground sets, the rewards under the best response add (\cref{lem:ops}), so the gaps of two combs add. If both combs have the width~$\delta$ and their contacts lie in $\delta\Z$, so that contacts closer than $\delta$ coincide, the sum is again a comb, and its contacts are the common contacts (\cref{lem:combsum}, \cref{fig:combsum}(c)).

The main construction, the \emph{comb gadget} (\cref{thm:gadget}), realizes the subset sums of a vector as contacts. Its inputs $\muin\in\Z^k_{>0}$ are perturbed powers of $8$ (\cref{def:family}). For every large enough integer $N$, the gadget is an instance of $\Ccut$, computed with its line in polynomial time, that is a comb of width $\delta=1/(3\muin_1)$ with the $2^k$ contacts $\delta(N+\chi^\top\muin)$, $\chi\in\{0,1\}^k$. The gadget is built in $k$ steps, a base case and $k-1$ doubling steps. We call the instance after the base case and $j$ doubling steps \emph{level}~$j$. Level~$0$, the base, is a three-element polymatroid whose reward touches a line at two shares, $1/3$ apart (\cref{prop:base}). Doubling step~$j$ adds two auxiliary elements and one new element whose marginal value is a constant minus a modular function. On the sets without the new element, level $j$ is level $j-1$ plus the auxiliary elements, and the modular function is chosen such that, on the sets that contain the new element, level $j-1$ reappears translated along the share axis (\cref{lem:translation,lem:dmt}). So level $j-1$ appears twice, in a first window of shares and translated in a second, its gap is copied unchanged into both windows, and the contacts double (\cref{fig:gadget} in \cref{sec:D-ideas}). The work lies between the two windows, where the new element crosses the whole old structure. There the gap first rises and then falls, and an explicit parameter rule keeps it at least a fixed $\varepsilon_0>0$ in every step. Proving that one rule works in every step is the bulk of \cref{app:hard}. Finally, a modular function turns the last level, on $3k$ elements, into a polymatroid instance, and one more element ensures the gap to the left of the contacts (\cref{thm:gadget}).

The reduction uses two combs.
Let~$(a,T)$ be an instance of Subset Sum, with $0\le T\le\sum_i a_i$, as otherwise it is a no-instance, and let $H:=10\sum_i a_i+1$. The first comb has the inputs $\smash{H\,8^{i-1}+a_i}$, which carry the items as small perturbations, and the second has the inputs $\smash{H\,8^{i-1}}$ and $N+T$ in place of~$N$. After a dilation (replacing $m$ by $m/\sigma$, \cref{lem:ops}(b)) that puts both on one grid, their contacts coincide exactly at the solutions of $(a,T)$ (\cref{lem:coincidence}), so their direct sum is a comb whose contacts encode the solutions. A further dilation of the rewards moves the window of all possible contacts into $[0,1/4]$, and one more element, the \emph{centering summand}, raises reward and line on this window by a constant and so moves the top of the parabola to the center of the window. The dilation also makes the gap of a no-instance exceed the drop of the parabola across this window. Then $\UP(\as)\ge\vartheta$, for the value $\vartheta$ of the parabola at the ends of the window, if and only if $(a,T)$ is a yes-instance (\cref{prop:correct}).

This hardness result concerns exact optimization and is compatible with the FPTAS in \cref{sec:fptas}.

\section{Approximating the optimal contract}
\label{sec:fptas}

In this section, we give approximation schemes for the principal's utility that apply beyond polymatroids, in particular to arbitrary compact convex action sets $K\subseteq\R^E_{\ge0}$ containing zero. Their running time is measured in best-response oracle calls; efficient implementations are given below for quadratic and general convex costs.
Let~$\mathrm{OPT}:=\sup_{\alpha\in[0,1]}\UP(\alpha)$ and call a contract $(1-\varepsilon)$-optimal if its utility is at least $(1-\varepsilon)\mathrm{OPT}$.

The scheme needs a \emph{share bound}, i.e., a number $\bar\alpha\in[0,1)$ such that $\sup_{\alpha\in[0,\bar\alpha]}\UP(\alpha)=\mathrm{OPT}$. This bounds how close to $1$ the search must go. We discretize $1-\alpha$ geometrically and round shares upward: reward monotonicity ensures that rounding loses utility only through the principal's retained fraction. This argument originates in D\"utting, Ezra, Feldman, and Kesselheim~\cite{Dut+21} and applies to any action set.
The formal proof is in \cref{app:halfrule}.
\begin{restatable}{theorem}{rsFptas}\label{thm:fptas}
Given an exact best-response oracle, a share bound $\bar\alpha\in[0,1)$, and $\varepsilon\in(0,1)$, a $(1-\varepsilon)$-optimal contract can be found with at most $\lceil2\varepsilon^{-1}\ln(1/(1-\bar\alpha))\rceil+1$ oracle calls. The queried shares are rationals with $\Oh(1+\log(1/\varepsilon)+\log(1/(1-\bar\alpha)))$ bits.
\end{restatable}

\Cref{thm:fptas} assumes that a share bound is supplied. The next corollary obtains one from responses at a given share $a<1$ with positive reward and at share $1$. The contract $a$ provides a positive benchmark utility $\UP(a)$, while reward monotonicity bounds the utility of any contract $\alpha$ by $(1-\alpha)r(\xs(1))$. Once this upper bound falls below $\UP(a)$, larger shares cannot improve on $a$ and can be discarded. Solving $(1-\bar\alpha_a)r(\xs(1))=\UP(a)$ gives the bound below; applying \cref{thm:fptas} then gives the approximation guarantee, with two additional oracle calls.
The proof is in \cref{app:halfrule}.
\begin{restatable}{corollary}{rsProbe}\label{cor:probe}
Let $\varepsilon\in(0,1)$. For every $a\in[0,1)$ with $r(\xs(a))>0$, the number $\bar\alpha_a:=1-(1-a)\,r(\xs(a))/r(\xs(1))$ is a share bound. So a $(1-\varepsilon)$-optimal contract is found with at most $\lceil2\varepsilon^{-1}\ln(1/(1-\bar\alpha_a))\rceil+3$ best-response calls.
\end{restatable}

The dependence on the share bound is necessary. Even for one-dimensional convex costs, deterministic algorithms require $\Omega(\varepsilon^{-1}\log(1/(1-\bar\alpha)))$ calls when this quantity is sufficiently large (\cref{prop:querylb} in \cref{sec:E-lower}).

\subsection{Homogeneous costs}\label{sec:fptas-quadratic}

We now turn to costs that are \emph{homogeneous of degree} $p>1$, i.e., $\smash{c\colon\R^E_{\ge0}\to\R_{\ge0}}$ is continuous and convex and $c(tx)=t^pc(x)$ for $t>0$. In the unconstrained setting, homogeneity determines how effort scales with the share. When $\smash{K = \mathbb{R}_{\geq 0}^E}$, if the best response exists and is unique, then $\smash{\xs(\alpha)=\alpha^{1/(p-1)}\xs(1)}$ for $\alpha>0$. Indeed, substituting $\smash{x=\alpha^{1/(p-1)}y}$ in the agent's objective gives $\smash{\alpha r(x)-c(x)=\alpha^{p/(p-1)}(r(y)-c(y))}$. The principal's utility is therefore proportional to $(1-\alpha)\alpha^{1/(p-1)}$, which is maximized at $\alpha=1/p$.

Capacity constraints can prevent the scaled allocation from being feasible, so the unconstrained formula no longer applies, and the optimal share may be smaller than $1/p$. Nevertheless, convexity and homogeneity imply that $r(\xs(\alpha))/\alpha^{1/(p-1)}$ is nonincreasing for $\alpha>0$ as constraints may slow the growth of reward but cannot make it exceed the unconstrained scaling rate. The proof obtains this bound by comparing optimal allocations along feasible segments, without differentiability or strict convexity. Since $(1-\alpha)\alpha^{1/(p-1)}$ is decreasing beyond $1/p$, so is the principal's utility there. Thus the unconstrained optimal share remains a tight upper bound on the shares that need to be searched.
The proof is in \cref{sec:E-prule}.
\begin{restatable}{theorem}{rsPrule}\label{thm:prule}
For homogeneous costs of degree $p>1$, $\bar\alpha=1/p$ is a share bound. This bound is tight.
\end{restatable}
Combining \cref{thm:fptas,thm:prule} gives the following guarantee; see \cref{app:halfrule} for the proof.
\begin{restatable}{corollary}{rsFptashom}\label{cor:fptashom}
For homogeneous costs of degree $p>1$ on a compact convex set $K\subseteq\R^E_{\ge0}$ containing zero, a $(1-\varepsilon)$-optimal contract can be found with $\Oh(1+\varepsilon^{-1}\log(p/(p-1)))$ exact best-response calls, for every $\varepsilon\in(0,1)$. For fixed $p$, this is $\Oh(1/\varepsilon)$ calls.
\end{restatable}

\Cref{cor:fptashom} bounds the number of oracle calls but does not bound the cost of computing a best response. For quadratic costs, convex quadratic programming on rational polytopes and water-filling on polymatroids provide efficient implementations. Taking $p=2$ therefore gives the following runtime guarantees; see \cref{app:halfrule} for a proof.
\begin{restatable}{corollary}{rsFptasquad}\label{cor:fptasquad}
For quadratic costs with rational data and $\varepsilon\in(0,1)$, a $(1-\varepsilon)$-optimal contract can be computed by an FPTAS on explicitly described rational polytopes $K\subseteq\R^E_{\ge0}$ containing zero. On value-oracle polymatroids, its runtime is $\Oh(n^7\log n/\varepsilon)\cdot\FO+\Oh(n^8\log n/\varepsilon)$.
\end{restatable}

The $1/\varepsilon$ dependence for fixed grids is optimal up to constants, already for one quadratic activity (\cref{prop:sharesets}).

\subsection{General convex costs}\label{sec:convex}

For continuous convex costs $c\colon K\to\R_{\ge0}$ with $c(0)=0$, exact best responses can be transcendental (\cref{app:convex}). We therefore use approximate versions. Formally, for a share $\alpha\in[0,1]$ and $\delta\ge0$, an action $\tilde x\in K$ is a \emph{$\delta$-optimal response} if $\UA(\tilde x\mid\alpha)\ge\max_{x\in K}\UA(x\mid\alpha)-\delta$.
An approximate best-response oracle obtains $\delta$ as input and returns such an action, together with its exact reward and cost.
Let $W:=\max_{x\in K}(r(x)-c(x))$ be the welfare and $\rbar>0$ a known reward bound. A lower bound $w$ on positive welfare allows the scheme to distinguish zero welfare from positive welfare.
The proof is in \cref{app:convex}.
\begin{restatable}{proposition}{rsApproxresponses}\label{prop:approxresponses}
Given an approximate best-response oracle, $\varepsilon\in(0,1)$, and $w\in(0,\rbar]$ such that $W=0$ or $W\ge w$, one can determine whether $W=0$ and otherwise compute a $(1-\varepsilon)$-optimal contract using $\Oh(\varepsilon^{-1}\log(3\rbar/w))$ oracle calls.
\end{restatable}

The scheme estimates welfare to obtain a share bound and slightly increases the sampled shares to obtain guarantees from approximate responses. This step is related to the conversion of approximately incentive-compatible contracts in D\"utting, Roughgarden, and Talgam-Cohen~\cite{DRT21}.

\Cref{prop:approxresponses} applies whenever approximate best responses can be computed. For $K=\Pp{f}$, the ellipsoid method provides an efficient implementation using a cost oracle that returns $c(x)$ and a subgradient $g(x)$ with $\|g(x)\|_\infty\le G$ for a known $G$. We use exact real arithmetic and comparisons, with square roots counted as single operations. The cost oracle accepts feasible real inputs and returns exact values, as does the rank oracle on subsets of $E$; rank queries outside submodular minimizations count as arithmetic operations.
The proof is in \cref{sec:F-ellipsoid}.
\begin{restatable}{theorem}{rsConvexfptas}\label{thm:convexfptas}
For convex costs on $\Pp{f}$, given $\varepsilon\in(0,1)$ and $w\in(0,\rbar]$ with $W=0$ or $W\ge w$, one can detect $W=0$ or compute a $(1-\varepsilon)$-optimal contract with numbers of submodular minimizations, cost-oracle queries, and arithmetic operations polynomial in $n$, $1/\varepsilon$, and $\log((\rbar+Gf(E))/w)$.
\end{restatable}

For rational piecewise linear costs, a suitable $w$ can be obtained from bounds on the input coefficients, without first computing $W$. Suppose $c(x)=\max_{j\in J}(a_j^\top x+b_j)$ for a finite family of rational affine functions, and a known bound $L$ limits the bit lengths of their coefficients, the rewards $m_e$, and every rank value $f(S)$. Welfare is then the optimal value of the linear program
\[
  \max_{x,z}\;m^\top x-z
  \quad\text{subject to}\quad
  x\in\Pp{f},\qquad z\ge a_j^\top x+b_j \quad \text{ for all } j\in J.
\]
This program has an optimal vertex. Its $n+1$ coordinates are determined by $n+1$ linearly independent tight constraints. Clearing denominators and applying Cramer's rule and determinant bounds therefore bounds the denominator of the optimal value by $2^{q(n,L)}$ for an explicitly computable polynomial $q$~\cite{GLS88}. The bound depends on the dimension and coefficient sizes, even though the polymatroid description may have exponentially many inequalities. Since a positive rational with denominator at most $2^{q(n,L)}$ is at least $2^{-q(n,L)}$, we may take $w=\min\{\rbar,2^{-q(n,L)}\}$. Thus, known coefficient bounds supply the required gap between zero and positive welfare. This justifies the choice of $w$ in \cref{thm:convexfptas}.

\section{Conclusion and open problems}\label{sec:conclusion}

Continuous contract design combines tractable agent optimization with a potentially difficult principal's problem. On polymatroids, linear costs permit polynomial-time exact optimization, whereas quadratic costs can produce exponentially many response pieces and make optimal contract design NP-hard. Approximation remains possible beyond polymatroids. For homogeneous costs, $\Oh(1/\varepsilon)$ best-response calls for fixed $p$ allow computing a $(1-\varepsilon)$-optimal contract. General convex costs can be handled through approximate responses.

For exact optimization, it remains open whether the hardness is strong and which additional polymatroids have polynomially many breakpoints. Our bounds cover partition polymatroids and laminar families of logarithmic height; the general laminar case remains unresolved. A bound polynomial in the numerical data would also give a pseudo-polynomial algorithm through \cref{thm:pieces}.
It remains open whether adaptive queries can beat the $\Omega(1/\varepsilon)$ bound for fixed grids under quadratic costs. The lower bound for adaptive algorithms uses more general convex costs (\cref{prop:querylb}). Further questions concern the structure of response curves beyond quadratic costs, share bounds without homogeneity, and richer outcome spaces, where linear contracts need not be optimal and the principal must optimize over more than one parameter.

\section*{Acknowledgments}
Funded by the Deutsche Forschungsgemeinschaft (DFG, German Research Foundation) under Germany's Excellence Strategy -- The Berlin Mathematics Research Center MATH+ (EXC-2046/1, EXC-2046/2, project ID: 390685689).

\clearpage
\appendix
\crefalias{section}{appendix}\crefalias{subsection}{subappendix}\crefalias{subsubsection}{subsubappendix}
\section{Omitted proofs for Section~\ref{sec:prelim}}
\label{app:proofs-prelim}

\subsection{Linear contracts}

\begin{lemma}\label{lem:linear-contracts}
Every contract with nonnegative outcome payments that admits a best response can be replaced by a linear contract with share $\alpha\in[0,1]$ without decreasing the principal's expected utility, with ties broken in favor of the principal.
\end{lemma}

\begin{proof}
Let the original contract pay $t_0\ge0$ for failure and $t_1\ge0$ for success, and set $\alpha':=(t_1-t_0)/\rbar$. Since success occurs with probability $r(x)/\rbar$, the agent's and the principal's expected utilities from an action $x\in\Pp{f}$ are, respectively,
\[
  t_0+\alpha'r(x)-c(x),\qquad (1-\alpha')r(x)-t_0.
\]
If $\alpha'\in[0,1]$, the linear contract with share $\alpha'$ has the same set of best responses, since the agent's utilities differ only by the constant $t_0$. For each such response, the principal's utility under the linear contract is larger by $t_0$. Principal-favoring tie-breaking therefore preserves this improvement.

If $\alpha'>1$, the principal's utility under the original contract is at most $0$, since $r(x)\ge0$ and $t_0\ge0$. If $\alpha'<0$, every best response $x$ satisfies $\alpha'r(x)-c(x)\ge0$, since choosing $x=0$ gives the agent utility $t_0$. As $r(x)\ge0$ and $c(x)\ge0$, this implies $r(x)=0$, so the principal's utility is $-t_0\le0$. In both cases, the linear contract with share $0$ is at least as good, since it gives the principal utility $r(\xs(0))\ge0$.
\end{proof}

\subsection{Linear costs}\label{sec:A-linear}

Throughout this subsection the cost function is $c(x)=\sum_{e\in E}c_ex_e$ with
$c_e\ge 0$ for all $e\in E$, and $n\ge 1$ (for $E=\emptyset$, the only response is $0$ and every share is optimal).

We use Edmonds' greedy theorem~\cite[(16)--(19)]{Edm03}, allowing activities with zero coefficient to receive their greedy allocation, since this does not change the objective value.

\paragraph{The greedy response.}
We apply Edmonds' greedy theorem~\cite[(19)]{Edm03} to compute the agent's optimal response with ties
broken in favor of the principal. Fix an enumeration $E=\{e_1,\dots,e_n\}$,
independent of the share $\alpha$. For $\alpha\in[0,1]$ let
$\rho_e(\alpha):=\alpha m_e-c_e$, and write $\rho_e=\rho_e(\alpha)$ when
$\alpha$ is fixed. An element $e$ is \emph{weakly profitable} (at share $\alpha$) if
$\rho_e(\alpha)\ge 0$. Let $\pi=\pi(\alpha)$ be the permutation of $[n]$
that orders the elements by nonincreasing $\rho_e$, breaking ties by
nonincreasing $m_e$ and any remaining ties by the fixed enumeration. In
particular,
\begin{align*}
  &\rho_{e_{\pi(1)}}\ge\rho_{e_{\pi(2)}}\ge\dots\ge\rho_{e_{\pi(n)}},\\
  &\rho_{e_{\pi(i)}}=\rho_{e_{\pi(i+1)}}\implies m_{e_{\pi(i)}}\ge m_{e_{\pi(i+1)}}
  \qquad\text{for } i\in[n-1].
\end{align*}
Set $S^\pi_0:=\emptyset$ and $S^\pi_i:=\{e_{\pi(1)},\dots,e_{\pi(i)}\}$
for $i\in[n]$. The \emph{greedy response} is
\begin{equation}\label{eq:A-greedy}
  \xs_{e_{\pi(i)}}(\alpha):=
  \begin{cases}
    f(S^\pi_i)-f(S^\pi_{i-1}) & \text{if } e_{\pi(i)} \text{ is weakly profitable, that is, } \rho_{e_{\pi(i)}}\ge 0,\\
    0 & \text{otherwise}.
  \end{cases}
\end{equation}
By Edmonds' greedy theorem~\cite{Edm03}, $\xs(\alpha)$ maximizes $\rho^\top x$ over $\Pp{f}$ and is
therefore agent-optimal. The next lemma establishes the required tie-breaking.

\begin{lemma}\label{lem:A-tiebreak}
Let $\alpha\in[0,1]$ be fixed. Then
\begin{enumerate}[label=(\roman*)]
  \item $\xs(\alpha)$ is agent-optimal, that is,
        $\xs(\alpha)\in\argmax\{\rho^\top x: x\in\Pp{f}\}$, \label{it:tiebreak-i}
  \item $\xs(\alpha)$ maximizes the principal's reward $r(x)=m^\top x$ among all agent-optimal solutions. \label{it:tiebreak-ii}
\end{enumerate}
\end{lemma}

\begin{proof}
Part \ref{it:tiebreak-i} follows from Edmonds' greedy theorem~\cite{Edm03}. For part \ref{it:tiebreak-ii}, consider the perturbed
coefficients
\[
  \rho^s_e:=\rho_e+s\,m_e+s^2\qquad\text{for } e\in E .
\]
Since $E$ is finite, for all sufficiently small $s>0$ every strict comparison
between the original coefficients is preserved and every negative coefficient
remains negative. When $\rho_e=\rho_{e'}$, the perturbed coefficients are
ordered by $m_e$ and $m_{e'}$, consistently with $\pi$. Moreover,
$\rho^s_e>0$ whenever $\rho_e\ge 0$, since $m_e\ge 0$. The greedy vector in
\eqref{eq:A-greedy} is determined solely by the order $\pi$ and the set
$\{e\in E:\rho_e\ge 0\}$ of weakly profitable elements. Since
$\pi$ is a nonincreasing order also for $\rho^s$, and
$\{e\in E:\rho^s_e\ge 0\}=\{e\in E:\rho_e\ge 0\}$, the greedy construction for
$\rho^s$ yields the same vector $\xs(\alpha)$. By Edmonds' greedy theorem~\cite{Edm03}, it
satisfies
\begin{equation}\label{eq:A-perturb}
  \rho^\top(\xs(\alpha)-y)+s\,m^\top(\xs(\alpha)-y)+s^2\,\mathbf 1^\top(\xs(\alpha)-y)\ge 0
\end{equation}
for every $y\in\Pp{f}$ and all sufficiently small $s>0$, where $\mathbf 1$ is
the all-ones vector. If $y$ is agent-optimal, then $\rho^\top(\xs(\alpha)-y)=0$
by part (i). Dividing \eqref{eq:A-perturb} by $s$ and letting $s\to 0$ gives
$m^\top\xs(\alpha)\ge m^\top y$, proving part \ref{it:tiebreak-ii}.
\end{proof}

\paragraph{Candidate breakpoints.}
We write $\UP(\alpha):=\UP(\xs(\alpha)\mid\alpha)=(1-\alpha)\,m^\top\xs(\alpha)$
and now regard $\rho_e(\alpha)=\alpha m_e-c_e$ as a function of $\alpha$. The \emph{Type~1 candidate breakpoints}, where an element
becomes weakly profitable, are
\[
  \alpha_e:=\frac{c_e}{m_e}\qquad\text{for } e\in E\text{ with } m_e>0 .
\]
An element with $m_e>0$ is weakly profitable exactly for the shares at or above
its Type~1 candidate breakpoint:
\begin{equation}\label{eq:A-type1}
  \alpha\ge\alpha_e\iff\rho_e(\alpha)\ge 0\qquad\text{for } e\in E\text{ with } m_e>0 .
\end{equation}
An element with $m_e=0$ is weakly profitable for every share if $c_e=0$, and
for no share if $c_e>0$. The \emph{Type~2 candidate breakpoints}, where two
coefficients exchange order, are
\[
  \alpha_{e,e'}:=\frac{c_e-c_{e'}}{m_e-m_{e'}}\qquad\text{for } e,e'\in E\text{ with } m_e>m_{e'} .
\]
For each such pair, $\rho_e-\rho_{e'}=(m_e-m_{e'})(\alpha-\alpha_{e,e'})$, so
\begin{equation}\label{eq:A-type2}
  \alpha\ge\alpha_{e,e'}\iff\rho_e(\alpha)\ge\rho_{e'}(\alpha)\qquad\text{for } e,e'\in E\text{ with } m_e>m_{e'} .
\end{equation}
Elements with equal rewards never change their relative order under the
tie-breaking rule, as $\rho_e-\rho_{e'}=c_{e'}-c_e$ is then constant. The two
types therefore include all shares at which the set of weakly profitable
elements or the ordering can change. A candidate breakpoint need not change the
response or the principal's utility. The candidate breakpoints in $[0,1]$ are the
candidate shares of \cref{sec:linear}. We add $0$ and $1$, remove duplicates, and sort the resulting shares as
\[
  0=\alpha^{(0)}<\alpha^{(1)}<\dots<\alpha^{(K)}=1,\qquad K\ge 1 .
\]

\begin{lemma}\label{lem:A-candidates}
For every $i\in\{0,\dots,K-1\}$, the response $\xs(\alpha)$ is constant on
$[\alpha^{(i)},\alpha^{(i+1)})$, and
\[
  \UP(\alpha^{(i)})\ge\UP(\alpha)\qquad\text{for all }\alpha\in[\alpha^{(i)},\alpha^{(i+1)}).
\]
There is an optimal contract in $\{\alpha^{(0)},\dots,\alpha^{(K-1)}\}$, so
evaluating the utility at these shares yields an optimal contract.
\end{lemma}

\begin{proof}
Fix $i\in\{0,\dots,K-1\}$. The interval $(\alpha^{(i)},\alpha^{(i+1)})$
contains no candidate breakpoint. By \eqref{eq:A-type1} and
\eqref{eq:A-type2}, and since elements with $m_e=0$ and pairs with equal
rewards behave the same for every share, the set of weakly profitable elements
and the ordering $\pi$ are constant on this interval.

We next include the left endpoint $\alpha^{(i)}$. Every coefficient $\rho_e$
that is zero at $\alpha^{(i)}$ remains nonnegative immediately to its right,
since $m_e\ge 0$. All nonzero coefficients retain their signs in a
sufficiently small neighborhood. Thus the set of weakly profitable elements at
$\alpha^{(i)}$ agrees with that on $(\alpha^{(i)},\alpha^{(i+1)})$. For any
pair of coefficients tied at $\alpha^{(i)}$ with unequal rewards, the element
with the larger reward precedes the other immediately to the right of
$\alpha^{(i)}$, and the tie-breaking rule gives the same order at
$\alpha^{(i)}$. Ties between elements with equal rewards persist and are
resolved by the fixed enumeration, while strict comparisons persist in a
sufficiently small neighborhood. Hence the ordering at $\alpha^{(i)}$ also
agrees with that on $(\alpha^{(i)},\alpha^{(i+1)})$. This argument applies to
simultaneous events and to $\alpha^{(0)}=0$, whether or not it is a candidate
breakpoint of either type.

By \eqref{eq:A-greedy}, $\xs(\alpha)=\xs(\alpha^{(i)})$ throughout
$[\alpha^{(i)},\alpha^{(i+1)})$. Consequently,
\[
  \UP(\alpha)=(1-\alpha)\,m^\top\xs(\alpha^{(i)})\qquad\text{for }\alpha\in[\alpha^{(i)},\alpha^{(i+1)}).
\]
This is affine in $\alpha$ with nonpositive slope $-m^\top\xs(\alpha^{(i)})$,
since $m\ge 0$ and $\xs(\alpha^{(i)})\ge 0$. It follows that
$\UP(\alpha^{(i)})\ge\UP(\alpha)$ on this interval. Finally,
$\UP(1)=0\le\UP(0)$, so the share $1$ need not be evaluated.
\end{proof}

\begin{lemma}\label{lem:A-lineartime}
An optimal contract $\as$ can be computed in time $\Oh(n^3)\cdot\FO+\Oh(n^3\log n)$.
\end{lemma}

\begin{proof}
There are at most $n$ Type~1 and $n(n-1)/2$ Type~2 candidate breakpoints.
Including $\alpha^{(0)}=0$ and omitting $\alpha^{(K)}=1$, as justified in
\cref{lem:A-candidates}, we evaluate at most
\[
  1+n+\frac{n(n-1)}{2}=\frac{n(n+1)}{2}+1
\]
shares. Generating the candidate breakpoints and keeping those in $[0,1)$ requires
$\Oh(n^2)$ arithmetic operations. Their order does not matter, and duplicates
do no harm. For each share, the greedy construction
\eqref{eq:A-greedy} takes $\Oh(n\log n)+\Oh(n)\cdot\FO$ time, and computing the
principal's utility takes an additional $\Oh(n)$ time. Evaluating all these
shares and retaining one with maximum utility therefore gives the stated
running time.
\end{proof}

We are now in position to prove the main result for linear costs.

\restate{\rsLinear}{thm:linear}
\begin{proof}[Proof of \cref{thm:linear}]
By \cref{lem:A-tiebreak}, the greedy response maximizes the principal's reward among the agent's best responses. By \cref{lem:A-candidates}, evaluating these responses at the candidate shares finds an optimal contract. The running time follows from \cref{lem:A-lineartime}.
\end{proof}

\paragraph{Relation to combinatorial contracts.} For an integral rank function $f$ and positive costs $c_e>0$, linear costs are a special case of the combinatorial contracts of D\"utting, Ezra, Feldman, and Kesselheim~\cite{Dut+21}, which assume positive costs. Split every element $e$ into $f(\{e\})$ copies. A set of copies is independent if its count vector, the number of its copies of each element, lies in $\Pp{f}$. This gives a matroid on the copies~\cite[(15), (70)--(72)]{Edm03}, \cite[\S44.6b]{Sch03}, and the count vectors of the independent sets are exactly the integral points of $\Pp{f}$. The agent chooses a set of copies, the success reward is $\rbar$, the success probability is the weighted rank of the set, with weight $m_e$ on every copy of $e$, divided by $\rbar$, and each copy of $e$ costs $c_e$. An independent set of copies then gives the same utilities as its count vector, and the agent may restrict itself to independent sets. As~$f$ is integral, the greedy response in \eqref{eq:A-greedy} is integral, and in the polymatroid instance it attains both the agent's optimal utility and the largest reward among the agent's best responses. So at every share the two instances have the same optimal utility of the agent and the same reward under the principal-favorable best responses, hence the same optimal contracts and utilities. Weighted matroid rank functions are gross substitutes~\cite[\S12.1]{PL17}, for which~\cite{Dut+21} compute an optimal contract in polynomial time. The reduction has $\sum_ef(\{e\})$ actions, so it is only pseudo-polynomial, while \cref{thm:linear} works on the polymatroid itself, with a value oracle, real values, and costs $c_e\ge0$.

\section{Omitted proofs for Section~\ref{sec:bestresponse}}\label{app:proofs-quadratic}

This appendix follows the order of the results in \cref{sec:bestresponse}: best responses, the structure of the response curve, bit lengths, piece enumeration, exponential lower bounds, and polynomial subclasses. The submodular tools in \cref{app:tools} precede the polynomial-size construction that uses them; they are also used in the hardness proof.

\subsection{The best response for quadratic costs}\label{sec:A-bestresponse}

Fix $\alpha\ge 0$. Recall the marginals $\rho_e(x)=\alpha m_e-\beta_ex_e$ and
the unconstrained maximizers $z_e(\alpha)=\alpha m_e/\beta_e$ from
\eqref{eq:marg}, and the marginal level sets
$A_\theta(x)=\{e\in E:\rho_e(x)\ge\theta\}$ for $x\in\Pp{f}$ and $\theta>0$.

\restate{\rsOptchar}{prop:optchar}
\begin{proof}[Proof of \cref{prop:optchar}]
At an optimum, $x_e\le z_e(\alpha)$: otherwise reducing $x_e$ to $z_e(\alpha)$
preserves feasibility and increases the utility. Hence $\rho(x)\ge 0$. By
concavity, $x$ is optimal if and only if it maximizes the linear objective
$\rho(x)^\top y$ over $y\in\Pp{f}$: necessity follows by taking directional
derivatives towards feasible points, and sufficiency from concavity, which
yields
\[
  \UA(y\mid\alpha)-\UA(x\mid\alpha)\le\rho(x)^\top(y-x).
\]
For $\rho(x)\ge 0$, let $\theta_1>\dots>\theta_k>0$ be its distinct positive
values, set $\theta_{k+1}:=0$, and put $A_i:=A_{\theta_i}(x)$. Then
$\rho(x)^\top y=\sum_{i=1}^k(\theta_i-\theta_{i+1})\,y(A_i)$. By
Edmonds' greedy theorem~\cite[(19)]{Edm03}, the maximum over $\Pp{f}$ is
$\sum_{i=1}^k(\theta_i-\theta_{i+1})f(A_i)$, so it is attained at $x$ exactly
when every $A_i$ is tight. These are all nonempty positive level sets. If
$k=0$, the condition is vacuous.
\end{proof}

\begin{proposition}[{\cite[Thm.~44.2]{Sch03}}]\label{prop:tight}
Let $(E,f)$ be a polymatroid. For every $x\in\Pp{f}$, the tight sets are closed under union and intersection.
\end{proposition}

\paragraph{Water-filling.}
Let $\hat E:=\{e\in E:\alpha m_e>0\}$ be the set of elements with positive
initial marginals, that is, the elements that are profitable at zero
allocation. We keep $x_e=0$ for the unprofitable elements outside $\hat E$. By
monotonicity of $f$, feasibility of such a nonnegative vector is determined by
the constraints on subsets of $\hat E$. For a feasible $x$, let
\[
  T(x):=\bigcup_{\substack{S\subseteq\hat E:\\ x(S)=f(S)}}S
\]
be its maximal tight subset of $\hat E$, which is tight by \cref{prop:tight}.
For nonempty $S\subseteq\hat E$, define $v^S_e:=\beta_e^{-1}$ for $e\in S$ and
$v^S_e:=0$ otherwise. If for some feasible $x\in\Pp{f}$ the marginals~$\rho_e(x)$ coincide for all $e\in S$, then the same holds at $x+dv^S$ for every
feasible step $d\ge 0$: since $\rho_e(x+dv^S)=\rho_e(x)-d$ for $e\in S$, these
marginals decrease uniformly as $d$ increases. Let
\begin{equation}\label{eq:A-dmax}
  d(x,S):=\max\{d\ge 0: x+dv^S\in\Pp{f}\}
  =\min_{\substack{S'\subseteq\hat E:\\ S'\cap S\neq\emptyset}}\frac{f(S')-x(S')}{v^S(S')} .
\end{equation}
The minimum is attained, so the maximal tight set $T(x+d(x,S)v^S)$ at
$x+d(x,S)v^S$ has a nonempty intersection with $S$.

Starting from $x=0$ and the \emph{saturated set} $Y:=\emptyset$, define
\begin{equation}\label{eq:A-level}
  \lambda:=\max\bigl(\{\rho_e(x):e\in\hat E\setminus Y\}\cup\{0\}\bigr)
  \quad\text{and}\quad
  S^\star:=\{e\in\hat E\setminus Y:\rho_e(x)=\lambda\}.
\end{equation}
We call $\lambda$ the \emph{water level}: it is the highest marginal of all
unsaturated elements. The \emph{active set} $S^\star$ contains all of these
elements, that is, the currently most profitable ones. If $\lambda=0$, no
unsaturated element has a positive marginal and we stop. Otherwise, we grow $x$
in direction $v^{S^\star}$, decreasing the common marginal of the active
elements at unit rate, until a capacity constraint becomes tight, another
element becomes active, or the level reaches zero. Precisely, with the
\emph{admission level} $\lambda'$ and the \emph{step} $\Delta$, set
\begin{equation}\label{eq:A-step}
\begin{aligned}
  \lambda'&:=\max\bigl(\{\alpha m_e:e\in\hat E\setminus(Y\cup S^\star)\}\cup\{0\}\bigr),\\
  \Delta&:=\min\{d(x,S^\star),\,\lambda-\lambda'\},\qquad x^+:=x+\Delta v^{S^\star},\\
  Y^+&:=\begin{cases}T(x^+) & \text{if }\Delta=d(x,S^\star),\\ Y & \text{otherwise,}\end{cases}
\end{aligned}
\end{equation}
and continue from $(x^+,Y^+)$. Thus a simultaneous capacity event is processed
as well, and a zero step saturates elements without changing $x$. If all active
elements are saturated, \eqref{eq:A-level} lowers the level to the next
unsaturated marginal. The \emph{blocking level} $\bar\lambda_e$ of $e$ is
$\lambda-\Delta$ at the step at which $e$ first enters $Y$, and we set
$\bar\lambda_e:=0$ if $e$ never enters $Y$.

\begin{lemma}\label{lem:wfstructure}
The construction remains feasible, its saturated sets are nested and tight, and
its water level, including the level $\lambda-\Delta$ reached at the end of each
step, never increases. Throughout, every unsaturated element $e$ has
$\rho_e(x)=\min\{\alpha m_e,\lambda\}$, and every saturated element $e$ has
$\rho_e(x)=\min\{\alpha m_e,\bar\lambda_e\}$ with $\bar\lambda_e\ge\lambda$. It
stops after at most $2|\hat E|$ steps at a point $\bar x$ satisfying
\begin{equation}\label{eq:A-terminal}
  \rho_e(\bar x)=\min\{\alpha m_e,\bar\lambda_e\}\ge 0\qquad\text{for every } e\in E .
\end{equation}
\end{lemma}

\begin{proof}
We maintain $\rho_e(x)=\min\{\alpha m_e,\lambda\}$ for unsaturated elements and
$\rho_e(x)=\min\{\alpha m_e,\bar\lambda_e\}$ with $\bar\lambda_e\ge\lambda$ for
saturated elements. Initially these identities hold, and $Y=\emptyset$ is
tight. They imply $S^\star=\{e\in\hat E\setminus Y:\alpha m_e\ge\lambda\}$. All
remaining unsaturated elements have $x_e=0$ and $\alpha m_e\le\lambda'<\lambda$.
Since $0\le\Delta\le d(x,S^\star)$, the new point is feasible. The coordinates
of $Y$ do not change, so $Y$ remains tight and, at a capacity event, is
contained in $T(x^+)$. Active marginals become $\lambda-\Delta\ge\lambda'$,
while waiting elements stay at zero allocation. Thus the identities hold at
level $\lambda-\Delta$, also for newly saturated elements. Recomputing the
largest unsaturated marginal can only lower this level. If it does, all
remaining unsaturated marginals equal their initial values, so the identities
still hold. This proves the induction and the monotonicity of the level.

Every capacity step saturates an active element by \eqref{eq:A-dmax}, so there
are at most $|\hat E|$ such steps. Every other step with $\lambda'>0$ admits a
previously inactive element: its initial marginal $\lambda'$ is strictly below
the current level and hence below every earlier active level. There are at most
$|\hat E|-1$ such steps when $\hat E\neq\emptyset$, since the first active set
is nonempty. A remaining step has $\lambda'=0$ and terminates the construction.
If $\hat E=\emptyset$, no step is taken. At termination all unsaturated
marginals are zero, giving \eqref{eq:A-terminal}.
\end{proof}

\begin{lemma}\label{lem:ray}
Let $x\in\Pp{f}$ with $x_e=0$ for $e\notin\hat E$, and let
$\emptyset\neq S\subseteq\hat E$. The step size $d(x,S)$ and the maximal tight
set at its endpoint can be computed with at most $|S|+1$ submodular
minimizations, $\Oh(|S|)$ further evaluations of $f$, and $\Oh(n|S|)$ further
arithmetic operations.
\end{lemma}

\begin{proof}
Write $v=v^S$ and start at $d=\min_{e\in S}(f(\{e\})-x_e)/v_e$. For each $d$,
find the maximal minimizer $\tilde S$ of the submodular function
$g_d(S'):=f(S')-x(S')-d\,v(S')$ on $2^{\hat E}$. If $g_d(\tilde S)=0$, return
$d$ and $\tilde S$. Otherwise replace $d$ by
\[
  d':=\frac{f(\tilde S)-x(\tilde S)}{v(\tilde S)} .
\]
Initially $d\ge d(x,S)$ by the singleton constraints. If $g_d(\tilde S)<0$,
feasibility of $x$ gives $v(\tilde S)>0$ and $d(x,S)\le d'<d$. If the minimum
is zero, the current step is feasible, hence equals $d(x,S)$, and its maximal
minimizer is exactly the maximal tight set at the endpoint.

To bound the number of calls, consider consecutive negative minima with
maximal minimizers $\tilde S,\tilde S'$ at $d>d'$. Since $\tilde S'$ minimizes
$g_{d'}=g_d+(d-d')v$ and $v\ge 0$,
\[
  g_d(\tilde S')-g_d(\tilde S\cap\tilde S')\le(d-d')\bigl(v(\tilde S\cap\tilde S')-v(\tilde S')\bigr)\le 0 .
\]
Submodularity gives $g_d(\tilde S\cup\tilde S')\le g_d(\tilde S)$, so
maximality of $\tilde S$ implies $\tilde S'\subseteq\tilde S$. If
$\tilde S'\cap S=\tilde S\cap S$, then
\[
  g_{d'}(\tilde S')=g_d(\tilde S')+(d-d')v(\tilde S)\ge g_d(\tilde S)+(d-d')v(\tilde S)=g_{d'}(\tilde S)=0,
\]
a contradiction. Thus the sets $\tilde S\cap S$ at negative minima form a
strictly decreasing chain of nonempty subsets of $S$. There are at most $|S|$
such calls and one final call with minimum zero. Each call is followed by one evaluation of $f$, at $\tilde S$, for the test and the update, and by $\Oh(n)$ arithmetic operations. With the $|S|$ singleton evaluations of the initialization, this makes at most $2|S|+1$ further evaluations of $f$.
\end{proof}

\restate{\rsWaterfilling}{thm:waterfilling}
\begin{proof}[Proof of \cref{thm:waterfilling}]
We first show that water-filling ends at the best response. Let $\bar x$ be the terminal point of water-filling. Fix
$\theta>0$ and let $C_\theta:=\{e:\bar\lambda_e\ge\theta\}$. As the water level never increases, elements saturated earlier have larger blocking levels, so $C_\theta$ is the set saturated after the last step with $\lambda-\Delta\ge\theta$, or empty. By
\cref{lem:wfstructure}, this is a cumulative saturated set, and as its coordinates no longer change, it
stays tight at $\bar x$. Equation \eqref{eq:A-terminal} gives
$A_\theta(\bar x)\subseteq C_\theta$, and every element of
$C_\theta\setminus A_\theta(\bar x)$ has $\rho_e(\bar x)=\alpha m_e$, hence
$\bar x_e=0$. Consequently,
\[
  \bar x(A_\theta(\bar x))=\bar x(C_\theta)=f(C_\theta)\ge f(A_\theta(\bar x))\ge\bar x(A_\theta(\bar x)).
\]
Thus every positive marginal level set is tight, $\rho(\bar x)\ge 0$ by
\eqref{eq:A-terminal}, and \cref{prop:optchar} shows that $\bar x$ is the
unique best response $\xs(\alpha)$.

It remains to count the operations. There are at most $2|\hat E|\le 2n$ steps by
\cref{lem:wfstructure}. Each uses one ray computation and $\Oh(n)$ additional
work to determine the active set, the admission level, and the updates. By
\cref{lem:ray}, this gives at most $2n(n+1)$ submodular minimizations,
$\Oh(n^2)$ further evaluations of $f$, and $\Oh(n^3)$ further arithmetic
operations. Throughout our operation counts, a submodular minimization returns the unique maximal minimizer. The strongly polynomial algorithm of Iwata and Orlin~\cite{IO09} finds it in time $\Oh(n^5\log n)\cdot\mathrm{EO}+\Oh(n^6\log n)$, where $\mathrm{EO}$ is the cost of evaluating the function being minimized. Here each such function is $f$ minus a modular function, so $\mathrm{EO}=\FO+\Oh(n)$. Multiplying by the $\Oh(n^2)$ minimizations gives the stated bound.
\end{proof}

\Cref{alg:waterfilling} summarizes the construction.

\begin{algorithm}[ht]
\caption{Water-filling for a fixed share}\label{alg:waterfilling}
\begin{algorithmic}[1]
\Require $f$, $m\ge 0$, $\beta>0$, $\alpha\ge 0$, and a value oracle for $f$ and submodular minimization returning the maximal minimizer (\cref{sec:bestresponse})
\Ensure the best response $\xs(\alpha)$
\State $x\gets 0$, $Y\gets\emptyset$, and determine $\lambda,S^\star$ by \eqref{eq:A-level}
\While{$\lambda>0$}
  \State compute $d(x,S^\star)$ and $T(x+d(x,S^\star)v^{S^\star})$ by \cref{lem:ray}
  \State update $(x,Y)\gets(x^+,Y^+)$ by \eqref{eq:A-step}
  \State determine $\lambda,S^\star$ by \eqref{eq:A-level}
\EndWhile
\State \Return $x$
\end{algorithmic}
\end{algorithm}

\paragraph{Level sets and monotone marginals.}
From now on, $\rho_e(\alpha):=\alpha m_e-\beta_e\xs_e(\alpha)$ and
$A_\theta(\alpha):=\{e\in E:\rho_e(\alpha)\ge\theta\}$ denote the optimal
marginals and their level sets, with $t_e=m_e/\beta_e$.

\restate{\rsLevelsets}{thm:levelsets}
\begin{proof}
Let $x:=\xs(\alpha)$, $\rho:=\rho(\alpha)$, and $A:=A_\theta(\alpha)$. As
$x_e=\alpha t_e-\beta_e^{-1}\rho_e$ for every $e\in E$, every $S\subseteq E$
satisfies
\begin{align*}
  F_{\alpha,\theta}(S)&=f(S)-x(S)+\sum_{e\in S}\beta_e^{-1}(\theta-\rho_e)\ge\sum_{e\in S}\beta_e^{-1}(\theta-\rho_e)\ \ge\sum_{e\in E:\,\rho_e>\theta}\beta_e^{-1}(\theta-\rho_e).
\end{align*}
The first inequality holds as $x\in\Pp{f}$, with equality if and only if $S$ is
tight, and the second holds with equality if and only if
$\{e:\rho_e>\theta\}\subseteq S\subseteq A$. As $A$ is tight
(\cref{prop:optchar}), it attains the lower bound, so it is a minimizer, and
every minimizer is contained in $A$.
\end{proof}

Topkis's monotonicity theorem~\cite[Thms.~6.1--6.2]{Top78} also gives monotone marginals.

\begin{theorem}\label{thm:monmarg}
For every $e\in E$, the marginal $\alpha\mapsto\rho_e(\alpha)$ is nondecreasing on $[0,\infty)$.
\end{theorem}
\begin{proof}
Let $0\le\alpha<\alpha'$ and $\theta>0$. The difference
$F_{\alpha',\theta}-F_{\alpha,\theta}=-(\alpha'-\alpha)t$ is modular and
nonincreasing under inclusion, so $A_\theta(\alpha)\subseteq A_\theta(\alpha')$
by \cref{thm:levelsets} and Topkis's monotonicity theorem~\cite[Thms.~6.1--6.2]{Top78}. If
$\rho_e(\alpha)>\rho_e(\alpha')$ for some $e$, then a $\theta$ with
$\rho_e(\alpha)>\theta>\rho_e(\alpha')\ge 0$ gives
$e\in A_\theta(\alpha)\setminus A_\theta(\alpha')$, a contradiction.
\end{proof}

\subsection{Structure of the best-response curve}\label{sec:A-structure}

\restate{\rsPwaffine}{lem:pwaffine}
\begin{proof}[Proof of \cref{lem:pwaffine}]
For the inner product $y^\top Dz$ with $D:=\operatorname{diag}(\beta)$ and its
norm $\|y\|_\beta:=(y^\top Dy)^{1/2}$, completing the square gives
\[
  \UA(x\mid\alpha)=\frac{\alpha^2}{2}\|t\|_\beta^2-\frac12\|x-\alpha t\|_\beta^2 .
\]
Thus $\xs(\alpha)$ is the projection of $\alpha t$ onto $\Pp{f}$.
Nonexpansiveness of projections \cite[Prop.~4.16]{BC17} gives
$\|\xs(\alpha)-\xs(\alpha')\|_\beta\le|\alpha-\alpha'|\,\|t\|_\beta$. For
$x\in\Pp{f}$, define
$N(x):=\{u:u^\top D(y-x)\le 0\text{ for all }y\in\Pp{f}\}$. Then
$x$ is the projection of $z$ exactly when $z-x\in N(x)$~\cite[Prop.~6.47]{BC17}. For a face $F$, this
cone is constant on $\relint F$: a normal at $x\in\relint F$ is orthogonal to
every direction parallel to $F$, and hence is normal at every other point of
$\relint F$. Denote it by $N(F)$. The shares for which
$\xs(\alpha)\in\relint F$ are exactly those with
$\alpha t\in\relint F+N(F)$, and they form an interval, since this set is
convex. On this interval, $\xs(\alpha)$ is the projection onto $\aff F$, hence
affine in $\alpha$. There are finitely many faces.
\end{proof}

\restate{\rsMonreward}{lem:monreward}
\begin{proof}
The argument uses only the optimality of $x$ and $x'$, not the form of $c$.
As $x$ maximizes $\UA(\cdot\mid\alpha)$ and $x'$ maximizes
$\UA(\cdot\mid\alpha')$ over $\Pp{f}$, we have
\[
  \alpha r(x)-c(x)\ge\alpha r(x')-c(x')
  \quad\text{and}\quad
  \alpha' r(x')-c(x')\ge\alpha' r(x)-c(x).
\]
Adding the two inequalities gives $(\alpha'-\alpha)\bigl(r(x')-r(x)\bigr)\ge 0$,
so $r(x')\ge r(x)$.
\end{proof}

Several constructions of this paper combine and rescale instances by the following operations. The \emph{direct sum} $I_1\oplus I_2$ of instances $I_i=(f_i,m_i,\beta_i)$ on disjoint sets $E_i$ has the rank function $(f_1\oplus f_2)(S):=f_1(S\cap E_1)+f_2(S\cap E_2)$, rewards $(m_1,m_2)$, and curvatures $(\beta_1,\beta_2)$.

\begin{lemma}\label{lem:ops}
Let $(f,m,\beta)$ be an instance with best response $\xs$.
\begin{enumerate}[label=(\alph*)]
\item If $f=f_1\oplus f_2$ for rank functions $f_1,f_2$, then $\xs=(\xs_1,\xs_2)$ and $r(\xs(\alpha))=r_1(\xs_1(\alpha))+r_2(\xs_2(\alpha))$, where $\xs_i$ and $r_i$ belong to $(f_i,m|_{E_i},\beta|_{E_i})$.\label{it:ops-a}
\item For $\sigma>0$, the instance $(f,m/\sigma,\beta)$ has the best response $\alpha\mapsto\xs(\alpha/\sigma)$ and the reward $r(\xs(\alpha/\sigma))/\sigma$ under it.\label{it:ops-b}
\item Let $C,\aknob>0$. Adding an element $e$ as a direct summand with $f(\{e\}):=C$, $m_e:=1$, and $\beta_e:=\aknob/C$, a \emph{capped linear summand}, adds the capped linear function $C\min(\alpha/\aknob,1)$ to the reward under the best response.\label{it:ops-c}
\end{enumerate}
\end{lemma}

\begin{proof}
We first show \ref{it:ops-a}. Let $E=E_1\cup E_2$. If $x\in\Pp{f}$ and
$S\subseteq E_i$, then $x(S)\le f(S)=f_i(S)$, as the other summand vanishes on
the empty set, so $x|_{E_i}\in\Pp{f_i}$. Conversely, if $x|_{E_i}\in\Pp{f_i}$
for $i=1,2$, then $x\ge 0$ and
$x(S)=x(S\cap E_1)+x(S\cap E_2)\le f_1(S\cap E_1)+f_2(S\cap E_2)=f(S)$ for
every $S\subseteq E$. So $\Pp{f}=\Pp{f_1}\times\Pp{f_2}$. The agent's utility
$\UA(x\mid\alpha)$ is separable: it is the sum of the agent's utilities in the
two instances $(f_i,m|_{E_i},\beta|_{E_i})$ at $x|_{E_1}$ and $x|_{E_2}$,
respectively. So, for every share $\alpha\ge 0$, its maximizer over the product
$\Pp{f}$ is the pair of the two maximizers, that is,
$\xs(\alpha)=(\xs_1(\alpha),\xs_2(\alpha))$. As
$r(x)=m^\top x=(m|_{E_1})^\top x|_{E_1}+(m|_{E_2})^\top x|_{E_2}$, this gives
$r(\xs(\alpha))=r_1(\xs_1(\alpha))+r_2(\xs_2(\alpha))$.

We next show \ref{it:ops-b}. At the share $\alpha\ge 0$, the agent's utility in the
instance $(f,m/\sigma,\beta)$ is
\[
  \sum_{e\in E}\Bigl(\alpha\frac{m_e}{\sigma}x_e-\frac{\beta_e}{2}x_e^2\Bigr)
  =\UA\Bigl(x\Bigm|\frac{\alpha}{\sigma}\Bigr),
\]
the agent's utility in $(f,m,\beta)$ at the share $\alpha/\sigma$, over the
same polyhedron $\Pp{f}$. So the best response of $(f,m/\sigma,\beta)$ at
$\alpha$ is $\xs(\alpha/\sigma)$. The reward of the new instance is measured
with $m/\sigma$, so its reward under the best response is
$(m/\sigma)^\top\xs(\alpha/\sigma)=r(\xs(\alpha/\sigma))/\sigma$. This is the
rescaling used with $\sigma=1/(\Lambda+1)$ at the end of the proof of
\cref{thm:expbp}.

Finally, we show \ref{it:ops-c}. By \ref{it:ops-a}, the new coordinate $y$ maximizes
$\alpha y-\aknob y^2/(2C)$ over $[0,C]$, so
$y=\min(\alpha C/\aknob,C)=C\min(\alpha/\aknob,1)$. As $m_e=1$, the new element
adds $y$ to the reward.
\end{proof}

At a share $\alpha>0$, let $\Phi(\alpha):=\{e:\rho_e(\alpha)=0\}$ and
$Z(\alpha):=\{e:\xs_e(\alpha)=0\}$ be the \emph{free set} and the \emph{zero
set}. Partition $E\setminus\Phi(\alpha)$ into \emph{blocks}
$D\in\mathcal P(\alpha)$ of equal positive marginal. Blocks may contain
elements of $Z(\alpha)$. The triple $(\Phi(\alpha),Z(\alpha),\mathcal P(\alpha))$
is the \emph{state}. A share is an \emph{event point} if the state is not
constant on any neighborhood of it.

\subsection{Bit lengths of breakpoints and optimal contracts}\label{sec:C-bitlength}

In this subsection we prove \cref{thm:bitlength}, with $\nu$ as in
\cref{thm:bitlength}: every $m_e$, every $\beta_e$, and every value of $f$ is a
quotient of two integers of absolute value less than $2^\nu$, and $n\le\nu$. We
show that the numerators and denominators of all breakpoints have absolute
value at most
\[
  Q := (2n+1)^{n+1}\,4^{(2n+1)\nu},
\]
and $\log_2 Q=(n+1)\log_2(2n+1)+2(2n+1)\nu=\Oh(n\nu)$.

\restate{\rsBitlength}{thm:bitlength}
\begin{proof}
  Recall from \cref{sec:A-structure} that a state records the elements of zero marginal utility, the elements of zero effort, and the blocks of elements with equal positive marginal utility. We first write down its linear equations. By \cref{lem:pwaffine},
  $[0,\infty)$ is the union of finitely many intervals on each of which $\xs$,
  and hence every $\rho_e$, is affine. On each of them, the state changes at
  most $\binom n2$ times, by the argument of \cref{lem:eventcount}, so
  there are finitely many event points. Let $J$ be an open interval on which
  the state is constant, with free elements $\Phi$, zero set $Z$, and partition
  $\mathcal P=\{D_1,\dots,D_k\}$ into blocks. The values $\rho_e(\alpha)$ of
  different blocks differ for every $\alpha\in J$ and depend continuously on
  $\alpha$, so the blocks are in the same order on all of $J$, and we number them
  such that their values satisfy $\rho_1(\alpha)>\dots>\rho_k(\alpha)>0$.
  Then $S_j:= D_1\cup\dots\cup D_j$ is the level set
  $A_{\rho_j(\alpha)}(\alpha)$, so it is tight with respect to $\xs(\alpha)$.
  Hence, for every $\alpha\in J$, the vector
  $(\alpha,\xs(\alpha),\rho_1(\alpha),\dots,\rho_k(\alpha))$ solves the linear
  equations
  \[
    \begin{array}{r@{\qquad}l}
      \beta_e x_e - m_e\alpha = 0 & \text{for } e\in\Phi,\\
      \beta_e x_e + \rho_j - m_e\alpha = 0 & \text{for } e\in D_j\setminus Z \text{ and } j=1,\dots,k,\\
      x_e = 0,\quad \rho_j - m_e\alpha = 0 & \text{for } e\in D_j\cap Z \text{ and } j=1,\dots,k,\\
      x(S_j) = f(S_j) & \text{for } j=1,\dots,k,
    \end{array}
  \]
  in the $N:= 1+n+k\le 2n+1$ unknowns $(\alpha,x,\rho_1,\dots,\rho_k)$.
  For fixed $\alpha$, these equations determine $x$ and $\rho_1,\dots,\rho_k$.
  Indeed, the first line fixes $x_e$ for $e\in\Phi$. If $D_j\cap Z\ne\emptyset$,
  the third line fixes $\rho_j$. Otherwise, subtracting the equations for
  $S_{j-1}$ and $S_j$, with $S_0:=\emptyset$, gives
  $\sum_{e\in D_j}(m_e\alpha-\rho_j)/\beta_e=f(S_j)-f(S_{j-1})$, which fixes
  $\rho_j$. Then every $x_e$ is fixed. By \cref{lem:velocity}, $\xs$ and
  $\rho_1,\dots,\rho_k$ are affine on $J$. Let $x^J$ and
  $\rho^J=(\rho^J_1,\dots,\rho^J_k)$ denote their affine extensions to $\R$.
  The solutions of the equations form an affine subspace that contains
  $(\alpha,x^J(\alpha),\rho^J(\alpha))$ for all $\alpha\in J$, hence for all
  $\alpha\in\R$, and that contains at most one point for each $\alpha$. So the
  solutions form exactly the line
  $\{(\alpha,x^J(\alpha),\rho^J(\alpha))\colon\alpha\in\R\}$.

  Next we single out a breakpoint by one more equation. Let $b$ be a breakpoint. As
  there are finitely many event points, there are open intervals $J_-$ with
  right endpoint $b$ and $J_+$ with left endpoint $b$ on each of which the
  state is constant. The two states differ: otherwise, \cref{lem:velocity}
  gives $\xs$ the same slope on $J_-$ and on $J_+$, and as $\xs$ is
  continuous, it would be affine on $J_-\cup\{b\}\cup J_+$. We find an
  interval $J\in\{J_-,J_+\}$ and an equation with coefficients in
  $\{0,\pm1\}$ and right-hand side $0$ in the unknowns of $J$ that, along the
  line of solutions above, holds at $\alpha=b$ and nowhere else. If the sets of free
  elements differ, some element $e$ is free on one of the two intervals and not
  on the other one, $J$, where it lies in a block $D_j$. Then $\rho_e$
  vanishes on the first interval, hence at $b$, and $\rho^J_j=\rho_e$ is
  positive on $J$, so $\rho^J_j$ is a nonconstant affine function that vanishes
  at $b$, and the equation $\rho_j=0$ works. If the zero sets differ, some
  element $e$ has $x_e^\star=0$ on one interval and $x_e^\star>0$ on the other
  one, $J$, and the equation $x_e=0$ works in the same way. Otherwise, the
  partitions into blocks differ, so there are elements $e,e'$ that lie in a
  common block on one interval and in different blocks $D_j\ne D_{j'}$ on the
  other one, $J$. Then $\rho_e=\rho_{e'}$ at $b$ by continuity and
  $\rho_e\ne\rho_{e'}$ on $J$, so $\rho^J_j-\rho^J_{j'}$ is a nonconstant
  affine function that vanishes at $b$, and the equation $\rho_j-\rho_{j'}=0$
  works. Adding this equation to the equations of $J$ gives a system whose
  only solution is $(b,\xs(b),\rho^J(b))$. So some $N$ of its equations form a
  square system $Gz=h$ with a nonsingular matrix $G$ and the same solution,
  and by Cramer's rule $b=\det G_\alpha/\det G$, where $G_\alpha$ arises from
  $G$ by replacing the column of $\alpha$ with $h$.

  Next we bound the size of $b$. Every equation contains at most two of the
  numbers $m_e$, $\beta_e$, or one value of $f$ on its right-hand side, and
  its other coefficients lie in $\{0,\pm1\}$. Multiplying each equation by the
  product of the denominators of these numbers multiplies $\det G$ and
  $\det G_\alpha$ by the same nonzero factor, and turns $G$ and $G_\alpha$
  into integer matrices whose entries have absolute value less than $4^\nu$.
  By Hadamard's inequality, both determinants then have absolute value at most
  $(\sqrt N\,4^\nu)^N\le Q$. So $b$ is a quotient of two integers of absolute
  value at most $Q$. The equations depend only on the state on $J$ and on the
  data, not on the breakpoints before $b$, so this bound holds for every
  breakpoint.

  Finally, we turn to an optimal contract. Let $J$ be an open interval with constant
  state in the interior of a piece $I$. Then the affine extension $x^J$ above is the affine form of $\xs$ on
  $I$. As the equations of $J$ determine $x$ and $\rho$ for fixed $\alpha$, some $n+k$ of them have a nonsingular matrix in
  the columns of $(x,\rho)$. We scale them as above, move the terms with
  $\alpha$ to the right-hand side, and apply Cramer's rule. This gives
  $x_e^\star(\alpha)=(u_e+\alpha v_e)/\zeta$ on the piece, with integers $u_e$,
  $v_e$, and $\zeta\ne 0$ of absolute value at most $Q$. Writing
  $m_e=a_e/\bar a_e$ with integers $a_e$ and $\bar a_e>0$, $|a_e|,\bar a_e<2^\nu$, the reward on
  $I$ is $r^0_I+r^1_I\alpha$, in the notation after \cref{lem:pwaffine}, where
  $r^0_I=P/(\zeta\prod_{e\in E}\bar a_e)$ and $r^1_I=P'/(\zeta\prod_{e\in E}\bar a_e)$ with
  $P:=\sum_{e\in E}a_eu_e\prod_{e'\ne e}\bar a_{e'}$ and
  $P':=\sum_{e\in E}a_ev_e\prod_{e'\ne e}\bar a_{e'}$, and the integers $P$, $P'$, and
  $\zeta\prod_{e\in E}\bar a_e$ have absolute value at most $nQ2^{n\nu}$. The maximum of $\UP$ over $I$ is
  attained at an endpoint, which is $0$, $1$, or a breakpoint, or, if $r^1_I>0$ and
  it lies in $I$, at the vertex $(r^1_I-r^0_I)/(2r^1_I)=(P'-P)/(2P')$ of the parabola
  $(1-\alpha)(r^0_I+r^1_I\alpha)$. So some optimal contract is a quotient of two integers of
  absolute value at most $2nQ2^{n\nu}$, and its bit length is $\Oh(n\nu)$.
\end{proof}

\subsection{Enumerating the pieces}\label{sec:A-pieces}

Recall that a \emph{breakpoint} is a share $\alpha>0$ at which $\xs$ is not
affine on any neighborhood, that $B$ denotes the number of breakpoints in $(0,1)$, and that a
\emph{piece} is the closure of a connected component of the non-breakpoint
shares in $(0,1)$. There are at most $B+1$ pieces, and they cover $[0,1]$, as there are
finitely many breakpoints (\cref{lem:pwaffine}). We write $t(S):=\sum_{e\in S}t_e$ and
$\binv(S):=\sum_{e\in S}\beta_e^{-1}$.

For a state $(\Phi,Z,\mathcal P)$ as in \cref{sec:A-structure}, define
\begin{equation}\label{eq:velocity}
  v_e:=\begin{cases}
    t_e & e\in\Phi,\\
    0 & e\in Z\setminus\Phi,\\
    \beta_e^{-1}(m_e-\bar m_D) & e\in D\setminus Z,\ D\in\mathcal P,
  \end{cases}
  \qquad
  \bar m_D:=\frac{t(D\setminus Z)}{\binv(D\setminus Z)},
\end{equation}
where the mean $\bar m_D$ is needed only when $D\setminus Z\neq\emptyset$.

\begin{lemma}\label{lem:velocity}
On every open interval with constant state, $\xs$ is affine with slope $v$ from
\eqref{eq:velocity}. For $e\in D\setminus Z$, the marginal $\rho_e$ has slope
$\bar m_D$.
\end{lemma}

\begin{proof}
Continuity keeps the blocks in the same order of marginal values throughout the
interval. Their cumulative unions are tight level sets, so $\xs(D)$ is constant
for every block $D$. Writing $\rho_D$ for its common marginal gives
\[
  \xs(D)=\alpha\,t(D\setminus Z)-\rho_D(\alpha)\,\binv(D\setminus Z).
\]
If $D\setminus Z\neq\emptyset$, this makes $\rho_D$ affine with slope $\bar m_D$
and gives the stated slope of $\xs$. For free and zero coordinates, the claim
follows from $\xs_e=\alpha t_e$ and $\xs_e=0$, respectively.
\end{proof}

\begin{lemma}\label{lem:eventcount}
Every breakpoint is an event point, and $(0,1)$ contains at most
$B+(B+1)\binom n2$ event points.
\end{lemma}

\begin{proof}
The first claim follows from \cref{lem:velocity}. Inside a piece, every $\xs_e$
and every $\rho_e$ is nonnegative and affine, so each vanishes either
identically or nowhere in its interior. Hence the free elements and the zero
set are constant there, and the blocks change only at an isolated zero of some
$\rho_e-\rho_g$, at most $\binom n2$ points per piece, in addition to the
breakpoints.
\end{proof}

\paragraph{Affine continuation.}
For $\hat\alpha\in(0,1)$, compute $\hat x:=\xs(\hat\alpha)$ and let
$v(\hat\alpha)$ be the vector assigned to its state by \eqref{eq:velocity}.
Define the \emph{affine continuation}
\begin{equation}\label{eq:A-cont}
  y_{\hat\alpha}(\alpha):=\hat x+(\alpha-\hat\alpha)\,v(\hat\alpha).
\end{equation}
Let $I(\hat\alpha)$ be the connected component containing $\hat\alpha$ of
$\{\alpha\in[0,1]:y_{\hat\alpha}(\alpha)=\xs(\alpha)\}$. It is a closed
interval, possibly a singleton, by \cref{lem:pwaffine}.

\begin{lemma}\label{lem:continuation}
Given $\hat x$ and $v(\hat\alpha)$, the interval $I(\hat\alpha)$ can be computed
with $\Oh(n^2)$ submodular minimizations, $\Oh(n^3)$ further evaluations of $f$,
and $\Oh(n^4)$ further arithmetic operations. Every nondegenerate
$I(\hat\alpha)$ is a piece. If $\hat\alpha$ is not an event point, it lies in
the interior of $I(\hat\alpha)$.
\end{lemma}

\begin{proof}
Write $y=y_{\hat\alpha}$, $v=v(\hat\alpha)$, and
$\tilde\rho_e(\alpha)=\alpha m_e-\beta_ey_e(\alpha)$. By \cref{prop:optchar},
$y=\xs$ exactly where $y$ is feasible, $\tilde\rho\ge 0$, and every positive
level set of $\tilde\rho$ is tight. Feasibility along $y$ is an interval:
coordinate nonnegativity gives ratio bounds, and the rank constraints are
handled by ray computations from $\hat x$ in the directions $v$ and $-v$. The
discrete Newton method for arbitrary directions uses $n^2+\Oh(n\log^2n)$
minimizations and $\Oh(n^2)$ evaluations of $f$ per ray~\cite{GGJ17}. Intersect the feasible interval with
$[0,1]$ and the inequalities $\tilde\rho_e\ge 0$, obtaining a closed interval
$K\ni\hat\alpha$.

Cut $K$ at $\hat\alpha$ and at the crossings of distinct affine functions
$\tilde\rho_e,\tilde\rho_g$. On each of the $\Oh(n^2)$ resulting open
intervals $J$, the positive level sets form a fixed chain
$L_1\subsetneq\dots\subsetneq L_q$ with $q\le n$: an affine function
nonnegative on $K$ cannot have an isolated zero in its interior. For the
midpoint $c$ of $J$, tightness throughout $J$ is equivalent to
\[
  y(c)(L_i)=f(L_i),\qquad v(L_i)=0\qquad(i=1,\dots,q).
\]
If these tests fail, tightness can hold at at most one point of $J$. Starting
at $\hat\alpha$, retain the consecutive intervals on either side that pass the
tests. By continuity, the union of their closures with $\{\hat\alpha\}$ is
exactly $I(\hat\alpha)$, including when $K$ is a singleton. The ray
computations and at most $n$ set tests on each of the $\Oh(n^2)$ intervals
give the stated costs.

If $I(\hat\alpha)$ is nondegenerate, any endpoint in $(0,1)$ must be a
breakpoint: otherwise the affine response there agrees with $y$ on an interval
and extends the component, a contradiction. Thus $I(\hat\alpha)$ is a piece. At
a non-event point, \cref{lem:velocity} gives $\xs=y$ on a neighborhood,
proving the last claim.
\end{proof}

\paragraph{Enumerating the pieces.}
Start with the covered set $C_0:=\emptyset$ and the collection of pieces
$\mathcal I_0:=\emptyset$. While $C_k\neq[0,1]$, choose the midpoint
$\hat\alpha_k$ of any connected component of $[0,1]\setminus C_k$ and set
\begin{equation}\label{eq:A-enum}
  C_{k+1}:=C_k\cup I(\hat\alpha_k),\qquad
  \mathcal I_{k+1}:=\begin{cases}
    \mathcal I_k\cup\{I(\hat\alpha_k)\} & \text{if } I(\hat\alpha_k)\text{ is nondegenerate},\\
    \mathcal I_k & \text{otherwise.}
  \end{cases}
\end{equation}
For each stored piece, retain its affine map $y_{\hat\alpha_k}$. After covering
$[0,1]$, maximize the corresponding quadratic utility
$\UP(\alpha)=(1-\alpha)\,m^\top y_{\hat\alpha_k}(\alpha)$ on each piece and take
the best share.

\restate{\rsPieces}{thm:pieces}
\begin{proof}
Each $C_k$ is a finite union of closed intervals, so every nonempty
complementary component has positive length and its midpoint lies in $(0,1)$.
By \cref{lem:continuation}, each step discovers a new piece or a singleton
event point that is never tried again. By \cref{lem:eventcount}, the number of
steps is at most
\[
  (B+1)\Biggl(1+\binom n2\Biggr)+B\le(n+2)^2(B+1).
\]
At termination, finitely many closed pieces and singletons cover $[0,1]$. The
pieces alone cover it, since a nonempty complement of their union would contain
an interval, so every piece has been found. Maximizing on each piece
therefore gives an optimal contract. Each step uses one best-response computation
and one affine continuation, costing $\Oh(n^2)$ minimizations, $\Oh(n^3)$
further rank evaluations, and $\Oh(n^4)$ further arithmetic operations by
the proof of \cref{thm:waterfilling} and \cref{lem:continuation}. Maintaining $C_k$ as a
sorted interval list costs at most $\Oh(n^2(B+1))$ operations per step.
Multiplying gives $\Oh(n^4(B+1))$ submodular
minimizations, $\Oh(n^5(B+1))$ further evaluations of $f$, and
$\Oh(n^6(B+1)+n^4(B+1)^2)$ further arithmetic operations.
As in the proof of \cref{thm:waterfilling}, each submodular minimization takes time $\Oh(n^5\log n)\cdot\FO+\Oh(n^6\log n)$ using Iwata and Orlin's algorithm~\cite{IO09}: its objective is $f$ minus a modular function, evaluated in time $\FO+\Oh(n)$. Substituting this bound gives the stated total runtime.
\end{proof}

\Cref{alg:pieces} summarizes this construction.

\begin{algorithm}[ht]
\caption{Computing an optimal contract from the affine pieces}\label{alg:pieces}
\begin{algorithmic}[1]
\Require $f$, $m\ge 0$, $\beta>0$, and a value oracle for $f$ and submodular minimization returning the maximal minimizer (\cref{sec:bestresponse})
\Ensure an optimal contract $\as$ and its best response
\State $C\gets\emptyset$, $\mathcal I\gets\emptyset$
\While{$C\neq[0,1]$}
  \State choose the midpoint $\hat\alpha$ of a component of $[0,1]\setminus C$
  \State compute $\xs(\hat\alpha)$ and $v(\hat\alpha)$ by \cref{thm:waterfilling} and \eqref{eq:velocity}
  \State compute $I(\hat\alpha)$ by \cref{lem:continuation} and update $C,\mathcal I$ by \eqref{eq:A-enum}
\EndWhile
\State \Return the best share over the pieces in $\mathcal I$, and its affine response
\end{algorithmic}
\end{algorithm}

\subsection{Exponentially many breakpoints}\label{sec:A-expbp}

\restate{\rsAlm}{thm:alm}
\begin{proof}
The network of Allman, Lo, and McCormick has a source $s$, a sink $s'$, and the
nodes $E$ (we avoid the letter $t$, which denotes the vector $m_e/\beta_e$).
The arcs $s\to j$ have capacity $a_j\lambda+b_j\theta$ with $a_j,b_j>0$, and
all other arcs have constant nonnegative capacities $u_{ij}$. Every
$S\subseteq E$ is the unique minimum cut $\{s\}\cup S$ at some point
$(\lambda_S,\theta_S)\in(0,1)^2$ \cite[Theorem~1.1, with the network of
Section~2]{ALM22}. The capacity of the cut $\{s\}\cup S$ is
$\lambda a(E\setminus S)+\theta b(E\setminus S)+\tilde f(S)$ with
\[
  \tilde f(S):=\sum_{i\in S}u_{is'}+\sum_{i\in S,\,j\in E\setminus S}u_{ij},
\]
that is, $\lambda a(E)+\theta b(E)$ plus $\tilde f(S)-\lambda a(S)-\theta b(S)$.
So $S$ is the unique minimizer of $\tilde f-\lambda_Sa-\theta_Sb$. The function
$\tilde f$ is a cut function of a directed graph with nonnegative capacities,
so it is nonnegative and submodular~\cite[(44.16), (44.77)]{Sch03}, and $\tilde f(\emptyset)=0$.
\end{proof}

\begin{lemma}\label{lem:embed}
Let $\smash{\tilde f\colon 2^E\to\R}$ be submodular with $\tilde f(\emptyset)=0$, let $a,b\in\R^E_{>0}$, $\Theta>0$, and
$\Lambda\ge\smash{\max\{0,\max_{S\subseteq E,\,e\in E\setminus S}(\tilde f(S)-\tilde f(S\cup\{e\})+\Theta b_e)/a_e\}}$.
Then $\smash{f(S):=\tilde f(S)+\sum_{e\in S}(\Lambda a_e-\Theta b_e)}$ is a rank function, and in the instance with $f$, $m_e:=a_e/b_e$, and $\beta_e:=1/b_e$, the level set $A_{\Theta-\theta}(\Lambda+\lambda)$ is the maximal minimizer of $\tilde f-\lambda a-\theta b$ for all $\lambda\ge-\Lambda$ and $\theta<\Theta$.
\end{lemma}
\begin{proof}
The function $f$ is submodular, as $\tilde f$ is and the second term is
modular, and $f(\emptyset)=0$. By the choice of $\Lambda$, every marginal
\[
  f(S\cup\{e\})-f(S)=\tilde f(S\cup\{e\})-\tilde f(S)+\Lambda a_e-\Theta b_e,
  \qquad S\subseteq E,\ e\in E\setminus S,
\]
is nonnegative, so $f$ is monotone. In the instance, $t_e=m_e/\beta_e=a_e$ and
$\beta_e^{-1}=b_e$, so for all $\lambda\ge-\Lambda$ and $\theta<\Theta$,
\[
  F_{\Lambda+\lambda,\,\Theta-\theta}(S)=\tilde f(S)-\lambda a(S)-\theta b(S)
  \qquad\text{for all } S\subseteq E,
\]
and \cref{thm:levelsets} applies with $\Lambda+\lambda\ge 0$ and
$\Theta-\theta>0$.
\end{proof}

\restate{\rsExpbp}{thm:expbp}
\begin{proof}
Take $a$, $b$, and $\tilde f$ from \cref{thm:alm}, and the instance of
\cref{lem:embed} with $\Theta:=2$ and $\Lambda$ as there. For $S\subseteq E$,
let $\alpha_S:=\Lambda+\lambda_S\in(\Lambda,\Lambda+1)$ and
$\bar\theta_S:=\Theta-\theta_S\in(1,2)$. As $\lambda_S>0\ge-\Lambda$ and
$\theta_S<\Theta$, \cref{lem:embed} gives $A_{\bar\theta_S}(\alpha_S)=S$, that
is, $\rho_e(\alpha_S)\ge\bar\theta_S$ for $e\in S$ and
$\rho_e(\alpha_S)<\bar\theta_S$ for $e\notin S$. For a level $\bar\theta$ with
$\max\{1,\rho_e(\alpha_S):e\notin S\}<\bar\theta<\bar\theta_S$, these
inequalities hold strictly at $(\alpha_S,\bar\theta)$, and by continuity of
$\rho$ (\cref{lem:pwaffine}) they hold strictly near $(\alpha_S,\bar\theta)$.
So there is a point $(\alpha'_S,\bar\theta'_S)$ with
$A_{\bar\theta'_S}(\alpha'_S)=S$, where $\alpha'_S\in(\Lambda,\Lambda+1)$ is
not a breakpoint and $\rho_e(\alpha'_S)\neq\bar\theta'_S$ for all $e\in E$.

Let $k$ be the number of breakpoints in $(\Lambda,\Lambda+1)$. They cut this
interval into $k+1$ open intervals, on each of which $\xs$, and hence every
$\rho_e$, is affine. Fix one of them, $I$. For $e\in E$, the graph of $\rho_e$
over $I$ lies on a line $\ell_e$ in the $(\alpha,\theta)$-plane, and for
$\alpha\in I$ the level set $A_\theta(\alpha)$ is the set of elements $e$ whose
line $\ell_e$ passes through or above $(\alpha,\theta)$. So all points
$(\alpha'_S,\bar\theta'_S)$ with $\alpha'_S\in I$ that lie in the same cell of
the arrangement of the lines $\ell_e$ have the same set $S$. As $n$ lines cut
the plane into at most $1+n+\binom n2=1+n(n+1)/2$ cells, we get
$2^n\le(k+1)\bigl(1+n(n+1)/2\bigr)$.

Finally, let $\sigma:=1/(\Lambda+1)$ and replace $m$ by $m/\sigma$. By
\cref{lem:ops}(b), the new best response is $\alpha\mapsto\xs(\alpha/\sigma)$.
Its breakpoints are the old breakpoints multiplied by $\sigma$, and the $k$
breakpoints in $(\Lambda,\Lambda+1)$ move into
$(\sigma\Lambda,1)\subseteq(0,1)$.
\end{proof}

\subsection{Tools for submodular functions}\label{app:tools}

This subsection collects general facts about submodular functions that are not necessarily monotone.
They are used both for the construction with polynomial bit length (\cref{app:bits}) and for the comb gadget (\cref{app:hard}).
\Cref{sec:B-fullplane} extends the best response and its marginals to the base polyhedron without sign constraints, \cref{sec:B-translation} adds one element whose marginal value is a constant minus a translation, \cref{sec:B-extensions} describes what a one-element extension does to the best response, and \cref{sec:B-dmt} specializes this to the extension that builds combs.

\subsubsection{The full plane}\label{sec:B-fullplane}

The functions to which we apply the tools of this subsection are submodular but need not be monotone, so in general they are not polymatroid rank functions and define no instance of their own.
We therefore extend the best response and its marginals to such a function $g$, on its base polyhedron without sign constraints, which we call the \emph{full plane}.
Two features make the full plane the right setting.
First, as in a base polyhedron, every point $y$ of the full plane satisfies $y(E)=g(E)$. Adding the same constant $c$ to all marginals amounts to adding $c\,y(E)=c\,g(E)$ to the agent's utility, a constant. So the level criterion (\cref{lem:fullplane}(a)) relaxes \cref{prop:optchar} by dropping the condition $\rho\ge0$, and it asks for tight upper level sets at every level $\theta\in\R$ instead of every $\theta>0$.
Second, there is no sign constraint.
Both are needed for the shift (\cref{lem:translation}(c)): subtracting a suitable modular function translates the best response along the share axis and raises all marginals by the same constant, and the translated point may have negative coordinates.
The one-element extensions of \cref{sec:B-extensions} lower marginal values by modular functions, and the shift subtracts a modular function of any sign, so the functions of the construction need not be monotone.
Part (g) of \cref{lem:fullplane} brings the result back to a polymatroid: after a suitable modular function is added, the full-plane best response is the best response $\xs$ of an instance, as in \cref{lem:embed}.

\paragraph{Full-plane best responses and marginals.}
Let $g\colon 2^E\to\R$ be submodular with $g(\emptyset)=0$, not necessarily monotone, let $m,\beta\in\R^E_{>0}$, and let $t_e=m_e/\beta_e$, $t(S)=\sum_{e\in S}t_e$, and $\binv(S)=\sum_{e\in S}1/\beta_e$ be as in \cref{sec:bestresponse}.
We also write $t$ and $\binv$ for the vectors $(t_e)_{e\in E}$ and $(\beta_e^{-1})_{e\in E}$, and we identify a vector $u\in\R^E$ with the modular function $S\mapsto u(S):=\sum_{e\in S}u_e$.
Let
\[
  \Bpm{g}:=\{y\in\R^E\colon y(S)\le g(S)\text{ for all }S\subseteq E,\ y(E)=g(E)\}
\]
be the base polyhedron of $g$ without a sign constraint.
The subscript $(\pm)$ indicates the missing sign constraint, in contrast to the $(+)$ of $\Pp{f}$. For a polymatroid rank function $f$, we write $B(f):=\{x\in\Pp{f}:x(E)=f(E)\}=\Bpm{f}\cap\R^E_{\ge0}$ for its base polyhedron.
For $y\in\R^E$ with $y(S)\le g(S)$ for all $S\subseteq E$, a set $S$ is \emph{tight} for $y$ if $y(S)=g(S)$, as for polymatroids.

The full-plane best response below is the maximizer of a strictly concave function on $\Bpm{g}$, so we first show that $\Bpm{g}$ is nonempty and compact.
For an order $e_1,\dots,e_n$ of $E$ and $S_i:=\{e_1,\dots,e_i\}$, the greedy vector $y_{e_i}:=g(S_i)-g(S_{i-1})$ satisfies $y(E)=g(E)$ (compare \cite[Thm.~44.3]{Sch03}).
Let $S\subseteq E$. For $e_i\in S$, we have $S\cap S_i=(S\cap S_{i-1})\cup\{e_i\}$ with $S\cap S_{i-1}\subseteq S_{i-1}$ and $e_i\notin S_{i-1}$, so submodularity gives $y_{e_i}\le g(S\cap S_i)-g(S\cap S_{i-1})$. For $e_i\notin S$, this right-hand side is $0$. So summing over $e_i\in S$ gives $y(S)\le\sum_{i=1}^n\bigl(g(S\cap S_i)-g(S\cap S_{i-1})\bigr)=g(S)$, and $y\in\Bpm{g}$.
The full plane is also compact, as it is closed and its points satisfy $g(E)-g(E\setminus\{e\})\le y_e\le g(\{e\})$, where the lower bound follows from $y_e=y(E)-y(E\setminus\{e\})$.
So for every $\alpha\in\R$ the strictly concave function $\UA(\cdot\mid\alpha)$, given on all of $\R^E$ by the same formula as on $\Pp{f}$, has a unique maximizer $\xi^g(\alpha)$ on $\Bpm{g}$, the \emph{full-plane best response} of $g$ (for $m$ and $\beta$).
Completing the square as in the proof of \cref{lem:pwaffine} shows that $\xi^g(\alpha)$ is the projection of $\alpha t$ onto $\Bpm{g}$ with respect to $\|\cdot\|_\beta$.
That proof applies verbatim with the polyhedron $\Bpm{g}$ in place of $\Pp{f}$ and with $\alpha\in\R$, so $\xi^g$ is continuous and piecewise affine on $\R$, with finitely many breakpoints.

We call $\rho^g(\alpha):=\alpha m-\beta\xi^g(\alpha)$ (componentwise), that is, $\rho^g_e(\alpha)=\alpha m_e-\beta_e\xi^g_e(\alpha)$, the \emph{full-plane marginals} of $g$, and $\rho^g_e(\alpha)$ the \emph{marginal} of $e$.
We extend the reward function $r(x)=m^\top x$ to all $x\in\R^E$ and consider the reward $r(\xi^g(\alpha))$ of the full-plane best response.
For a polymatroid rank function $g$ and a share $\alpha\ge 0$ at which all full-plane marginals are nonnegative, $\xi^g(\alpha)=\xs(\alpha)$, and $\rho^g(\alpha)$ is the vector of the marginals $\rho_e(\alpha)$ of \cref{sec:bestresponse} (\cref{lem:fullplane}(g) below, with $\Lambda=\Theta=0$).

For a share $\alpha\in\R$ and a vector $y\in\R^E$, let $\rho(y):=\alpha m-\beta y$ be the vector of marginals at $y$, as in \eqref{eq:marg}, and let $A_\theta(y):=\{e\in E\colon\rho_e(y)\ge\theta\}$ be its upper level set at the level $\theta$, as in \cref{sec:bestresponse} but now for every $\theta\in\R$.
So $\rho^g(\alpha)=\rho(\xi^g(\alpha))$ at the share $\alpha$, and we write for $\theta \in \mathbb{R}$
\[
  A^g_\theta(\alpha):=A_\theta(\xi^g(\alpha))=\{e\in E\colon\rho^g_e(\alpha)\ge\theta\}.
\]
For $\alpha,\theta\in\R$ and $S\subseteq E$, let
\[
  F^g_{\alpha,\theta}(S):=g(S)-\alpha\,t(S)+\theta\,\binv(S).
\]
For a polymatroid rank function $g$, $\alpha\ge 0$, and $\theta>0$, this is the function $F_{\alpha,\theta}$ of \cref{thm:levelsets}, and for $\beta_e=1$ for all $e\in E$ it is the function $F^g_{\alpha,\theta}$ of \cref{app:bits}.
A function $\phi\colon\R\to\R$ \emph{grows at least} (\emph{at most}) \emph{at the rate} $c$ if $\phi(\alpha+\Delta)\ge\phi(\alpha)+c\Delta$ (respectively $\phi(\alpha+\Delta)\le\phi(\alpha)+c\Delta$) for all $\alpha\in\R$ and $\Delta\ge 0$.
For an affine function, this compares its slope with $c$.

\paragraph{Weighted means.}
For nonempty $D\subseteq E$, let
\[
  \bar m_D:=\frac{t(D)}{\binv(D)}=\frac{\sum_{e\in D}m_e/\beta_e}{\sum_{e\in D}1/\beta_e}
\]
be the mean reward of $D$, weighted by $1/\beta$.
The mean of \eqref{eq:velocity}, written $\bar m_D$ there, is $\bar m_{D\setminus Z}$ in the present notation.
We also use the weighted sum of squares
\[
  \SSq(D):=\sum_{e\in D}\frac{(m_e-\bar m_D)^2}{\beta_e}.
\]
The function $v\mapsto\sum_{e\in D}(m_e-v)^2/\beta_e$ is a convex quadratic with derivative $-2\bigl(t(D)-v\,\binv(D)\bigr)$, so it is minimized at $v=\bar m_D$, and
\[
  \SSq(D)=\min_{v\in\R}\sum_{e\in D}\frac{(m_e-v)^2}{\beta_e}.
\]
Hence $\SSq$ is monotone under inclusion, as for $D\subseteq D'$ the choice $v=\bar m_{D'}$ gives $\SSq(D)\le\SSq(D')$.
For distinct $e,e'\in E$, the mean $\bar m:=\bar m_{\{e,e'\}}$ divides the gap between $m_e$ and $m_{e'}$ in the ratio of the curvatures, $m_e-\bar m=\beta_e(m_e-m_{e'})/(\beta_e+\beta_{e'})$ and $m_{e'}-\bar m=-\beta_{e'}(m_e-m_{e'})/(\beta_e+\beta_{e'})$, so
\begin{equation}\label{eq:B-pair}
  \SSq(\{e,e'\})=\frac{(m_e-m_{e'})^2}{\beta_e+\beta_{e'}}.
\end{equation}
For disjoint nonempty $D_1,D_2\subseteq E$, expanding about the two group means, the cross terms vanish, and the term between the groups is \eqref{eq:B-pair} for two elements with rewards $\bar m_{D_i}$ and curvatures $1/\binv(D_i)$. This gives
\begin{equation}\label{eq:B-twogroups}
  \SSq(D_1\cup D_2)=\SSq(D_1)+\SSq(D_2)+\frac{(\bar m_{D_1}-\bar m_{D_2})^2}{1/\binv(D_1)+1/\binv(D_2)}.
\end{equation}

\begin{lemma}\label{lem:fullplane}
Let $g\colon 2^E\to\R$ be submodular with $g(\emptyset)=0$, let $m,\beta\in\R^E_{>0}$, and let $\alpha,\theta\in\R$.
\begin{enumerate}[label=(\alph*)]
  \item A vector $y\in\R^E$ is $\xi^g(\alpha)$ if and only if $y(S)\le g(S)$ for all $S\subseteq E$ and every upper level set $A_{\theta'}(y)$, $\theta'\in\R$, of the marginals at the share $\alpha$ is tight for $y$.
  In particular, the sets $A^g_\theta(\alpha)$ and $\{e\in E\colon\rho^g_e(\alpha)>\theta\}$ are tight for $\xi^g(\alpha)$. \label{it:fullplane-a}
  \item The set $A^g_\theta(\alpha)$ minimizes $F^g_{\alpha,\theta}$ and contains every minimizer. \label{it:fullplane-b}
  \item Let $b\in E$ and $c\in\R$ with $c\,\binv(S)\le t(S)$ for every $S\subseteq E$ with $b\in S$.
  Then $\rho^g_b$ grows at least at the rate $c$, and every $\rho^g_e$ grows at most at the rate $\max_{e'\in E}m_{e'}$.
  In particular, every $\rho^g_e$ grows at least at the rate $\min_{e'\in E}m_{e'}>0$ and at most at the rate $\max_{e'\in E}m_{e'}$, so it is nondecreasing, and $\rho^g$, $\xi^g$, and $\alpha\mapsto r(\xi^g(\alpha))$ are continuous.
  Moreover, $\rho^g_b$ grows at least at the rate $t_b/\binv(E)$. \label{it:fullplane-c}
  \item If $U\subseteq E$ is tight for $\xi^g(a)$ and for $\xi^g(a')$, where $a\le a'$, then $\sum_{e\in U}m_e\xi^g_e$ and $\sum_{e\in E\setminus U}m_e\xi^g_e$ are both at least as large at $a'$ as at $a$.
  In particular, $\alpha\mapsto r(\xi^g(\alpha))$ is nondecreasing. \label{it:fullplane-d}
  \item \label{it:fullplane-e} If the elements of a nonempty $D\subseteq E$ have equal marginals at $\alpha$, then $\xi:=\xi^g(\alpha)$ satisfies
  \[
    \sum_{e\in D}m_e\xi_e=\alpha\,\SSq(D)+\xi(D)\,\bar m_D.
  \]
  \item Let $g=g_1\oplus g_2$, that is, $g(S)=g_1(S\cap E_1)+g_2(S\cap E_2)$ for $S\subseteq E$, for submodular $g_i\colon 2^{E_i}\to\R$ with $g_i(\emptyset)=0$ on disjoint sets $E_1,E_2$ with $E=E_1\cup E_2$, and let $\xi^{g_i}$ and $\rho^{g_i}$ be taken with respect to $m|_{E_i}$ and $\beta|_{E_i}$.
  Then $\xi^g=(\xi^{g_1},\xi^{g_2})$ and $\rho^g=(\rho^{g_1},\rho^{g_2})$.
  A modular singleton $\pi_F$ on $\{\tau_1\}$ with $\pi_F(\{\tau_1\})=F$ (as in \cref{lem:cutforms}(c)(ii), now with $F\in\R$) has $\xi^{\pi_F}_{\tau_1}(\alpha)=F$ and $\rho^{\pi_F}_{\tau_1}(\alpha)=\alpha m_{\tau_1}-\beta_{\tau_1}F$ for every $\alpha\in\R$. \label{it:fullplane-f}
  \item If $f:=g+\Lambda t-\Theta\binv$ is a polymatroid rank function for some $\Lambda,\Theta\in\R$, then the instance $(f,m,\beta)$ has
  \[
    \xs(\alpha+\Lambda)=\xi^g(\alpha)+\Lambda t-\Theta\binv
  \]
  at every $\alpha\ge-\Lambda$ with $\rho^g_e(\alpha)+\Theta\ge 0$ for all $e\in E$. \label{it:fullplane-g}
\end{enumerate}
\end{lemma}

Parts \ref{it:fullplane-a}--\ref{it:fullplane-d} are full-plane forms of \cref{prop:optchar}, \cref{thm:levelsets}, \cref{thm:monmarg}, and \cref{lem:monreward}, and \ref{it:fullplane-c} and \ref{it:fullplane-d} are stronger, with rates and with parts of the reward. Part \ref{it:fullplane-a} is in substance Fujishige's criterion for separable convex objectives over a base polyhedron~\cite[Thm.~9.1]{Fuj05}, and its exchange step is Groenevelt's~\cite[Thm.~8.1]{Fuj05}.
Part \ref{it:fullplane-e} is a pointwise form of the slopes \eqref{eq:velocity}: for a block $D$ with $D\cap Z=\emptyset$ on an interval with constant state, $\sum_{e\in D}m_ev_e=\SSq(D)$.
Part \ref{it:fullplane-g} gives the best response itself under the embedding of \cref{lem:embed}. The translation identities used later are stated separately in \cref{lem:translation}(c).

\begin{proof}
\ref{it:fullplane-a} We first show that $y:=\xi^g(\alpha)$ has the two properties.
It satisfies $y(S)\le g(S)$ for all $S$, as $y\in\Bpm{g}$.
The gradient of $\UA(\cdot\mid\alpha)$ at $y$ is $\rho(y)$.
Let $e,e'\in E$ with $\rho_e(y)>\rho_{e'}(y)$, and suppose that no set that is tight for $y$ contains $e$ but not $e'$.
Moving a little mass from $e'$ to $e$, that is, increasing $y_e$ and decreasing $y_{e'}$ by the same small amount, changes $y(S)$ only for the sets $S$ that contain exactly one of $e$ and $e'$.
It increases $y(S)$ only for the sets that contain $e$ but not $e'$, which are finitely many and not tight, so for a small enough amount the new vector stays in $\Bpm{g}$.
As $\rho_e(y)>\rho_{e'}(y)$, it increases $\UA(\cdot\mid\alpha)$, which contradicts the maximality of $y$.
So for an upper level set $U:=A_{\theta'}(y)$, every $e\in U$ and every $e'\in E\setminus U$, which satisfy $\rho_e(y)\ge\theta'>\rho_{e'}(y)$, have a tight set $S_{ee'}$ with $e\in S_{ee'}$ and $e'\notin S_{ee'}$.
Then
\[
  U=\bigcup_{e\in U}\ \bigcap_{e'\in E\setminus U}S_{ee'},
\]
an empty intersection being $E$ and an empty union being $\emptyset$, because each intersection contains $e$ and no element of $E\setminus U$.
The tight sets of a vector $y$ with $y(S)\le g(S)$ for all $S$ are closed under union and intersection, by the argument of \cref{prop:tight}~\cite[Thm.~44.2]{Sch03}, which needs only the submodularity of $g$ and $y(S)\le g(S)$ for all $S$.
Indeed, for tight sets $S,S'$,
\begin{align*}
  g(S\cup S')+g(S\cap S')&\ge y(S\cup S')+y(S\cap S')=y(S)+y(S')\\
  &=g(S)+g(S')\ge g(S\cup S')+g(S\cap S'),
\end{align*}
so equality holds throughout, and as $y(S\cup S')\le g(S\cup S')$ and $y(S\cap S')\le g(S\cap S')$, both sets are tight.
Hence $U$ is tight.

Conversely, let $y$ satisfy the two conditions.
As $E=A_{\theta'}(y)$ for every $\theta'\le\min_{e\in E}\rho_e(y)$, the set~$E$ is tight, so $y\in\Bpm{g}$.
We use the concavity step and the decomposition into level sets of the proof of \cref{prop:optchar}, with $\Bpm{g}$ in place of $\Pp{f}$ and with a level below all marginals in place of $0$.
By concavity, $\UA(x\mid\alpha)-\UA(y\mid\alpha)\le\rho(y)^\top(x-y)$ for every $x\in\R^E$.
Let $x\in\Bpm{g}$, let $\varrho_1>\dots>\varrho_k$ be the distinct values of $\rho(y)$, let $\varrho_{k+1}<\varrho_k$ be arbitrary, and let $A_i:=A_{\varrho_i}(y)$ for $i=1,\dots,k$, so that $A_k=E$.
Since $x(E)=y(E)$,
\[
  \rho(y)^\top(x-y)=\sum_{e\in E}\bigl(\rho_e(y)-\varrho_{k+1}\bigr)(x_e-y_e)=\sum_{i=1}^k(\varrho_i-\varrho_{i+1})\bigl(x(A_i)-y(A_i)\bigr)\le 0,
\]
as every $A_i$ is tight for $y$ and $x(A_i)\le g(A_i)$.
So $y$ maximizes $\UA(\cdot\mid\alpha)$ on $\Bpm{g}$, that is, $y=\xi^g(\alpha)$.
Here $x(E)=y(E)$ takes the place of the condition $\rho(y)\ge 0$ of \cref{prop:optchar} and allows levels of any sign, and the upper bound needs only $x(A_i)\le g(A_i)$, not the value of the maximum.
That proof uses the value of the maximum, from Edmonds' greedy theorem~\cite[(19)]{Edm03} for rank functions and $\Pp{f}$, in both directions, so we have shown the other direction by the exchange argument above.

Finally, the set $\{e\in E\colon\rho^g_e(\alpha)>\theta\}$ is empty or equals $A^g_{\theta''}(\alpha)$, where $\theta''$ is the smallest marginal of its elements.
So both sets of the last claim are upper level sets of $\rho(\xi^g(\alpha))$ at the share $\alpha$, and they are tight for $\xi^g(\alpha)$.

\ref{it:fullplane-b} The proof of \cref{thm:levelsets} applies verbatim, with $g$, $\xi^g(\alpha)$, $\rho^g(\alpha)$, $A^g_\theta(\alpha)$, and $\Bpm{g}$ in place of $f$, $\xs(\alpha)$, $\rho(\alpha)$, $A_\theta(\alpha)$, and $\Pp{f}$, and with (a) in place of \cref{prop:optchar} for the tightness of $A^g_\theta(\alpha)$.
It uses neither $\alpha\ge 0$ nor $\theta>0$.
In short, $\xi:=\xi^g(\alpha)$ has $\xi_e=\alpha t_e-\rho^g_e(\alpha)/\beta_e$, so
\[
  F^g_{\alpha,\theta}(S)=\bigl(g(S)-\xi(S)\bigr)+\sum_{e\in S}\frac{\theta-\rho^g_e(\alpha)}{\beta_e}\qquad(S\subseteq E).
\]
The first term is nonnegative, and it vanishes at $S=A^g_\theta(\alpha)$ by \ref{it:fullplane-a}.
The sum is smallest exactly for the sets that contain every $e$ with $\rho^g_e(\alpha)>\theta$ and no $e$ with $\rho^g_e(\alpha)<\theta$, among them $A^g_\theta(\alpha)$.
So $A^g_\theta(\alpha)$ minimizes $F^g_{\alpha,\theta}$, and every minimizer makes the sum smallest, hence is contained in $A^g_\theta(\alpha)$.

\ref{it:fullplane-c} Let $\Delta\ge 0$, $\rho_0:=\rho^g_b(\alpha)$, $A:=A^g_{\rho_0}(\alpha)$, and $A':=A^g_{\rho_0+c\Delta}(\alpha+\Delta)$, so that $b\in A$, and suppose that $b\notin A'$.
Write $G:=F^g_{\alpha,\rho_0}$ and $G':=F^g_{\alpha+\Delta,\rho_0+c\Delta}$.
Then $G'=G-\Delta(t-c\binv)$, and by submodularity of $G'$ and the minimality of $A$ for $G$ (by (b)),
\begin{align*}
  G'(A\cup A')-G'(A')&\le G'(A)-G'(A\cap A')\\
  &=G(A)-G(A\cap A')-\Delta\,(t-c\binv)(A\setminus A')\le 0,
\end{align*}
as $b\in A\setminus A'$ gives $c\,\binv(A\setminus A')\le t(A\setminus A')$.
So $A\cup A'$ minimizes $G'$, and $A\subseteq A\cup A'\subseteq A'$ by (b), a contradiction.
Hence $\rho^g_b(\alpha+\Delta)\ge\rho_0+c\Delta$.

For the upper rate, let $e\in E$, $m_{\max}:=\max_{e'\in E}m_{e'}$, $\rho_1:=\rho^g_e(\alpha+\Delta)$, $A':=A^g_{\rho_1}(\alpha+\Delta)$, and $A:=A^g_{\rho_1-m_{\max}\Delta}(\alpha)$, so that $e\in A'$, and suppose that $e\notin A$.
Now $G:=F^g_{\alpha+\Delta,\rho_1}$ and $G':=F^g_{\alpha,\rho_1-m_{\max}\Delta}$ satisfy $G'=G-\Delta(m_{\max}\binv-t)$ with $m_{\max}\binv-t\ge 0$, and by \ref{it:fullplane-b}, $A'$ minimizes $G$ and $A$ minimizes $G'$.
The same computation with the roles of $A$ and $A'$ exchanged,
\begin{align*}
  G'(A\cup A')-G'(A)&\le G'(A')-G'(A\cap A')\\
  &=G(A')-G(A\cap A')-\Delta\,(m_{\max}\binv-t)(A'\setminus A)\le 0,
\end{align*}
shows that $A\cup A'$ minimizes $G'$, so $e\in A'\subseteq A\cup A'\subseteq A$ by \ref{it:fullplane-b}, a contradiction.
Hence $\rho^g_e(\alpha)\ge\rho_1-m_{\max}\Delta$, that is, $\rho^g_e(\alpha+\Delta)\le\rho^g_e(\alpha)+m_{\max}\Delta$.

The two lower rates hold because $c=\min_{e'\in E}m_{e'}$ satisfies $c\,\binv(S)\le t(S)$ for every $S\subseteq E$, and $c=t_b/\binv(E)$ satisfies $c\,\binv(S)\le t_b\le t(S)$ for every $S\ni b$.
These rates give $|\rho^g_e(\alpha')-\rho^g_e(\alpha)|\le m_{\max}|\alpha'-\alpha|$ for all $\alpha,\alpha'\in\R$, so $\rho^g$ is continuous, and so are $\xi^g_e(\alpha)=\alpha t_e-\rho^g_e(\alpha)/\beta_e$ and $r(\xi^g(\alpha))=m^\top\xi^g(\alpha)$.
When the modular difference $G'-G$ is signed on all sets, as for $c=\min_{e'}m_{e'}$ and for the upper rate, it is nonincreasing under inclusion, and the argument above is Topkis's monotonicity theorem~\cite[Thms.~6.1--6.2]{Top78}, as in the proof of \cref{thm:monmarg}.
For a general $c$, such as $t_b/\binv(E)$, the function $t-c\binv$ is known to be nonnegative only on the sets that contain $b$, and the argument uses only that.

\ref{it:fullplane-d} Let $x:=\xi^g(a)$ and $x':=\xi^g(a')$.
The vector $y$ equal to $x'$ on $U$ and to $x$ on $E\setminus U$ lies in $\Bpm{g}$.
Indeed, as $U$ is tight for $x$ and for $x'$, $y(E)=g(U)+g(E)-g(U)=g(E)$, and for every $S\subseteq E$, by submodularity,
\[
  y(S)=x'(S\cap U)+x(S\cup U)-x(U)\le g(S\cap U)+g(S\cup U)-g(U)\le g(S).
\]
In the same way, the vector $y'$ equal to $x$ on $U$ and to $x'$ on $E\setminus U$ lies in $\Bpm{g}$.
As in the proof of \cref{lem:monreward}, we add two optimality inequalities, here $\UA(x\mid a)\ge\UA(y\mid a)$ and $\UA(x'\mid a')\ge\UA(y'\mid a')$.
The vectors in each of them differ only on $U$, the quadratic terms cancel, and we obtain $(a'-a)\sum_{e\in U}m_e(x'_e-x_e)\ge 0$.
The comparisons of $x$ with $y'$ at $a$ and of $x'$ with $y$ at $a'$ give the same for $E\setminus U$.
The set $U=\emptyset$ is tight for every vector of the full plane, and for it this is the proof of \cref{lem:monreward} in the full plane, which shows that $\alpha\mapsto r(\xi^g(\alpha))$ is nondecreasing.

\ref{it:fullplane-e} Let $\rho_D$ be the common marginal, so $\xi_e=\alpha t_e-\rho_D/\beta_e$ for $e\in D$.
Summing over $D$ gives $\rho_D=\bigl(\alpha t(D)-\xi(D)\bigr)/\binv(D)$, and therefore
\begin{align*}
  \sum_{e\in D}m_e\xi_e&=\alpha\sum_{e\in D}\frac{m_e^2}{\beta_e}-\rho_D\,t(D)=\alpha\Bigl(\sum_{e\in D}\frac{m_e^2}{\beta_e}-\frac{t(D)^2}{\binv(D)}\Bigr)+\xi(D)\,\bar m_D\\
  &=\alpha\,\SSq(D)+\xi(D)\,\bar m_D,
\end{align*}
the last step by expanding the square in $\SSq(D)$.

\ref{it:fullplane-f} The claimed vectors satisfy the conditions of \ref{it:fullplane-a}.
The vector $y:=(\xi^{g_1}(\alpha),\xi^{g_2}(\alpha))$ satisfies $y(S)=y(S\cap E_1)+y(S\cap E_2)\le g(S)$ for every $S\subseteq E$.
Its marginals at the share $\alpha$ are those of the two summands, so its upper level sets are the unions of the upper level sets of the two summands at the same level, which are tight for the summands, hence for $y$.
So $y=\xi^g(\alpha)$ by (a), and $\rho^g(\alpha)=\rho(y)=(\rho^{g_1}(\alpha),\rho^{g_2}(\alpha))$.
For a modular singleton, $\Bpm{\pi_F}=\{F\}$.

\ref{it:fullplane-g} The vector $y:=\xi^g(\alpha)+\Lambda t-\Theta\binv$ satisfies $y(S)\le g(S)+\Lambda t(S)-\Theta\binv(S)=f(S)$ for every $S\subseteq E$, with equality exactly for the sets that are tight for $\xi^g(\alpha)$ with respect to $g$, and its marginals at the share $\alpha+\Lambda$ are
\[
  \rho_e(y)=(\alpha+\Lambda)m_e-\beta_e\xi^g_e(\alpha)-\Lambda m_e+\Theta=\rho^g_e(\alpha)+\Theta.
\]
So the upper level sets of $\rho(y)$ are those of $\rho^g(\alpha)$, which are tight for $\xi^g(\alpha)$ by (a), hence for $y$ with respect to $f$.
By (a), applied to $f$, $y=\xi^f(\alpha+\Lambda)$, and its marginals $\rho^g(\alpha)+\Theta$ are nonnegative by assumption.
As $f$ is monotone, every $\tilde{y}\in\Bpm{f}$ is nonnegative, since $\tilde{y}_e\ge f(E)-f(E\setminus\{e\})\ge 0$, so $\Bpm{f}=B(f)$ and $y\in\Pp{f}$.
At the share $\alpha+\Lambda\ge 0$, the marginals $\rho(y)$ are nonnegative and every level set $A_{\theta'}(y)$ with $\theta'>0$ is tight by (a).
By \cref{prop:optchar}, $y=\xs(\alpha+\Lambda)$.
\end{proof}

\subsubsection{Translation extensions}\label{sec:B-translation}

The following lemma adds a new element $b$ whose marginal value at $S$ is a constant minus a modular function $d$ of a special form, a translation.
Up to a common additive constant, the values of $F^{g'}_{\alpha,\theta}$ on the sets that contain $b$ are then the values of $F^g$ at a fixed translate of $(\alpha,\theta)$ on the sets without $b$.

\begin{lemma}\label{lem:translation}
Let $g\colon 2^E\to\R$ be submodular with $g(\emptyset)=0$, let $m,\beta\in\R^E_{>0}$ and $t:=m/\beta$.
Let $(\mu_\alpha,\mu_\theta)\in\R^2$ be such that $d_e:=(\mu_\theta-\mu_\alpha m_e)/\beta_e\ge 0$ for all $e\in E$,
let $K\in\R$, and let $b\notin E$ be a new element with $m_b,\beta_b>0$ and $t_b:=m_b/\beta_b$.
Define $g'\colon 2^{E\cup\{b\}}\to\R$ by
\[
  g'(S):=g(S),\qquad g'(S\cup\{b\}):=g(S)+K-d(S)\qquad(S\subseteq E).
\]
Then the following hold, where $F^{g'}$ is taken with respect to $t$ and $\binv$ extended by $t_b$ and $1/\beta_b$.
\begin{enumerate}[label=(\alph*)]
  \item The function $g'$ is submodular with $g'(\emptyset)=0$.
  If $g$ is given by a cut form and $K$ and $d$ are rational, then $g'$ is given by a cut form (\cref{lem:cutforms}(c)(iii)).
  \item For all $\alpha,\theta\in\R$ and $T\subseteq E$,
  \begin{align*}
    F^{g'}_{\alpha,\theta}(T)&=F^g_{\alpha,\theta}(T),\\
    F^{g'}_{\alpha,\theta}(T\cup\{b\})&=F^g_{\alpha-\mu_\alpha,\theta-\mu_\theta}(T)+K-\alpha t_b+\theta/\beta_b.
  \end{align*}
  \item $F^{g-d}_{\alpha,\theta}=F^g_{\alpha-\mu_\alpha,\theta-\mu_\theta}$ on $2^E$ for all $\alpha,\theta\in\R$, and for every $\alpha\in\R$,
  \[
    \xi^{g-d}(\alpha)=\xi^g(\alpha-\mu_\alpha)-d\qquad\text{and}\qquad\rho^{g-d}(\alpha)=\rho^g(\alpha-\mu_\alpha)+\mu_\theta,
  \]
  with the full-plane best responses and marginals of $g-d$ and $g$ taken with respect to $m$ and $\beta$.
  Moreover, $A^{g-d}_\theta(\alpha)=A^g_{\theta-\mu_\theta}(\alpha-\mu_\alpha)$ for all $\alpha,\theta\in\R$.
\end{enumerate}
Parts (b) and (c) do not use $d\ge 0$. In particular, (c) holds for every $(\mu_\alpha,\mu_\theta)\in\R^2$.
\end{lemma}

\begin{proof}
(a) We show that the marginal values of $g'$ do not increase as the set grows.
By the definition of $g'$, every $X\subseteq E$ and $e\in E\setminus X$ satisfy
\begin{align*}
  g'(X\cup\{e\})-g'(X)&=g(X\cup\{e\})-g(X),\\
  g'(X\cup\{b,e\})-g'(X\cup\{b\})&=g(X\cup\{e\})-g(X)-d_e,\\
  g'(X\cup\{b\})-g'(X)&=K-d(X).
\end{align*}
Let $X\subseteq Y\subseteq E\cup\{b\}$ and $e\in(E\cup\{b\})\setminus Y$.
If $e=b$, the marginal values of $b$ at $X$ and at~$Y$ are $K-d(X)\ge K-d(Y)$, as $d\ge 0$.
If $e\in E$, let $X':=X\setminus\{b\}$ and $Y':=Y\setminus\{b\}$.
The marginal value of $e$ at $X$ is $g(X'\cup\{e\})-g(X')$, minus $d_e$ if $b\in X$, and its marginal value at $Y$ is $g(Y'\cup\{e\})-g(Y')$, minus $d_e$ if $b\in Y$.
As $g$ is submodular, $X'\subseteq Y'$, $d_e\ge 0$, and $b\in X$ implies $b\in Y$, the marginal value at $X$ is at least the marginal value at $Y$.
As $g'(\emptyset)=g(\emptyset)=0$, this proves that $g'$ is submodular with $g'(\emptyset)=0$.
This argument uses only that $g$ is submodular with $g(\emptyset)=0$ and that $d\ge 0$, not the form of $d$.
The statement on cut forms is \cref{lem:cutforms}(c)(iii): add the node $b$ with weight $K-d(E)$ and the arcs $b\to e$ of capacity $d_e$ for $e\in E$.

(b) The first identity holds as $g'=g$ on $2^E$.
For the second, note first that, for every $T\subseteq E$,
$d(T)=\mu_\theta\,\binv(T)-\mu_\alpha\,t(T)$, so
\[
  F^g_{\alpha,\theta}(T)-d(T)=g(T)-(\alpha-\mu_\alpha)t(T)+(\theta-\mu_\theta)\binv(T)=F^g_{\alpha-\mu_\alpha,\theta-\mu_\theta}(T).
\]
Hence
\begin{align*}
  F^{g'}_{\alpha,\theta}(T\cup\{b\})&=g(T)+K-d(T)-\alpha\bigl(t(T)+t_b\bigr)+\theta\bigl(\binv(T)+1/\beta_b\bigr)\\
  &=F^g_{\alpha,\theta}(T)-d(T)+K-\alpha t_b+\theta/\beta_b\\
  &=F^g_{\alpha-\mu_\alpha,\theta-\mu_\theta}(T)+K-\alpha t_b+\theta/\beta_b.
\end{align*}

(c) The identity $F^{g-d}_{\alpha,\theta}=F^g_{\alpha,\theta}-d=F^g_{\alpha-\mu_\alpha,\theta-\mu_\theta}$ on $2^E$ is the computation in (b).
The function $g-d$ is submodular with $(g-d)(\emptyset)=0$, as $d$ is modular.
The vector $y:=\xi^g(\alpha-\mu_\alpha)-d$ satisfies $y(S)\le g(S)-d(S)$ for every $S\subseteq E$, with equality exactly for the sets that are tight for $\xi^g(\alpha-\mu_\alpha)$ with respect to $g$, and as $\beta_ed_e=\mu_\theta-\mu_\alpha m_e$, its marginals at the share $\alpha$ are
\begin{align*}
  \rho_e(y)&=\alpha m_e-\beta_e\xi^g_e(\alpha-\mu_\alpha)+\beta_ed_e\\
  &=(\alpha-\mu_\alpha)m_e-\beta_e\xi^g_e(\alpha-\mu_\alpha)+\mu_\theta=\rho^g_e(\alpha-\mu_\alpha)+\mu_\theta.
\end{align*}
So its upper level sets are those of $\rho^g(\alpha-\mu_\alpha)$, which are tight for $\xi^g(\alpha-\mu_\alpha)$ with respect to $g$ by \cref{lem:fullplane}(a), hence tight for $y$ with respect to $g-d$.
By \cref{lem:fullplane}(a), $y=\xi^{g-d}(\alpha)$, and $\rho^{g-d}(\alpha)=\rho(y)=\rho^g(\alpha-\mu_\alpha)+\mu_\theta$.
Finally, the identity $F^{g-d}_{\alpha,\theta}=F^g_{\alpha-\mu_\alpha,\theta-\mu_\theta}$ and \cref{lem:fullplane}(b), applied to $g-d$ and to $g$, give $A^{g-d}_\theta(\alpha)=A^g_{\theta-\mu_\theta}(\alpha-\mu_\alpha)$.
None of these steps uses $d\ge 0$.
\end{proof}

\subsubsection{One-element extensions and the splice}\label{sec:B-extensions}

In the comb gadget, combs are built by adding, again and again, one element whose marginal value is a constant minus a nonnegative modular function.
This subsection describes exactly what such an extension does to the full-plane best response.
The marginal of an old element for the old function is at most its marginal for the corrected function, the old function minus the modular correction, and its new marginal is the median of these two and of the marginal of the new element (\cref{thm:splice}).
So the old elements above and below the new element keep one of their two marginals, and the new element is merged with the old elements in between.
We then compute the reward of the extension in windows of shares on which this pattern is fixed (\cref{lem:windows}), and in \cref{sec:B-dmt} we specialize to the extension that builds combs, the doubly merged translation (\cref{lem:dmt}).

\paragraph{Setting.}
Let $f\colon 2^E\to\R$ be submodular with $f(\emptyset)=0$, not necessarily monotone, let $m,\beta\in\R^E_{>0}$, let $d\in\R^E_{\ge 0}$, which we also view as the modular function $S\mapsto d(S)$, let $h:=f-d$ and $K\in\R$, and let $b\notin E$ be a new element with $m_b,\beta_b>0$ and $t_b:=m_b/\beta_b$.
The \emph{extension} $f'\colon 2^{E\cup\{b\}}\to\R$ of $f$ by $b$ is
\[
  f'(S):=f(S),\qquad f'(S\cup\{b\}):=K+h(S)=f(S)+K-d(S)\qquad(S\subseteq E).
\]
The marginal value $f'(S\cup\{b\})-f'(S)$ of $b$ at $S\subseteq E$ is $K-d(S)$.
This is a marginal value of the set function $f'$, as in \cref{lem:translation}, and not a marginal $\rho_e$ of the agent.
If $d=\mu_\theta\binv-\mu_\alpha t$ for some $(\mu_\alpha,\mu_\theta)\in\R^2$, then $f'$ is the translation extension $g'$ of \cref{lem:translation} with $g:=f$, but everything in this subsection holds for an arbitrary $d\in\R^E_{\ge 0}$.
The proof of \cref{lem:translation}(a) uses only that $f$ is submodular with $f(\emptyset)=0$ and that $d\ge 0$, not the form of $d$, so it shows that $f'$ is submodular with $f'(\emptyset)=0$.
This can also be checked directly: among the inequalities $f'(X\cup\{u\})+f'(X\cup\{v\})\ge f'(X\cup\{u,v\})+f'(X)$ for $X\subseteq E\cup\{b\}$ and distinct $u,v\notin X$, those with $u,v\in E$ are inequalities of $f$ (if $b\notin X$) or of $h$ (if $b\in X$), and those with $u\in E$ and $v=b$ read $d_u\ge 0$.
We write $\rho':=\rho^{f'}$ and $\xi':=\xi^{f'}$, taken with respect to $m$ and $\beta$ extended by $m_b$ and $\beta_b$, and $\theta_b:=\rho'_b$ for the marginal of the new element.

\begin{theorem}\label{thm:splice}
Let $f\colon 2^E\to\R$ be submodular with $f(\emptyset)=0$, let $m,\beta\in\R^E_{>0}$, $d\in\R^E_{\ge 0}$, and $K\in\R$, let $b\notin E$ with $m_b,\beta_b>0$, let $h:=f-d$, and let $f'$ be the extension of $f$ by $b$ as in the setting above.
Fix $\alpha\in\R$, and let $\rho^f$, $\rho^h$, $\rho'$, and $\theta_b$ denote the marginals at $\alpha$.
\begin{enumerate}[label=(\alph*)]
  \item $\rho^f_e\le\rho^h_e$ for every $e\in E$, and $\sum_{e\in E}(\rho^h_e-\rho^f_e)/\beta_e=d(E)$.
  \item $\rho'_e=\med(\rho^f_e,\theta_b,\rho^h_e)$, the median of the three numbers, for every $e\in E$.
  \item $\theta_b$ is the unique solution $\theta$ of
  \begin{equation}\label{eq:splice}
    \frac{\alpha m_b-\theta}{\beta_b}=K-d(E)+Y(\theta),
  \end{equation}
  where
  \[
    Y(\theta):=\sum_{e\in E}\frac{\clamp\bigl(\theta-\rho^f_e,\,0,\,\rho^h_e-\rho^f_e\bigr)}{\beta_e}\in[0,d(E)]
  \]
  and $\clamp(v,a,a'):=\min(\max(v,a),a')$ for $a\le a'$.
\end{enumerate}
\end{theorem}

\begin{proof}
We first prove property (a).
Fix $\theta\in\R$, and let $A:=A^f_\theta(\alpha)$ and $A':=A^h_\theta(\alpha)$.
As $F^h_{\alpha,\theta}=F^f_{\alpha,\theta}-d$ and $d\ge 0$, the submodularity of $F^f_{\alpha,\theta}$ and the minimality of $A$ (\cref{lem:fullplane}(b)) give
\begin{align*}
  F^h_{\alpha,\theta}(A\cup A')-F^h_{\alpha,\theta}(A')&=F^f_{\alpha,\theta}(A\cup A')-F^f_{\alpha,\theta}(A')-d(A\setminus A')\\
  &\le F^f_{\alpha,\theta}(A\cup A')-F^f_{\alpha,\theta}(A')\\
  &\le F^f_{\alpha,\theta}(A)-F^f_{\alpha,\theta}(A\cap A')\le 0.
\end{align*}
So $A\cup A'$ minimizes $F^h_{\alpha,\theta}$, as $A'$ does, and $A\subseteq A'$ by \cref{lem:fullplane}(b).
As this holds for every $\theta$, $\rho^f\le\rho^h$: an element $e$ with $\rho^f_e>\rho^h_e$ would lie in $A$ but not in $A'$ for $\theta:=\rho^f_e$.
The set $E$ is tight for $\xi^f(\alpha)$ and for $\xi^h(\alpha)$, and $\rho^h_e-\rho^f_e=\beta_e(\xi^f_e-\xi^h_e)$, so
\[
  \sum_{e\in E}\frac{\rho^h_e-\rho^f_e}{\beta_e}=\sum_{e\in E}(\xi^f_e-\xi^h_e)=f(E)-h(E)=d(E).
\]
This proves (a).

Now we prove properties (b) and (c).
The difference of the two sides of \eqref{eq:splice} is continuous in $\theta$, as $Y$ is a sum of clamps of affine functions, and strictly decreasing, as $Y$ is nondecreasing.
As $0\le Y\le d(E)$ by (a), the difference is nonnegative at $\theta=\alpha m_b-\beta_bK$ and nonpositive at $\theta=\alpha m_b-\beta_b(K-d(E))$.
So, by the intermediate value theorem, \eqref{eq:splice} has a solution $\theta$, and it is unique, as the difference is strictly decreasing.
For $e\in E$, write $c_e:=\clamp(\theta-\rho^f_e,0,\rho^h_e-\rho^f_e)$, which is well defined by (a).
Define $y\in\R^{E\cup\{b\}}$ by $y_b:=(\alpha m_b-\theta)/\beta_b$ and $y_e:=\xi^f_e-c_e/\beta_e$ for $e\in E$.
At the share $\alpha$, its marginals \eqref{eq:marg} are $\rho_b(y)=\theta$ and $\rho_e(y)=\rho^f_e+c_e=\med(\rho^f_e,\theta,\rho^h_e)$ for $e\in E$.
Let $\pi_e:=(\rho^h_e-\rho^f_e)/\beta_e=\xi^f_e-\xi^h_e$ for $e\in E$.
Then $\pi_e\ge c_e/\beta_e\ge 0$, and $\sum_{e\in E}\pi_e=d(E)$ by (a).
By \eqref{eq:splice}, $y_b=K-d(E)+\sum_{e\in E}c_e/\beta_e$.
With $\xi^f(S)=\xi^h(S)+\sum_{e\in S}\pi_e$, every $S\subseteq E$ therefore satisfies
\begin{equation}\label{eq:B-feas}
\begin{aligned}
  y(S)&\le\xi^f(S)\le f(S),\\
  y(S)+y_b&=K+\xi^h(S)-\sum_{e\in E\setminus S}\Bigl(\pi_e-\frac{c_e}{\beta_e}\Bigr)\le K+h(S),
\end{aligned}
\end{equation}
so $y(S')\le f'(S')$ for every $S'\subseteq E\cup\{b\}$.

We now show that every upper level set $\{e\in E\cup\{b\}\colon\rho_e(y)\ge\theta'\}$, $\theta'\in\R$, is tight for $y$ with respect to $f'$, using $\rho^f\le\rho^h$ throughout.
Let $\theta'>\theta$.
For $e\in E$, the median $\med(\rho^f_e,\theta,\rho^h_e)$ exceeds $\theta$ exactly if $\rho^f_e>\theta$, and then it equals $\rho^f_e$.
As $\rho_b(y)=\theta<\theta'$, the level set is therefore $\{e\in E\colon\rho^f_e\ge\theta'\}$, on which $c=0$, that is, $y=\xi^f$.
This set is tight for $\xi^f(\alpha)$ by \cref{lem:fullplane}(a), hence for $y$, as $f'=f$ on subsets of $E$.
Now let $\theta'\le\theta$.
For $e\in E$, the median $\med(\rho^f_e,\theta,\rho^h_e)=\min(\max(\rho^f_e,\theta),\rho^h_e)$ is at least $\theta'$ exactly if $\rho^h_e\ge\theta'$, and $\rho_b(y)=\theta\ge\theta'$.
So the level set is $A\cup\{b\}$ with $A:=\{e\in E\colon\rho^h_e\ge\theta'\}$.
Every $e\in E\setminus A$ has $\rho^h_e<\theta'\le\theta$, so $c_e=\rho^h_e-\rho^f_e$, that is, $c_e/\beta_e=\pi_e$.
So the sum in \eqref{eq:B-feas} vanishes for $S=A$, and as $A$ is tight for $\xi^h(\alpha)$ by \cref{lem:fullplane}(a),
\[
  y(A)+y_b=K+\xi^h(A)=K+h(A)=f'(A\cup\{b\}).
\]
By \cref{lem:fullplane}(a), $y=\xi'(\alpha)$.
This proves (b), and (c) with $\theta_b=\rho_b(y)=\theta$.
\end{proof}

\paragraph{Configurations.}
We call the pair $(U,W)$ with $U:=\{e\in E\colon\rho^f_e(\alpha)>\theta_b(\alpha)\}$ and $W:=\{e\in E\colon\rho^h_e(\alpha)\ge\theta_b(\alpha)\}$ the \emph{configuration} at $\alpha$.
By \cref{thm:splice}, $U\subseteq W$, and $\rho'=\rho^f>\theta_b$ on $U$, $\rho'=\theta_b$ on $W\setminus U$, and $\rho'=\rho^h<\theta_b$ on $E\setminus W$.
So the block of $b$, the set of elements of $E\cup\{b\}$ with the marginal $\theta_b$, is $D:=(W\setminus U)\cup\{b\}$: the new element is merged with the old elements whose two marginals $\rho^f_e$ and $\rho^h_e$ enclose $\theta_b$.
For a function $\phi$ of the share, we write $[\phi]_a^{a'}:=\phi(a')-\phi(a)$.

\begin{lemma}\label{lem:windows}
In the setting of \cref{thm:splice}, fix $\alpha\in\R$, let $U\subseteq W\subseteq E$ and $D:=(W\setminus U)\cup\{b\}$, and let
\[
  X_{U,W}:=K+h(W)-f(U)\qquad\text{and}\qquad\theta_{U,W}:=\frac{\alpha t(D)-X_{U,W}}{\binv(D)}.
\]
We call $(U,W)$ a \emph{weak configuration} at $\alpha$ if, at $\alpha$, $\rho^f_e\ge\theta_b$ for $e\in U$, $\rho^f_e\le\theta_b\le\rho^h_e$ for $e\in W\setminus U$, and $\rho^h_e\le\theta_b$ for $e\in E\setminus W$, and if $U$ is tight for $\xi^f(\alpha)$ and $W$ is tight for $\xi^h(\alpha)$.
The last two conditions hold if $U$ is an upper level set of $\rho^f(\alpha)$ and $W$ one of $\rho^h(\alpha)$, strict or not (\cref{lem:fullplane}(a)).
In particular, the configuration at $\alpha$ is a weak configuration.
At a weak configuration, $X_{U,W}=\xi'(D)$ is the mass of $D$ and $\theta_{U,W}=\theta_b$ is the common marginal of its elements, by (a), and for the configuration, $D$ is the block of $b$.
\begin{enumerate}[label=(\alph*)]
  \item If $(U,W)$ is a weak configuration at $\alpha$, then $\theta_b(\alpha)=\theta_{U,W}$, $\xi'(\alpha)=\xi^f(\alpha)$ on $U$, $\xi'(\alpha)=\xi^h(\alpha)$ on $E\setminus W$, and
  \begin{equation}\label{eq:windows}
    r(\xi^{f'}(\alpha))=\sum_{e\in U}m_e\xi^f_e(\alpha)+\sum_{e\in E\setminus W}m_e\xi^h_e(\alpha)+\alpha\,\SSq(D)+X_{U,W}\,\bar m_D.
  \end{equation}
  \item Conversely, if $\rho^f_e(\alpha)>\theta_{U,W}$ for $e\in U$, $\rho^f_e(\alpha)\le\theta_{U,W}$ for $e\in E\setminus U$, $\rho^h_e(\alpha)\ge\theta_{U,W}$ for $e\in W$, and $\rho^h_e(\alpha)<\theta_{U,W}$ for $e\in E\setminus W$, then $(U,W)$ is the configuration at $\alpha$.
  \item If $(U,W)$ is a weak configuration at $a$ and at $a'\ge a$, then
  \[
    r(\xi^{f'}(a'))-r(\xi^{f'}(a))=\Bigl[\sum_{e\in U}m_e\xi^f_e\Bigr]_a^{a'}+\Bigl[\sum_{e\in E\setminus W}m_e\xi^h_e\Bigr]_a^{a'}+(a'-a)\,\SSq(D),
  \]
  where the first bracket lies in $[0,r(\xi^f(a'))-r(\xi^f(a))]$ and the second in $[0,r(\xi^h(a'))-r(\xi^h(a))]$.
\end{enumerate}
\end{lemma}

\begin{proof}
(a) By \cref{thm:splice}(b), $\rho'_e=\med(\rho^f_e,\theta_b,\rho^h_e)$.
By the inequalities of a weak configuration and as $\rho^f\le\rho^h$, this is $\rho^f_e$ for $e\in U$, $\theta_b$ for $e\in W\setminus U$, and $\rho^h_e$ for $e\in E\setminus W$.
As $\xi_e=\alpha t_e-\rho_e/\beta_e$ for each of the three best responses $\xi'$, $\xi^f$, $\xi^h$ and their marginals, $\xi'=\xi^f$ on $U$ and $\xi'=\xi^h$ on $E\setminus W$.
The set $E\cup\{b\}$ is tight for $\xi'$, $U$ is tight for $\xi^f$, and $W$ and $E$ are tight for $\xi^h$, so
\begin{align*}
  \xi'(D)&=f'(E\cup\{b\})-\xi^f(U)-\xi^h(E\setminus W)\\
  &=K+h(E)-f(U)-h(E)+h(W)=X_{U,W}.
\end{align*}
All elements of $D$ have the marginal $\theta_b$, so $\xi'(D)=\alpha t(D)-\theta_b\binv(D)$, which gives $\theta_b=\theta_{U,W}$.
By \cref{lem:fullplane}(e), $\sum_{e\in D}m_e\xi'_e=\alpha\,\SSq(D)+X_{U,W}\,\bar m_D$, and adding the sums of $m_e\xi'_e$ over $U$ and over $E\setminus W$ gives \eqref{eq:windows}.

(b) We show that $\theta:=\theta_{U,W}$ solves \eqref{eq:splice}.
Then $\theta_b=\theta$ by the uniqueness in \cref{thm:splice}(c), and the hypotheses say that $U=\{e\in E\colon\rho^f_e>\theta_b\}$ and $W=\{e\in E\colon\rho^h_e\ge\theta_b\}$.
By the hypotheses, the clamp of $e$ in $Y(\theta)$ is $0$ for $e\in U$, $\theta-\rho^f_e$ for $e\in W\setminus U$, and $\rho^h_e-\rho^f_e$ for $e\in E\setminus W$.
With $\rho^f_e/\beta_e=\alpha t_e-\xi^f_e$ and $\rho^h_e/\beta_e=\alpha t_e-\xi^h_e$, and as $(W\setminus U)\cup(E\setminus W)=E\setminus U$, this gives
\[
  Y(\theta)=\theta\,\binv(W\setminus U)-\alpha\,t(W\setminus U)+\xi^f(E\setminus U)-\xi^h(E\setminus W).
\]
The sets $E$, $U=\{e\in E\colon\rho^f_e>\theta\}$, and $W=\{e\in E\colon\rho^h_e\ge\theta\}$ are upper level sets, so $\xi^f(E\setminus U)=f(E)-f(U)$ and $\xi^h(E\setminus W)=h(E)-h(W)$ by \cref{lem:fullplane}(a), and $f(E)-h(E)=d(E)$.
Hence $K-d(E)+Y(\theta)=X_{U,W}+\theta\,\binv(W\setminus U)-\alpha\,t(W\setminus U)$.
The left-hand side of \eqref{eq:splice} is $\alpha t_b-\theta/\beta_b$, so \eqref{eq:splice} becomes $\alpha t(D)-\theta\,\binv(D)=X_{U,W}$, which is the definition of $\theta_{U,W}$.

(c) Subtract \eqref{eq:windows} at $a$ from \eqref{eq:windows} at $a'$.
The numbers $X_{U,W}$ and $\bar m_D$ do not depend on the share, so the last two terms give $(a'-a)\,\SSq(D)$.
By the definition of a weak configuration at $a$ and at $a'$, the set $U$ is tight for $\xi^f(a)$ and for $\xi^f(a')$.
So, by \cref{lem:fullplane}(d), the sums $\sum_{e\in U}m_e\xi^f_e$ and $\sum_{e\in E\setminus U}m_e\xi^f_e$ are both at least as large at $a'$ as at $a$.
Their increments add up to $r(\xi^f(a'))-r(\xi^f(a))$, so the first bracket lies in $[0,r(\xi^f(a'))-r(\xi^f(a))]$.
Likewise, the set $W$ is tight for $\xi^h(a)$ and for $\xi^h(a')$, so, by \cref{lem:fullplane}(d), the sums $\sum_{e\in W}m_e\xi^h_e$ and $\sum_{e\in E\setminus W}m_e\xi^h_e$ are both at least as large at $a'$ as at $a$.
The second bracket is the increment of the sum over the complement $E\setminus W$, so it lies in $[0,r(\xi^h(a'))-r(\xi^h(a))]$.
\end{proof}

\subsubsection{The doubly merged translation}\label{sec:B-dmt}

We now specialize to the extension that builds combs.
Two modular singletons on new elements $\tau_1$ and $\tau_2$ are added to the old function, and the correction $d$ is a translation (\cref{lem:translation}).
In one window, the new element is merged at the bottom with $\tau_1$, and in a window translated by $\mu$, it is merged at the top with $\tau_2$.
A suitable difference $F_{\tau_1}-F_{\tau_2}$ then puts the reward of the old best response and its translate under one line.

\paragraph{Setting.}
Let $E=C\cup\{\tau_1,\tau_2\}$ and $f:=f_C\oplus\pi_{F_{\tau_1}}\oplus\pi_{F_{\tau_2}}$, where $f_C\colon 2^C\to\R$ is submodular with $f_C(\emptyset)=0$, and $\pi_{F_{\tau_1}}$ and $\pi_{F_{\tau_2}}$ are the modular singletons on $\{\tau_1\}$ and $\{\tau_2\}$ with $\pi_{F_{\tau_1}}(\{\tau_1\})=F_{\tau_1}$ and $\pi_{F_{\tau_2}}(\{\tau_2\})=F_{\tau_2}$, with $m_{\tau_1}=m_{\tau_2}=m_\delta$ and $\beta_{\tau_1}=\beta_{\tau_2}=\beta_\delta$, and with $m|_C,\beta|_C\in\R^C_{>0}$.
For the reward of the full-plane best response of $f_C$, we write $r(\xi^{f_C}(\alpha)):=\sum_{e\in C}m_e\xi^{f_C}_e(\alpha)$.
Let $\mu>0$ and $m_*\ge\max_{e\in E}m_e$, let $d_e:=\mu(m_*-m_e)/\beta_e\ge 0$ for $e\in E$, that is, $d(S)=\mu\bigl(m_*\binv(S)-t(S)\bigr)$ for $S\subseteq E$, and let $w:=\mu(m_*-m_\delta)$, so that $d_{\tau_1}=d_{\tau_2}=w/\beta_\delta$.
So $d$ is the vector $d$ of \cref{lem:translation} with $(\mu_\alpha,\mu_\theta)=(\mu,m_*\mu)$.
The condition $m_*\ge\max_{e\in E}m_e$ makes $d$ nonnegative, so $f'$ is the translation extension of \cref{lem:translation} with $g:=f$, and an extension as in \cref{sec:B-extensions}.
The new element $b$ has $m_b>m_\delta$ and $\beta_b:=\beta_\delta/12$.
Put
\[
  \bar m:=\frac{12m_b+m_\delta}{13},\qquad s:=\bar m-m_\delta,\qquad\psi:=\frac{12(m_b-m_\delta)^2}{13\beta_\delta},
\]
and $\Delta F:=F_{\tau_1}-F_{\tau_2}$.
Then $\bar m$ is the weighted mean reward $\bar m_{\{b,\tau_1\}}=\bar m_{\{b,\tau_2\}}$ of $\{b,\tau_1\}$ and of $\{b,\tau_2\}$, as $1/\beta_b=12/\beta_\delta$.
Moreover, $\psi=\SSq(\{b,\tau_1\})=\SSq(\{b,\tau_2\})$, as $\SSq(\{b,\tau_1\})=(m_b-m_\delta)^2/(\beta_b+\beta_\delta)$ by \eqref{eq:B-pair}.

\begin{lemma}\label{lem:dmt}
In the setting of this subsection, let $\Delta_p:=m_\delta F_{\tau_2}+(K-d(E)+F_{\tau_1})\bar m$.
\begin{enumerate}[label=(\alph*)]
  \item If $(E\setminus\{\tau_1\},E)$ is a weak configuration at $\alpha$, then
  \[
    r(\xi^{f'}(\alpha))=r(\xi^{f_C}(\alpha))+\psi\alpha+\Delta_p.
  \]
  \item If $(\emptyset,\{\tau_2\})$ is a weak configuration at $\alpha$, then
  \[
    r(\xi^{f'}(\alpha))=r(\xi^{f_C}(\alpha-\mu))+\psi\alpha+\Delta_p+\sum_{e\in C}(\bar m-m_e)d_e+s\Bigl(\frac{w}{\beta_\delta}-\Delta F\Bigr).
  \]
  \item Let $L_C(\alpha)=p_C+q_C\alpha$, and suppose that
  \[
    \Delta F=\frac{w}{\beta_\delta}+d(C)-\frac{q_C\mu-\sum_{e\in C}(m_\delta-m_e)d_e}{s}.
  \]
  Then $L(\alpha):=L_C(\alpha)+\psi\alpha+\Delta_p$ satisfies $L(\alpha)-r(\xi^{f'}(\alpha))=L_C(\alpha)-r(\xi^{f_C}(\alpha))$ in case (a) and $L(\alpha)-r(\xi^{f'}(\alpha))=L_C(\alpha-\mu)-r(\xi^{f_C}(\alpha-\mu))$ in case (b).
\end{enumerate}
\end{lemma}

\begin{proof}
(a) Apply \cref{lem:windows}(a) with $U=E\setminus\{\tau_1\}$ and $W=E$.
Then $D=\{b,\tau_1\}$, and $X_{U,W}=K+h(E)-f(E\setminus\{\tau_1\})=K-d(E)+F_{\tau_1}$, as $f(E)-f(E\setminus\{\tau_1\})=F_{\tau_1}$.
By \cref{lem:fullplane}(f), $\xi^f(\alpha)=(\xi^{f_C}(\alpha),F_{\tau_1},F_{\tau_2})$, so the first sum in \eqref{eq:windows} is $r(\xi^{f_C}(\alpha))+m_\delta F_{\tau_2}$, while the second is empty.
With $\SSq(\{b,\tau_1\})=\psi$ and $\bar m_{\{b,\tau_1\}}=\bar m$, \eqref{eq:windows} becomes $r(\xi^{f'}(\alpha))=r(\xi^{f_C}(\alpha))+m_\delta F_{\tau_2}+\psi\alpha+(K-d(E)+F_{\tau_1})\bar m$, which is the claim.

(b) Apply \cref{lem:windows}(a) with $U=\emptyset$ and $W=\{\tau_2\}$.
Then $D=\{b,\tau_2\}$, and $X_{U,W}=K+h(\{\tau_2\})=K+F_{\tau_2}-w/\beta_\delta$, as $d_{\tau_2}=w/\beta_\delta$.
The function $h=f-d$ is the direct sum of $f_C-d|_C$ and the two modular singletons with the values $F_{\tau_1}-w/\beta_\delta$ and $F_{\tau_2}-w/\beta_\delta$, and $d_e=\mu(m_*-m_e)/\beta_e$ for $e\in C$.
So, by \cref{lem:fullplane}(f) and \cref{lem:translation}(c) (applied to $f_C$ with $(\mu_\alpha,\mu_\theta)=(\mu,m_*\mu)$), $\xi^h_e(\alpha)=\xi^{f_C}_e(\alpha-\mu)-d_e$ for $e\in C$, hence
\[
  \sum_{e\in C}m_e\xi^h_e(\alpha)=r(\xi^{f_C}(\alpha-\mu))-\sum_{e\in C}m_ed_e,
\]
and $\xi^h_{\tau_1}(\alpha)=F_{\tau_1}-w/\beta_\delta$.
The first sum in \eqref{eq:windows} is empty, and the second, over $E\setminus W=C\cup\{\tau_1\}$, is $r(\xi^{f_C}(\alpha-\mu))-\sum_{e\in C}m_ed_e+m_\delta(F_{\tau_1}-w/\beta_\delta)$.
We insert this into \eqref{eq:windows}, with $\SSq(\{b,\tau_2\})=\psi$ and $\bar m_{\{b,\tau_2\}}=\bar m$, and subtract $r(\xi^{f_C}(\alpha-\mu))+\psi\alpha+\Delta_p$.
Using $d(E)=d(C)+2w/\beta_\delta$, the remaining constant is
\begin{align*}
  &-\sum_{e\in C}m_ed_e+m_\delta\Bigl(F_{\tau_1}-\frac{w}{\beta_\delta}\Bigr)+\Bigl(K+F_{\tau_2}-\frac{w}{\beta_\delta}\Bigr)\bar m\\
  &\qquad-m_\delta F_{\tau_2}-\Bigl(K-d(C)-\frac{2w}{\beta_\delta}+F_{\tau_1}\Bigr)\bar m\\
  &=-\sum_{e\in C}m_ed_e+\bar m\,d(C)-(\bar m-m_\delta)\Delta F+(\bar m-m_\delta)\frac{w}{\beta_\delta}\\
  &=\sum_{e\in C}(\bar m-m_e)d_e+s\Bigl(\frac{w}{\beta_\delta}-\Delta F\Bigr).
\end{align*}

(c) With the given $\Delta F$, we have $s(w/\beta_\delta-\Delta F)=-s\,d(C)+q_C\mu-\sum_{e\in C}(m_\delta-m_e)d_e$.
As $\bar m=m_\delta+s$, also $\sum_{e\in C}(\bar m-m_e)d_e=s\,d(C)+\sum_{e\in C}(m_\delta-m_e)d_e$.
So the constant of (b) is $q_C\mu$.
In case (b), therefore,
\[
  L(\alpha)-r(\xi^{f'}(\alpha))=L_C(\alpha)-q_C\mu-r(\xi^{f_C}(\alpha-\mu))=L_C(\alpha-\mu)-r(\xi^{f_C}(\alpha-\mu)).
\]
In case (a), the claim follows from (a) directly, as the terms $\psi\alpha+\Delta_p$ cancel.
\end{proof}

So if $\Delta F$ is as in \cref{lem:dmt}(c) and $r(\xi^{f_C}(\gamma))=L_C(\gamma)$, then $r(\xi^{f'}(\alpha))=L(\alpha)$ both at $\alpha=\gamma$ and at $\alpha=\gamma+\mu$, provided the configurations are as in \cref{lem:dmt}(a) at $\gamma$ and as in \cref{lem:dmt}(b) at $\gamma+\mu$.
More generally, the gap $L_C(\alpha)-r(\xi^{f_C}(\alpha))$ is copied unchanged into both windows.
The lemma says nothing about where the two configurations hold, nor about the gap $L(\alpha)-r(\xi^{f'}(\alpha))$ between and outside the windows.
This is the work of the comb-gadget part of \cref{app:hard}.

\subsection{Polynomial-size instances with exponentially many breakpoints}\label{app:bits}\label{sec:C-polybits}

In this subsection we prove \cref{thm:polybits}. We use the instance of
\cref{lem:embed}, but instead of the networks of Allman, Lo, and McCormick~\cite{ALM22} we
construct the submodular function directly, adding the elements one at a time
(\cref{cor:witness}). When an element is added, every old set keeps a point at
which it is the unique minimizer, and every new set receives the point of an
old set translated by a fixed vector. The region containing these points
therefore grows only additively, and all numbers have $\Oh(n\log n)$ bits
(\cref{prop:witness}\ref{it:C-witness-i} and the end of the proof of \cref{thm:polybits}). The construction also gives networks of the kind for
which Allman, Lo, and McCormick pose their question~\cite[\S5]{ALM22} (the proof of part~(ii) of \cref{thm:polybits}).

For a finite set $E$, possibly empty, a vector $t\in\R^E_{>0}$, a function
$g\colon 2^E\to\R$, and $w=(\lambda,\theta)\in\R^2$, let
\[
  F^g_w(S) = F^g_{\lambda,\theta}(S) := g(S) - \lambda t(S) + \theta\,|S|
  \qquad\text{for } S\subseteq E,
\]
that is, the function $F^g_{\lambda,\theta}$ of \cref{sec:B-fullplane} with
$\beta=\mathbf 1$, and let $\min F^g_w := \min_{S\subseteq E} F^g_w(S)$.
For an instance with $\beta_e=1$ for all $e\in E$, the function
$F_{\alpha,\theta}$ of \cref{thm:levelsets} is $F^f_{\alpha,\theta}$. We call
$w$ a \emph{witness} of $S\subseteq E$ for $g$, with respect to $t$, if $S$ is
the unique minimizer of $F^g_w$.

The following corollary of \cref{lem:translation} adds a new element $b$ to
$E$. Its marginal value at $S\subseteq E$ is $K-d(S)$ for a constant $K$ and a
nonnegative modular function $d$, chosen such that, up to a common additive
constant, the values of $F$ at a point $w$ on the sets that contain $b$ are the
values at a fixed translate of $w$ on the sets without $b$ (see
\eqref{eq:C-extend} below).

\begin{corollary}\label{cor:witness}
  Let $E$ be a finite set, possibly empty, let $t\in\R^E_{>0}$, and let
  $g\colon 2^E\to\R$ be submodular with $g(\emptyset)=0$. Let
  $\mathcal F\subseteq 2^E$ be a family such that every $S\in\mathcal F$ has a
  witness $w_S\in[0,\bar\lambda]\times[0,\bar\theta]$ for $g$ with respect to
  $t$, where $\bar\lambda,\bar\theta\ge 0$. Let $b\notin E$ be a new element
  with $t_b>0$, and let $\hat w=(\hat\lambda,\hat\theta)\in\R^2$ be such that
  $d:=\hat\theta\mathbf 1-\hat\lambda t\ge 0$, where $\mathbf 1\in\R^E$ is
  the all-ones vector. Let $K\in\R$ satisfy
  \begin{equation}\label{eq:C-window}
    t_b\bar\lambda + d(E) < K < t_b\hat\lambda - \bar\theta - \hat\theta .
  \end{equation}
  Define $g'\colon 2^{E\cup\{b\}}\to\R$ by $g'(S):= g(S)$ and
  $g'(S\cup\{b\}):= g(S)+K-d(S)$ for $S\subseteq E$. Then $g'$ is
  submodular with $g'(\emptyset)=0$. Moreover, for every $S\in\mathcal F$, with
  respect to $g'$ and the vector $t$ extended by $t_b$, the point $w_S$ is a
  witness of $S$ and the point $w_S+\hat w$ is a witness of
  $S\cup\{b\}$.
\end{corollary}

\begin{proof}
  We apply \cref{lem:translation} to $g$ with $\beta:=\mathbf 1$,
  $m:= t$, $m_b:= t_b$, $\beta_b:= 1$, the constant $K$,
  and $(\mu_\alpha,\mu_\theta):=(\hat\lambda,\hat\theta)$. Then
  $\mu_\theta\binv - \mu_\alpha t = \hat\theta\mathbf 1-\hat\lambda t = d\ge 0$, and
  the function $g'$ of the lemma is the function $g'$ above. By
  \cref{lem:translation}(a), $g'$ is submodular with $g'(\emptyset)=0$.

  Next, we relate $F^{g'}$ to $F^g$. Let $w=(\lambda,\theta)\in\R^2$ and let
  $\ell(w):=\lambda t_b-\theta-K$. For every $T\subseteq E$, we have
  $F^{g'}_w(T)=F^g_w(T)$ by \cref{lem:translation}(b). By
  \cref{lem:translation}(c), $F^g_{w-\hat w}=F^{g-d}_w$ on $2^E$. Equivalently,
  as $d(T)=\hat\theta|T|-\hat\lambda t(T)$,
  \begin{equation}\label{eq:C-shift}
    F^g_{w-\hat w}(T) = F^g_w(T) - d(T).
  \end{equation}
  By \cref{lem:translation}(b) with $\beta_b=1$, or directly from the
  definition of $g'$ and \eqref{eq:C-shift},
  \begin{equation}\label{eq:C-extend}
    \begin{aligned}
      F^{g'}_w(T\cup\{b\}) &= g(T)+K-d(T)-\lambda\bigl(t(T)+t_b\bigr)+\theta\bigl(|T|+1\bigr)\\
      &= F^g_w(T)-d(T)-(\lambda t_b-\theta-K) = F^g_{w-\hat w}(T)-\ell(w).
    \end{aligned}
  \end{equation}
  As $0\le d(T)\le d(E)$ for all $T\subseteq E$, taking the minimum over $T$
  in \eqref{eq:C-shift} gives
  \begin{equation}\label{eq:C-minbounds}
    \min F^g_w - d(E) \le \min F^g_{w-\hat w} \le \min F^g_w .
  \end{equation}

  We now show that $w_S$ is a witness of $S$ for $g'$, for every
  $S\in\mathcal F$. Let $w:= w_S=(\lambda,\theta)$. As
  $\lambda\le\bar\lambda$, $\theta\ge 0$, and $t_b>0$, the left inequality in
  \eqref{eq:C-window} gives $\ell(w)\le t_b\bar\lambda-K<-d(E)$. Hence, by
  \eqref{eq:C-extend} and \eqref{eq:C-minbounds}, every $T\subseteq E$
  satisfies
  \[
    F^{g'}_w(T\cup\{b\}) = F^g_{w-\hat w}(T)-\ell(w) > \min F^g_{w-\hat w}+d(E)
    \ge \min F^g_w = F^g_w(S) = F^{g'}_w(S),
  \]
  and every $T\subseteq E$ with $T\ne S$ satisfies
  $F^{g'}_w(T)=F^g_w(T)>F^g_w(S)$. So $S$ is the unique minimizer of
  $F^{g'}_w$, as claimed.

  Finally, we show that $w':= w_S+\hat w$ is a witness of $S\cup\{b\}$ for
  $g'$, for every $S\in\mathcal F$. Write $w_S=(\lambda,\theta)$. As
  $\lambda\ge 0$ and $\theta\le\bar\theta$, the right inequality in
  \eqref{eq:C-window} gives
  \[
    \ell(w') = (\lambda+\hat\lambda)t_b-(\theta+\hat\theta)-K
    \ge t_b\hat\lambda-\bar\theta-\hat\theta-K > 0 .
  \]
  Hence, by \eqref{eq:C-minbounds} at the point $w'$ and by
  \eqref{eq:C-extend}, every $T\subseteq E$ satisfies
  \[
    F^{g'}_{w'}(T) = F^g_{w'}(T) \ge \min F^g_{w'} \ge \min F^g_{w_S}
    = F^g_{w_S}(S) > F^g_{w_S}(S)-\ell(w') = F^{g'}_{w'}(S\cup\{b\}),
  \]
  and every $T\subseteq E$ with $T\ne S$ satisfies
  \[
    F^{g'}_{w'}(T\cup\{b\}) = F^g_{w_S}(T)-\ell(w') > F^g_{w_S}(S)-\ell(w')
    = F^{g'}_{w'}(S\cup\{b\}).
  \]
  So $S\cup\{b\}$ is the unique minimizer of $F^{g'}_{w'}$.
\end{proof}

Starting from the empty ground set, we apply \cref{cor:witness} $n$ times. In
the $k$th application, the first coordinate $\lambda_k=2\bar\lambda_k+1$ of the
translation exceeds the first coordinates of all witnesses so far, and for
$k\ge 2$ its second coordinate $\theta_k$ is the least one for which $d\ge 0$.
Then $t_k$ is the least integer for which the window \eqref{eq:C-window}
contains an integer, and $K_k$ is the least integer in it.

\begin{proposition}\label{prop:witness}
  Let $n\ge 1$, and define integers recursively for $k=1,\dots,n$ by
  \begin{align*}
    \lambda_k &:= 3^{k-1}, &
    \bar\lambda_k &:= \sum_{j<k}\lambda_j = \frac{3^{k-1}-1}{2},\\
    \theta_1 &:= 0,\quad \theta_k:=\lambda_k t_{k-1}\ \text{for } k\ge 2, &
    \bar\theta_k &:= \sum_{j<k}\theta_j,\\
    d_i^{(k)} &:= \theta_k-\lambda_k t_i\ \text{for } i<k, &
    D_k &:= \sum_{i<k} d_i^{(k)},\\
    t_k &:= \biggl\lfloor\frac{D_k+1+\theta_k+\bar\theta_k}{\bar\lambda_k+1}\biggr\rfloor+1, &
    K_k &:= t_k\bar\lambda_k+D_k+1.
  \end{align*}
  Let $E:=\{1,\dots,n\}$, and let
  \[
    g(S) := \sum_{k\in S}K_k - \sum_{i,k\in S,\ i<k} d_i^{(k)}
    \qquad\text{for } S\subseteq E.
  \]
  Then the following hold.
  \begin{enumerate}[label=(\roman*)]
    \item\label{it:C-witness-i} $2=t_1<t_2<\dots<t_n\le 2(2n+2)^{n-1}$, and
      $d_i^{(k)}\ge 0$ for all $i<k$.
    \item\label{it:C-witness-ii} The function $g$ is submodular with
      $g(\emptyset)=0$, and every $S\subseteq E$ has the witness
      $(\lambda_S,\theta_S):=\sum_{k\in S}(\lambda_k,\theta_k)$ for $g$
      with respect to $t=(t_1,\dots,t_n)$.
  \end{enumerate}
\end{proposition}

\begin{proof}
  We first prove~\ref{it:C-witness-i}, by induction on $k$. For $k=1$, we have
  $D_1=\theta_1=\bar\theta_1=\bar\lambda_1=0$, so $t_1=\lfloor 1\rfloor+1=2$.
  Let $k\in\{2,\dots,n\}$ and assume that $t_1<\dots<t_{k-1}$. Then
  $d_i^{(k)}=\lambda_k(t_{k-1}-t_i)\ge 0$ for all $i<k$, so $D_k\ge 0$, and
  $\bar\theta_k\ge 0$ as all $\theta_j\ge 0$. As $\lfloor x\rfloor+1>x$ for all
  $x\in\R$, we obtain
  \[
    t_k > \frac{D_k+1+\theta_k+\bar\theta_k}{\bar\lambda_k+1}
    > \frac{\theta_k}{\bar\lambda_k+1}
    = \frac{2\cdot 3^{k-1}}{3^{k-1}+1}\,t_{k-1} \ge t_{k-1}.
  \]
  This shows that $t$ is strictly increasing and that $d_i^{(k)}\ge 0$ for all
  $i<k$. For the upper bound, let $k\in\{2,\dots,n\}$. Then
  $0\le d_i^{(k)}\le\theta_k=3^{k-1}t_{k-1}$ for all $i<k$, so
  $D_k\le(k-1)3^{k-1}t_{k-1}$, and as $t$ is increasing,
  \[
    \bar\theta_k = \sum_{j=2}^{k-1}3^{j-1}t_{j-1}
    \le t_{k-1}\sum_{j=2}^{k-1}3^{j-1} = \frac{3^{k-1}-3}{2}\,t_{k-1}.
  \]
  Adding these bounds and using $t_{k-1}\ge 2$, we get
  \[
    D_k+1+\theta_k+\bar\theta_k
    \le \Bigl(k+\frac12\Bigr)3^{k-1}t_{k-1}+1-\frac32\,t_{k-1}
    \le \Bigl(k+\frac12\Bigr)3^{k-1}t_{k-1}.
  \]
  As $\bar\lambda_k+1=(3^{k-1}+1)/2$ and $t_{k-1}\ge 1$, it follows that
  \[
    t_k \le \frac{(2k+1)\,3^{k-1}}{3^{k-1}+1}\,t_{k-1}+1
    < (2k+1)t_{k-1}+1 \le (2k+2)t_{k-1}.
  \]
  Hence $t_n\le t_1\prod_{k=2}^n(2k+2)\le 2(2n+2)^{n-1}$, which completes the
  proof of~\ref{it:C-witness-i}.

  We now prove~\ref{it:C-witness-ii}. For $k\in\{0,\dots,n\}$, let
  $E_k:=\{1,\dots,k\}$, let $g_k$ be the restriction of $g$ to $2^{E_k}$, and
  let $(\mathrm C_k)$ be the claim that $g_k$ is submodular with
  $g_k(\emptyset)=0$ and that every $S\subseteq E_k$ has the witness
  $(\lambda_S,\theta_S)$ for $g_k$, with respect to $(t_1,\dots,t_k)$. We prove
  $(\mathrm C_k)$ for all $k\in\{0,\dots,n\}$ by induction on $k$. For
  $k=0$, the only set is $\emptyset$, $g_0(\emptyset)=0$, and
  $(\lambda_\emptyset,\theta_\emptyset)=(0,0)$ is a witness of $\emptyset$. Let
  $k\in\{1,\dots,n\}$ and assume $(\mathrm C_{k-1})$. We apply \cref{cor:witness} to
  $E_{k-1}$, $g_{k-1}$, the family $\mathcal F:= 2^{E_{k-1}}$ with the
  witnesses $(\lambda_S,\theta_S)$, which lie in
  $[0,\bar\lambda_k]\times[0,\bar\theta_k]$ as $S\subseteq E_{k-1}$, the new
  element $b:= k$ with $t_b:= t_k$,
  $\hat w:=(\lambda_k,\theta_k)$, and $K:= K_k$. Its
  hypotheses hold. Indeed, $d=\theta_k\mathbf 1-\lambda_k t$ is the vector
  $(d_i^{(k)})_{i<k}$, which is nonnegative by~\ref{it:C-witness-i} and
  satisfies $d(E_{k-1})=D_k$. So the window \eqref{eq:C-window} reads
  $t_k\bar\lambda_k+D_k<K_k<t_k\lambda_k-\bar\theta_k-\theta_k$. Its left
  inequality holds as $K_k=t_k\bar\lambda_k+D_k+1$. As
  $\lambda_k=2\bar\lambda_k+1$, its right inequality is equivalent to
  $D_k+1+\theta_k+\bar\theta_k<t_k(\bar\lambda_k+1)$, which holds by the
  definition of $t_k$, as $\lfloor x\rfloor+1>x$. Moreover, the function $g'$
  of the corollary is $g_k$: for $S\subseteq E_{k-1}$, the pairs of elements of
  $S\cup\{k\}$ that contain $k$ are the pairs $\{i,k\}$ with $i\in S$, all with
  $i<k$, so $g(S\cup\{k\})=g(S)+K_k-\sum_{i\in S}d_i^{(k)}=g(S)+K_k-d(S)$. By
  the corollary, $g_k$ is submodular with $g_k(\emptyset)=0$, every
  $S\subseteq E_{k-1}$ has the witness $(\lambda_S,\theta_S)$ for $g_k$, and
  $S\cup\{k\}$ has the witness
  $(\lambda_S,\theta_S)+(\lambda_k,\theta_k)=(\lambda_{S\cup\{k\}},\theta_{S\cup\{k\}})$.
  This proves $(\mathrm C_k)$ for all $k\in\{0,\dots,n\}$. As $E_n=E$ and
  $g_n=g$, the claim $(\mathrm C_n)$ is~\ref{it:C-witness-ii}.
\end{proof}

Now that every subset of $E$ has a witness, the proof of \cref{thm:polybits}
follows the proof of \cref{thm:expbp}: we embed $g$ by \cref{lem:embed}, count
the breakpoints, and dilate the rewards so that all breakpoints lie in $(0,1)$.

\restate{\rsPolybits}{thm:polybits}
\begin{proof}
  Let $E=\{1,\dots,n\}$, let $t$, $g$, and the other numbers be as in
  \cref{prop:witness}, let $\Lambda:= 1+\sum_{k\in E}D_k$, and let
  \[
    f(S) := g(S)+\sum_{e\in S}(\Lambda t_e-1) \qquad\text{for } S\subseteq E.
  \]
  We first prove part~(i), that the instance with the rank function $f$, the
  rewards $m:= t/\sigma$ for the dilation factor
  $\sigma:= 1/(f(E)+1)$ and $\beta:=\mathbf 1$ has the claimed
  properties, and then part~(ii), for networks, the setting in which Allman,
  Lo, and McCormick pose their question. Until we pass to the rewards
  $t/\sigma$, we consider the instance $(f,t,\mathbf 1)$ with the rewards $t$
  instead.

  We first show that $f$ is a polymatroid rank function and that every
  $S\subseteq E$ is a level set. We apply \cref{lem:embed} with $\tilde f:= g$,
  $a:= t$, the all-ones vector $\mathbf 1$ as the vector $b$ of the
  lemma, and its parameters $(\Lambda,\Theta):=(\Lambda,1)$. The
  instance of the lemma then has the rewards $m_e=t_e$ and $\beta_e=1$ for all
  $e\in E$. To check the condition on $\Lambda$, let $e\in E$ and
  $S\subseteq E\setminus\{e\}$. By submodularity and by the definition of $g$,
  \[
    g(S\cup\{e\})-g(S) \ge g(E)-g(E\setminus\{e\}) = K_e-D_e-\sum_{k>e}d_e^{(k)}.
  \]
  Here $K_e-D_e=t_e\bar\lambda_e+1\ge 1$ and $0\le d_e^{(k)}\le D_k$ for $k>e$,
  so $g(S)-g(S\cup\{e\})+1\le\sum_{k>e}D_k<\Lambda\le\Lambda t_e$. As also
  $\Lambda\ge 0$, the condition on $\Lambda$ holds, and $f$ is a polymatroid
  rank function, with integer values as all numbers involved are integers. Now
  let $S\subseteq E$. By \cref{lem:embed} with the offsets
  $(\lambda,\theta):=(\lambda_S,-\theta_S)$, where
  $\lambda_S\ge 0>-\Lambda$ and $-\theta_S\le 0<1=\Theta$, the level set
  $A_{1+\theta_S}(\Lambda+\lambda_S)$ is the maximal minimizer of
  $g-\lambda_S t+\theta_S|\cdot|=F^g_{\lambda_S,\theta_S}$. By
  \cref{prop:witness}\ref{it:C-witness-ii}, this function has the unique
  minimizer $S$, so $A_{1+\theta_S}(\Lambda+\lambda_S)=S$.

  Next we show that the number $B_\infty$ of breakpoints of $(f,t,\mathbf 1)$
  in $(0,\infty)$ satisfies $2^n\le(B_\infty+1)\bigl(1+n(n+1)/2\bigr)$. By
  \cref{lem:pwaffine}, $B_\infty$ is finite. Let $S\subseteq E$ and
  $\alpha_S:=\Lambda+\lambda_S>0$. As $S=A_{1+\theta_S}(\alpha_S)$,
  $\rho_e(\alpha_S)\ge 1+\theta_S$ for $e\in S$ and
  $\rho_e(\alpha_S)<1+\theta_S$ for $e\notin S$. Choose $\theta'_S$ with
  $\max\{0,\rho_e(\alpha_S)\colon e\notin S\}<\theta'_S<1+\theta_S$. Then
  $\rho_e(\alpha_S)>\theta'_S$ for $e\in S$ and $\rho_e(\alpha_S)<\theta'_S$
  for $e\notin S$, and by continuity of $\rho$ (\cref{lem:pwaffine}), these
  strict inequalities hold at every share near $\alpha_S$. As $\theta'_S>0$
  and there are finitely many breakpoints, there is a share $\alpha'_S>0$ that
  is not a breakpoint such that $A_{\theta'_S}(\alpha'_S)=S$ and
  $\rho_e(\alpha'_S)\ne\theta'_S$ for all $e\in E$. With these points
  $(\alpha'_S,\theta'_S)$, the line-arrangement argument of the second
  paragraph of the proof of \cref{thm:expbp}, with $(0,\infty)$ in place of
  the bounded interval of shares considered there and $B_\infty$ in place of the
  number of breakpoints in it, gives
  $2^n\le(B_\infty+1)\bigl(1+n(n+1)/2\bigr)$.

  Next we show that all breakpoints lie in $(0,f(E)]$. Let
  $v_k:= f(\{k,\dots,n\})-f(\{k+1,\dots,n\})$ for $k\in E$. We show that
  $\xs(\alpha)=v$ for every $\alpha\ge f(E)$. As $t_n>\dots>t_1>0$ by
  \cref{prop:witness}\ref{it:C-witness-i}, Edmonds' greedy theorem~\cite[(19)]{Edm03} for the order
  $n,n-1,\dots,1$ shows that $v$ maximizes $t^\top x$ over $\Pp{f}$, so in
  particular $v\in\Pp{f}$. Every set $\{k,\dots,n\}$ is tight with respect to
  $v$, as the differences telescope, and $0\le v_k\le f(\{k\})\le f(E)$ by
  monotonicity and submodularity of $f$. Let $\alpha\ge f(E)$. The marginals
  at $v$ are $\rho_k(v)=\alpha t_k-v_k$. For $k\ge 2$, the integers $t_k$
  satisfy $t_k\ge t_{k-1}+1$, so
  \[
    \rho_k(v)-\rho_{k-1}(v) = \alpha(t_k-t_{k-1})-v_k+v_{k-1} \ge \alpha-f(E) \ge 0,
  \]
  and $\rho_1(v)\ge\alpha-f(E)\ge 0$ as $t_1\ge 1$. So
  $0\le\rho_1(v)\le\dots\le\rho_n(v)$, and every level set $A_\theta(v)$ with
  $\theta>0$ is empty or of the form $\{k,\dots,n\}$, hence tight. By
  \cref{prop:optchar}, $v$ is the best response at the share $\alpha$. So
  $\xs$ is constant on $[f(E),\infty)$, and every breakpoint lies in
  $(0,f(E)]$.

  Now we pass to the rewards $t/\sigma$ and show that the instance
  $(f,t/\sigma,\mathbf 1)$ has at least $\frac{2^n}{1+n(n+1)/2}-1$ breakpoints
  in $(0,1)$. By the dilation \cref{lem:ops}(b) with $\sigma=1/(f(E)+1)$, the
  best response of this instance is $\alpha\mapsto\xs(\alpha/\sigma)$, and its
  breakpoints are the $B_\infty$ breakpoints of $(f,t,\mathbf 1)$ multiplied by
  $\sigma$. As those lie in $(0,f(E)]$, these lie in
  $(0,\sigma f(E)]=(0,f(E)/(f(E)+1)]\subseteq(0,1)$, and by the bound on
  $B_\infty$ shown above, $B_\infty\ge\frac{2^n}{1+n(n+1)/2}-1$.

  Finally, for part~(i), we show that all numbers have $\Oh(n\log n)$ bits and
  that $f$ can be evaluated in polynomial time. By
  \cref{prop:witness}\ref{it:C-witness-i}, every $t_k$ is at most
  $\tau:= 2(2n+2)^{n-1}$. For $k\ge 2$, we have
  $\theta_k=3^{k-1}t_{k-1}\le 3^n\tau$, so $D_k\le(k-1)\theta_k\le n3^n\tau$
  and $K_k=t_k\bar\lambda_k+D_k+1\le(n+2)3^n\tau$, and these bounds also hold
  for $k=1$. Hence $\Lambda\le 1+n^23^n\tau$, and as
  $g(E)\le\sum_{k\in E}K_k$,
  \[
    f(E) \le \sum_{k\in E}(K_k+\Lambda t_k)
    \le n(n+2)3^n\tau + n(1+n^23^n\tau)\tau \le (n+2)^3 9^n\tau^2 .
  \]
  As $f$ is monotone, every value of $f$ is an integer in $[0,f(E)]$, and
  every reward $m_e=t_e/\sigma=(f(E)+1)t_e$ is an integer in
  $(0,(f(E)+1)\tau]$. So all these numbers are at most $2(n+2)^39^n\tau^3$,
  and as $\log_2\tau=\Oh(n\log n)$, they have $\Oh(n\log n)$ bits. The
  recursion of \cref{prop:witness} computes all $t_k$, $\theta_k$,
  $d_i^{(k)}$, $D_k$, $K_k$, and $\Lambda$ with $\Oh(n^2)$ arithmetic
  operations on integers of this size, and
  \[
    f(S) = \sum_{k\in S}(K_k+\Lambda t_k-1) - \sum_{i,k\in S,\ i<k} d_i^{(k)}
  \]
  takes $\Oh(n^2)$ further operations, so $f$ and $1/\sigma=f(E)+1$ can be
  computed in polynomial time. This completes the proof of part~(i).

  It remains to prove part~(ii), for networks. Let
  $\Xi:= 1+\sum_{k\in E}\theta_k$, and consider the network with a
  source $s$, a sink $s'$, and the nodes $E$, with an arc $k\to i$ of capacity
  $d_i^{(k)}$ for all $i<k$, an arc $k\to s'$ of capacity
  $K_k-D_k+\Xi+t_k=t_k(\bar\lambda_k+1)+\Xi+1$ for all $k\in E$, and an arc
  $s\to j$ of capacity $t_j\lambda+\mu$ for all $j\in E$. All constant
  capacities are nonnegative integers. As $D_k=\sum_{i<k}d_i^{(k)}$, every
  $S\subseteq E$ satisfies
  \[
    \sum_{k\in S}(K_k-D_k) + \sum_{\substack{k\in S,\ i\in E\setminus S,\\ i<k}} d_i^{(k)}
    = \sum_{k\in S}K_k - \sum_{\substack{i,k\in S,\\ i<k}} d_i^{(k)} = g(S).
  \]
  So the capacity of the cut $\{s\}\cup S$, the total capacity of the arcs
  that leave $\{s\}\cup S$, is
  \begin{align*}
    \lambda t(E\setminus S)+\mu|E\setminus S|+g(S)+\Xi|S|+t(S)
    &= \lambda t(E)+\mu n+g(S)-(\lambda-1)t(S)+(\Xi-\mu)|S|\\
    &= \lambda t(E)+\mu n+F^g_{\lambda-1,\Xi-\mu}(S).
  \end{align*}
  By \cref{prop:witness}\ref{it:C-witness-ii}, the cut $\{s\}\cup S$ is
  therefore the unique minimum cut at
  $(\lambda,\mu)=(\lambda_S+1,\Xi-\theta_S)$. Both coordinates of this point
  lie in $[1,C)$ for $C:= 3^n+\Xi$. So after the coefficients $t_j$ and
  $1$ of the arcs $s\to j$ are multiplied by $C$ (that is, $a_j:= Ct_j$
  and $b_j:= C$), every cut $\{s\}\cup S$ is the unique minimum cut at
  the point $\bigl((\lambda_S+1)/C,(\Xi-\theta_S)/C\bigr)\in(0,1)^2$. By the
  bounds shown for part~(i), all capacities and all coordinates of these points
  have $\Oh(n\log n)$ bits. This completes the proof.
\end{proof}

\subsection{Polymatroids with polynomially many breakpoints}\label{sec:A-subclasses}

We prove \cref{rem:subclasses} through the more detailed statements below. Here we bound the number of
breakpoints of $\xs$ in all of $(0,\infty)$, not only in $(0,1)$. In this subsection, $B$ denotes this
number. It is at least the number of breakpoints in $(0,1)$, so the bounds below also hold for the $B$ of
\cref{sec:bestresponse}. Recall from \cref{sec:bestresponse} that a polymatroid is laminar if
$\Pp{f}=\{x\ge0:x(L)\le f(L)\text{ for }L\in\mathcal L\}$ for a family $\mathcal L$ of nonempty sets, the
constraint sets, of which any two are disjoint or nested, and a partition polymatroid if the constraint
sets are pairwise disjoint. Ordered by inclusion, the constraint sets form a forest, whose height $h$ is the
largest number of constraint sets in a chain. The state, the free set $\Phi$, the zero set $Z$, and the
blocks are those of \cref{sec:A-structure}. We first show that elements leave the support of $\xs$ for good,
for every polymatroid.

\begin{lemma}\label{lem:supportmon}
Let $0<\alpha<\alpha'$ and $e\in E$. If $\xs_e(\alpha)=0$, then $\xs_e(\alpha')=0$. So the support of $\xs$ is nonincreasing in $\alpha$ and changes at no more than $n$ shares.
\end{lemma}

\begin{proof}
By \cref{lem:pwaffine}, $\xs$ is continuous and has finitely many affine
pieces. Recall $\rho_e(\tilde\alpha)=\tilde\alpha m_e-\beta_e\xs_e(\tilde\alpha)$, so that
$\xs_e(\tilde\alpha)=0$ if and only if $\rho_e(\tilde\alpha)=\tilde\alpha m_e$.

Suppose that $\xs_e(\alpha)=0$ and $\xs_e(\alpha')>0$ for some
$0<\alpha<\alpha'$. Then $m_e>0$, as $0<\xs_e(\alpha')\le\alpha't_e$ by
\cref{prop:optchar}. Let
$\underline\alpha:=\inf\{\tilde\alpha>\alpha:\xs_e(\tilde\alpha)>0\}$. By continuity,
$\xs_e(\underline\alpha)=0$, that is,
$\rho_e(\underline\alpha)=\underline\alpha m_e>0$. Choose $h>0$ such that $\xs$,
and hence $\rho$, is affine on $J:=(\underline\alpha,\underline\alpha+h)$.
After shrinking $h$, the chain of positive level sets of $\rho$, the zero set
$Z$, and the free set $\Phi$ are constant on $J$ (so the state of
\cref{sec:A-structure} is constant on $J$), and $\rho_e>0$ on $J$. Moreover,
$\xs_e>0$ on $J$, because $\xs_e$ is affine on $J$, vanishes at
$\underline\alpha$, and is positive somewhere in $J$. Let $D$ be the block of
$e$ on $J$, that is, the set of elements $g$ with $\rho_g=\rho_e$ on $J$. By
\cref{lem:velocity}, $\rho_e=\tilde\alpha m_e-\beta_e\xs_e$ has slope
$\bar m_D=t(D\setminus Z)/\binv(D\setminus Z)$ on $J$, which is the mean of the
rewards $m_g$, $g\in D\setminus Z$, weighted by $\beta_g^{-1}$. Each
$g\in D\setminus Z$ satisfies $\xs_g>0$, hence $\rho_e=\rho_g<\tilde\alpha m_g$ for
$\tilde\alpha\in J$. Letting $\tilde\alpha\downarrow\underline\alpha$ gives
$\underline\alpha m_e\le\underline\alpha m_g$, that is, $m_g\ge m_e$. So the
slope of $\rho_e$ on $J$ is at least $m_e$, and
$\rho_e(\tilde\alpha)\ge\underline\alpha m_e+(\tilde\alpha-\underline\alpha)m_e=\tilde\alpha m_e$ for
$\tilde\alpha\in J$. This means $\xs_e(\tilde\alpha)\le 0$ on $J$, a contradiction. The last
statement holds since every element leaves the support at most once.
\end{proof}

\subsubsection{Partition polymatroids}

In this subsection, we show the following result.

\begin{proposition}
\label{prop:partition}
For every partition polymatroid, $B\le n$.
\end{proposition}

\begin{proof}
Let $E_1,\dots,E_k$ be the constraint sets of the partition polymatroid, so
that $\Pp{f}=\{x\ge0:x(E_i)\le f(E_i)\text{ for }i=1,\dots,k\}$ with pairwise
disjoint sets $E_i$. They cover $E$ since $\Pp{f}$ is bounded, so they
partition $E$, and the response decomposes over these parts. Fix a part
$E_i$. If $f(E_i)=0$ or $t(E_i)=0$, the response is zero on $E_i$, so assume
that both are positive and set $\alpha_i:=f(E_i)/t(E_i)$. For
$\alpha\le\alpha_i$, the unconstrained response $\xs_e=\alpha t_e$, $e\in E_i$,
is feasible, as $\alpha t(E_i)\le f(E_i)$, and hence optimal.

Now let $\alpha>\alpha_i$. For $\lambda\ge0$ and $e\in E_i$, let
$x_e(\lambda):=(\alpha m_e-\lambda)^+/\beta_e$. Each $x_e(\lambda)$ is
continuous and nonincreasing in $\lambda$, and strictly decreasing while it is
positive, that is, for $\lambda<\alpha m_e$. Hence the sum
$s_i(\lambda):=\sum_{e\in E_i}x_e(\lambda)$ is continuous, equals
$\alpha t(E_i)$ at $\lambda=0$ as $m\ge0$, decreases strictly until
$\lambda=\alpha\max_{e\in E_i}m_e$, and is zero from there on. As
$\alpha>\alpha_i$, we have $\alpha t(E_i)>f(E_i)>0$. So there is exactly one
$\lambda\ge0$ with $s_i(\lambda)=f(E_i)$. We call it $\lambda_i(\alpha)$, and
it is positive since $s_i(0)>f(E_i)$. Let $x_e:=x_e(\lambda_i(\alpha))$ for
$e\in E_i$, so that
\[
  x_e=\frac{(\alpha m_e-\lambda_i(\alpha))^+}{\beta_e},\qquad x(E_i)=f(E_i).
\]
This $x$ is optimal on $E_i$. Its marginals $\alpha m_e-\beta_e x_e$ are at
most $\lambda_i(\alpha)$, with equality whenever $x_e>0$. For every feasible
$y$ on $E_i$, as $y\ge0$, this gives
$(\alpha m_e-\beta_e x_e)(y_e-x_e)\le\lambda_i(\alpha)(y_e-x_e)$ for every
$e\in E_i$. Hence, by concavity of the per-element utility
$x_e\mapsto\alpha m_e x_e-\frac{\beta_e}{2}x_e^2$, every feasible $y$ on $E_i$
satisfies
\[
  \sum_{e\in E_i}\Bigl(\alpha m_e(y_e-x_e)-\tfrac{\beta_e}{2}\bigl(y_e^2-x_e^2\bigr)\Bigr)
  \le\sum_{e\in E_i}(\alpha m_e-\beta_e x_e)(y_e-x_e)
  \le\lambda_i(\alpha)\bigl(y(E_i)-f(E_i)\bigr)\le 0.
\]
Dividing $s_i(\lambda_i(\alpha))=f(E_i)$ by $\alpha$ shows that
$\mu_i(\alpha):=\lambda_i(\alpha)/\alpha$ satisfies
\[
  q_i\bigl(\mu_i(\alpha)\bigr)=\frac{f(E_i)}{\alpha},\qquad
  \text{where }q_i(\mu):=\sum_{e\in E_i}\frac{(m_e-\mu)^+}{\beta_e}.
\]
The left-hand side depends only on $\mu_i(\alpha)$, and the right-hand side
only on $\alpha$. As $f(E_i)>0$, the right-hand side strictly decreases as
$\alpha$ increases. The function $q_i$ is continuous and strictly decreasing on
$[0,\max_{e\in E_i}m_e]$, which contains $\mu_i(\alpha)$ because
$0<\lambda_i(\alpha)<\alpha\max_{e\in E_i}m_e$. So when $\alpha$ increases,
$q_i(\mu_i(\alpha))$ decreases, and this forces $\mu_i(\alpha)$ to increase.
As $q_i$ has a continuous inverse on this interval,
$\mu_i(\alpha)=q_i^{-1}(f(E_i)/\alpha)$ is moreover continuous in $\alpha$. As
$\mu_i(\alpha)$ increases and stays below $\max_{e\in E_i}m_e$, the support
$W_i:=\{e\in E_i: m_e>\mu_i(\alpha)\}$ only loses elements, never those with
the largest reward, so it changes at most $|E_i|-1$ times. Between these
changes, $W_i$ is constant. As $x_e=(\alpha m_e-\lambda_i(\alpha))/\beta_e$ for
$e\in W_i$ and $x_e=0$ for the other elements of $E_i$, the equation
$x(E_i)=f(E_i)$ reads $\alpha t(W_i)-\lambda_i(\alpha)\binv(W_i)=f(E_i)$.
Hence $\lambda_i(\alpha)=(\alpha t(W_i)-f(E_i))/\binv(W_i)$ is affine in
$\alpha$, and so is the response. Including $\alpha_i$ gives at most $|E_i|$
breakpoints on $E_i$, hence at most $|E_1|+\dots+|E_k|=n$ overall.
\end{proof}

\subsubsection{Laminar polymatroids}

In this subsection, we show the following result.

\begin{theorem}\label{thm:laminar}
If $(E,f)$ is laminar with a family $\mathcal L$ of height $h$, then $B\le n+2h(n+1)|\mathcal L|\,3^{h-1}$. As $|\mathcal L|\le 2n-1$, this is $\Oh(n^2)$ for fixed $h$ and polynomial for $h=\Oh(\log n)$.
\end{theorem}

Before we are in a position to prove \cref{thm:laminar}, we need two technical lemmas, which need the following notation. Every element $e$ lies in some constraint set, since otherwise $x_e$
would be unbounded on $\Pp{f}$. If $m_e=0$, then $\xs_e(\alpha)=0$ for all
$\alpha$, since lowering $x_e$ to $0$ keeps a point in $\Pp{f}$ and does not
decrease $\UA(\cdot\mid\alpha)$.

By \cref{lem:supportmon}, which holds for every polymatroid, the support
changes at no more than $n$ shares. These shares cut $(0,\infty)$ into at most
$n+1$ open intervals on which the support is constant. We call these intervals
\emph{pinning intervals}. Fix a pinning interval $I$ with support $W$. We
remove from $\mathcal L$ the constraint sets that contain no element of the
support $W$, that is, $\xs_e=0$ on $I$ for all their elements $e$. We define
$\mathcal L_W$ to be the remaining ones. We regard $\mathcal L_W$ as a forest.
The \emph{parent} $\operatorname{par}(L)$ of $L$ is the smallest constraint set
in $\mathcal L_W$ that properly contains $L$, the \emph{children} of $L$ are
the constraint sets $K$ with $\operatorname{par}(K)=L$, and for $e\in W$ we
denote by $\ell(e)$ the smallest constraint set in $\mathcal L_W$ that contains
$e$. For $L\in\mathcal L_W$ let $E_L:=\{e\in W:\ell(e)=L\}$ be the elements
whose smallest constraint set is $L$. For $L\in\mathcal L_W$ and
$(\alpha,\theta)\in\R^2$ define, recursively over the forest,
\begin{equation}\label{eq:E-D}
  D_L(\alpha,\theta):=\sum_{e\in E_L}\frac{\alpha m_e-\theta}{\beta_e}
  +\sum_{K\text{ child of }L}\min\bigl\{f(K),D_K(\alpha,\theta)\bigr\}.
\end{equation}
This is the amount that the elements of $L\cap W$ take if each of them has
marginal $\theta$, except that the elements inside a child $K$ take at most
$f(K)$ together, and so on down the forest. By induction over the forest,
$D_L$ is the minimum of finitely many affine functions
$(\alpha,\theta)\mapsto\alpha\,t(U)-\theta\,\binv(U)+c$ with $U\subseteq L\cap W$
and $c\in\R$, one of them with $U=L\cap W$. So $D_L$ is continuous, concave, and
piecewise affine, nondecreasing in $\alpha$ and nonincreasing in $\theta$, and
$D_L(\alpha,\theta)\to-\infty$ for $\theta\to\infty$, as $L$ contains an
element of $W$. Let
\[
  \pi_L(\alpha):=\max\{\theta\in\R : D_L(\alpha,\theta)\ge f(L)\}
\]
be the water level at which $L$ is exactly full, with $\pi_L(\alpha):=-\infty$
if $D_L(\alpha,\theta)<f(L)$ for all $\theta$. The maximum exists, as $D_L$ is
continuous and nonincreasing in $\theta$ and tends to $-\infty$. So
$\{\theta:D_L(\alpha,\theta)\ge f(L)\}=(-\infty,\pi_L(\alpha)]$. In
particular, $D_L(\alpha,\theta)<f(L)$ for $\theta>\pi_L(\alpha)$ and, by
continuity, $D_L(\alpha,\pi_L(\alpha))=f(L)$ if $\pi_L(\alpha)>-\infty$.
Finally, for $V\in\mathcal L_W$ let
$\Pi_V:=\max\{0,\max_{L\supseteq V}\pi_L\}$, where $L$ runs over the
constraint sets in $\mathcal L_W$ that contain $V$. Since the constraint sets
in $\mathcal L_W$ that contain $\operatorname{par}(V)$ are those that contain
$V$, except $V$ itself, we have $\Pi_V=\max\{\Pi_{\operatorname{par}(V)},\pi_V\}$,
where $\Pi_{\operatorname{par}(V)}:=0$ for maximal $V$.

\begin{lemma}\label{lem:E-laminarformula}
Let $I$ be a pinning interval with support $W$. For every $\alpha\in I$ and
$e\in W$, we have
$\xs_e(\alpha)=\bigl(\alpha m_e-\Pi_{\ell(e)}(\alpha)\bigr)/\beta_e$. Moreover,
every $\pi_L\not\equiv-\infty$ is concave, piecewise linear, and strictly
increasing.
\end{lemma}

\begin{proof}
We first show the second claim. As $D_L$ is the minimum of finitely many
affine functions $\alpha\,t(U)-\theta\,\binv(U)+c$ with $U\subseteq L\cap W$,
the set $\{(\alpha,\theta):D_L(\alpha,\theta)\ge f(L)\}$ is the intersection of
the half-planes on which these functions are at least $f(L)$. For $U=\emptyset$
such a half-plane is the whole plane or empty, and in the latter case
$\pi_L\equiv-\infty$. Otherwise, $\pi_L(\alpha)$ is the minimum of
$(\alpha\,t(U)+c-f(L))/\binv(U)$ over the functions with $U\ne\emptyset$, a
minimum of finitely many affine functions of $\alpha$ whose slopes
$t(U)/\binv(U)$ are positive, as $m_e>0$ for every $e\in W$. So $\pi_L$ is
concave, piecewise linear, and strictly increasing.

Now fix $\alpha\in I$ and drop $\alpha$ from the notation. Since all
constraints are linear, $\xs$ satisfies the KKT conditions, and the
multipliers of the constraints $x_e\ge0$, $e\in W$, vanish by complementary
slackness. The removed constraint sets contain no element of $W$, so their
multipliers do not enter the stationarity conditions of the elements of $W$.
So the restriction of $\xs$ to $W$ satisfies the KKT conditions of the
problem of maximizing $\sum_{e\in W}\bigl(\alpha m_e x_e-\frac{\beta_e}{2}x_e^2\bigr)$
over $x\in\R^W$ subject to $x(L\cap W)\le f(L)$ for $L\in\mathcal L_W$,
without sign constraints. This problem has a strictly concave objective and
linear constraints, so its KKT conditions are sufficient and its optimum is
unique. Hence it suffices to show that
$x_e:=(\alpha m_e-\Pi_{\ell(e)})/\beta_e$, $e\in W$, together with the
multipliers $\lambda_V:=\Pi_V-\Pi_{\operatorname{par}(V)}\ge0$,
$V\in\mathcal L_W$, satisfies these conditions. Stationarity holds by
construction, as $\alpha m_e-\beta_e x_e=\Pi_{\ell(e)}=\sum_{L\ni e}\lambda_L$,
where the sum telescopes along the constraint sets in $\mathcal L_W$ that
contain $\ell(e)$.

For feasibility and complementary slackness, let, for $V\in\mathcal L_W$ and
$\theta\in\R$,
\[
  Y_V(\theta):=\sum_{e\in V\cap W}\frac1{\beta_e}
  \Bigl(\alpha m_e-\max\bigl\{\theta,\max_{V\supseteq L\supseteq\ell(e)}\pi_L\bigr\}\Bigr).
\]
We claim that $Y_V(\theta)=\min\{f(V),D_V(\theta)\}$ for all $\theta\in\R$.
We proceed by induction over the forest, so let the claim hold for the
children $K$ of $V$, and put $\theta':=\max\{\theta,\pi_V\}$. The elements $e$
with $\ell(e)=V$ contribute $\alpha t(E_V)-\theta'\binv(E_V)$ to
$Y_V(\theta)$, and the elements inside a child $K$ contribute
$Y_K(\theta')=\min\{f(K),D_K(\theta')\}$. By \eqref{eq:E-D}, this gives
$Y_V(\theta)=D_V(\theta')$. If $\theta\ge\pi_V$, then $\theta'=\theta$ and
$D_V(\theta)\le f(V)$. Otherwise, $D_V(\theta')=D_V(\pi_V)=f(V)\le D_V(\theta)$.
In both cases, the claim follows.

For $e\in V\cap W$, we have
$\Pi_{\ell(e)}=\max\{\Pi_{\operatorname{par}(V)},\max_{V\supseteq L\supseteq\ell(e)}\pi_L\}$,
hence $x(V\cap W)=Y_V(\Pi_{\operatorname{par}(V)})$ and, by the claim,
\[
  x(V\cap W)=\min\bigl\{f(V),D_V(\Pi_{\operatorname{par}(V)})\bigr\}\le f(V).
\]
This is feasibility. If $\lambda_V>0$, then $\pi_V>\Pi_{\operatorname{par}(V)}$,
hence $D_V(\Pi_{\operatorname{par}(V)})\ge f(V)$ and $x(V\cap W)=f(V)$, which is
complementary slackness.
\end{proof}

\begin{lemma}\label{lem:E-piececount}
Let $I$ be a pinning interval with support $W$. For $L\in\mathcal L_W$, let
$\#(L)$ be the number of linear pieces of $\pi_L$ on $\R$, and $\#(L):=0$ if
$\pi_L\equiv-\infty$. Then $\#(L)\le1+2\sum_{J\subsetneq L}\#(J)$, where $J$
runs over the constraint sets in $\mathcal L_W$ properly contained in $L$.
Consequently, $\sum_{L\in\mathcal L_W}\#(L)\le|\mathcal L|\,3^{h-1}$.
\end{lemma}

\begin{proof}
The set $T_L:=\{(\alpha,\theta):D_L(\alpha,\theta)\ge f(L)\}$ is the hypograph
of $\pi_L$, and $D_L$ is concave and piecewise affine on $\R^2$. Let
$\operatorname{brk}(F)$ denote the set of points near which a function $F$ is
not affine. Near interior points of $T_K$ the function $\min\{f(K),D_K\}$
equals $f(K)$, and near points outside $T_K$ it equals $D_K$. So
$\operatorname{brk}(\min\{f(K),D_K\})\subseteq\partial T_K\cup\operatorname{brk}(D_K)$.
By \eqref{eq:E-D}, $\operatorname{brk}(D_L)$ is contained in the union of the
sets $\operatorname{brk}(\min\{f(K),D_K\})$ over the children $K$ of $L$, and
by induction $\operatorname{brk}(D_L)\subseteq\bigcup_{J\subsetneq L}\partial T_J$.

Let $z$ be a vertex of the graph of $\pi_L$. Then $D_L(z)=f(L)$, and $D_L$ is
not affine near $z$. Otherwise, near $z$, the set $T_L$ would be a half-plane
bounded by a line through $z$, or a whole neighborhood of $z$, and in neither
case would $z$ be a vertex of the graph of $\pi_L$. So $z\in\partial T_J$ for
some $J\subsetneq L$, that is, $z$ lies on the graph of $\pi_J$. Now let
$\gamma$ be a linear piece of the graph of some $\pi_J$, over a closed
interval~$O$. On $O$, the function $\pi_L-\pi_J$ is concave, so its zero set
in $O$ consists of at most two points or is an interval, on whose interior
$\pi_L=\pi_J$ is affine. Hence $\gamma$ contains at most two vertices of the
graph of $\pi_L$. As this graph has $\#(L)-1$ vertices, the first inequality
follows.

For the second inequality, let $\Sigma(L):=\sum_{J\subseteq L}\#(J)$, where
$J$ runs over the constraint sets in $\mathcal L_W$ contained in $L$. The first
inequality gives $\Sigma(L)\le1+3\sum_K\Sigma(K)$, where $K$ runs over the
children of $L$, and by induction
$\Sigma(L)\le\sum_{J\subseteq L}3^{\dist(L,J)}$, where
$\dist(L,J)\le h-1$ is the number of edges of the forest between $L$ and $J$.
Summing over the maximal constraint sets in $\mathcal L_W$ and using
$|\mathcal L_W|\le|\mathcal L|$ gives the claim.
\end{proof}

\begin{proof}[Proof of \cref{thm:laminar}]\phantomsection\label{prf:laminar}
Let $I$ be a pinning interval with support $W$, and let
$N_I:=\sum_{L\in\mathcal L_W}\#(L)\le|\mathcal L|\,3^{h-1}$
(\cref{lem:E-piececount}).
By \cref{lem:E-laminarformula}, on $I$ the vector $\xs$ vanishes outside $W$
and is determined on $W$ by the functions $\Pi_V=\max\mathcal F_V$, where
$\mathcal F_V:=\{0\}\cup\{\pi_L: L\supseteq V,\ \pi_L\not\equiv-\infty\}$.
Hence, near every breakpoint $\alpha_0\in I$ of $\xs$, some $\Pi_V$ is not
affine. If $\alpha_0$ is neither a vertex of a function in $\mathcal F_V$ nor a
boundary point of $\{\varphi=\psi\}$ for $\varphi,\psi\in\mathcal F_V$, then
near $\alpha_0$ all functions in $\mathcal F_V$ are affine and every difference
$\varphi-\psi$ either vanishes identically or has constant sign, so one of them
is maximal on a whole neighborhood of $\alpha_0$ and $\Pi_V$ is affine there.
Therefore, every breakpoint in $I$ is (a) a vertex of some $\pi_L$, (b) a zero
of some $\pi_L$, or (c) a boundary point of $\{\pi_L=\pi_{L'}\}$ that is not a
vertex of $\pi_L$, for constraint sets $L\subsetneq L'$ in $\mathcal L_W$.

There are at most $N_I$ points of type (a) and, as the functions $\pi_L$ are
strictly increasing, at most $|\mathcal L|$ points of type (b). For (c), fix
$L\subsetneq L'$ and an open linear piece $O$ of $\pi_L$. On $O$, the
difference $\pi_L-\pi_{L'}$ is convex, so its zero set in $O$ either consists
of at most two points or is an interval. In both cases, it has at most two
boundary points in $O$. So the pair $L\subsetneq L'$ yields at most $2\#(L)$
points of type (c). As every $L$ is contained in at most $h-1$ other
constraint sets in $\mathcal L_W$, there are at most $2(h-1)N_I$ points of type
(c). Altogether, $I$ contains at most
\[
  (2h-1)N_I+|\mathcal L|\le 2h\,|\mathcal L|\,3^{h-1}
\]
breakpoints. Combining this with at most $n+1$ pinning intervals and $n$
support changes (\cref{lem:supportmon}) gives
$B\le n+2h(n+1)|\mathcal L|\,3^{h-1}$.
\end{proof}

\section{Omitted proofs for Section~\ref{sec:hardness}}\label{app:hard}

This appendix gives the precise statements and the proofs for \cref{sec:hardness}.
\Cref{sec:D-ideas} gives the definition of combs, a picture of one step of the comb gadget, and the centering of the parabola, the last step of the reduction, which \cref{sec:hardness} only sketches.
\Cref{sec:D-cutforms} shows that cut forms define a class $\Ccut$ that is recognized in polynomial time and is closed under the operations used in the construction, and it proves the rules for sums and dilations of combs (\cref{lem:combsum}).
\Cref{sec:D-correct} states the comb gadget (\cref{thm:gadget}) and the reduction (\cref{def:reduction}), and proves the correctness of the reduction (\cref{prop:correct}) assuming the gadget theorem.
The following subsections construct the combs and prove \cref{thm:gadget} with the tools of \cref{app:tools}: an overview (\cref{sec:D-overview}), the base and the steps with their parameter rule (\cref{sec:D-base}), one level (\cref{sec:D-onelevel}), and the proof that every level works (\cref{sec:D-alllevels}).
Finally, \cref{sec:D-npcomplete} establishes exact best-response computation and verification in polynomial time (\cref{lem:exactbr}) and combines these facts with the reduction to prove \cref{thm:nphard}.
A reader who wants the ideas first can read \cref{sec:D-ideas}, the reduction in \cref{sec:D-correct}, and the overview in \cref{sec:D-overview}, and skip the rest on a first reading.

\subsection{Combs}\label{sec:D-ideas}

This subsection gives the definition of combs, a picture of one step of the comb gadget, and the last step of the reduction, the centering of the parabola, which \cref{sec:hardness} only sketches. The following subsections give the precise statements and the proofs.

The following definition makes the combs of \cref{sec:hardness} precise.

\begin{definition}\label{def:comb}
Let $\Gamma\subseteq[0,\infty)$ be finite, $L(\alpha)=p+q\alpha$ affine, and $\kappa,\delta>0$. An instance is a \emph{comb} with \emph{contact set} $\Gamma$, \emph{line} $L$, \emph{dip} $\kappa$, and \emph{width} $\delta$ if $r(\xs(\gamma))=L(\gamma)$ for $\gamma\in\Gamma$ and
\[
L(\alpha)-r(\xs(\alpha))\ \ge\ \kappa\min\big(\dist(\alpha,\Gamma),\delta\big)\qquad\text{for all }\alpha\ge0,
\]
where $\dist(\alpha,\emptyset):=\infty$. We call $L(\alpha)-r(\xs(\alpha))$ the \emph{gap} and the points of $\Gamma$ the \emph{contacts}.
\end{definition}

\begin{figure}[!ht]
\centering
\begin{tikzpicture}[xscale=0.9,yscale=0.7,>=Stealth]
\draw[->] (0,2.6)--(13.2,2.6) node[right,font=\scriptsize]{$\alpha$};
\draw[thick,blue!70!black] (0.2,2.75)--(0.5,2.6)--(1.1,3.0)--(1.7,2.6)--(2.0,3.5)--(12.8,4.0);
\foreach \x in {0.5,1.7} \fill[blue!70!black] (\x,2.6) circle (2pt);
\draw[decorate,decoration={brace,amplitude=3pt}] (0.2,3.6)--(2.0,3.6) node[midway,above=2pt,font=\scriptsize]{window $R$};
\node[font=\scriptsize,anchor=west] at (4.4,4.2) {gap of level $j-1$};
\draw[->] (0,0)--(13.2,0) node[right,font=\scriptsize]{$\alpha$};
\draw[thick] (0.2,0.15)--(0.5,0)--(1.1,0.4)--(1.7,0)--(2.0,0.9)--(4.3,1.5)--(7.5,0.15)--(7.8,0)--(8.4,0.4)--(9.0,0)--(9.3,0.9)--(11,1.3)--(12.8,1.6);
\foreach \x in {0.5,1.7,7.8,9.0} \fill (\x,0) circle (2pt);
\draw[gray,dashed] (2.0,0.15)--(7.5,0.15);
\node[font=\scriptsize,text=gray,above] at (3.2,0.15) {$1/30$};
\draw[decorate,decoration={brace,amplitude=3pt,mirror}] (0.2,-0.25)--(2.0,-0.25) node[midway,below=2pt,font=\scriptsize]{$R$};
\draw[decorate,decoration={brace,amplitude=3pt,mirror}] (7.5,-0.25)--(9.3,-0.25) node[midway,below=2pt,font=\scriptsize]{$R+\delta\muin_{j+1}$};
\node[font=\scriptsize] at (4.75,-0.55) {transition: gap rises, then falls};
\node[font=\scriptsize,anchor=west] at (3.0,1.9) {gap of level $j$};
\end{tikzpicture}
\caption{Schematic of one step of the gadget. The gap of level $j-1$ is copied into the window $R$ and into its translate, which doubles the contacts, and between the windows the gap stays at least $1/30$.}\label{fig:gadget}
\end{figure}
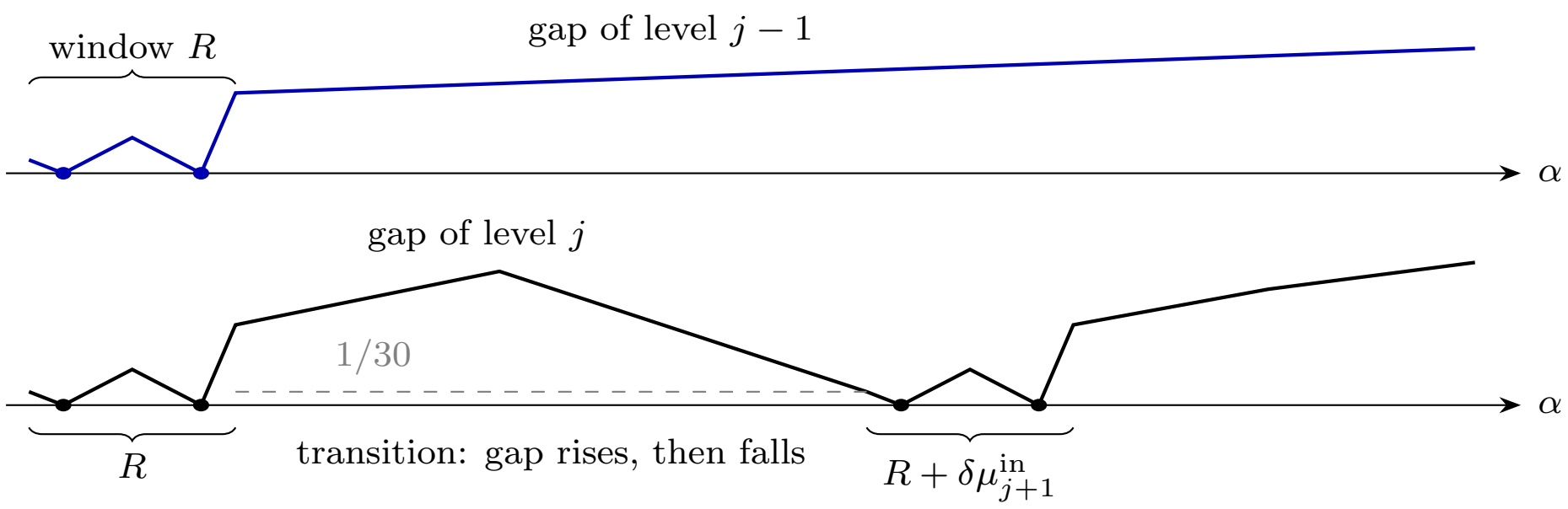

\Cref{fig:gadget} shows one step of the construction described in \cref{sec:hardness}.

\begin{figure}[t]
\centering
\begin{tikzpicture}[scale=0.8,>=Stealth]
\begin{scope}
\fill[gray!15] (0.55,0)--(0.95,0)--(0.95,2)--(0.55,2)--cycle;
\draw[->] (0,0)--(3.4,0) node[below,font=\scriptsize]{$\alpha$};
\draw[->] (0,0)--(0,2.3);
\draw[thick,domain=0:3,samples=40] plot (\x,{1.9-0.21*(\x-0.75)^2*1.0});
\draw[thick,dashed,domain=0:3,samples=40] plot (\x,{max(0,1.0-0.25*(\x-1.6)^2)});
\draw[->] (0.75,1.0)--(0.75,1.85);
\fill[orange!85!black] (0.75,1.9) circle (2pt);
\node[font=\scriptsize] at (2.4,2.05) {$(1-\alpha)(L+C)$};
\node[font=\scriptsize] at (3.0,0.3) {$(1-\alpha)L$};
\node[font=\scriptsize,below] at (0.75,-0.02) {center};
\node[font=\scriptsize] at (1.6,-0.95) {(a) centering};
\end{scope}
\begin{scope}[xshift=4.3cm]
\fill[gray!15] (0.6,0)--(2.6,0)--(2.6,2)--(0.6,2)--cycle;
\draw[->] (0,0)--(3.4,0) node[below,font=\scriptsize]{$\alpha$};
\draw[thick,domain=0.1:3.2,samples=40] plot (\x,{1.9-0.45*(\x-1.6)^2});
\draw[gray,dashed] (0,1.45)--(3.3,1.45) node[right,font=\scriptsize,text=black]{$\vartheta$};
\draw[thick,purple!70] (0.2,0.2)--(0.6,0.7)--(0.9,0.9)--(1.2,1.05)--(1.4,1.25)--(1.6,1.9)--(1.8,1.25)--(2.1,1.05)--(2.4,0.85)--(2.8,0.5)--(3.1,0.1);
\fill[orange!85!black] (1.6,1.9) circle (2pt);
\node[font=\scriptsize] at (1.6,-0.95) {(b) yes-instance};
\end{scope}
\begin{scope}[xshift=8.9cm]
\fill[gray!15] (0.6,0)--(2.6,0)--(2.6,2)--(0.6,2)--cycle;
\draw[->] (0,0)--(3.4,0) node[below,font=\scriptsize]{$\alpha$};
\draw[thick,domain=0.1:3.2,samples=40] plot (\x,{1.9-0.45*(\x-1.6)^2});
\draw[gray,dashed] (0,1.45)--(3.3,1.45) node[right,font=\scriptsize,text=black]{$\vartheta$};
\draw[thick,purple!70] (0.2,0.2)--(0.6,0.65)--(0.9,0.85)--(1.2,1.0)--(1.4,1.05)--(1.6,1.15)--(1.8,1.05)--(2.1,1.0)--(2.4,0.8)--(2.8,0.45)--(3.1,0.1);
\node[font=\scriptsize] at (1.6,-0.95) {(c) no-instance};
\end{scope}
\end{tikzpicture}
\caption{Schematic. (a) The centering summand raises reward and line on the window of contacts (shaded) by a constant $C$ and moves the top of the parabola to the center of the window. (b), (c) The window enlarged. (b) At a common contact, $\UP$ touches the parabola above $\vartheta$. (c) Without one, $\UP$ stays below the parabola by $(1-\alpha)$ times the gap, which exceeds the drop of the parabola across the window, so $\UP$ stays below $\vartheta$.}\label{fig:knob}
\end{figure}
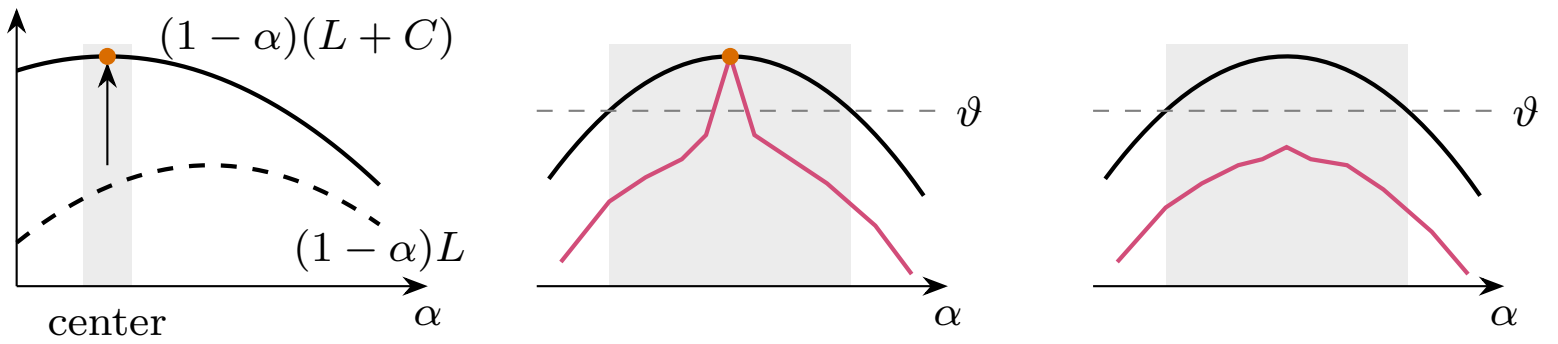

In the reduction of \cref{sec:hardness}, a dilation of the rewards of the direct sum of the two combs by a small power of two $\sigma$ moves the window that contains all possible contacts into $[0,1/4]$ and multiplies the gap of a no-instance by $1/\sigma$. One more element, the \emph{centering summand}, a capped linear summand (\cref{lem:ops}(c)), raises reward and line on this window by a constant $C$ such that, for the line $L$ of the dilated sum, the top of the parabola $(1-\alpha)(L+C)$ lies at the center of the window (\cref{fig:knob}). The drop of this parabola across the window does not depend on $\sigma$. Let $\vartheta$ be its value at the ends of the window. At a common contact, $\UP$ touches the parabola inside the window, so $\UP(\as)\ge\vartheta$. Without one, $\UP$ stays below the parabola by $(1-\alpha)$ times the gap, which for small $\sigma$ exceeds the drop, so $\UP(\as)<\vartheta$ (\cref{prop:correct}). All numbers of the reduction have polynomial bit length.

\subsection{Cut forms}\label{sec:D-cutforms}

This subsection defines $\Ccut$ (\cref{def:cutform}) and shows that it is a well-defined class of inputs, recognized in polynomial time, and that cut forms are preserved by the operations used in the reduction and in the construction of combs.
It then states and proves the rules for sums and dilations of combs (\cref{lem:combsum}). \Cref{lem:ops} is stated and proved in \cref{sec:A-structure}.

\begin{definition}\label{def:cutform}
A \emph{cut form} on $E$ consists of a loopless digraph $D$ on $E$, capacities $\mathrm{cap}\colon D\to\Q_{\ge0}$, and weights $\omega\in\Q^E$, and defines
\begin{equation}\label{eq:cutform}
f(S):=\cut_D(S)+\omega(S),\qquad \cut_D(S):=\sum_{(u,v)\in D:\,u\in S,\,v\notin S}\mathrm{cap}(u,v).
\end{equation}
Let $\Ccut$ be the class of instances $(f,m,\beta)$ with $m,\beta\in\Q^E_{>0}$ for which $f$ is a rank function given by a cut form, encoded by the cut form, $m$, and $\beta$ in binary.
\end{definition}

\begin{lemma}\label{lem:cutforms}
\begin{enumerate}[label=(\alph*)]
\item Every function given by a cut form as in \eqref{eq:cutform} is submodular.
\item The cut form $(D,\mathrm{cap},\omega)$ defines a polymatroid rank function if and only if, for every $e\in E$, $\omega_e$ is at least the total capacity of the arcs of $D$ that enter $e$. In particular, membership in $\Ccut$ is decided in polynomial time.
\item Let $g$ be given by a cut form $(D,\mathrm{cap},\omega)$ on $E$, that is, by \eqref{eq:cutform} with $g$ in place of $f$. Then each of the following functions has a cut form.
  \begin{enumerate}[label=(\roman*)]
  \item The function $g$ plus a modular function $S\mapsto y(S)$ with $y\in\Q^E$: replace $\omega$ by $\omega+y$.
  \item The direct sum of $g$ and a \emph{modular singleton}, that is, the function $\pi_F$ on a one-element set $\{b\}$ with $b\notin E$, $\pi_F(\emptyset)=0$, and $\pi_F(\{b\})=F$ for some $F\in\Q$: add $b$ as an isolated node of weight $F$.
  \item The \emph{one-element extension} $g'\colon 2^{E\cup\{b\}}\to\R$ with $b\notin E$, given by
  \[
    g'(S):= g(S)\quad\text{and}\quad g'(S\cup\{b\}):= g(S)+K-d(S)\qquad(S\subseteq E)
  \]
  for some $K\in\Q$ and $d\in\Q^E_{\ge0}$, where $d(S)=\sum_{e\in S}d_e$: add the node $b$ with weight $K-d(E)$ and the arcs $b\to e$ of capacity $d_e$ for $e\in E$.
  \end{enumerate}
  The direct sum of two functions given by cut forms has the disjoint union of the two forms as a cut form.
\end{enumerate}
\end{lemma}

The one-element extension in part (c)(iii) is the extension of \cref{lem:translation} and \cref{cor:witness}, with an arbitrary $d\in\Q^E_{\ge0}$ and an arbitrary $K\in\Q$ (\cref{cor:witness} takes a special nonnegative $d$ and a $K$ in a window).
In particular, when $d$ has the form $d=\mu_\theta\binv-\mu_\alpha t\ge0$ of \cref{lem:translation} (with rational data), part (c)(iii) is the cut form of the translation extension, as used in \cref{lem:translation}(a).
The marginal value of $b$ at $S\subseteq E$ in $g'$ is $K-d(S)$.
By (a) and (c), $g'$ is submodular, as \cref{lem:translation}(a) also shows directly.

\begin{proof}
\textit{(a)} A directed cut function with nonnegative capacities is submodular~\cite[(44.16)]{Sch03}, as used in the proof of \cref{thm:alm}, and so is its sum with a modular function.
Indeed, each arc $(u,v)\in D$ contributes $\mathrm{cap}(u,v)$ to $\cut_D(S)$ if $u\in S$ and $v\notin S$, and nothing otherwise.
In this contribution, the marginal value of $u$ at a set $S\not\ni u$ is $\mathrm{cap}(u,v)$ if $v\notin S$ and $0$ otherwise, the marginal value of $v$ at a set $S\not\ni v$ is $-\mathrm{cap}(u,v)$ if $u\in S$ and $0$ otherwise, and the marginal value of every other element is $0$.
As $\mathrm{cap}(u,v)\ge0$, all of them are nonincreasing in $S$, so each contribution is submodular, and so is their sum $\cut_D$.

\textit{(b)} Always $f(\emptyset)=0$, as $\cut_D(\emptyset)=0$, and $f$ is submodular by (a).
A submodular $f$ with $f(\emptyset)=0$ is monotone if and only if $f(E)-f(E\setminus\{e\})\ge0$ for every $e\in E$, because $f(S\cup\{e\})-f(S)\ge f(E)-f(E\setminus\{e\})$ for every $S\subseteq E\setminus\{e\}$.
A monotone $f$ with $f(\emptyset)=0$ is nonnegative.
No arc leaves $E$, and as $D$ has no loops, the arcs that leave $E\setminus\{e\}$ are exactly those that enter $e$.
So $f(E)-f(E\setminus\{e\})$ is $\omega_e$ minus the total capacity of the arcs entering $e$, which proves the equivalence.
To decide whether an encoded instance $(f,m,\beta)$ lies in $\Ccut$, it therefore suffices to check that $D$ has no loops, that all capacities are nonnegative, and that $m$ and $\beta$ are positive, and to compare, for each $e\in E$, the weight $\omega_e$ with the sum of the capacities of the arcs entering $e$.
These are polynomially many additions and comparisons of input numbers.

\textit{(c)} For the extension, let $g''$ be the function of the extended form on $E\cup\{b\}$, and let $S\subseteq E$.
The arcs of $D$ join nodes of $E$, so they contribute $\cut_D(S)$ to the cuts of both $S$ and $S\cup\{b\}$.
The new arcs $b\to e$ leave $b$, so they contribute nothing to the cut of $S$.
To the cut of $S\cup\{b\}$ they contribute $\sum_{e\in E\setminus S}d_e=d(E)-d(S)$, and the weight $K-d(E)$ of $b$ is counted there as well.
Hence $g''(S)=g(S)=g'(S)$ and $g''(S\cup\{b\})=g(S)+d(E)-d(S)+K-d(E)=g'(S\cup\{b\})$.
The other operations add modular terms or add two forms on disjoint sets.
An isolated node $b$ of weight $F$ adds nothing to any cut and adds $F$ to the weight of exactly the sets that contain $b$, which gives $(g\oplus\pi_F)(S)=g(S\cap E)+\pi_F(S\cap\{b\})$.
For two cut forms on disjoint sets $E_1$ and $E_2$, no arc joins $E_1$ and $E_2$, so the cut and the weight of $S\subseteq E_1\cup E_2$ in the disjoint union are those of $S\cap E_1$ plus those of $S\cap E_2$.
\end{proof}

\Cref{lem:cutforms} ensures that the construction preserves cut representations. We next describe how rescaling rewards and taking direct sums affect the combs themselves. Rescaling lets us place their contacts in the desired interval of shares. In a direct sum, the two nonnegative gaps add, so the reward touches the sum of the upper lines exactly at common contacts (\cref{fig:combsum}(c)). The next lemma quantifies these operations; its common-grid assumption ensures a lower bound on the gap away from the common contacts.

\begin{lemma}\label{lem:combsum}
\begin{enumerate}[label=(\alph*)]
\item \label{it:combsum-a} Replacing $m$ by $m/\sigma$, $\sigma>0$, in a comb with contact set $\Gamma$, line $p+q\alpha$, dip $\kappa$, and width $\delta$ gives a comb with contact set $\sigma\Gamma$, line $p/\sigma+q\alpha/\sigma^2$, dip $\kappa/\sigma^2$, and width $\sigma\delta$.
\item \label{it:combsum-b} If $I_1$ and $I_2$ are combs on disjoint ground sets with the same width $\delta$, contact sets $\Gamma_1,\Gamma_2\subseteq\delta\Z$, lines $L_1,L_2$, and dips $\kappa_1,\kappa_2$, then $I_1\oplus I_2$ is a comb with contact set $\Gamma_1\cap\Gamma_2$, line $L_1+L_2$, dip $\min(\kappa_1,\kappa_2)$, and width $\delta/2$.
\end{enumerate}
\end{lemma}

\begin{proof}
\ref{it:combsum-a} Let $L(\alpha):= p+q\alpha$ be the line of the comb.
By \cref{lem:ops}(b), at the share $\alpha$ the reward of the instance with $m/\sigma$ in place of $m$ under its best response is $r(\xs(\alpha/\sigma))/\sigma$.
The new line is $L'(\alpha):= L(\alpha/\sigma)/\sigma=p/\sigma+q\alpha/\sigma^2$.
Since $\dist(\alpha/\sigma,\Gamma)=\dist(\alpha,\sigma\Gamma)/\sigma$, the inequality of \cref{def:comb} at the share $\alpha/\sigma$ gives, for every $\alpha\ge0$,
\begin{align*}
  L'(\alpha)-\frac{r(\xs(\alpha/\sigma))}{\sigma}
  &=\frac{L(\alpha/\sigma)-r(\xs(\alpha/\sigma))}{\sigma}
   \ge\frac{\kappa}{\sigma}\min\Bigl(\frac{\dist(\alpha,\sigma\Gamma)}{\sigma},\delta\Bigr)\\
  &=\frac{\kappa}{\sigma^2}\min\bigl(\dist(\alpha,\sigma\Gamma),\sigma\delta\bigr).
\end{align*}
At a contact $\sigma\gamma$ with $\gamma\in\Gamma$, the left-hand side is $\bigl(L(\gamma)-r(\xs(\gamma))\bigr)/\sigma=0$.
So the new instance is a comb with contact set $\sigma\Gamma$, line $L'$, dip $\kappa/\sigma^2$, and width $\sigma\delta$.

\ref{it:combsum-b} By \cref{lem:ops}(a), the reward of $I_1\oplus I_2$ is $r_1(\xs_1(\alpha))+r_2(\xs_2(\alpha))$, so the gaps add. Fix $\alpha\ge0$, let $d_i:=\dist(\alpha,\Gamma_i)$ and $\kappa:=\min(\kappa_1,\kappa_2)$. The gap of the sum is at least $\kappa_1\min(d_1,\delta)+\kappa_2\min(d_2,\delta)$. If $d_1\ge\delta/2$ or $d_2\ge\delta/2$, this is at least $\kappa\delta/2$. Otherwise the points $\gamma_i\in\Gamma_i$ nearest to $\alpha$ satisfy $|\gamma_1-\gamma_2|<\delta$, and as both lie in $\delta\Z$, $\gamma_1=\gamma_2\in\Gamma_1\cap\Gamma_2$ and $d_1=d_2=|\alpha-\gamma_1|\ge\dist(\alpha,\Gamma_1\cap\Gamma_2)$. Then the gap is at least $(\kappa_1+\kappa_2)|\alpha-\gamma_1|\ge\kappa\min(\dist(\alpha,\Gamma_1\cap\Gamma_2),\delta/2)$. On $\Gamma_1\cap\Gamma_2$ both gaps vanish.
\end{proof}

\subsection{The gadget}\label{sec:D-correct}

This subsection states the comb gadget and the reduction precisely and proves \cref{prop:correct}, assuming the gadget theorem. We give the reduction first to explain what the gadget must accomplish; its construction and proof follow in \cref{sec:D-overview,sec:D-base,sec:D-onelevel,sec:D-alllevels}. The complete proof of \cref{thm:nphard} is in \cref{sec:D-npcomplete}, after all its ingredients have been established.

The inputs of the gadget are perturbed powers of $8$.

\begin{definition}\label{def:family}
Let $\Fin$ be the set of all $\mu\in\Z^k_{>0}$, $k\ge1$, for which there is an integer $H\ge1$ with $H\,8^{i-1}\le\mu_i<\frac{11}{10}H\,8^{i-1}$ for $i\in[k]$.
\end{definition}

In every $\mu\in\Fin$, each entry exceeds the sum of the earlier ones, as $\sum_{j<i}\mu_j\le\frac{11}{10}H\sum_{j<i}8^{j-1}=\frac{11}{70}H(8^{i-1}-1)<H\,8^{i-1}\le\mu_i$. So the entries increase fast enough for the $2^k$ numbers $\mu^\top\chi$, $\chi\in\{0,1\}^k$, to be distinct. Together, the two bounds keep every input within the factor $11/10$ of $H\,8^{i-1}$. So every input exceeds the sum of the earlier ones by a wide margin, which separates the two windows of every level $j\ge1$ of the gadget (\cref{def:tower}) by a transition, the ratios of consecutive inputs lie in $(80/11,44/5)$ (\cref{lem:inputs}), and one explicit parameter rule (\cref{def:ruleP}) works at every level (\cref{thm:alllevels}).

The construction and proof occupy \cref{sec:D-overview,sec:D-base,sec:D-onelevel,sec:D-alllevels}; the proof is completed in \cref{sec:D-alllevels}.

\begin{theorem}\label{thm:gadget}
There is a polynomial-time algorithm that computes, for $\muin\in\Fin$ of length $k$, an integer $N_{\min}(\muin)\ge1$ and, for every integer $N\ge N_{\min}(\muin)$, an instance $I(\muin,N)\in\Ccut$ with $3k+1$ elements, a cut form of it, and rationals $p$ and $q>0$, such that $I(\muin,N)$ is a comb with contact set $\Gamma$, line $p+q\alpha$, dip $\kappa$, and width $\delta$ given by
\[
\Gamma=\big\{\delta(N+\chi^\top\muin):\chi\in\{0,1\}^k\big\},\qquad\kappa=\min\Big(1,\frac{1}{30\delta}\Big),\qquad\delta=\frac{1}{3\muin_1}.
\]
All numbers computed have $\Oh(k+\log\muin_k+\log N)$ bits.
\end{theorem}

The guarantees of \cref{thm:gadget} cannot extend to arbitrary positive integer inputs unless $\mathrm{P}=\mathrm{NP}$. For $\muin=a$, one exact best response at the share $\delta(N+T)$, compared with the line, would decide whether $T$ is a subset sum of $a$. \Cref{thm:gadget} guarantees the construction for perturbed powers of $8$. Their subset sums are distinct, and Subset Sum is easy on such inputs, greedily from the largest, so the argument above gives no contradiction. The reduction therefore needs two combs, one whose inputs carry the items as small perturbations and one for the target.

The reduction combines two combs of \cref{thm:gadget}: the first encodes the items, the second the target. Let $(a,T)$ be an instance of Subset Sum with $k\ge1$ and $a\in\Z^k_{>0}$.

\begin{lemma}\label{lem:coincidence}
Let $0\le T\le\sum_i a_i$, let $N_0\in\Z$ and $\delta_1>0$, and let $H:=10\sum_i a_i+1$. Define $\mu,\mu'\in\Z^k_{>0}$ by
\[
  \mu_i:=H\,8^{i-1}+a_i\qquad\text{and}\qquad\mu'_i:=H\,8^{i-1}\qquad\text{for }i\in[k],
\]
and the sets
\[
  \Gamma_1:=\bigl\{\delta_1(N_0+\mu^\top\chi):\chi\in\{0,1\}^k\bigr\}\qquad\text{and}\qquad
  \Gamma_2:=\bigl\{\delta_1(N_0+T+\mu'^\top\chi'):\chi'\in\{0,1\}^k\bigr\}.
\]
Then the common points of $\Gamma_1$ and $\Gamma_2$ are the points $\delta_1(N_0+\mu^\top\chi)$ with $a^\top\chi=T$. In particular, $\Gamma_1$ and $\Gamma_2$ meet if and only if $(a,T)$ is a yes-instance.
\end{lemma}

\begin{figure}[t]
\centering
\begin{tikzpicture}[x=1.4mm,y=1cm]
\draw[gray] (-4,0)--(37,0) (44,0)--(85,0) node[right,font=\scriptsize,text=black]{$\alpha/\delta_1-N_0$};
\draw[gray] (36.4,-0.1)--(37.6,0.1) (43.4,-0.1)--(44.6,0.1);
\draw[thick,blue!70!black] (-4,0.45)--(-1,0.45)--(0,0)--(1,0.45)--(31,0.45)--(32,0)--(33,0.45)--(37,0.45)
  (44,0.45)--(47,0.45)--(48,0)--(49,0.45)--(79,0.45)--(80,0)--(81,0.45)--(85,0.45);
\draw[thick,orange!85!black] (-4,-0.45)--(1,-0.45)--(2,0)--(3,-0.45)--(32,-0.45)--(33,0)--(34,-0.45)--(37,-0.45)
  (44,-0.45)--(47,-0.45)--(48,0)--(49,-0.45)--(78,-0.45)--(79,0)--(80,-0.45)--(85,-0.45);
\foreach \x/\p in {0/0,32/32,80/282} {
  \fill[blue!70!black] (\x,0) circle (1.4pt);
  \node[above,font=\scriptsize,text=blue!70!black] at (\x,0.45) {$\p$};}
\foreach \x/\p in {2/2,33/33,79/281} {
  \fill[orange!85!black] (\x,0) circle (1.4pt);
  \node[below,font=\scriptsize,text=orange!85!black] at (\x,-0.45) {$\p$};}
\fill[red!70!black] (48,0) circle (2.2pt);
\node[above,font=\scriptsize,text=blue!70!black] at (48,0.45) {$250$};
\node[below,font=\scriptsize,text=orange!85!black] at (48,-0.45) {$250$};
\foreach \x/\c/\d in {1/{(0,0)}/2,32.5/{(1,0)}/1,48/{(0,1)}/0,79.5/{(1,1)}/-1} {
  \node[font=\scriptsize,text=gray!70!black] at (\x,1.1) {$\c$};
  \node[font=\scriptsize,text=gray!70!black] at (\x,-1.1) {$\d$};}
\node[font=\scriptsize,anchor=east] at (-6,1.1) {$\chi$};
\node[font=\scriptsize,anchor=east,text=blue!70!black] at (-6,0.45) {$\Gamma_1$};
\node[font=\scriptsize,anchor=east,text=orange!85!black] at (-6,-0.45) {$\Gamma_2$};
\node[font=\scriptsize,anchor=east] at (-6,-1.1) {$T-a^\top\chi$};
\end{tikzpicture}
\caption{The sets of \cref{lem:coincidence} for the items $a=(1,2)$ and the target $T=2$, so that $H=31$, $\mu=(32,250)$, and $\mu'=(31,248)$, in units of $\delta_1$ and after the offset $N_0$, with the axis broken between $37$ and $246$. Schematically, the gap of a comb with the contact set $\Gamma_1$ is drawn above the axis, and that of a comb with the contact set $\Gamma_2$, mirrored, below it. The two contacts with the same $\chi$ differ by $T-a^\top\chi$, which vanishes only for $\chi=(0,1)$, the subset with sum $T$. Only there do both combs touch their lines (red).}\label{fig:coincidence}
\end{figure}
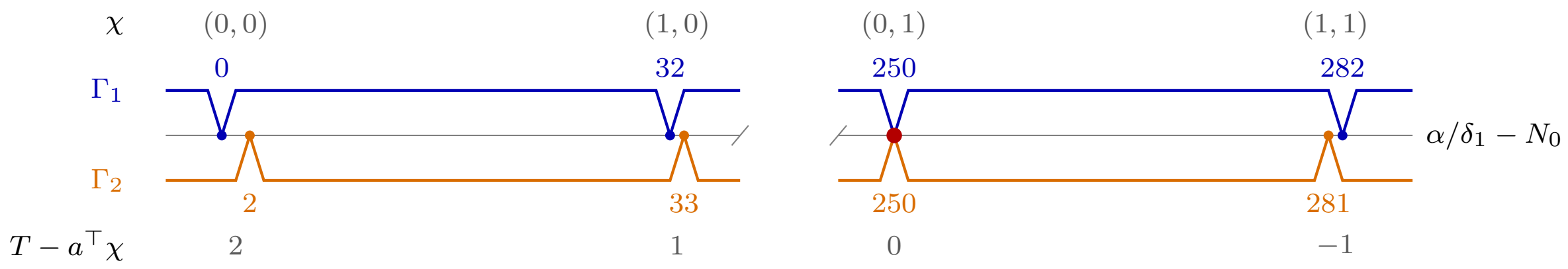

\Cref{fig:coincidence} shows the two sets for a small instance.

\begin{proof}
As $\mu^\top\chi=H\sum_i\chi_i8^{i-1}+a^\top\chi$ and $\mu'^\top\chi'=H\sum_i\chi'_i8^{i-1}$, two points coincide if and only if $H\sum_i(\chi_i-\chi'_i)8^{i-1}=T-a^\top\chi$. The right-hand side has absolute value less than~$H$, as $T,a^\top\chi\in[0,\sum_i a_i]$, and the left-hand side is a multiple of~$H$, so both vanish, and $a^\top\chi=T$. Conversely, $\chi':=\chi$ gives a common point if $a^\top\chi=T$.
\end{proof}

Both $\mu$ and $\mu'$ lie in $\Fin$, as $a_i\le\sum_j a_j<H/10$.

\begin{definition}\label{def:reduction}
Given $(a,T)$, proceed as follows.
\begin{enumerate}[label=(\arabic*),leftmargin=*]
\item\label{red:1} If $T<0$ or $T>\sum_i a_i$, replace $(a,T)$ by the no-instance $a=(2)$, $T=1$.
\item\label{red:2} With $H,\mu,\mu'$ as in \cref{lem:coincidence}, let $N_0:=\max(N_{\min}(\mu),N_{\min}(\mu'))$, $I_1:=I(\mu,N_0)$, and $I_2:=I(\mu',N_0+T)$ (\cref{thm:gadget}), with widths $\delta_1,\delta_2$, dips $\kappa_1,\kappa_2$, and lines $p_1+q_1\alpha$, $p_2+q_2\alpha$.
\item\label{red:3} Let $I_2'$ be $I_2$ with $m$ replaced by $(\delta_2/\delta_1)m$, let $I^*:=I_1\oplus I_2'$, $p:=p_1+(\delta_2/\delta_1)p_2$, $q:=q_1+(\delta_2/\delta_1)^2q_2$, and $\kappa^*:=\min(\kappa_1,(\delta_2/\delta_1)^2\kappa_2)$, and let $c$ and $\ell$ be the center and the half-length of $[\delta_1N_0,\delta_1(N_0+\sum_i\mu_i)]$.
\item\label{red:4} Let $\sigma$ be the largest power of $2$ with $\sigma\le1$, $\sigma\le\kappa^*\delta_1/(4q\ell^2)$, $\sigma(c+\ell)\le1/4$, and $\sigma(p+2qc)<q$. Let $\hat I$ be $I^*$ with $m$ replaced by $m/\sigma$, plus a capped linear summand (\cref{lem:ops}(c)), the \emph{centering summand}, with $\aknob:=\sigma\delta_1N_0$ and $C:=\big(q(1-2\sigma c)-\sigma p\big)/\sigma^2$.
\item\label{red:5} Output $\hat I$ and $\vartheta:=\frac{q}{\sigma^2}\big((1-\sigma c)^2-\sigma^2\ell^2\big)$.
\end{enumerate}
\end{definition}

Every condition in step~\ref{red:4} is an upper bound on $\sigma$ with a positive right-hand side, or vacuous, so $\sigma$ exists and is at least half the smallest of these right-hand sides. By the last condition $C>0$, and by the vertex formula of \cref{sec:bestresponse}, the centering summand moves the top of the parabola to the center $\sigma c$ of the window of contacts (\cref{fig:knob}).

\begin{proposition}\label{prop:correct}
Let $(\hat I,\vartheta)$ be the output for $(a,T)$. If $(a,T)$ is a yes-instance, then $\UP(\as)\ge\vartheta$ for $\hat I$. Otherwise $\UP(\as)\le\vartheta-q\ell^2/2<\vartheta$.
\end{proposition}

\begin{proof}
We first show that step~\ref{red:1} of \cref{def:reduction} does not change the answer.
If it applies, then $(a,T)$ is a no-instance, as $0\le a^\top\chi\le\sum_i a_i$ for every $\chi\in\{0,1\}^k$.
It is replaced by $a=(2)$, $T=1$, which is a no-instance as well, as $2\chi_1\ne1$, and which satisfies $0\le T\le\sum_i a_i=2$.
So the steps \ref{red:2} to \ref{red:5} are applied to a pair $(a,T)$ with $0\le T\le\sum_i a_i$ that is a yes-instance if and only if the given pair is one.
From now on, $(a,T)$ denotes this pair, and $0\le T\le\sum_i a_i$.

Next we show that $I^*$ is a comb whose contacts lie in the window $[c-\ell,c+\ell]$ and exist if and only if $(a,T)$ is a yes-instance.
Both $\mu$ and $\mu'$ lie in $\Fin$, and $N_0\ge N_{\min}(\mu)$ and $N_0+T\ge N_0\ge N_{\min}(\mu')$, so \cref{thm:gadget} applies to both calls in step \ref{red:2}.
So $I_1$ is a comb with the contact set $\Gamma_1$ of \cref{lem:coincidence}, line $p_1+q_1\alpha$, dip $\kappa_1$, and width $\delta_1$, and $I_2$ is a comb with contact set $\{\delta_2(N_0+T+\mu'^\top\chi')\colon\chi'\in\{0,1\}^k\}$, line $p_2+q_2\alpha$, dip $\kappa_2$, and width $\delta_2$, where $q_1,q_2>0$.
By \cref{lem:combsum}(a), applied with $\delta_1/\delta_2$ in place of $\sigma$, the instance $I_2'$ is a comb with the contact set $\Gamma_2$ of \cref{lem:coincidence}, line $(\delta_2/\delta_1)p_2+(\delta_2/\delta_1)^2q_2\alpha$, dip $(\delta_2/\delta_1)^2\kappa_2$, and width $\delta_1$.
Both contact sets lie in $\delta_1\Z$, as $N_0$, $T$, $\mu$, and $\mu'$ are integral.
So, by \cref{lem:combsum}(b), the instance $I^*=I_1\oplus I_2'$ is a comb with line $p+q\alpha$, dip $\kappa^*$, width $\delta_1/2$, and contact set $\Gamma^*:=\Gamma_1\cap\Gamma_2$.
By \cref{lem:coincidence}, $\Gamma^*\ne\emptyset$ if and only if $(a,T)$ is a yes-instance, and
\[
  \Gamma^*\subseteq\Gamma_1\subseteq\Bigl[\delta_1N_0,\ \delta_1\Bigl(N_0+\sum_i\mu_i\Bigr)\Bigr]=[c-\ell,c+\ell].
\]
Moreover, $q\ge q_1>0$ and $\ell=\tfrac12\delta_1\sum_i\mu_i>0$, and $c-\ell=\delta_1N_0>0$, as $N_0\ge1$.

Then we show that the reward of $\hat I$, the rescaled sum plus the centering summand, stays below a line and touches it at the contacts.
Let $I'$ be $I^*$ with $m$ replaced by $m/\sigma$.
We write $r'(x'^\star(\alpha))$ for the reward of $I'$ under its best response, and $\hat r(\hat x^\star(\alpha))$ for that of $\hat I$.
By \cref{lem:combsum}(a), $I'$ is a comb with contact set $\sigma\Gamma^*$, line $L'(\alpha):= p/\sigma+q\alpha/\sigma^2$, dip $\kappa^*/\sigma^2$, and width $\sigma\delta_1/2$.
The centering summand has $\aknob=\sigma\delta_1N_0>0$ and $C>0$, by the condition $\sigma(p+2qc)<q$ in step \ref{red:4}, so, by \cref{lem:ops}(c),
\[
  \hat r(\hat x^\star(\alpha))=r'(x'^\star(\alpha))+C\min(\alpha/\aknob,1)\qquad\text{for }\alpha\ge0.
\]
Put $\tilde L:= L'+C$, $c':=\sigma c$, and $\ell':=\sigma\ell$.
As $C\min(\alpha/\aknob,1)\le C$, we have $\tilde L(\alpha)-\hat r(\hat x^\star(\alpha))\ge L'(\alpha)-r'(x'^\star(\alpha))$ for all $\alpha\ge0$.
As $\Gamma^*\subseteq[c-\ell,c+\ell]$, the contacts satisfy $\sigma\Gamma^*\subseteq[c'-\ell',c'+\ell']$, and $c'-\ell'=\sigma\delta_1N_0=\aknob$.
So the centering summand adds $C$ at every contact, and $\tilde L(\gamma)=L'(\gamma)+C=\hat r(\hat x^\star(\gamma))$ for every $\gamma\in\sigma\Gamma^*$.

We now show that $(1-\alpha)\tilde L(\alpha)$ is a parabola with its top at the center $c'$ of the window of contacts.
Put $q':= q/\sigma^2$.
By the choice of $C$ in step \ref{red:4},
\[
  \tilde L(\alpha)=\frac p\sigma+\frac{q\alpha}{\sigma^2}+\frac q{\sigma^2}-\frac{2qc}\sigma-\frac p\sigma=q'(1-2c'+\alpha),
\]
so
\begin{equation}\label{eq:D-parabola}
  (1-\alpha)\tilde L(\alpha)=q'\bigl((1-c')^2-(\alpha-c')^2\bigr)=\vartheta+q\ell^2-q'(\alpha-c')^2,
\end{equation}
as $\vartheta=q'\bigl((1-c')^2-\ell'^2\bigr)$ and $q'\ell'^2=q\ell^2$.
Moreover, $c'+\ell'\le1/4$ by step \ref{red:4}.

With this, we prove the claim for yes-instances.
Let $(a,T)$ be a yes-instance.
Then $\Gamma^*\ne\emptyset$ by \cref{lem:coincidence}, and we let $\gamma\in\sigma\Gamma^*$.
Then $|\gamma-c'|\le\ell'$ and $0\le\gamma\le c'+\ell'\le1/4$, so $\tilde L(\gamma)=\hat r(\hat x^\star(\gamma))$ and \eqref{eq:D-parabola} give
\[
  \UP(\gamma)=(1-\gamma)\,\hat r(\hat x^\star(\gamma))=(1-\gamma)\tilde L(\gamma)\ge\vartheta+q\ell^2-q'\ell'^2=\vartheta.
\]
As $\gamma\in[0,1]$, we get $\UP(\as)\ge\UP(\gamma)\ge\vartheta$.

Finally, we prove the claim for no-instances.
Let $(a,T)$ be a no-instance.
Then $\Gamma^*=\emptyset$, so $\dist(\alpha,\sigma\Gamma^*)=\infty$ for every $\alpha$, and \cref{def:comb} for $I'$ gives $L'(\alpha)-r'(x'^\star(\alpha))\ge(\kappa^*/\sigma^2)(\sigma\delta_1/2)=\kappa^*\delta_1/(2\sigma)$ for all $\alpha\ge0$.
As $\tilde L(\alpha)-\hat r(\hat x^\star(\alpha))\ge L'(\alpha)-r'(x'^\star(\alpha))$, and by the condition $\sigma\le\kappa^*\delta_1/(4q\ell^2)$ in step \ref{red:4}, we get
\[
  \tilde L(\alpha)-\hat r(\hat x^\star(\alpha))\ge\frac{\kappa^*\delta_1}{2\sigma}\ge2q\ell^2\qquad\text{for all }\alpha\ge0.
\]
For $\alpha\in[0,1]$, therefore,
\[
  \UP(\alpha)=(1-\alpha)\,\hat r(\hat x^\star(\alpha))\le(1-\alpha)\tilde L(\alpha)-(1-\alpha)\,2q\ell^2,
\]
and we distinguish three cases.
If $|\alpha-c'|<\ell'$, then $\alpha<c'+\ell'\le1/4$, and \eqref{eq:D-parabola} gives $\UP(\alpha)\le\vartheta+q\ell^2-\frac34\cdot2q\ell^2$.
If $|\alpha-c'|\ge\ell'$ and $\alpha\le3/4$, then $(1-\alpha)\tilde L(\alpha)\le\vartheta+q\ell^2-q'\ell'^2=\vartheta$ by \eqref{eq:D-parabola}, and $\UP(\alpha)\le\vartheta-\frac14\cdot2q\ell^2$.
If $\alpha>3/4$, then $\alpha-c'\ge1/2$, as $c'\le1/4$, so \eqref{eq:D-parabola} gives
\[
  \UP(\alpha)\le(1-\alpha)\tilde L(\alpha)\le\vartheta+q\ell^2-\frac{q'}4\le\vartheta-3q\ell^2,
\]
as $q'/4=q/(4\sigma^2)\ge4q\ell^2$ by $\sigma\ell\le\sigma(c+\ell)\le1/4$.
In all cases $\UP(\alpha)\le\vartheta-q\ell^2/2$, so $\UP(\as)\le\vartheta-q\ell^2/2<\vartheta$, as $q>0$ and $\ell>0$.
\end{proof}

So the output instance has $\UP(\as)\ge\vartheta$ if $(a,T)$ is a yes-instance and $\UP(\as)\le\vartheta-q\ell^2/2$ otherwise.

\subsection{Overview of the gadget}\label{sec:D-overview}

This subsection and \cref{sec:D-base,sec:D-onelevel,sec:D-alllevels} prove
\cref{thm:gadget}. Before the details, we explain the idea.

As explained after \cref{thm:gadget}, some restriction of the inputs is
needed unless $\mathrm{P}=\mathrm{NP}$ (comparing reward and line at one share
is done exactly by \cref{lem:exactbr}). The inputs of the gadget therefore come from the family
$\Fin$ of \cref{def:family}: base-$8$ digits $H8^{i-1}\le\muin_i<\frac{11}{10}H8^{i-1}$
perturbed by small offsets.

The gadget is built from a three-element base in $k-1$ steps. We call the
instance after $j$ steps \emph{level}~$j$, so the base is level~$0$. Each step copies
the gap of the previous level into two windows of shares and so doubles the
contacts, from two on the base to $2^k$ on level $k-1$. The
base $f_0$ on $E_0=\{C_0,C_1,C_2\}$ is a polymatroid rank function whose reward
under the full-plane best response touches the line $L_0(\alpha)=10+7\alpha$ on
$[-1/6,\infty)$ at exactly two shares, the \emph{contacts} $\zeta_1=1/3$ and $\zeta_2=2/3$
(\cref{prop:base}). Step $j$ adds three elements: two \emph{dummies} $\tau_{1,j}$
and $\tau_{2,j}$, each a direct summand (a modular singleton with value
$F_{\tau_1}^{(j)}$, resp.\ $F_{\tau_2}^{(j)}$), and one element $b_j$ whose marginal value
$g_j(S\cup\{b_j\})-g_j(S)$, for the set function $g_j$ of level $j$, is
$K_j-d_j(S)$ with a constant $K_j$ and the modular function
\[
  d_j(S):=\mu_j\bigl(m_{*,j}\binv(S)-t(S)\bigr),\qquad \mu_j=\delta\muin_{j+1},
\]
which is nonnegative on its domain $E_{j-1}\cup\{\tau_{1,j},\tau_{2,j}\}$, as the constant $m_{*,j}$ is at least the reward of every element of this set. This is the
translation extension of \cref{lem:translation} with the translation vector
$(\mu_\alpha,\mu_\theta)=(\mu_j,m_{*,j}\mu_j)$, so $g_j$ is submodular
(\cref{lem:translation}(a)) and keeps a cut form (\cref{lem:cutforms}(c)).
By \cref{lem:translation}(b), for every set $S$ without $b_j$, the value
$F^{g_j}_{\alpha,\theta}(S\cup\{b_j\})$ is, up to a term independent of $S$,
the value at $S$ of the same function for level $j-1$ with the dummies, at the
point $(\alpha-\mu_j,\theta-m_{*,j}\mu_j)$. In particular
(\cref{lem:translation}(c)), $\xi^{h_j}(\alpha)=\xi^{f_j}(\alpha-\mu_j)-d_j$ for $h_j:=f_j-d_j$, where $f_j$ is level $j-1$ plus the dummies (\cref{def:tower}).

So the old best response reappears in two windows of shares
(\cref{thm:splice,lem:windows}). In the first window $\mathcal R_{j-1}=[R_0,R_1^{(j-1)}]$, which starts at
$R_0=\zeta_1-1/30$ and ends $1/30$ above the largest contact of level $j-1$
(\cref{def:tower}), the new element $b_j$ is merged at the bottom with $\tau_{1,j}$ (configuration
$(E\setminus\{\tau_{1,j}\},E)$), and the reward is the old one plus an affine function.
In the second window $\mathcal R_{j-1}+\mu_j$, translated by $\mu_j$, $b_j$ is
merged at the top with $\tau_{2,j}$ (configuration $(\emptyset,\{\tau_{2,j}\})$), and
the reward is the translated old one plus an affine function. A suitable
difference $\Delta F=F_{\tau_1}-F_{\tau_2}$ of the dummy values puts both copies of the old
gap under one new line $L_j=L_{j-1}+\psi_j\alpha+\Delta_p$ (\cref{lem:dmt}), so
that the old gap is copied unchanged into both windows and the contacts double:
$\Gamma_j=\Gamma_{j-1}\cup(\Gamma_{j-1}+\mu_j)$. After $k-1$ steps there are
$2^k$ contacts, the points $\zeta_1+\sum_{i\in Q}\mu_i$ with $Q\subseteq\{0,\dots,k-1\}$ and $\mu_0=1/3$ (\cref{def:tower}).

Between the two windows, on the \emph{transition} $[R_1^{(j-1)},R_0+\mu_j]$,
neither window description applies: $b_j$ crosses the whole old structure from
the bottom to the top, and old elements join and leave its block. There we show
that the gap first rises and then falls, and that it stays at least
$1/30$ (\cref{thm:transition}), under explicit conditions
(T1)--(T3) on the marginals and on weighted sums of squares. Together with the
window conditions (W1)--(W6) of \cref{lem:onelevel}, these are the only
requirements of one step.

All parameters of level $j$ are fixed by one explicit rule, the \emph{parameter rule}
(\cref{def:ruleP}, summarized in \cref{tab:ruleP}), which uses no best response
and has the same form at every level. The rewards are $\frac12M_{j-1}$ (dummies) and $M_j=12\cdot208^j$,
the reciprocal curvatures are $1$ and $1/4$ on the base, $\frac34\cdot8^{j-1}$ on
the dummies, and $9\cdot8^{j-1}$ on $b_j$, and the remaining numbers are given by
explicit formulas in the data of the previous levels. The work is to show that the rule
succeeds at \emph{every} level for \emph{every} input in $\Fin$. The two dummies
of a level weigh exactly as much as all old elements, and this invariant
(\cref{lem:weights}) bounds the numbers of every step uniformly in the level. We
reduce (W1)--(W6) and (T1)--(T3) to explicit inequalities in the input
(\cref{lem:inputs}) through this invariant, and conclude by induction
(\cref{thm:onestep,prop:level1,thm:alllevels}). The levels are not polymatroid
rank functions, and we analyze them on the base polyhedron $\Bpm{g}$ without
sign constraints (\cref{lem:fullplane}).

Finally, the last level $g_{k-1}$ on $3k$ elements has the right contacts, but
only on $[R_0,\infty)$. It is embedded into a polymatroid instance by adding
the modular function $\Lambda't-\Theta\binv$ for a suitable constant $\Theta>0$ (\cref{lem:embed} and
\cref{lem:fullplane}(g), with $\Theta$ fixed in the proof of \cref{thm:gadget}), which shifts the share by $\Lambda'=\delta N-\zeta_1$
and moves the contacts to $\Gamma=\{\delta(N+\chi^\top\muin)\colon\chi\in\{0,1\}^k\}$.
One capped linear summand (\cref{lem:ops}(c)) with the share $\aknob:=\alpha_\Gamma=\delta N$ repairs
the reward to the left of the contacts. This gives $3k+1$ elements, and all
numbers have polynomial bit length (\cref{lem:sizes} and the proof of \cref{thm:gadget}).

\subsection{The base and the steps}\label{sec:D-base}

\subsubsection{The base}

\begin{proposition}\label{prop:base}
Let $E_0:=\{C_0,C_1,C_2\}$, $m:=(5,12,4)$, and $\beta:=(1,4,4)$. We write $C_0$
for $\{C_0\}$, $C_0C_1$ for $\{C_0,C_1\}$, and so on. Let $f_0(\emptyset):=0$ and
\[
\begin{aligned}
  &f_0(C_0):=\tfrac13, && f_0(C_1):=\tfrac76, && f_0(C_2):=\tfrac23, &&\\
  &f_0(C_0C_1):=\tfrac76, && f_0(C_0C_2):=1, && f_0(C_1C_2):=\tfrac43, && f_0(E_0):=\tfrac43.
\end{aligned}
\]
Then $f_0$ is a polymatroid rank function, given by the cut form with the arcs
$C_0\to C_1$ of capacity $1/3$ and $C_1\to C_2$ of capacity $1/2$ and
the weights $0,2/3,2/3$ on $C_0,C_1,C_2$. Its full-plane best response
$\xi^{(0)}:=\xi^{f_0}$ has the marginals $\rho:=\rho^{f_0}$ and the reward
$r(\xi^{(0)}(\alpha))$ given by the following table.
\begin{center}
\begin{tabular}{lllll}
\toprule
piece & $\rho_{C_0}$ & $\rho_{C_1}$ & $\rho_{C_2}$ & $r(\xi^{(0)}(\alpha))$\\
\midrule
$(-\infty,-1/6]$ & $5\alpha-1/3$ & $12\alpha-4/3$ & $4\alpha-8/3$ & $25/3$\\[2pt]
$[-1/6,1/3]$ & $5\alpha-1/3$ & $8\alpha-2$ & $8\alpha-2$ & $8\alpha+29/3$\\[2pt]
$[1/3,3/7]$ & $5\alpha-1/3$ & $12\alpha-10/3$ & $4\alpha-2/3$ & $37/3$\\[2pt]
$[3/7,2/3]$ & $32\alpha/5-14/15$ & $32\alpha/5-14/15$ & $4\alpha-2/3$ & $49\alpha/5+122/15$\\[2pt]
$[2/3,\infty)$ & $5\alpha$ & $12\alpha-14/3$ & $4\alpha-2/3$ & $44/3$\\
\bottomrule
\end{tabular}
\end{center}
Let $\zeta_1:=1/3$, $\zeta_2:=2/3$, $L_0(\alpha):=p_0+q_0\alpha$ with
$p_0:=10$ and $q_0:=7$, $\varepsilon_0:=1/30$ and $R_0:=\zeta_1-\varepsilon_0=3/10$. Then the
following hold.
\begin{enumerate}[label=(B\arabic*),leftmargin=*]
\item The slopes of $\alpha\mapsto r(\xi^{(0)}(\alpha))$ on the five pieces are
  $0,8,0,49/5,0$.
\item On $[-1/6,\infty)$, $L_0(\alpha)\ge r(\xi^{(0)}(\alpha))$, with
  equality exactly at $\zeta_1$ and $\zeta_2$. On $[-1/6,\zeta_1]$,
  $L_0(\alpha)-r(\xi^{(0)}(\alpha))=\zeta_1-\alpha$, and
  $L_0(\alpha)-r(\xi^{(0)}(\alpha))\ge\dist(\alpha,\{\zeta_1,\zeta_2\})$ for every
  $\alpha\in[R_0,\infty)$. In particular, $L_0(\alpha)-r(\xi^{(0)}(\alpha))$ is
  $\varepsilon_0$ at $\alpha=R_0$ and $7\varepsilon_0$ at $\alpha=\zeta_2+\varepsilon_0$.
\item On $(-7/3,\zeta_1]$, $\rho_{C_0}$ is the strict maximum of the three
  marginals. At $R_0$, the marginals are $\rho(R_0)=(7/6,2/5,2/5)$,
  so $\max_{e\in E_0}\rho_e(R_0)=\rho_{C_0}(R_0)=7/6$.
\end{enumerate}
\end{proposition}

\begin{proof}
We first show that the cut form gives $f_0$. For $S\subseteq E_0$, the form gives
the value
$\omega(S)+\frac13\iverson{C_0\in S,\ C_1\notin S}+\frac12\iverson{C_1\in S,\ C_2\notin S}$,
where a bracket is $1$ if the conditions in it hold and $0$ otherwise. These
eight values agree with $f_0$, for instance $f_0(C_0C_2)=0+2/3+1/3=1$.
The weights $0,2/3,2/3$ are at least the capacities $0,1/3,1/2$
of the arcs entering $C_0,C_1,C_2$, so $f_0$ is a polymatroid rank function by
\cref{lem:cutforms}(b).

Next we show that the rows of the table are the full-plane marginals. Fix a piece and $\alpha$ in it, let $\rho(\alpha)$ be
the row of the table, and let $y:=\alpha t-\rho(\alpha)/\beta$
(componentwise). The following table gives $y$ and the upper level sets of
$\rho(\alpha)$ in the interior of the piece. At the ends of a piece, two
marginals may coincide, which only merges two level sets. On the first piece,
the smallest level set is $C_2$ left of $-7/3$ and $C_0$ right of it, and at
$-7/3$ itself it is $C_0C_2$.
\begin{center}
\begin{tabular}{lll}
\toprule
piece & $y=(y_{C_0},y_{C_1},y_{C_2})$ & upper level sets\\
\midrule
$(-\infty,-1/6]$ & $(1/3,1/3,2/3)$ & $C_2$ or $C_0$, $C_0C_2$, $E_0$\\[2pt]
$[-1/6,1/3]$ & $(1/3,\alpha+1/2,1/2-\alpha)$ & $C_0$, $E_0$\\[2pt]
$[1/3,3/7]$ & $(1/3,5/6,1/6)$ & $C_0$, $C_0C_1$, $E_0$\\[2pt]
$[3/7,2/3]$ & $(14/15-7\alpha/5,7\alpha/5+7/30,1/6)$ & $C_0C_1$, $E_0$\\[2pt]
$[2/3,\infty)$ & $(0,7/6,1/6)$ & $C_1$, $C_0C_1$, $E_0$\\
\bottomrule
\end{tabular}
\end{center}
On every piece, $y(E_0)=4/3$, the listed level sets are tight, and
$y(S)\le f_0(S)$ for all $S\subseteq E_0$. On the bounded pieces $y$ is affine,
so it suffices to check this at the end points, and on the two unbounded pieces
$y$ is constant. By \cref{lem:fullplane}(a), $y=\xi^{(0)}(\alpha)$ and
$\rho(\alpha)=\rho^{f_0}(\alpha)$, and $r(\xi^{(0)}(\alpha))=m^\top y$, which is
the last column of the first table.

Finally, we derive (B1)--(B3). (B1) is read off the first table. On the four pieces right
of $-1/6$, the difference $L_0(\alpha)-r(\xi^{(0)}(\alpha))$ equals
$\zeta_1-\alpha$, $7(\alpha-\zeta_1)$, $\frac{14}5(\zeta_2-\alpha)$ and
$7(\alpha-\zeta_2)$, respectively. For example,
$L_0(\alpha)-r(\xi^{(0)}(\alpha))=10+7\alpha-(8\alpha+29/3)=1/3-\alpha$
on $[-1/6,1/3]$. All four coefficients are at least $1$, which gives
(B2), with the values $\varepsilon_0$ at $R_0$ and $7\varepsilon_0$ at
$\zeta_2+\varepsilon_0$. (B3) is read off the first table, with
$\rho_{C_0}(R_0)=5R_0-1/3=7/6$ and
$\rho_{C_1}(R_0)=\rho_{C_2}(R_0)=8R_0-2=2/5$.
\end{proof}

\subsubsection{The steps and the parameter rule}

\begin{definition}\label{def:tower}
Let $\muin\in\Fin$ have length $k$, put $\delta:=1/(3\muin_1)$ and
$\mu_j:=\delta\muin_{j+1}$ for $0\le j\le k-1$, so that
$\mu_0=1/3=\zeta_2-\zeta_1$. Level $0$ is $g_0:=f_0$ on $E_0$. For $j\ge1$,
step $j$ adds three elements $\tau_{1,j},\tau_{2,j},b_j$ to level $j-1$: the \emph{dummies} $\tau_{1,j}$ and
$\tau_{2,j}$ with the reward $m_\delta^{(j)}$ and the curvature $\beta_\delta^{(j)}$,
and $b_j$ with the reward $M_j$ and the curvature $\beta_\delta^{(j)}/12$. (The subscript $\delta$ of $m_\delta$ and $\beta_\delta$ is a label, unrelated to the width $\delta$.) With
$E_j:=E_{j-1}\cup\{\tau_{1,j},\tau_{2,j},b_j\}$, let $f_j$ be the direct sum of $g_{j-1}$
and of the modular singletons on $\{\tau_{1,j}\}$ and on $\{\tau_{2,j}\}$ with the values
$F_{\tau_1}^{(j)}$ and $F_{\tau_2}^{(j)}$, that is,
\[
  f_j(S):=g_{j-1}(S\cap E_{j-1})+F_{\tau_1}^{(j)}\iverson{\tau_{1,j}\in S}+F_{\tau_2}^{(j)}\iverson{\tau_{2,j}\in S}
  \qquad(S\subseteq E_{j-1}\cup\{\tau_{1,j},\tau_{2,j}\}),
\]
where, as in the proof of \cref{prop:base}, a bracket is $1$ if the condition in
it holds and $0$ otherwise, and let
\[
  d_j(e):=\frac{\mu_j(m_{*,j}-m_e)}{\beta_e}\quad\text{for }e\in E_{j-1}\cup\{\tau_{1,j},\tau_{2,j}\}
  \qquad\text{and}\qquad h_j:=f_j-d_j.
\]
Let $g_j$ be the extension of $f_j$ by $b_j$ with $d_j$ and a constant $K_j$, as
in the setting of \cref{thm:splice}, that is, $g_j(S):=f_j(S)$ and
$g_j(S\cup\{b_j\}):=f_j(S)+K_j-d_j(S)$ for $S\subseteq E_{j-1}\cup\{\tau_{1,j},\tau_{2,j}\}$.
All numbers with index $j$ are fixed by the parameter rule (\cref{def:ruleP} below). We
write $\xi^{(j)}:=\xi^{g_j}$ and $\rho^{(j)}:=\rho^{g_j}$ for the full-plane best
response of level $j$ and its marginals, so that $\xi^{(0)}$ is the one of
\cref{prop:base}, and
\[
\begin{aligned}
  R_1^{(j)}&:=\zeta_1+\sum_{i=0}^{j}\mu_i+\varepsilon_0,\qquad
  \mathcal R_j:=[R_0,R_1^{(j)}],\\
  \Gamma_j&:=\Bigl\{\zeta_1+\sum_{i\in Q}\mu_i\colon Q\subseteq\{0,\dots,j\}\Bigr\},
\end{aligned}
\]
and $L_j(\alpha):=p_j+q_j\alpha$, with $p_0$ and $q_0$ as in \cref{prop:base}.
We call $L_j(\alpha)-r(\xi^{(j)}(\alpha))$ the \emph{gap} of level $j$ at
$\alpha$. For $j\ge1$, we call $\mathcal R_{j-1}$ and $\mathcal R_{j-1}+\mu_j$
the \emph{first} and the \emph{second window} of level $j$.
\end{definition}

So $\Gamma_{j+1}=\Gamma_j\cup(\Gamma_j+\mu_{j+1})$ and
$\mathcal R_{j+1}=\mathcal R_j\cup[R_1^{(j)},R_0+\mu_{j+1}]\cup(\mathcal R_j+\mu_{j+1})$.
The middle interval is the \emph{transition} between the two windows of level
$j+1$. Every level is a doubly merged translation of the previous one. We will
show that the contacts of level $j$ with $L_j$ on $[R_0,\infty)$, the shares at
which its gap vanishes, are exactly the points of $\Gamma_j$, which has $2^{j+1}$
points.

The parameter rule below fixes, by explicit formulas in the data of the levels
built so far, the rewards and curvatures of the three new elements, the constant
$m_*$ of the translation, the values $F_{\tau_1}$ and $F_{\tau_2}$ of the dummies, the
constant $K$, and the new line $L_{k+1}$. These are the numbers with which the
gadget works. Their roles are listed after the rule
and in \cref{tab:ruleP}, and \cref{sec:D-onelevel,sec:D-alllevels} prove that
every step succeeds with them. Besides the data of the previous levels, (P6)
uses the constant $\iota:=1/100$, the length of a short stretch
$[R_0-\iota,R_0]$ to the left of the windows. On it, level $j$ keeps the gap
$\zeta_1-\alpha$ of the base (\cref{lem:twosides}(b)), which the transition of
level $j+1$ needs at its right end (\cref{thm:transition}). (P6) and condition
(W1) of \cref{lem:onelevel} leave room for this stretch.

\begin{definition}\label{def:ruleP}
Let $k\ge0$, suppose that levels $0,\dots,k$ have been built, and write
$\mu:=\mu_{k+1}$, $R_1:=R_1^{(k)}$, $\mathcal R:=\mathcal R_k=[R_0,R_1]$,
$|\mathcal R|:=R_1-R_0$, $\tau_1:=\tau_{1,k+1}$, $\tau_2:=\tau_{2,k+1}$, $b:=b_{k+1}$, and
$E:=E_k\cup\{\tau_1,\tau_2\}$. Quantities of level $k+1$ are written without index. Up
to \cref{prop:level1}, $k$ is the level before the step, not the length of
$\muin$.
\begin{enumerate}[label=(P\arabic*),leftmargin=*]
\item $M:=M_k$, $m_\delta:=M/2$, and $m_*:=m_\delta+64M$.
\item $1/\beta_\delta:=\frac34\cdot8^k$ and $\beta_b:=\beta_\delta/12$.
\item $w:=64M\mu$, $d_e:=\mu(m_*-m_e)/\beta_e$ for $e\in E_k$, and
  $d_{\tau_1}=d_{\tau_2}:=w/\beta_\delta$, so that $d=m_*\mu\binv-\mu t$ on $E$, and
  $\bar d:=d(E_k)/d_{\tau_1}$.
\item $\Delta_b:=\frac{415}2M$. Put
  $m_b:=m_\delta+\Delta_b$, $s:=\frac{12}{13}\Delta_b$, $\bar m:=m_\delta+s$, and
  $\psi_{k+1}:=\frac{12}{13}\Delta_b^2/\beta_\delta$.
\item With $\Pi:=\sum_{e\in E_k}(m_\delta-m_e)d_e$, let
  $\Delta F:=w/\beta_\delta+d(E_k)-(q_k\mu-\Pi)/s$.
\item Let $\bar\rho_0:=7/6$, let
  $\bar\rho_k:=R_0m_\delta^{(k)}-\beta_\delta^{(k)}F_{\tau_2}^{(k)}$ for $k\ge1$, and let
  \[
    u^0:=a_2\Delta_b-\bigl(\tfrac12+\bar d\bigr)w+\frac{a_0}{\Delta_b},
  \]
  where $a_2:=\frac{12}{13}\bigl(\mu-|\mathcal R|/2\bigr)$ and
  $a_0:=\beta_\delta\bigl(q_k\mu-\Pi\bigr)$. Let
  $F_{\tau_2}:=(R_0m_\delta-\bar\rho_k-w+u^0-s\iota)/\beta_\delta$, so that
  $R_0m_\delta-\beta_\delta F_{\tau_2}=\bar\rho_k+w-u^0+s\iota$, and let
  $F_{\tau_1}:=F_{\tau_2}+\Delta F$.
\item $K:=d(E)+12F_{\tau_1}-13\bigl(w/2-s(R_0+R_1)/2\bigr)/\beta_\delta$.
\item $q_{k+1}:=q_k+\psi_{k+1}$ and $p_{k+1}:=p_k+\Delta_p$ with
  $\Delta_p:=m_\delta F_{\tau_2}+(K-d(E)+F_{\tau_1})\bar m$.
\end{enumerate}
Level $k+1$ is then given by \cref{def:tower} with $m_\delta^{(k+1)}:=m_\delta$,
$\beta_\delta^{(k+1)}:=\beta_\delta$, $M_{k+1}:=m_b$, $m_{*,k+1}:=m_*$,
$F_{\tau_1}^{(k+1)}:=F_{\tau_1}$, $F_{\tau_2}^{(k+1)}:=F_{\tau_2}$, and $K_{k+1}:=K$, so $d_{k+1}=d$. We
also set $M_0:=12$.
\end{definition}

\begin{table}[htbp]
\centering
\small
\renewcommand{\arraystretch}{1.15}
\begin{tabularx}{\textwidth}{@{}ll>{\raggedright\arraybackslash}p{0.35\textwidth}>{\raggedright\arraybackslash}X@{}}
\toprule
item & parameter & value at the step $k\to k+1$ & role\\
\midrule
(P1) & $M$ & $M_k$ & largest reward of $E_k$ (that of $b_k$ for $k\ge1$, of $C_1$ for $k=0$), $=12\cdot208^k$\\
 & $m_\delta$ & $M/2$ & reward of the two dummies $\tau_1,\tau_2$\\
 & $m_*$ & $m_\delta+64M$ & slope of the translation vector $(\mu,m_*\mu)$, by which the old level is shifted in the $(\alpha,\theta)$-plane (\cref{lem:translation}(b)). As $m_*\ge m_e$ for all $e\in E$, $d\ge0$\\
(P2) & $1/\beta_\delta$ & $\frac34\cdot8^k$ & reciprocal curvature of the dummies, which together weigh as much as the old elements: $\binv(\{\tau_1,\tau_2\})=\binv(E_k)$ (\cref{lem:weights}(a))\\
 & $\beta_b$ & $\beta_\delta/12$ & curvature of $b$, with $1/\beta_b+1/\beta_\delta=13/\beta_\delta$\\
(P3) & $w$ & $64M\mu$ & width of the bands of $\tau_1$ and $\tau_2$, $=(m_*-m_\delta)\mu$\\
 & $d_e$, $d_{\tau_1}=d_{\tau_2}$ & $\mu(m_*-m_e)/\beta_e$ ($e\in E_k$), $w/\beta_\delta$ & translation correction, $d=m_*\mu\binv-\mu t$ (\cref{lem:translation}(c))\\
 & $\bar d$ & $d(E_k)/d_{\tau_1}$ & normalized correction of the old elements\\
(P4) & $\Delta_b$ & $\frac{415}2M$ & places $b$ in the band of $\tau_2$ in the second window at a fraction $u^0/w$ of its width between $0.23$ and $0.52$ (\cref{lem:weights}(d) and \cref{prop:level1})\\
 & $m_b$ & $m_\delta+\Delta_b$ & reward of $b$, $=208M=M_{k+1}$\\
 & $s$ & $\frac{12}{13}\Delta_b$ & $=\bar m-m_\delta$, the slope of $u_1$ and $u_2$\\
 & $\bar m$ & $m_\delta+s$ & weighted mean reward of $\{b,\tau_1\}$ and of $\{b,\tau_2\}$\\
 & $\psi_{k+1}$ & $\frac{12}{13}\Delta_b^2/\beta_\delta$ & $=\SSq(\{b,\tau_1\})=\SSq(\{b,\tau_2\})$, the slope added to the line\\
(P5) & $\Pi$ & $\sum_{e\in E_k}(m_\delta-m_e)d_e$ & auxiliary sum for $\Delta F$ and $a_0$\\
 & $\Delta F$ & $w/\beta_\delta+d(E_k)-(q_k\mu-\Pi)/s$ & $=F_{\tau_1}-F_{\tau_2}$, aligns the two windows on one line (\cref{lem:dmt}(c))\\
(P6) & $\bar\rho_k$ & $7/6$ if $k=0$, and $R_0m_\delta^{(k)}-\beta_\delta^{(k)}F_{\tau_2}^{(k)}$ if $k\ge1$ & largest marginal of level $k$ at $R_0$\\
 & $a_2$, $a_0$ & $\frac{12}{13}(\mu-|\mathcal R|/2)$, $\beta_\delta(q_k\mu-\Pi)$ & auxiliary coefficients of $u^0$\\
 & $u^0$ & $a_2\Delta_b-(1/2+\bar d)w+a_0/\Delta_b$ & $=u_2(R_0+\mu)$: position of $b$ above the lower end of the band of $\tau_2$ at the start of the second window\\
 & $F_{\tau_2}$ & $(R_0m_\delta-\bar\rho_k-w+u^0-s\iota)/\beta_\delta$ & value of the dummy $\tau_2$. It sets the marginal of $\tau_2$ at $R_0$ to $\bar\rho_k+w-u^0+s\iota$, so that $b$ enters the second window just above the translated old elements, with margin $s\iota$\\
 & $F_{\tau_1}$ & $F_{\tau_2}+\Delta F$ & value of the dummy $\tau_1$\\
(P7) & $K$ & $d(E)+12F_{\tau_1}-13(w/2-s(R_0+R_1)/2)/\beta_\delta$ & centers $b$ in the band of $\tau_1$ in the first window ($u_1=w/2$ at the midpoint of $\mathcal R$)\\
(P8) & $q_{k+1}$ & $q_k+\psi_{k+1}$ & slope of the new line $L_{k+1}$\\
 & $p_{k+1}$ & $p_k+\Delta_p$, $\Delta_p:=m_\delta F_{\tau_2}+(K-d(E)+F_{\tau_1})\bar m$ & intercept of the new line $L_{k+1}$\\
\bottomrule
\end{tabularx}
\caption{The parameter rule at the step $k\to k+1$. The bands of $\tau_1$ and $\tau_2$ and the positions $u_1,u_2$ of $b$ in them are defined in \cref{sec:D-onelevel}.}\label{tab:ruleP}
\end{table}

The rule is \emph{trace-free}: it uses the data of the levels built so far, but
no best response. It is also explicit, and it has the same form at every level.
We have $M_{k+1}=\frac12M_k+\frac{415}2M_k=208M_k$, so $M_k=12\cdot208^k$. The
rewards added in step $j\ge1$ are $\frac12M_{j-1}$ (the dummies) and $M_j$, so all
rewards are integers and $M_k$ is the largest reward of $E_k$. So, at the step
$k\to k+1$, $m_*=m_\delta+64M\ge M\ge m_e$ for every $e\in E$, hence $d\ge0$, and
$g_{k+1}$ is an extension of $f_{k+1}$ as in the setting of \cref{thm:splice}.
Moreover, $1/\beta_\delta^{(k)}=\frac34\cdot8^{k-1}$ for $k\ge1$. The items of the rule have a
geometric meaning, which \cref{sec:D-onelevel} makes precise with the bands of $\tau_1$ and $\tau_2$ and the positions $u_1,u_2$ of $b$ in them (\cref{lem:closedforms}).
\begin{itemize}
\item (P2) makes the new dummies comparable to the old elements: together, $\tau_1$
  and $\tau_2$ weigh exactly as much as $E_k$, that is,
  $\binv(\{\tau_1,\tau_2\})=\binv(E_k)$ (\cref{lem:weights}(a)).
\item (P4) chooses the reward of $b$ such that $b$ enters the band of $\tau_2$ in
  the second window at a fraction $u^0/w$ of its width between $0.23$ and
  $0.52$ (\cref{lem:weights}(d) and \cref{prop:level1}).
\item (P5) aligns the two windows on one line (\cref{lem:dmt}(c)).
\item (P6) places $\tau_2$ such that $b$, merged with $\tau_2$, enters the second
  window just above the translated old elements, with a margin $s\iota$.
\item (P7) centers $b$ in the band of $\tau_1$ in the first window.
\end{itemize}
Together, (P4) and (P6) put the lower end of the band of $\tau_2$ about $u^0$
below the largest translated old marginal at the start of the second window,
which is more than the spread of the old marginals and their drift across the
transition. This gives the order (T2) of the marginals on the transition
(\cref{thm:transition}). The rule has no roundings, and its numbers still have
polynomial bit length (\cref{lem:sizes}).

\subsection{One level}\label{sec:D-onelevel}

\subsubsection{The two windows}

Consider the step $k\to k+1$, and write $f:=f_{k+1}$, $h:=f-d$,
$\rho':=\rho^{(k+1)}$, and $\theta_b$ for the marginal of $b$, as in the setting
of \cref{thm:splice} with $f'=g_{k+1}$. By \cref{lem:fullplane}(f) and
\cref{lem:translation}(c) with the translation vector $(\mu,m_*\mu)$, every
$e\in E_k$ satisfies
\begin{equation}\label{eq:D-bands}
  \rho^f_e(\alpha)=\rho^{(k)}_e(\alpha),\qquad
  \rho^h_e(\alpha)=\rho^{(k)}_e(\alpha-\mu)+m_*\mu,
\end{equation}
and the dummies have $\rho^f_{\tau_1}(\alpha)=\alpha m_\delta-\beta_\delta F_{\tau_1}$,
$\rho^f_{\tau_2}=\rho^f_{\tau_1}+\varphi w$ with $\varphi:=\beta_\delta\Delta F/w$, and
$\rho^h_{\tau_1}=\rho^f_{\tau_1}+w$ and $\rho^h_{\tau_2}=\rho^f_{\tau_2}+w$, since
$m_*\mu-\mu m_\delta=w$. So $\tau_1$ and $\tau_2$ have the \emph{bands}
$[\rho^f_{\tau_1},\rho^h_{\tau_1}]$ and $[\rho^f_{\tau_2},\rho^h_{\tau_2}]$ of width $w$, and the first
lies below the second if $\varphi>1$. The aim is to merge $b$ with $\tau_1$ on the
window $\mathcal R$ and with $\tau_2$ on $\mathcal R+\mu$. By
\cref{lem:windows}(a), $b$ then has the marginal $\theta_1$ on $\mathcal R$ and
$\theta_2$ on $\mathcal R+\mu$, where
\[
\begin{aligned}
  \theta_1(\alpha)&:=\frac{\alpha(t_b+t_{\tau_1})-(K-d(E)+F_{\tau_1})}{1/\beta_b+1/\beta_\delta},\\
  \theta_2(\alpha)&:=\frac{\alpha(t_b+t_{\tau_2})-(K+F_{\tau_2}-w/\beta_\delta)}{1/\beta_b+1/\beta_\delta}
\end{aligned}
\]
are the numbers $\theta_{U,W}$ of \cref{lem:windows} for
$(U,W)=(E\setminus\{\tau_1\},E)$ and for $(U,W)=(\emptyset,\{\tau_2\})$, respectively.
We measure the position of $b$ in the bands by $u_1:=\theta_1-\rho^f_{\tau_1}$ and
$u_2:=\theta_2-\rho^f_{\tau_2}$.

\begin{lemma}\label{lem:closedforms}
At the step $k\to k+1$, the following hold.
\begin{enumerate}[label=(\alph*)]
\item $d(E)=(2+\bar d)w/\beta_\delta$ and $\varphi=1+\bar d-a_0/(sw)$.
\item $u_1(\alpha)=s\bigl(\alpha-\frac{R_0+R_1}2\bigr)+\frac w2$ and
  $u_2(\alpha)=u_1(\alpha)-(1+\bar d)w+a_0/\Delta_b$ for every $\alpha\in\R$. In
  particular, $u_2(R_0+\mu)=u^0$, and $u_1$ and $u_2$ increase with slope $s$.
\item $\rho^f_{\tau_2}(\alpha)=(\alpha-R_0)m_\delta+\bar\rho_k+w-u^0+s\iota$ for every
  $\alpha\in\R$. In particular, $\bar\rho_{k+1}=\rho^f_{\tau_2}(R_0)=\bar\rho_k+w-u^0+s\iota$.
\end{enumerate}
\end{lemma}

\begin{proof}
(a) By (P3), $d(E)=d(E_k)+2w/\beta_\delta=(2+\bar d)w/\beta_\delta$, and (P5) with
the definition of $a_0$ in (P6) gives $\varphi=\beta_\delta\Delta F/w=1+\bar d-a_0/(sw)$.

(b) As $1/\beta_b+1/\beta_\delta=13/\beta_\delta$ and
$t_b+t_{\tau_1}=t_b+t_{\tau_2}=(12m_b+m_\delta)/\beta_\delta$, we have
$\theta_1(\alpha)=\alpha\bar m-\beta_\delta(K-d(E)+F_{\tau_1})/13$. As
$\rho^f_{\tau_1}(\alpha)=\alpha m_\delta-\beta_\delta F_{\tau_1}$ and $\bar m-m_\delta=s$, this
gives $u_1(\alpha)=s\alpha-\beta_\delta(K-d(E)-12F_{\tau_1})/13$, and (P7) gives the
formula for $u_1$. In the same way,
$u_2(\alpha)=s\alpha-\beta_\delta(K-12F_{\tau_2}-w/\beta_\delta)/13$, so
\[
  u_2-u_1= \frac{\beta_\delta\bigl(w/\beta_\delta-d(E)-12\Delta F\bigr)}{13}.
\]
Inserting (a), (P5), and the definition of $a_0$ in (P6) gives
$$u_2-u_1=-w-\beta_\delta d(E_k)+\frac{12}{13}\cdot \frac{a_0}{s} =-(1+\bar d)w+a_0/\Delta_b,$$
as
$\beta_\delta d(E_k)=\bar dw$ and $s=\frac{12}{13}\Delta_b$. At $R_0+\mu$,
$u_1(R_0+\mu)=s(\mu-|\mathcal R|/2)+w/2=a_2\Delta_b+w/2$, which gives
$u_2(R_0+\mu)=u^0$.

(c) We have
$\rho^f_{\tau_2}(R_0)=R_0m_\delta-\beta_\delta F_{\tau_2}=R_0m_\delta-\beta_\delta(F_{\tau_1}-\Delta F)$,
which is $\bar\rho_k+w-u^0+s\iota$ by (P6), and $\rho^f_{\tau_2}$ has slope
$m_\delta$. The last claim is the definition of $\bar\rho_{k+1}$ in (P6), as
$m_\delta^{(k+1)}=m_\delta$, $\beta_\delta^{(k+1)}=\beta_\delta$, and
$F_{\tau_2}^{(k+1)}=F_{\tau_2}$.
\end{proof}

\subsubsection{One level}

The marginals of the old elements and of the dummies grow at most at the rate
$M$, by \eqref{eq:D-bands} and \cref{lem:fullplane}(c). The proofs below need
the marginal of $b$ to be faster than all of them.

\begin{lemma}\label{lem:fast}
Consider the step $k\to k+1$, and suppose that $t_b/\binv(E\cup\{b\})>M$. Then
$\bar m>M$, and $\theta_b-\rho^f_e$ and $\theta_b-\rho^h_e$ are continuous and
strictly increasing for every $e\in E$.
\end{lemma}

\begin{proof}
As $\binv(E\cup\{b\})\ge\binv(\{b,\tau_1\})=13/\beta_\delta$ and
$t_b=12m_b/\beta_\delta$, we have
$M<t_b/\binv(E\cup\{b\})\le\frac{12}{13}m_b\le\frac{12}{13}m_b+\frac1{13}m_\delta=\bar m$,
the last step by (P4). By
\cref{lem:fullplane}(c), applied to $g_{k+1}$ on $E\cup\{b\}$, $\theta_b$ is
continuous and grows at least at the rate $c:=t_b/\binv(E\cup\{b\})$. For
$e\in E_k$, $\rho^f_e$ and $\rho^h_e$ grow at most at the rate
$\max_{e'\in E_k}m_{e'}=M$, by \eqref{eq:D-bands} and \cref{lem:fullplane}(c),
and the marginals of $\tau_1$ and $\tau_2$ are affine with slope $m_\delta<M$. As
$c>M$, every difference $\theta_b-\rho^f_e$ and $\theta_b-\rho^h_e$ is
continuous and grows at least at the rate $c-M>0$.
\end{proof}

The next lemma gives conditions under which a level has the two windows of
\cref{lem:dmt}, and it describes the level to the right of its second window.

\begin{lemma}\label{lem:onelevel}
Consider the step $k\to k+1$, and assume the following.
\begin{enumerate}[label=(W\arabic*),leftmargin=*]
\item $s\iota<u_1(R_0)$ and $u_1(R_1)<w$.
\item $0<u_2(R_0+\mu)$ and $u_2(R_1+\mu)<w$.
\item $\rho^{(k)}_e(\alpha)>\rho^h_{\tau_1}(\alpha)$ for all $e\in E_k$ and
  $\alpha\in\mathcal R$.
\item $\rho^h_e(\alpha)<\theta_2(\alpha)$ for all $e\in E_k$ and
  $\alpha\in\mathcal R+\mu$.
\item $\varphi>1$.
\item $t_b/\binv(E\cup\{b\})>M$.
\end{enumerate}
Let $\alpha_{\tau_2}$ be the zero of $u_2-w$, the share at which $b$ reaches the top
of the band of $\tau_2$. Then $\alpha_{\tau_2}>R_1+\mu$, and the following hold.
\begin{enumerate}[label=(\alph*)]
\item The configuration of level $k+1$ is $(E\setminus\{\tau_1\},E)$, with
  $\theta_b=\theta_1$, on $\mathcal R$. It is $(\emptyset,\{\tau_2\})$, with
  $\theta_b=\theta_2$, on $[R_0+\mu,\alpha_{\tau_2}]$, and it is $(\emptyset,\emptyset)$
  on $(\alpha_{\tau_2},\infty)$.
\item The gap $L_{k+1}(\alpha)-r(\xi^{(k+1)}(\alpha))$ of level $k+1$ equals
  \[
  \begin{array}{ll}
    L_k(\alpha)-r(\xi^{(k)}(\alpha)) & \text{for }\alpha\in\mathcal R,\\
    L_k(\alpha-\mu)-r(\xi^{(k)}(\alpha-\mu)) & \text{for }\alpha\in[R_0+\mu,\alpha_{\tau_2}],\\
    L_k(\alpha-\mu)-r(\xi^{(k)}(\alpha-\mu))+\psi_{k+1}(\alpha-\alpha_{\tau_2}) & \text{for }\alpha\in[\alpha_{\tau_2},\infty).
  \end{array}
  \]
\item On $\mathcal R$, the smallest marginal of level $k+1$ is $\theta_1$ and the
  largest is $\rho^f_{\tau_2}$. On $\mathcal R+\mu$, the largest is $\theta_2$ and the
  smallest is $\rho^h_{\tau_1}$.
\end{enumerate}
\end{lemma}

In words, (W1) and (W2) keep $b$ inside the band of $\tau_1$ on the first window and
inside the band of $\tau_2$ on the second. (W3) and (W4) keep the old elements
above $b$ on the first window and below $b$ on the second. (W5) puts the band of
$\tau_1$ below that of $\tau_2$, and (W6) makes $b$ faster than every other element.
The margin $s\iota$ in (W1) is used only in \cref{lem:twosides}(b). By
\cref{lem:fast}, (W6) implies $\bar m>M$, and every difference
$\theta_b-\rho^f_e$ and $\theta_b-\rho^h_e$, $e\in E$, is strictly increasing.

\begin{proof}
We first show that $b$ is merged with $\tau_1$ on the first window $\mathcal R$. Let
$\alpha\in\mathcal R$. As $u_1$ is
affine and $s\iota>0$, (W1) gives $0<u_1(\alpha)<w$, that is,
$\rho^f_{\tau_1}<\theta_1<\rho^h_{\tau_1}$ at $\alpha$. We apply \cref{lem:windows}(b) with
$U=E\setminus\{\tau_1\}$ and $W=E$. Then $D=\{b,\tau_1\}$, $X_{U,W}=K-d(E)+F_{\tau_1}$, and
$\theta_{U,W}=\theta_1$. The conditions of the lemma hold. For $\tau_1$, we have just
seen it. For $e\in E_k$, (W3) and \cref{thm:splice}(a) give
$\rho^h_e\ge\rho^f_e=\rho^{(k)}_e>\rho^h_{\tau_1}>\theta_1$. For $\tau_2$, (W5) gives
$\rho^h_{\tau_2}\ge\rho^f_{\tau_2}=\rho^f_{\tau_1}+\varphi w>\rho^h_{\tau_1}>\theta_1$. So the
configuration is $(E\setminus\{\tau_1\},E)$, and $\theta_b=\theta_1$.

Next we show that $b$ is merged with $\tau_2$ on the second window
$\mathcal R+\mu$. Let $\alpha\in\mathcal R+\mu$. By
(W2), $\rho^f_{\tau_2}<\theta_2<\rho^h_{\tau_2}$. We apply \cref{lem:windows}(b) with
$U=\emptyset$ and $W=\{\tau_2\}$. Then $D=\{b,\tau_2\}$,
$X_{U,W}=K+F_{\tau_2}-w/\beta_\delta$, and $\theta_{U,W}=\theta_2$. For $e\in E_k$,
$\rho^f_e\le\rho^h_e<\theta_2$ by (W4). For $\tau_1$,
$\rho^f_{\tau_1}\le\rho^h_{\tau_1}<\rho^f_{\tau_2}<\theta_2$ by (W5). So the configuration is
$(\emptyset,\{\tau_2\})$, and $\theta_b=\theta_2$.

To complete (a), we follow $b$ to the right of the second window. By (W2) and
as $u_2$ increases,
$\alpha_{\tau_2}>R_1+\mu$ and $u_2>0$ on $[R_1+\mu,\infty)$. For $e\in E_k$,
$\theta_2-\rho^h_e$ is positive at $R_1+\mu$ by (W4), and it increases, as
$\theta_2$ has slope $\bar m>M$ (by (W6) and \cref{lem:fast}) while $\rho^h_e$
grows at most at the rate $M$. Moreover, $\rho^h_{\tau_1}<\rho^f_{\tau_2}<\theta_2$ by (W5)
and as $u_2>0$. So every $e\in E_k\cup\{\tau_1\}$ has $\rho^f_e\le\rho^h_e<\theta_2$
on $[R_1+\mu,\infty)$. On $[R_1+\mu,\alpha_{\tau_2}]$, where $0<u_2\le w$, also
$\rho^f_{\tau_2}<\theta_2\le\rho^h_{\tau_2}$, and \cref{lem:windows}(b) with
$U=\emptyset$ and $W=\{\tau_2\}$ shows that the configuration is
$(\emptyset,\{\tau_2\})$, with $\theta_b=\theta_2$. Let
$\theta_{\mathrm{out}}:=\alpha m_b-\beta_bK$, the marginal that $b$ would have on
its own, which is $\theta_{U,W}$ for $U=W=\emptyset$. From the definitions,
$13\theta_2=\alpha(12m_b+m_\delta)-\beta_\delta(K+F_{\tau_2})+w=12\theta_{\mathrm{out}}+\rho^h_{\tau_2}$,
as $\beta_\delta=12\beta_b$ and
$\rho^h_{\tau_2}(\alpha)=\alpha m_\delta-\beta_\delta F_{\tau_2}+w$. So
$12(\theta_{\mathrm{out}}-\theta_2)=u_2-w$ and
$12(\theta_{\mathrm{out}}-\rho^h_{\tau_2})=13(u_2-w)$. On $(\alpha_{\tau_2},\infty)$,
where $u_2>w$, every $e\in E$ therefore has
$\rho^f_e\le\rho^h_e<\theta_{\mathrm{out}}$, and \cref{lem:windows}(b) with
$U=W=\emptyset$ shows that the configuration is $(\emptyset,\emptyset)$. With the
two windows, this proves (a).

Now we prove (b). The setting of \cref{lem:dmt} holds with $f_C:=g_k$ and
$C:=E_k$, as $m_*\ge\max_{e\in E}m_e$ (see the paragraph after
\cref{def:ruleP}), $\beta_b=\beta_\delta/12$ by (P2) and $m_b>m_\delta$ by (P4),
and its numbers $w$, $\bar m$, $s$, and $\psi$ are those of the parameter rule, with
$\psi=\psi_{k+1}$. On $\mathcal R$ and on $[R_0+\mu,\alpha_{\tau_2}]$, the claim is
\cref{lem:dmt}(c), applied with $L_C:=L_k$, whose cases (a) and (b) hold there
by (a). Its hypothesis on $\Delta F$ is (P5), and its line $L$ is $L_{k+1}$ by
(P8). On $[\alpha_{\tau_2},\infty)$, the pair $(\emptyset,\emptyset)$ is a weak
configuration, also at $\alpha_{\tau_2}$, where $\rho^h_{\tau_2}=\theta_2=\theta_b$. So
\eqref{eq:windows}, with \cref{lem:fullplane}(f) and \cref{lem:translation}(c)
(translation vector $(\mu,m_*\mu)$), gives
\[
  r(\xi^{(k+1)}(\alpha))=r(\xi^{(k)}(\alpha-\mu))+m_\delta(F_{\tau_1}+F_{\tau_2})-\sum_{e\in E}m_ed_e+Km_b.
\]
Hence
$L_{k+1}(\alpha)-r(\xi^{(k+1)}(\alpha))-\bigl(L_k(\alpha-\mu)-r(\xi^{(k)}(\alpha-\mu))\bigr)$
is affine in $\alpha$ on $[\alpha_{\tau_2},\infty)$, with slope
$q_{k+1}-q_k=\psi_{k+1}$, and it vanishes at $\alpha_{\tau_2}$ by the case of
$[R_0+\mu,\alpha_{\tau_2}]$.

Finally, we prove (c). On $\mathcal R$, $b$ and $\tau_1$ have the marginal $\theta_1$,
and all other elements have $\rho'=\rho^f>\theta_1$ (as shown for the first window). For
$\alpha\in\mathcal R$ and $e\in E_k$, (W4) at $\alpha+\mu$ with
\eqref{eq:D-bands}, then (W2) and the slope $m_\delta$ of $\rho^f_{\tau_2}$ give
\[
\begin{aligned}
  \rho^{(k)}_e(\alpha)+m_*\mu=\rho^h_e(\alpha+\mu)&<\theta_2(\alpha+\mu)=\rho^f_{\tau_2}(\alpha+\mu)+u_2(\alpha+\mu)\\
  &\le\rho^f_{\tau_2}(\alpha)+\mu m_\delta+w=\rho^f_{\tau_2}(\alpha)+m_*\mu,
\end{aligned}
\]
so $\rho^f_{\tau_2}$ is the largest marginal. On $\mathcal R+\mu$, $b$ and $\tau_2$
have the marginal $\theta_2$, and all other elements have
$\rho'=\rho^h<\theta_2$ (as shown for the second window). For $\alpha\in\mathcal R+\mu$ and $e\in E_k$,
(W3) at $\alpha-\mu$ and the slope $m_\delta$ of $\rho^h_{\tau_1}$ give
\[
  \rho^h_e(\alpha)=\rho^{(k)}_e(\alpha-\mu)+m_*\mu>\rho^h_{\tau_1}(\alpha-\mu)+m_*\mu=\rho^h_{\tau_1}(\alpha)+w,
\]
so $\rho^h_{\tau_1}$ is the smallest marginal.
\end{proof}

To the left of its windows a level is simple, and the transition of the next
level uses this.

\begin{lemma}\label{lem:leftwindows}
Let $j\ge1$, and assume the hypotheses of \cref{lem:onelevel} at the step
$j-1\to j$. In this lemma, $f$, $h$, $\tau_1$, $\tau_2$, $b$, $u_1$, $\theta_1$, and
$\theta_b$ are the quantities of this step, written without the index $j$, and
$E:=E_{j-1}\cup\{\tau_1,\tau_2\}$. Then the following hold for every $\alpha\le R_0$.
\begin{enumerate}[label=(\alph*)]
\item Every $e\in E_{j-1}\cup\{\tau_2\}$ has $\rho^f_e(\alpha)>\theta_b(\alpha)$, and
  $\theta_b(\alpha)\le\rho^h_{\tau_1}(\alpha)$.
\item $\rho^{(j)}_e(\alpha)=\rho^{(j-1)}_e(\alpha)$ for $e\in E_{j-1}$,
  $\rho^{(j)}_{\tau_2}(\alpha)=\rho^f_{\tau_2}(\alpha)$, and $b$ has the smallest marginal
  of level $j$.
\item If $u_1(\alpha)\ge0$, then level $j$ has the configuration
  $(E\setminus\{\tau_1\},E)$, so the block of $b$ is $\{b,\tau_1\}$, and
  $\theta_b(\alpha)=\theta_1(\alpha)$.
\end{enumerate}
\end{lemma}

\begin{proof}
By \cref{lem:onelevel}(a), the configuration at $R_0$ is
$(E\setminus\{\tau_1\},E)$, with $\theta_b=\theta_1$. So $\rho^f_e>\theta_b$ at $R_0$
for $e\in E_{j-1}\cup\{\tau_2\}$, and $\theta_b<\rho^h_{\tau_1}$ at $R_0$ by (W1). By
(W6) and \cref{lem:fast}, the differences $\rho^f_e-\theta_b$ and
$\rho^h_{\tau_1}-\theta_b$ decrease, so these inequalities hold on $(-\infty,R_0]$.
This proves (a). As
$\rho^f\le\rho^h$ (\cref{thm:splice}(a)), \cref{thm:splice}(b) then gives
$\rho^{(j)}_e=\med(\rho^f_e,\theta_b,\rho^h_e)=\rho^f_e$ for
$e\in E_{j-1}\cup\{\tau_2\}$, which is $\rho^{(j-1)}_e$ for $e\in E_{j-1}$ by
\cref{lem:fullplane}(f), and
$\rho^{(j)}_{\tau_1}=\med(\rho^f_{\tau_1},\theta_b,\rho^h_{\tau_1})\ge\theta_b$. So $b$ has the
smallest marginal, which proves (b).

For (c), let $\alpha\le R_0$ with $u_1(\alpha)\ge0$. As $u_1$ increases and $u_1(R_1)<w$
by (W1), $\rho^f_{\tau_1}\le\theta_1<\rho^h_{\tau_1}$ at $\alpha$. The function $\theta_1$ has
slope $\bar m>M$, so $\rho^f_e-\theta_1$ decreases for
$e\in E_{j-1}\cup\{\tau_2\}$, and it is positive at $R_0$, where
$\theta_1=\theta_b$. So $\rho^h_e\ge\rho^f_e>\theta_1$ at $\alpha$ for these $e$.
By \cref{lem:windows}(b) with $U=E\setminus\{\tau_1\}$ and $W=E$, the configuration
at $\alpha$ is $(E\setminus\{\tau_1\},E)$, and $\theta_b=\theta_1$. So the block of
$b$ is $(W\setminus U)\cup\{b\}=\{b,\tau_1\}$.
\end{proof}

\subsubsection{The transition}

On the transition $[R_1,R_0+\mu]$ between the two windows, neither window
description applies. The new element $b$ crosses the whole old structure from
the bottom to the top, and old elements join and leave its block. We show that
the gap first rises and then falls there, and that it stays at least~$\varepsilon_0$. \Cref{lem:twosides} describes the old level to the right of its
contacts and just to the left of them. \Cref{thm:transition} then gives the
bound on the transition.

\begin{lemma}\label{lem:twosides}
Let $k\ge0$, and assume the hypotheses of \cref{lem:onelevel} at the steps
$0\to1,\dots,k-1\to k$.
\begin{enumerate}[label=(\alph*)]
\item On $[R_1^{(k)},\infty)$, the gap $L_k(\alpha)-r(\xi^{(k)}(\alpha))$ of
  level $k$ is nondecreasing in $\alpha$, with the value $7\varepsilon_0$ at
  $R_1^{(k)}$, as for the base at $R_1^{(0)}$. So
  $r(\xi^{(k)}(\alpha'))-r(\xi^{(k)}(\alpha))\le q_k(\alpha'-\alpha)$
  for $R_1^{(k)}\le\alpha\le\alpha'$.
\item $L_k(\alpha)-r(\xi^{(k)}(\alpha))=\zeta_1-\alpha$ for
  $\alpha\in[R_0-\iota,R_0]$.
\end{enumerate}
\end{lemma}

\begin{proof}
By induction on $k$. For $k=0$, the proof of \cref{prop:base} gives
$L_0(\alpha)-r(\xi^{(0)}(\alpha))=7(\alpha-\zeta_2)$ on $[\zeta_2,\infty)$, so the
gap is $7\varepsilon_0$ at $R_1^{(0)}=\zeta_2+\varepsilon_0$. It also gives
$L_0(\alpha)-r(\xi^{(0)}(\alpha))=\zeta_1-\alpha$ on $[-1/6,\zeta_1]$, which
contains $[R_0-\iota,R_0]$.

Let $k\ge1$. By \cref{lem:onelevel}(b) at the step $k-1\to k$, on
$[R_0+\mu_k,\infty)$, which contains $[R_1^{(k)},\infty)$, the gap of level $k$
at $\alpha$ is
\[
  L_{k-1}(\alpha-\mu_k)-r(\xi^{(k-1)}(\alpha-\mu_k))+\psi_k\max(\alpha-\alpha_{\tau_2},0),
\]
where $\alpha_{\tau_2}>R_1^{(k-1)}+\mu_k=R_1^{(k)}$ is the point $\alpha_{\tau_2}$ of that
step. For $\alpha\ge R_1^{(k)}$, the first two terms are the gap of level $k-1$
at $\alpha-\mu_k\ge R_1^{(k-1)}$, which is nondecreasing in $\alpha$ by
induction, with the value $7\varepsilon_0$ at $\alpha=R_1^{(k)}$. The last term is
nondecreasing and vanishes at $R_1^{(k)}$. This gives the first claim of (a),
and the second follows, as $L_k$ has slope $q_k$.

For (b), let $\alpha\in[R_0-\iota,R_0]$. As $u_1^{(k)}$ has slope $s_k$
(\cref{lem:closedforms}(b)), (W1) at the step $k-1\to k$ gives
$u_1^{(k)}(\alpha)\ge u_1^{(k)}(R_0)-s_k\iota>0$. So, by \cref{lem:leftwindows}(c),
level $k$ has the configuration $(E\setminus\{\tau_{1,k}\},E)$ of its first window at
$\alpha$, with $E=E_{k-1}\cup\{\tau_{1,k},\tau_{2,k}\}$. As in the proof of
\cref{lem:onelevel}(b), \cref{lem:dmt}(a) and (c) then show that the gap of level
$k$ at $\alpha$ is $L_{k-1}(\alpha)-r(\xi^{(k-1)}(\alpha))$, which is
$\zeta_1-\alpha$ by induction.
\end{proof}

The transition theorem holds for every level. Its hypotheses name two old
elements, that is, elements of $E_k$. The first is a heavy one, $e_t$, whose
arrival in the block of $b$ turns the gap from rising to falling. The second is
the old element $e_l$ with the largest marginal for $h$, the last to leave the
block. They are $b_k$ and $\tau_{2,k}$ for $k\ge1$, and both are $C_0$ at the step
$0\to1$.

\begin{theorem}\label{thm:transition}
Let $k\ge0$ and $e_t,e_l\in E_k$. Assume the hypotheses of \cref{lem:onelevel}
at the steps $0\to1,\dots,k\to k+1$, and assume the following.
\begin{enumerate}[label=(T\arabic*),leftmargin=*]
\item $\psi_{k+1}\ge\SSq\bigl((E_k\setminus\{e_t\})\cup\{b\}\bigr)$ and
  $\SSq(\{b,e_t\})\ge q_{k+1}$, and, if $e_l\ne e_t$, also
  $\SSq(\{b,\tau_2,e_l\})\ge q_{k+1}$.
\item On the transition,
  $\rho^h_{\tau_1}<\rho^f_e<\rho^f_{\tau_2}<\rho^h_{e'}\le\rho^h_{e_l}$ for all
  $e,e'\in E_k$.
\item $\theta_2\le\rho^h_{e_l}$ at $R_0+\mu-\iota$.
\end{enumerate}
Then the gap $L_{k+1}(\alpha)-r(\xi^{(k+1)}(\alpha))$ of level $k+1$ is at
least $\varepsilon_0$ for every $\alpha$ in the transition $[R_1,R_0+\mu]$.
\end{theorem}

\begin{proof}
By (W6) and \cref{lem:fast}, every difference $\theta_b-\rho^f_e$ and
$\theta_b-\rho^h_e$, $e\in E$, is continuous and strictly increasing, so it has
at most one zero. We cut the transition at these finitely many zeros into
pieces. On the interior of a piece $[c,c']$, the sets
$U:=\{e\in E\colon\rho^f_e>\theta_b\}$ and $W:=\{e\in E\colon\rho^h_e>\theta_b\}$
are constant, and $U\subseteq W$ by \cref{thm:splice}(a). As the differences
increase, $U$ and $W$ are also $\{e\colon\rho^f_e>\theta_b\}$ and
$\{e\colon\rho^h_e>\theta_b\}$ at $c$, and $\{e\colon\rho^f_e\ge\theta_b\}$ and
$\{e\colon\rho^h_e\ge\theta_b\}$ at $c'$. So, at every point of $[c,c']$, the
inequalities of a weak configuration hold for $(U,W)$ by continuity, and $U$
and $W$ are upper level sets of $\rho^f$ and of $\rho^h$. Hence $(U,W)$ is a
weak configuration at every point of $[c,c']$, and in the interior it is the
configuration. By \cref{lem:windows}(c), with $D=(W\setminus U)\cup\{b\}$, the
increment $r(\xi^{(k+1)}(a'))-r(\xi^{(k+1)}(a))$ over every interval
$[a,a']\subseteq[c,c']$ is
\begin{equation}\label{eq:D-increment}
  \Bigl[\sum_{e\in U}m_e\xi^f_e\Bigr]_a^{a'}+\Bigl[\sum_{e\in E\setminus W}m_e\xi^h_e\Bigr]_a^{a'}+(a'-a)\SSq(D).
\end{equation}
Here the second bracket is nonnegative, and the first lies in
$[0,r(\xi^{(k)}(a'))-r(\xi^{(k)}(a))]\subseteq[0,q_k(a'-a)]$, by
\cref{lem:fullplane}(f), which gives
$r(\xi^f(\alpha))=r(\xi^{(k)}(\alpha))+m_\delta(F_{\tau_1}+F_{\tau_2})$, and by
\cref{lem:twosides}(a), as $a\ge R_1$.

By \cref{lem:onelevel}(a), (W1), and (W2), $\rho^f_{\tau_1}<\theta_b$ at $R_1$ and
$\theta_b<\rho^h_{\tau_2}$ at $R_0+\mu$. As the differences increase,
$\rho^f_{\tau_1}<\theta_b<\rho^h_{\tau_2}$ on the whole transition. By (W4) and (T2),
$\theta_b\ge\rho^h_{e_l}\ge\rho^h_{e_t}\ge\rho^f_{e_t}$ at $R_0+\mu$, where
$\theta_b=\theta_2$, while $\theta_b<\rho^f_{e_t}\le\rho^h_{e_l}$ at $R_1$, where
the configuration is $(E\setminus\{\tau_1\},E)$. Let $\check\alpha$ and $\hat\alpha$
be the zeros of $\theta_b-\rho^f_{e_t}$ and of $\theta_b-\rho^h_{e_l}$ in the
transition, which exist by continuity. Then $\check\alpha<\hat\alpha$, as
$\theta_b=\rho^h_{e_l}>\rho^f_{e_t}$ at $\hat\alpha$ by (T2). Every comparison of
marginals below is (T2) at a single share.

We first show that the gap of level $k+1$ is nondecreasing on
$[R_1,\check\alpha]$, before $e_t$ arrives in the block of $b$. In the
interior of each piece, $\theta_b<\rho^f_{e_t}<\rho^f_{\tau_2}$, so $e_t$ and $\tau_2$ lie in $U$,
above the block of $b$. Moreover, $\rho^h_e>\rho^f_{\tau_2}>\theta_b$ for every old element
$e\in E_k$, so only $\tau_1$ can lie outside $W$. As $\xi^h_{\tau_1}$ is constant, the second
bracket of \eqref{eq:D-increment} vanishes. If $\tau_1$ is in the block, no old
element $e$ is, since $\rho^f_e>\rho^h_{\tau_1}\ge\theta_b$. So the block is $\{b,\tau_1\}$
or lies in $(E_k\setminus\{e_t\})\cup\{b\}$, and $\SSq(D)\le\psi_{k+1}$, as
$\SSq(\{b,\tau_1\})=\psi_{k+1}$, by (T1) and the monotonicity of $\SSq$. Hence, by
\eqref{eq:D-increment}, the reward $r(\xi^{(k+1)}(\alpha))$ grows by at most
$(q_k+\psi_{k+1})(a'-a)=q_{k+1}(a'-a)$ on every interval $[a,a']$ in a piece,
and the gap of level $k+1$ is nondecreasing on $[R_1,\check\alpha]$.

Next we show that the gap is nonincreasing on $[\check\alpha,\hat\alpha]$, from
the arrival of $e_t$ in the block until $e_l$ leaves it. In the interior of each
piece, $\rho^f_{e_t}<\theta_b<\rho^h_{e_l}$. If $\theta_b<\rho^h_{e_t}$ there, then
$e_t$ is in the block. Otherwise
\[
  \rho^f_{e_l}<\rho^f_{\tau_2}<\rho^h_{e_t}<\theta_b<\rho^h_{e_l}
  \qquad\text{and}\qquad\theta_b<\rho^h_{\tau_2},
\]
so $\tau_2$ and $e_l\ne e_t$ are in the block. By (T1) and the monotonicity of
$\SSq$, $\SSq(D)\ge q_{k+1}$ in both cases. So, by \eqref{eq:D-increment}, whose
two brackets are nonnegative, the reward $r(\xi^{(k+1)}(\alpha))$ grows by at
least $q_{k+1}(a'-a)$ on every interval $[a,a']$ in a piece, and the gap of level
$k+1$ is nonincreasing on $[\check\alpha,\hat\alpha]$.

Finally, we show that the gap is at least $\varepsilon_0$ on
$[\hat\alpha,R_0+\mu]$, where $b$ has passed every old element. Here every old
element $e\in E_k$ satisfies
\[
  \theta_b\ge\rho^h_{e_l}\ge\rho^h_e>\rho^f_e>\rho^h_{\tau_1},
  \qquad\text{and}\qquad
  \rho^f_{\tau_2}<\rho^h_{e_l}\le\theta_b<\rho^h_{\tau_2}.
\]
So $\{\tau_2\}=\{e\in E\colon\rho^h_e>\theta_b\}$, and $(\emptyset,\{\tau_2\})$ is a
weak configuration. As in the proof of \cref{lem:onelevel}(b),
\cref{lem:dmt}(b) and (c) then show that the gap of level $k+1$ at $\alpha$ is
$L_k(\alpha-\mu)-r(\xi^{(k)}(\alpha-\mu))$. By \cref{lem:windows}(a),
$\theta_b=\theta_2$ there, so $\theta_2=\rho^h_{e_l}$ at $\hat\alpha$. As
$\theta_2-\rho^h_{e_l}$ increases ($\theta_2$ has slope $\bar m>M$), (T3) gives
$\hat\alpha\ge R_0+\mu-\iota$. So $\alpha-\mu\in[R_0-\iota,R_0]$, and
\cref{lem:twosides}(b) gives
\[
  L_{k+1}(\alpha)-r(\xi^{(k+1)}(\alpha))=\zeta_1-(\alpha-\mu)\ge\zeta_1-R_0=\varepsilon_0.
\]

So, on $[R_1,\check\alpha]$, the gap of level $k+1$ is at least its value
$L_k(R_1)-r(\xi^{(k)}(R_1))=7\varepsilon_0$ at $R_1$, by \cref{lem:onelevel}(b) and
\cref{lem:twosides}(a). On $[\check\alpha,\hat\alpha]$, it is at least its value
at $\hat\alpha$, and on $[\hat\alpha,R_0+\mu]$, it is at least $\varepsilon_0$.
As $7\varepsilon_0>\varepsilon_0$ and the gap at $\hat\alpha$ is at least
$\varepsilon_0$, the gap of level $k+1$ is at least $\varepsilon_0$ on the whole
transition.
\end{proof}

Under the hypotheses of \cref{thm:transition}, the two windows copy the gap of
level $k$ exactly, so the gap of level $k+1$ is $7\varepsilon_0$ at $R_1$ and
$\varepsilon_0$ at $R_0+\mu$, by \cref{lem:onelevel}(b) and \cref{lem:twosides}(a)
and (b). So the gap must fall somewhere on the transition, for every choice of
the parameters that keeps these exact copies. In the proof above, it is
nondecreasing up to $\check\alpha$ and falls only afterwards.

\subsection{Every level works}\label{sec:D-alllevels}

This subsection shows that the hypotheses of \cref{lem:onelevel} and of
\cref{thm:transition} hold in every step of the construction by the parameter rule, for every
input in $\Fin$ (\cref{prop:level1,thm:onestep}), and that the reward of every
level then stays below its line on $[R_0,\infty)$ and touches it exactly at
$\Gamma_j$ (\cref{thm:alllevels}). It also bounds the sizes of all numbers of
the construction (\cref{lem:sizes}) and proves \cref{thm:gadget}. We first bound
the inputs, and then the curvatures and rewards of a level, which determine the
numbers of the next step. As in \cref{def:ruleP}, up to \cref{prop:level1} the
letter $k$ denotes the level before a step, not the length of $\muin$.

For $j\ge0$, let $|\mathcal R_j|:=R_1^{(j)}-R_0$ be the length of $\mathcal R_j$.

\begin{lemma}\label{lem:inputs}
Let $\muin\in\Fin$ have length at least $2$. Then
$\mu_1\in(80/33,44/15)$, and for every $j\ge1$ for which
$\mu_{j+1}$ is defined,
\[
  \hat\mu_j:=\frac{\mu_j}{\mu_{j+1}}\in\Bigl(\frac5{44},\frac{11}{80}\Bigr),\qquad
  \hat\ell_j:=\frac{|\mathcal R_j|}{\mu_{j+1}}<\frac{253}{1600}=0.158125,\qquad
  \frac1{\mu_{j+1}}<\frac{33}{640}.
\]
Moreover, $\mu_{j+1}>\sum_{i\le j}\mu_i$ and $|\mathcal R_j|<\mu_{j+1}$ for every
$j\ge0$ for which $\mu_{j+1}$ is defined. So $\Gamma_j$ has $2^{j+1}$ points for
every $j$ for which $\mu_j$ is defined, and for every $j$ for which $\mu_{j+1}$
is defined, the transition
$[R_1^{(j)},R_0+\mu_{j+1}]$ has positive length.
\end{lemma}

\begin{proof}
With $H$ as in the definition of $\Fin$,
$\muin_i\in[H8^{i-1},\frac{11}{10}H8^{i-1})$, so the ratio of two consecutive
inputs lies in $(80/11,44/5)$. This gives the ranges of
$\mu_1=\muin_2/(3\muin_1)$ and of $\hat\mu_j=\muin_{j+1}/\muin_{j+2}$. Also
$1/\mu_{j+1}=3\muin_1/\muin_{j+2}<\frac{33}{10}\cdot8^{-j-1}\le\frac{33}{640}$ for
$j\ge1$. Next, $|\mathcal R_j|=\sum_{i\le j}\mu_i+1/15$ and
$\mu_i=\muin_{i+1}/(3\muin_1)$, so
\[
  \frac{|\mathcal R_j|}{\mu_{j+1}}=\Bigl(\sum_{i=1}^{j+1}\muin_i+\frac15\muin_1\Bigr)\Big/\muin_{j+2}.
\]
For $j\ge1$, the numerator is less than
$\frac{11}{10}H\bigl(\frac17(8^{j+1}-1)+\frac15\bigr)$ and the denominator is at
least $H8^{j+1}$, so
$\hat\ell_j<\frac{11}{10}\bigl(\frac17+\frac2{35}\cdot8^{-j-1}\bigr)\le\frac{11}{10}\bigl(\frac17+\frac2{35}\cdot8^{-2}\bigr)=\frac{253}{1600}$.
For $j=0$, $|\mathcal R_0|/\mu_1=\frac2{5\mu_1}<\frac{33}{200}$. So
$|\mathcal R_j|/\mu_{j+1}<1$ for every $j\ge0$, that is,
$\sum_{i\le j}\mu_i<|\mathcal R_j|<\mu_{j+1}$. This gives the remaining claims,
as a sequence in which every entry exceeds the sum of the earlier ones has
distinct subset sums.
\end{proof}

For a level $j\ge1$, let $\Delta_j$, $\varphi_j$, $u^0_j$, and $w_j$ be the
quantities $\Delta_b$, $\varphi$, $u^0$, and $w$ of the step $j-1\to j$, so that
$\Delta_j=\frac{415}2M_{j-1}=2490\cdot208^{j-1}$. Let
$\upsilon^0_j:=u^0_j/w_j$ and $\varpi_j:=1/\beta_\delta^{(j)}=\frac34\cdot8^{j-1}$, and
$\varpi_0:=1$. By (P4), $s=\frac{12}{13}\Delta_b=\frac{2490}{13}M$ at every step, so
$\hat s:=s/(64M)=1245/416$ does not depend on the step. The next lemma
bounds all numbers of a step by an exact
invariant of the curvatures: the two dummies of a level weigh as much as all
older elements together.

\begin{lemma}\label{lem:weights}
\begin{enumerate}[label=(\alph*)]
\item $\binv(E_k)=2\varpi_{k+1}$ for every $k\ge0$. Hence
  $\binv(\{\tau_{1,k+1},\tau_{2,k+1}\})=\binv(E_k)$ and
  $\binv(E_k\cup\{\tau_{1,k+1},\tau_{2,k+1},b_{k+1}\})=16\varpi_{k+1}$ for $k\ge0$, and
  $\binv(E_k\setminus\{b_k\})=4\varpi_k$ for $k\ge1$.
\item For every $k\ge0$, all rewards of $E_k$ are integers in $(0,M_k]$, and
  $M_k=12\cdot208^k$ is the reward of $C_1$ if $k=0$ and of $b_k$ if $k\ge1$.
\item $q_0=7$, and $\frac9{10}\varpi_kM_k^2<q_k\le\varpi_kM_k^2$ for $k\ge1$.
\item Let $k\ge1$, and let $\muin\in\Fin$ have length at least $k+2$. At the step
  $k\to k+1$, $\bar d\in[127/64,129/64)$,
  $|a_0|/(\Delta_bw)\le517/106240<0.00487$, and
  $\varphi\in(2.9791,3.0209)$. Moreover, $u^0/w=A-\frac{1245}{832}\hat\ell_k$ with
  \[
    A:=\frac{1245}{416}-\frac12-\bar d+\frac{a_0}{\Delta_bw}\in(0.4722,0.5133),
  \]
  and $u^0/w\in(0.2354,0.5133)$.
\end{enumerate}
\end{lemma}

\begin{proof}
(a) By induction on $k$. First, $\binv(E_0)=1+1/4+1/4=3/2=2\varpi_1$.
By (P2), the step $k\to k+1$ adds $1/\beta_{\tau_1}=1/\beta_{\tau_2}=\varpi_{k+1}$ and
$1/\beta_b=12\varpi_{k+1}$, so
$\binv(E_{k+1})=\binv(E_k)+14\varpi_{k+1}=16\varpi_{k+1}=2\varpi_{k+2}$. This gives the
first two consequences. For $k\ge1$, $1/\beta_{b_k}=12\varpi_k$ and
$\binv(E_k)=2\varpi_{k+1}=16\varpi_k$, so $\binv(E_k\setminus\{b_k\})=4\varpi_k$.

(b) By induction on $k$. The rewards of $E_0$ are $5,12,4$, and $M_0=12$. By (P1)
and (P4), the step $k\to k+1$ adds the reward $\frac12M_k$ (of $\tau_1$ and $\tau_2$) and
$m_b=\frac12M_k+\frac{415}2M_k=208M_k=M_{k+1}$ (of $b$), and $\frac12M_k$ is an
integer, as $M_k$ is even.

(c) By (P2), (P4), and (P8), $q_{k+1}=q_k+\psi_{k+1}$ with
$\psi_{k+1}=\frac{12}{13}\varpi_{k+1}\Delta_b^2=\frac{12}{13}\bigl(\frac{415}{416}\bigr)^2\varpi_{k+1}M_{k+1}^2$,
and $\frac9{10}<\frac{12}{13}(\frac{415}{416})^2<\frac{12}{13}$. As $q_0=7>0$, this
gives $q_k\ge\psi_k>\frac9{10}\varpi_kM_k^2$ for $k\ge1$. For the upper bound,
$q_1=7+\psi_1<7+\frac{12}{13}\varpi_1M_1^2<\varpi_1M_1^2$, as
$\frac1{13}\varpi_1M_1^2=\frac1{13}\cdot\frac34\cdot2496^2>7$. For $k\ge2$, by
induction,
$q_{k-1}\le\varpi_{k-1}M_{k-1}^2=\varpi_kM_k^2/(8\cdot208^2)<\frac1{13}\varpi_kM_k^2$,
so $q_k=q_{k-1}+\psi_k<(1/13+12/13)\varpi_kM_k^2$.

(d) Write $M:=M_k$ and $\varpi:=\varpi_k$, so that $1/\beta_\delta=\varpi_{k+1}=8\varpi$,
$m_*=\frac{129}2M$, and $w=64M\mu$, and $\binv(E_k)=16\varpi$ by (a). By (b),
$m_*-m_e\in[\frac{127}2M,\frac{129}2M)$ for $e\in E_k$. As $d_e=\mu(m_*-m_e)/\beta_e$
and $d_{\tau_1}=w/\beta_\delta=512\varpi M\mu$, (P3) gives
$\bar d=\sum_{e\in E_k}(m_*-m_e)/(512\varpi M\beta_e)\in[127/64,129/64)$.
As $|m_\delta-m_e|\le\frac M2$ for $e\in E_k$, we get
$|\Pi|\le\frac M2d(E_k)<\frac M2\cdot\frac{129}2M\mu\binv(E_k)=516\varpi\mu M^2$.
With $0<q_k\mu\le\varpi\mu M^2$ by (c) and $a_0=\beta_\delta(q_k\mu-\Pi)=(q_k\mu-\Pi)/(8\varpi)$,
this gives $|a_0|\le\frac{517}8\mu M^2$, and as $\Delta_bw=13280\mu M^2$,
$|a_0|/(\Delta_bw)\le517/106240$. By \cref{lem:closedforms}(a),
$\varphi=1+\bar d-a_0/(sw)$, where $a_0/(sw)=\frac{13}{12}a_0/(\Delta_bw)$, which
gives the range of $\varphi$. Next,
$a_2\Delta_b=s(\mu-|\mathcal R|/2)$, so $a_2\Delta_b/w=\hat s(1-\hat\ell_k/2)$, as
$|\mathcal R|/\mu=\hat\ell_k$, and (P6) gives
$u^0/w=\hat s(1-\hat\ell_k/2)-\frac12-\bar d+a_0/(\Delta_bw)=A-\frac12\hat s\hat\ell_k$.
The range of $A$ follows from those of $\bar d$ and $a_0/(\Delta_bw)$, and
$0<\frac12\hat s\hat\ell_k<0.2367$ by \cref{lem:inputs} gives the range of $u^0/w$.
\end{proof}

\begin{theorem}\label{thm:onestep}
Let $k\ge1$, and let $\muin\in\Fin$ have length at least $k+2$. Assume that the
hypotheses of \cref{lem:onelevel} hold at the steps $0\to1,\dots,k-1\to k$.
Then at the step $k\to k+1$, the hypotheses of \cref{lem:onelevel} and those of
\cref{thm:transition} with $(e_t,e_l)=(b_k,\tau_{2,k})$ hold.
\end{theorem}

\begin{proof}
The numbers of a step grow from level to level. We first write them in terms of
$M:=M_k$, $\varpi:=\varpi_k$, and the band width $w$, so that the same
computation works at every level $k\ge1$. Let $m_o:=m_\delta^{(k)}=\frac12M_{k-1}$
and $\hat M:=M_{k-1}/M=1/208$ (\cref{lem:weights}(b)). By the parameter rule, $\varpi_{k+1}=8\varpi$, $m_\delta=\frac12M$,
$m_*=\frac{129}2M$, $m_b=208M$, $\Delta_b=\frac{415}2M$, $s=\frac{2490}{13}M$,
and $\bar m=\frac{4993}{26}M$, so $\bar m-M=\frac{4967}{26}M>191M$. The inputs
enter through $\hat\mu:=\hat\mu_k$
and $\hat\ell:=\hat\ell_k$ of \cref{lem:inputs}, so $\hat\mu\in(5/44,11/80)$,
$\hat\ell<0.158125$, and $1/\mu<33/640$. We use $w=64M\mu$ and
$w_k=64\hat MM\mu_k$, so $w_k/w=\hat M\hat\mu$. We take $\bar d$, $\varphi$, $A$,
and $\upsilon^0:=u^0/w$ from \cref{lem:weights}(d) at the step $k\to k+1$, and
recall that $\hat s=s/(64M)=1245/416\in(2.9927,2.9928)$. Decimals are
rounded in the safe direction. Three kinds of quantities are small compared
with $w$.
\begin{itemize}
\item The margin of (P6): $c_{\tau_2}:=s\iota/w=\hat s\iota/\mu<0.00155$.
\item The \emph{width} $c_w:=\varphi_kw_k/w=\varphi_k\hat M\hat\mu<0.00200$ of
  level $k$ at $R_0$, as $\varphi_1<3.000004$ (\cref{prop:level1}) and
  $\varphi_k<3.0209$ for $k\ge2$ (\cref{lem:weights}(d) at the step $k-1\to k$).
  By \cref{lem:onelevel}(c) for level $k$, the largest marginal of level $k$ at
  $R_0$ is $\rho^{f_k}_{\tau_{2,k}}(R_0)=\bar\rho_k$, and the smallest is $\theta_1$, which exceeds
  $\rho^{f_k}_{\tau_{1,k}}(R_0)=\bar\rho_k-c_ww$, as $u_1(R_0)>0$ by (W1) at the step $k-1\to k$.
\item The \emph{drifts} over one shift, $m_\delta\mu=(M-m_\delta)\mu=w/128$.
\end{itemize}
By \cref{lem:closedforms}(c), and as the marginals of level $k$ are
nondecreasing and grow at most at the rate $M$ (\cref{lem:fullplane}(c)), the
dummies and the old marginals lie as follows.
\begin{enumerate}[label=(Q\arabic*),leftmargin=*]
\item $\rho^f_{\tau_2}(\alpha)=\bar\rho_k+m_\delta(\alpha-R_0)+(1-\upsilon^0+c_{\tau_2})w$
  and $\rho^h_{\tau_1}=\rho^f_{\tau_2}-(\varphi-1)w$.
\item For $e\in E_k$,
  $\rho^{(k)}_e(\alpha)\in[\bar\rho_k-c_ww,\ \bar\rho_k+M(\alpha-R_0)]$ if
  $\alpha\ge R_0$, and
  $\rho^{(k)}_e(\alpha)\in[\bar\rho_k-c_ww-M(R_0-\alpha),\ \bar\rho_k]$ if
  $\alpha\le R_0$.
\end{enumerate}
Recall that $\rho^h_e(\alpha)=\rho^{(k)}_e(\alpha-\mu)+m_*\mu$ for $e\in E_k$ by
\eqref{eq:D-bands}, and that $m_*\mu=m_\delta\mu+w$. We now check the
hypotheses one by one.
\begin{itemize}
\item (T1): $\psi_{k+1}=\frac{12}{13}\Delta_b^2/\beta_\delta=\frac{96}{13}\varpi\Delta_b^2$.
  By \cref{lem:weights}(a), $\binv(E_k\setminus\{b_k\})=4\varpi$. Taking $v=m_b$ in
  the minimum form of $\SSq$, as all rewards of $E_k$ lie in $(0,m_b)$,
  \[
    \SSq\bigl((E_k\setminus\{b_k\})\cup\{b\}\bigr)\le\binv(E_k\setminus\{b_k\})\,m_b^2=4\varpi\Bigl(\frac{416}{415}\Delta_b\Bigr)^2
    <4.02\,\varpi\Delta_b^2<\psi_{k+1}.
  \]
  Next, by the formula for $\SSq$ of a pair and by \cref{lem:weights}(c),
  \[
  \begin{aligned}
    \SSq(\{b,b_k\})&=\frac{(m_b-M)^2}{\beta_b+\beta_{b_k}}=\frac{32}3\,\varpi(207M)^2=457056\,\varpi M^2,\\
    q_{k+1}&=q_k+\psi_{k+1}\le\varpi M^2+\frac{96}{13}\Bigl(\frac{415}2\Bigr)^2\varpi M^2<317955\,\varpi M^2.
  \end{aligned}
  \]
  Finally, we decompose $\SSq$ into the groups $\{b,\tau_2\}$, with
  $\binv(\{b,\tau_2\})=104\varpi$ and mean $\bar m$, and $\{\tau_{2,k}\}$, with
  $\binv(\{\tau_{2,k}\})=\varpi$ and mean $m_o$. As
  $\bar m-m_o\ge\bar m-M>191M$ and $q_k\le\varpi M^2$, this gives
  \[
    \SSq(\{b,\tau_2,\tau_{2,k}\})=\psi_{k+1}+\frac{104}{105}\,\varpi(\bar m-m_o)^2>\psi_{k+1}+\varpi M^2\ge q_{k+1}.
  \]
\item (W1), (W2): $s|\mathcal R|/w=\hat s\hat\ell<0.4733$. So, by
  \cref{lem:closedforms}(b),
  $u_1(R_0)=\frac12(w-s|\mathcal R|)>0.2633w>s\iota$,
  $u_1(R_1)=\frac12(w+s|\mathcal R|)<0.7367w$, $u_2(R_0+\mu)=\upsilon^0w>0.2354w$, and
  \[
    u_2(R_1+\mu)=(\upsilon^0+\hat s\hat\ell)w
    =\Bigl(A+\frac{1245}{832}\hat\ell\Bigr)w<0.7501w.
  \]
\item (W3) and the inequalities $\rho^h_{\tau_1}<\rho^f_e<\rho^f_{\tau_2}$ of (T2): for $e\in E_k$ and
  $\alpha\in[R_0,R_0+\mu]$, (Q1) and (Q2) give
  \[
  \begin{aligned}
    \rho^{(k)}_e-\rho^h_{\tau_1}&\ge(\varphi+\upsilon^0-2-c_{\tau_2}-c_w)w-m_\delta\mu\\
    &>\bigl(2.9791+0.2354-2-0.00155-0.002-\tfrac1{128}\bigr)w>1.2w,\\
    \rho^f_{\tau_2}-\rho^{(k)}_e&\ge(1-\upsilon^0)w-(M-m_\delta)\mu>\bigl(1-0.5133-\tfrac1{128}\bigr)w>0.47w.
  \end{aligned}
  \]
\item The inequality $\rho^f_{\tau_2}<\rho^h_{e'}$ of (T2): let $e'\in E_k$ and let $\alpha$ lie in the
  transition. As $\alpha\ge R_1$, we have
  $0\le R_0+\mu-\alpha\le\mu-|\mathcal R|$, and (Q1) at $\alpha$ and (Q2) at
  $\alpha-\mu\le R_0$ give
  \[
  \begin{aligned}
    \rho^h_{e'}(\alpha)-\rho^f_{\tau_2}(\alpha)&\ge(\upsilon^0-c_{\tau_2}-c_w)w-(M-m_\delta)(\mu-|\mathcal R|)\\
    &=\Bigl(A-\frac1{128}-\frac{2477}{1664}\hat\ell-c_{\tau_2}-c_w\Bigr)w\\
    &>\bigl(0.4722-\tfrac1{128}-0.2354-0.00155-0.002\bigr)w>0.225w.
  \end{aligned}
  \]
\item The inequality $\rho^h_{e'}\le\rho^h_{e_l}$ of (T2): let $\alpha$ lie in the transition, and let
  $\alpha':=\alpha-\mu$, so that $0\le R_0-\alpha'<\mu$. By
  \cref{lem:leftwindows}(b) for level $k$, $b_k$ has the smallest marginal of level
  $k$ at $\alpha'$, and
  \[
    \rho^{(k)}_{\tau_{1,k}}\le\rho^{h_k}_{\tau_{1,k}}=\rho^{f_k}_{\tau_{2,k}}-(\varphi_k-1)w_k<\rho^{f_k}_{\tau_{2,k}}=\rho^{(k)}_{\tau_{2,k}}.
  \]
  Moreover, $\rho^{(k)}_e(\alpha')=\rho^{(k-1)}_e(\alpha')\le\bar\rho_{k-1}$ for
  $e\in E_{k-1}$, as $\bar\rho_{k-1}$ is the largest marginal of level $k-1$ at
  $R_0$ (\cref{lem:onelevel}(c), or (B3) of \cref{prop:base} for $k=1$). By
  \cref{lem:closedforms}(c) at level $k$,
  $\bar\rho_k-\bar\rho_{k-1}\ge(1-\upsilon^0_k)w_k$, where $\upsilon^0_k<0.5133$
  (\cref{prop:level1} for $k=1$, and \cref{lem:weights}(d) at the step $k-1\to k$
  for $k\ge2$). The marginal $\rho^{f_k}_{\tau_{2,k}}$ has slope
  $m_o=\frac12M_{k-1}$, where $\frac12M_{k-1}\mu=w_k/(128\hat\mu)<\frac{11}{160}w_k$.
  So $\rho^{(k)}_{\tau_{2,k}}(\alpha')-\bar\rho_{k-1}>(1-0.5133-11/160)w_k>0.41w_k$,
  and $\tau_{2,k}$ has the largest of the marginals $\rho^h_e(\alpha)$, $e\in E_k$.
\item (W4): for $e\in E_k$, $\theta_2-\rho^h_e$ increases, as $\bar m>M$. At
  $R_0+\mu$ it is at least
  $\rho^f_{\tau_2}(R_0+\mu)+u_2(R_0+\mu)-\bar\rho_k-m_*\mu=s\iota>0$, by (Q1), (Q2),
  and $u_2(R_0+\mu)=u^0$.
\item (T3): on $(-\infty,R_0+\mu]$,
  $\rho^h_{\tau_{2,k}}(\alpha)=\rho^{f_k}_{\tau_{2,k}}(\alpha-\mu)+m_*\mu$
  (\cref{lem:leftwindows}(b)) has slope $m_o\le m_\delta$, so
  $\theta_2-\rho^h_{\tau_{2,k}}$ has slope at least $s$. At $R_0+\mu$ it is
  $s\iota$, as for (W4) and since $\rho^{f_k}_{\tau_{2,k}}(R_0)=\bar\rho_k$. So it is
  at most $0$ at $R_0+\mu-\iota$.
\item (W5): $\varphi>2.9791$.
\item (W6): $\binv(E\cup\{b\})=16\varpi_{k+1}=128\varpi$ (\cref{lem:weights}(a))
  and $t_b=12\varpi_{k+1}\cdot208M=96\varpi\cdot208M$, so
  $t_b/\binv(E\cup\{b\})=156M>M$.
\end{itemize}
This proves all the hypotheses, and the theorem.
\end{proof}

\begin{proposition}\label{prop:level1}
Let $\muin\in\Fin$ have length at least $2$. Then $\varphi_1\in(3,3.000004)$ and
$\upsilon^0_1\in(0.2458,0.2888)$, and the hypotheses of \cref{lem:onelevel} and
those of \cref{thm:transition} with $(e_t,e_l)=(C_0,C_0)$ hold at the step
$0\to1$.
\end{proposition}

\begin{proof}
We have $M=12$, $m_\delta=6$, $m_*=774$, $1/\beta_\delta=3/4$,
$1/\beta_b=9$, $\Delta_b=2490$, $m_b=2496$, and
$s=29880/13\in(2298.46,2298.47)$, and
$\sum_{e\in E_0}(m_*-m_e)/\beta_e=769+762/4+770/4=1152$. So
$\bar d=2$, $w=768\mu$, $m_*\mu=774\mu$, $\Pi=\mu(769-1143+385)=11\mu$, and
$a_0=\frac43(q_0-11)\mu=-\frac{16}3\mu$. By \cref{lem:inputs},
$\mu\in(80/33,44/15)$, so $w>1861$, and $|\mathcal R_0|=2/5$.
By (P6) and \cref{lem:closedforms}(a),
\[
  u^0=s\Bigl(\mu-\frac15\Bigr)-\frac52w-\frac{16\mu}{3\cdot2490}=c\mu-\frac s5,
\]
where $c:=s-1920-8/3735\in(378.45,378.46)$, and
$\varphi_1=3+\frac{16\mu/3}{sw}=3+\frac1{144s}$. So
$\varphi_1\in(3,3.000004)$, and $\upsilon^0_1=u^0/w=c/768-s/(3840\mu)$ increases
with $\mu$ and lies in $(0.2458,0.2888)$.

To check the hypotheses, we use $s\iota<22.99$, $R_0=3/10$,
$R_1=7/10$, $s|\mathcal R_0|<919.4$, $\frac s5<459.70$, and
$m_\delta\mu=(M-m_\delta)\mu=w/128$. By the first table of \cref{prop:base} and (B3),
the base marginals at $R_0$ lie in
$[2/5,7/6]=[\bar\rho_0-23/30,\bar\rho_0]$, and their slopes are
at most $12$. So (Q1) and (Q2) of the proof of \cref{thm:onestep} hold with
$k=0$, with $c_ww=23/30$, the spread of the base marginals at $R_0$, and
with $c_{\tau_2} w=s\iota<22.99$. That derivation uses the assumption $k\ge1$ only through the spread and the slopes of the old marginals at $R_0$, which these bounds replace.
\begin{itemize}
\item (T1): $\psi_1=\frac9{13}\cdot2490^2$ and $q_1=7+\psi_1<4.3\cdot10^6$.
  Taking $v=m_b$ in the minimum form of $\SSq$,
  $\SSq(\{b,C_1,C_2\})\le\frac14(2484^2+2492^2)<3.1\cdot10^6<\psi_1$. Moreover,
  $\SSq(\{b,C_0\})=2491^2/(\beta_b+1)=\frac9{10}\cdot2491^2>5.58\cdot10^6>q_1$.
\item (W1): $u_1(R_0)=\frac12(w-s|\mathcal R_0|)>\frac12(1861-919.4)>470>s\iota$,
  and $w-u_1(R_1)=u_1(R_0)>470$. (W2): $u_2(R_0+\mu)=u^0>0.2458w>0$ and
  \[
    w-u_2(R_1+\mu)=w-u^0-s|\mathcal R_0|=(768-c)\mu-\frac s5>389.54\cdot\frac{80}{33}-459.70>484.
  \]
\item (W3) and (T2): as in the proof of \cref{thm:onestep}, on
  $[R_0,R_0+\mu]$,
  \[
  \begin{aligned}
    \rho^{(0)}_e-\rho^h_{\tau_1}&\ge\Bigl(\varphi_1+\upsilon^0_1-2-\frac1{128}\Bigr)w-s\iota-\frac{23}{30}>1.2379\cdot1861-23.76>2200,\\
    \rho^f_{\tau_2}-\rho^{(0)}_e&\ge\Bigl(1-\upsilon^0_1-\frac1{128}\Bigr)w>0.7w.
  \end{aligned}
  \]
  On the transition, where the drift is
  $(M-m_\delta)(\mu-|\mathcal R_0|)=6\mu-12/5$,
  \[
  \begin{aligned}
    \rho^h_{e'}-\rho^f_{\tau_2}&\ge u^0-s\iota-\frac{23}{30}-6\mu+\frac{12}5
    =(c-6)\mu-\frac s5-s\iota-\frac{23}{30}+\frac{12}5\\
    &>372.45\cdot\frac{80}{33}-459.70-22.99-0.77+2.4>421.
  \end{aligned}
  \]
  By (B3), $C_0$ has the largest of the marginals $\rho^h_e$, $e\in E_0$, on the
  transition, as $\alpha-\mu\ge R_1-\mu>7/10-44/15=-67/30>-7/3$
  there.
\item (W4) and (T3): as in the proof of \cref{thm:onestep}. Here $\bar\rho_0$ is
  the largest base marginal at $R_0$ by (B3), and the marginal
  $\rho^h_{C_0}(\alpha)=5(\alpha-\mu)-1/3+m_*\mu$ on $(-\infty,R_0+\mu]$ has
  slope $5<m_\delta$.
\item (W5): $\varphi_1>3$. (W6): $\binv(E\cup\{b\})=3/2+3/2+9=12$, so
  $t_b/\binv(E\cup\{b\})=9\cdot2496/12=1872>12$.
\end{itemize}
This proves the proposition.
\end{proof}

\begin{theorem}\label{thm:alllevels}
Let $\muin\in\Fin$ have length $k$, build the levels $0,\dots,k-1$ by the parameter rule,
and let $\kappa:=\min(1,\varepsilon_0/\delta)$. Then every level $j\le k-1$
satisfies
\begin{enumerate}[label=(H$_j$),leftmargin=*]
\item $L_j(\alpha)-r(\xi^{(j)}(\alpha))\ge\kappa\min\bigl(\dist(\alpha,\Gamma_j),\delta\bigr)$
  for all $\alpha\ge R_0$, and $r(\xi^{(j)}(\gamma))=L_j(\gamma)$ for every
  $\gamma\in\Gamma_j$,
\end{enumerate}
and for $j\ge1$ the hypotheses of \cref{lem:onelevel} hold at the step
$j-1\to j$.
\end{theorem}

\begin{proof}
By induction on $j$. (H$_0$) is (B2) of \cref{prop:base}, since $\kappa\le1$
and $\Gamma_0=\{\zeta_1,\zeta_2\}$. The assertions on level $1$ are
\cref{prop:level1}, and for $j\ge1$ those on level $j+1$ follow from
\cref{thm:onestep} at the step $j\to j+1$. It remains to derive (H$_{j+1}$)
from (H$_j$). By \cref{thm:transition}, with $(e_t,e_l)=(C_0,C_0)$ for $j=0$
and $(b_j,\tau_{2,j})$ for $j\ge1$, whose hypotheses hold by \cref{prop:level1} and
\cref{thm:onestep},
\[
  L_{j+1}(\alpha)-r(\xi^{(j+1)}(\alpha))\ge\varepsilon_0\ge\kappa\delta
\]
on the transition $[R_1^{(j)},R_0+\mu_{j+1}]$. On $\mathcal R_j$,
\cref{lem:onelevel}(b) and (H$_j$) give
\[
  L_{j+1}(\alpha)-r(\xi^{(j+1)}(\alpha))=L_j(\alpha)-r(\xi^{(j)}(\alpha))\ge\kappa\min\bigl(\dist(\alpha,\Gamma_j),\delta\bigr).
\]
On $[R_0+\mu_{j+1},\infty)$, as $\psi_{j+1}>0$ and $\alpha-\mu_{j+1}\ge R_0$,
they give
\[
\begin{aligned}
  L_{j+1}(\alpha)-r(\xi^{(j+1)}(\alpha))&\ge L_j(\alpha-\mu_{j+1})-r(\xi^{(j)}(\alpha-\mu_{j+1}))\\
  &\ge\kappa\min\bigl(\dist(\alpha,\Gamma_j+\mu_{j+1}),\delta\bigr).
\end{aligned}
\]
As $\Gamma_{j+1}=\Gamma_j\cup(\Gamma_j+\mu_{j+1})$ and these three intervals
cover $[R_0,\infty)$, this gives the inequality of (H$_{j+1}$). Finally,
$\Gamma_j\subseteq\mathcal R_j$, and $\Gamma_j+\mu_{j+1}$ lies in
$\mathcal R_j+\mu_{j+1}$, which is contained in the interval
$[R_0+\mu_{j+1},\alpha_{\tau_2}]$ of \cref{lem:onelevel}(b). So, by the first two
cases of \cref{lem:onelevel}(b) and by (H$_j$),
$L_{j+1}(\alpha)-r(\xi^{(j+1)}(\alpha))$ vanishes on $\Gamma_j$ and on
$\Gamma_j+\mu_{j+1}$.
\end{proof}

The last level $g_{k-1}$ has the right contacts, but only on $[R_0,\infty)$, and in
general it is not a polymatroid rank function. An embedding (\cref{lem:embed}
and \cref{lem:fullplane}(g)) turns it into a polymatroid instance on a
half-line, and one capped linear summand repairs the reward to the left of the contacts. The
next lemma bounds the sizes.

\begin{lemma}\label{lem:sizes}
Let $\muin\in\Fin$ have length $k$. Every number computed by the parameter rule for the
levels $1,\dots,k-1$, and every value of $g_{k-1}$, is a rational with
$\Oh(k+\log\muin_k)$ bits, and the parameter rule runs in time polynomial in
$k+\log\muin_k$.
\end{lemma}

\begin{proof}
A rational number has $\Oh(k+\log\muin_k)$ bits if its magnitude and its
denominator are at most $2^{\Oh(k+\log\muin_k)}$. We first bound the magnitudes
of the numbers of a level, and then their denominators. By the parameter rule,
$M_j=12\cdot208^j$ and $\varpi_j=\frac34\cdot8^{j-1}$ for
$j\ge1$, so $\log M_j$ and $\log\varpi_j$ are $\Oh(j)$, and $\mu_j\le\muin_k$.
Put $\Omega_j:=\varpi_jM_j(1+\mu_j)$, which more than doubles from level to
level. For the step $j-1\to j$, let $\bar d_j$ and $\varphi_j$ be its numbers
$\bar d$ and $\varphi$, and write $d_j(E_j\setminus\{b_j\})$ for the sum of
$d_j$ over the ground set $E_j\setminus\{b_j\}=E_{j-1}\cup\{\tau_{1,j},\tau_{2,j}\}$ of
$f_j$. Then $\bar d_j<129/64$ and $\varphi_j<3.021$ (\cref{lem:weights}(d)
at the step $j-1\to j$ for $j\ge2$, and $\bar d_1=2$ and $\varphi_1<3.000004$ by
\cref{prop:level1} and its proof), and $w_j=64M_{j-1}\mu_j=\frac4{13}M_j\mu_j$. So
$d_j(E_j\setminus\{b_j\})=(2+\bar d_j)\varpi_jw_j\le2\Omega_j$ and
$\Delta F_j=\varphi_j\varpi_jw_j\le\Omega_j$, and the number $\Pi$ of (P5) at
the step to level $j$ satisfies $|\Pi|\le M_jd_j(E_j\setminus\{b_j\})$. By
\cref{lem:closedforms}(c), and as $0<u^0_j<w_j$ (\cref{lem:weights}(d) and
\cref{prop:level1}),
\[
  0<\bar\rho_j\le\bar\rho_0+\sum_{i\le j}(w_i+s_i\iota)\le\frac76+2w_j+\frac{M_j}{50}\le M_j(1+\mu_j),
\]
because $w_i$ more than doubles from level to level, and
$s_i\iota=\frac{2490}{1300}M_{i-1}<\frac{M_i}{100}$. Hence
$|F_{\tau_2}^{(j)}|=\varpi_j|R_0m_\delta^{(j)}-\bar\rho_j|\le2\Omega_j$ and
$|F_{\tau_1}^{(j)}|\le|F_{\tau_2}^{(j)}|+\Delta F_j\le3\Omega_j$. By (P7), as
$s_j<M_j$ and $R_0+R_1^{(j-1)}\le1+\mu_j$,
\[
  |K_j|\le d_j(E_j\setminus\{b_j\})+12|F_{\tau_1}^{(j)}|+13\varpi_j\Bigl(\frac{w_j}2+\frac{s_j(1+\mu_j)}2\Bigr)\le50\,\Omega_j.
\]
Then
$|\Delta_p^{(j)}|\le M_j\bigl(|F_{\tau_2}^{(j)}|+|K_j|+d_j(E_j\setminus\{b_j\})+|F_{\tau_1}^{(j)}|\bigr)\le60M_j\Omega_j$,
$q_j\le\varpi_jM_j^2$ (\cref{lem:weights}(c)), and
$|p_j|\le p_0+\sum_{i\le j}|\Delta_p^{(i)}|\le10+120M_j\Omega_j$. As
\[
\begin{aligned}
  g_j(S)&=g_{j-1}(S\cap E_{j-1})+F_{\tau_1}^{(j)}\iverson{\tau_{1,j}\in S}+F_{\tau_2}^{(j)}\iverson{\tau_{2,j}\in S}\\
  &\quad+\iverson{b_j\in S}\bigl(K_j-d_j(S\setminus\{b_j\})\bigr)\qquad(S\subseteq E_j),
\end{aligned}
\]
we get
$\max_{S\subseteq E_j}|g_j(S)|\le4/3+\sum_{i\le j}\bigl(|F_{\tau_1}^{(i)}|+|F_{\tau_2}^{(i)}|+|K_i|+d_i(E_i\setminus\{b_i\})\bigr)\le2+114\,\Omega_j$.
Every reward is at most $M_j$, and every $1/\beta_e$ is at most $12\varpi_j$.

Next we bound the denominators. Here $1/\delta=3\muin_1$ is an integer, and
$\mu_j\in\delta\Z$ for $j\ge0$. All rewards, $\Delta_j=2490\cdot208^{j-1}$, and
$\varpi_j\Delta_j^2$ are integers, $\varpi_j\Delta_j\in\frac12\Z$, and
$1/\beta_e\in\{1,\frac14\}\cup\{\frac34\cdot8^i,9\cdot8^i\colon i\ge0\}\subseteq\frac14\Z$.
As common denominators we use the positive integers $\mathfrak D_0:=6$ and,
for $j\ge1$,
\[
  \mathfrak E_j:=\frac{600\cdot13\cdot415\cdot208^{j-1}}\delta,\qquad
  \mathfrak D_j:=12\,\mathfrak E_j,\qquad
  \mathfrak T_j:=3\cdot8^{j-1}\mathfrak E_j,\qquad
  \mathfrak P:=\frac{200\cdot13\cdot6\cdot415}\delta.
\]
At the step to level $j\ge1$, this gives the following.
\begin{itemize}
\item $w_j,m_{*,j}\mu_j\in\delta\Z$, and $d_e$ for $e\in E_j\setminus\{b_j\}$,
  $d_j(E_{j-1})$, $d_j(E_j\setminus\{b_j\})$, and $\Pi$ lie in $\frac\delta4\Z$.
\item $s_j,\bar m,\psi_j,q_j\in\frac1{13}\Z$, as $13s_j=12\Delta_j$.
\item $\Delta F_j\in\frac{\delta}{48\cdot2490\cdot208^{j-1}}\Z$ by (P5), as
  $1/s_j=13/(12\Delta_j)$.
\item $a_2\in\frac\delta{130}\Z$, as $\mu_j-|\mathcal R_{j-1}|/2\in\frac\delta{10}\Z$
  by $\frac1{15}=\frac15\delta\muin_1$.
\end{itemize}
By (P6), with $\varpi_j\bar d_jw_j=d_j(E_{j-1})$ and $\varpi_ja_0=q_{j-1}\mu_j-\Pi$,
\begin{equation}\tag{$*$}
  F_{\tau_2}^{(j)}=\varpi_jR_0m_\delta^{(j)}-\varpi_j\bar\rho_{j-1}-\frac32\varpi_jw_j+\varpi_ja_2\Delta_j-d_j(E_{j-1})+\frac{q_{j-1}\mu_j-\Pi}{\Delta_j}-\varpi_js_j\iota,
\end{equation}
where $\varpi_1\bar\rho_0=7/8$ and
$\varpi_j\bar\rho_{j-1}=\varpi_jR_0m_\delta^{(j-1)}-8F_{\tau_2}^{(j-1)}$ for $j\ge2$.
The seven terms lie in $\frac1{40}\Z$, in $\frac18\Z$ or
$\frac1{40}\Z+\frac1{\mathfrak E_{j-1}}\Z$, in $\frac\delta8\Z$,
$\frac\delta{260}\Z$, $\frac\delta4\Z$, $\frac\delta{312\cdot415\cdot208^{j-1}}\Z$,
and $\frac1{1300}\Z$, so all of them lie in $\frac1{\mathfrak E_j}\Z$. By
induction on $j$, $F_{\tau_2}^{(j)}\in\frac1{\mathfrak E_j}\Z$, and then
$F_{\tau_1}^{(j)}=F_{\tau_2}^{(j)}+\Delta F_j\in\frac1{\mathfrak D_j}\Z$. By (P7),
\[
  K_j=d_j(E_j\setminus\{b_j\})+12F_{\tau_1}^{(j)}-\frac{13}2\varpi_jw_j+6\varpi_j\Delta_j\bigl(R_0+R_1^{(j-1)}\bigr),
\]
with $R_0+R_1^{(j-1)}=1+\sum_{1\le i<j}\mu_i\in\delta\Z$, so
$K_j\in\frac1{\mathfrak D_j}\Z$. By the formula for $g_j$ above and induction on
$j$, starting from the base values in $\frac16\Z=\frac1{\mathfrak D_0}\Z$, every
value of $g_j$ lies in $\frac1{\mathfrak D_j}\Z$. Next, (P7) gives
$K_j-d_j(E_j\setminus\{b_j\})+F_{\tau_1}^{(j)}=13\bigl(F_{\tau_1}^{(j)}-\frac12\varpi_jw_j\bigr)+6\varpi_j\Delta_j(R_0+R_1^{(j-1)})$,
and $13\bar m=\frac{4993}2M_{j-1}$, so (P8) becomes
\begin{equation}\tag{$\dagger$}
  \Delta_p^{(j)}=2497M_{j-1}F_{\tau_2}^{(j)}+\frac{4993}2M_{j-1}\Delta F_j-\frac{4993}4M_{j-1}\varpi_jw_j+6\bar m\varpi_j\Delta_j\bigl(R_0+R_1^{(j-1)}\bigr).
\end{equation}
As $M_{j-1}=12\cdot208^{j-1}$, the four terms lie in
$\frac\delta{50\cdot13\cdot415}\Z$, $\frac\delta{48\cdot415}\Z$, $\frac\delta4\Z$,
and $\frac\delta{13}\Z$, hence in $\frac1{\mathfrak P}\Z$. As $p_0=10$, we get
$\Delta_p^{(j)},p_j\in\frac1{\mathfrak P}\Z$. Finally,
$\bar\rho_j=R_0m_\delta^{(j)}-F_{\tau_2}^{(j)}/\varpi_j$ and
$u^0_j=\bar\rho_{j-1}+w_j+s_j\iota-\bar\rho_j$ (\cref{lem:closedforms}(c)) lie
in $\frac1{\mathfrak T_j}\Z$. The integers $\mathfrak D_j$, $\mathfrak T_j$, and
$\mathfrak P$ have $\Oh(j+\log\muin_1)$ bits. With the magnitudes above, every
value of $g_{k-1}$ and every number $F_{\tau_1}$, $F_{\tau_2}$, $K$, $\Delta F$, $d_e$,
$\Delta_p$, $p$, $q$, $\bar\rho$, and $u^0$ of the levels $1,\dots,k-1$ therefore
has $\Oh(k+\log\muin_k)$ bits. The sums $d_j(E_{j-1})$ and $\Pi$ lie in $\frac\delta4\Z$, with magnitudes at most $2\Omega_j$ and $2M_j\Omega_j$, and $a_2$ lies in $\frac\delta{130}\Z$, with magnitude at most $1+\mu_j$, as $|\mathcal R_{j-1}|\le R_0+R_1^{(j-1)}\le1+\mu_j$. So they too have $\Oh(k+\log\muin_k)$ bits, and so do the remaining quantities $\bar d$, $\varphi$, and $a_0$ of the parameter rule, which are obtained from these by $\Oh(1)$ arithmetic operations at each level.

Finally, we bound the running time. Per level, the parameter rule needs
$\Oh(|E_{k-1}|)$ arithmetic operations on
numbers of $\Oh(k+\log\muin_k)$ bits. Values of $g_{k-1}$ are computed
recursively, by the formula above, in $\Oh(k|E_{k-1}|)$ operations.
\end{proof}

\begin{proof}[Proof of \cref{thm:gadget}]
Let $\muin\in\Fin$ have length $k$, and let $g:=g_{k-1}$ on $E:=E_{k-1}$, with
$|E|=3k$.

We first define the output and the numbers that come with it. Put
$\Theta:=\max_{e\in E}\bigl(\beta_eg(\{e\})-\zeta_1m_e\bigr)$
and
\[
\begin{aligned}
  \Lambda&:=\max\Bigl(0,\ \max_{e\in E}\frac{\Theta/\beta_e-g(E)+g(E\setminus\{e\})}{t_e}\Bigr),\\
  N_{\min}&:=\max\bigl(1,\lceil(\zeta_1+\Lambda)/\delta\rceil\bigr).
\end{aligned}
\]
For an integer $N\ge N_{\min}$, put $\Lambda':=\delta N-\zeta_1\ge\Lambda$,
$f_c:=g+\Lambda't-\Theta\binv$, $q:=q_{k-1}$, and $\alpha_\Gamma:=\delta N$. The
output $I(\muin,N)$ is the instance $(f_c,m,\beta)$ plus a capped linear summand $e_g$
(\cref{lem:ops}(c)) with the capacity $C_g:=(q+\kappa)\alpha_\Gamma$ and the
share $\aknob:=\alpha_\Gamma$, so that $\beta_{e_g}=1/(q+\kappa)$. It comes with
$q$ and $p:=p_{k-1}-q\Lambda'+\Lambda'm^\top t-\Theta\,t(E)+C_g$.

Next we show that $f_c$ is a polymatroid rank function and that $(f_c,m,\beta)$
keeps the gap of level $k-1$, shifted by $\Lambda'$, to the right of
$\alpha_\Gamma$. By \cref{prop:base} and \cref{lem:cutforms}(c), the
function $g$ has a cut form, so it is submodular by \cref{lem:cutforms}(a), and
$g(\emptyset)=f_0(\emptyset)=0$. We apply \cref{lem:embed} with $\tilde f:=g$,
$a:=t$, $b:=\binv$, and the parameters $(\Lambda',\Theta)$. Then its function
$f$ is $f_c$, and its instance is $(f_c,m,\beta)$, as $a_e/b_e=m_e$ and
$1/b_e=\beta_e$. Its hypotheses hold. The vectors $t$ and $\binv$ are positive,
as $m,\beta>0$. Next, $\Theta>0$, as $g=f_0$ on $2^{E_0}$ by \cref{def:tower},
so $\Theta\ge\beta_{C_1}g(\{C_1\})-\zeta_1m_{C_1}=4\cdot\frac76-\frac{12}3=\frac23$.
Finally, for $e\in E$ and $S\subseteq E\setminus\{e\}$, submodularity gives
$g(S)-g(S\cup\{e\})\le g(E\setminus\{e\})-g(E)$, with equality for
$S=E\setminus\{e\}$. So the lower bound that \cref{lem:embed} imposes on its
first parameter is $\Lambda$, and $\Lambda'\ge\Lambda$. Equivalently, for
$e\in E$ and $S\subseteq E\setminus\{e\}$, the marginal value
$f_c(S\cup\{e\})-f_c(S)$ is at least
$f_c(E)-f_c(E\setminus\{e\})=g(E)-g(E\setminus\{e\})+\Lambda't_e-\Theta/\beta_e$,
which is nonnegative as $\Lambda'\ge\Lambda$. By \cref{lem:embed}, $f_c$ is a
polymatroid rank function.

\Cref{lem:embed} describes the level sets of $(f_c,m,\beta)$, and
\cref{lem:fullplane}(g), with $(\Lambda',\Theta)$ in place of $(\Lambda,\Theta)$,
gives its best response $\xs_c$ itself. For $\alpha\ge\zeta_1$,
$\xi^g_e(\alpha)\le g(\{e\})$ and $m_e>0$ give
\[
  \rho^g_e(\alpha)\ge\alpha m_e-\beta_eg(\{e\})\ge\zeta_1m_e-\beta_eg(\{e\})\ge-\Theta,
\]
and $\alpha+\Lambda'\ge0$. So, by \cref{lem:fullplane}(g),
$\xs_c(\alpha+\Lambda')=\xi^g(\alpha)+\Lambda't-\Theta\binv$ for
$\alpha\ge\zeta_1$, and as $m^\top\binv=t(E)$, the reward of $(f_c,m,\beta)$
under its best response satisfies
\[
  r(\xs_c(\alpha+\Lambda'))=r(\xi^{(k-1)}(\alpha))+\Lambda'm^\top t-\Theta\,t(E)\qquad(\alpha\ge\zeta_1).
\]
Let $L_c(\alpha):=L_{k-1}(\alpha-\Lambda')+\Lambda'm^\top t-\Theta\,t(E)$. For
$\alpha\ge\alpha_\Gamma=\zeta_1+\Lambda'$, so that
$\alpha-\Lambda'\ge\zeta_1>R_0$, (H$_{k-1}$) of \cref{thm:alllevels} gives
\[
\begin{aligned}
  L_c(\alpha)-r(\xs_c(\alpha))&=L_{k-1}(\alpha-\Lambda')-r(\xi^{(k-1)}(\alpha-\Lambda'))\\
  &\ge\kappa\min\bigl(\dist(\alpha,\Lambda'+\Gamma_{k-1}),\delta\bigr),
\end{aligned}
\]
with equality $0$ on
$\Lambda'+\Gamma_{k-1}=\{\delta(N+\chi^\top\muin)\colon\chi\in\{0,1\}^k\}=\Gamma$,
whose smallest point is $\alpha_\Gamma$.

Then we show that the output is a comb. The summand $e_g$ takes care of the gap
to the left of the contacts. By \cref{lem:ops}(c), the reward of the output under
its best response is $r(\xs_c(\alpha))+C_g\min(\alpha/\alpha_\Gamma,1)$, and
$L:=L_c+C_g$ is the line $p+q\alpha$. On $[\alpha_\Gamma,\infty)$, the summand $e_g$ adds
$C_g$ to the reward and to the line, so the gap of the output is
$L_c(\alpha)-r(\xs_c(\alpha))$. By the bound in the previous paragraph, it is at least $\kappa\min(\dist(\alpha,\Gamma),\delta)$. On $[0,\alpha_\Gamma)$, the
reward $r(\xs_c(\cdot))$ is nondecreasing (\cref{lem:monreward}), so
\[
  r(\xs_c(\alpha))\le r(\xs_c(\alpha_\Gamma))=L_c(\alpha_\Gamma)=L_c(\alpha)+q(\alpha_\Gamma-\alpha),
\]
and the gap of the output satisfies
\[
\begin{aligned}
  L(\alpha)-r(\xs_c(\alpha))-C_g\frac{\alpha}{\alpha_\Gamma}
  &\ge-q(\alpha_\Gamma-\alpha)+C_g\Bigl(1-\frac{\alpha}{\alpha_\Gamma}\Bigr)\\
  &=\kappa(\alpha_\Gamma-\alpha)=\kappa\dist(\alpha,\Gamma).
\end{aligned}
\]
So the output is a comb with the stated contact set, line, dip, and width
(\cref{def:comb}), and $q=q_{k-1}\ge q_0=7>0$ by (P8).

Finally, we show that the output lies in $\Ccut$, that its numbers have
polynomial size, and that it is computed in polynomial time. The function $f_c$ is
obtained from $f_0$
by one-element extensions, direct sums with modular singletons, and the addition
of a modular function, and adding $e_g$ is a direct sum with a modular singleton. By
\cref{prop:base} and \cref{lem:cutforms}(c), the output has a cut form, computed
along the construction. Its rank function is a polymatroid rank function, as
$f_c$ is one and $C_g>0$, and with $3k+1$ elements and positive rational $m$ and
$\beta$, the output lies in $\Ccut$. For the sizes, \cref{lem:sizes} applies to
$g$, $m$, and $\beta$. Moreover,
$|\Theta|\le M_{k-1}+4\max_{S\subseteq E}|g(S)|$, since $\zeta_1<1$ and
$\beta_e\le4$, and
$\Lambda\le12\varpi_{k-1}|\Theta|+2\max_{S\subseteq E}|g(S)|$, since
$1/\beta_e\le12\varpi_{k-1}$ and $t_e\ge1$. Also $\Lambda'\le\delta N$. So
$N_{\min}$ has $\Oh(k+\log\muin_k)$ bits. For the denominators,
$\zeta_1=1/3$, all rewards are integers, and
$\beta_e\in\frac1{9\cdot8^k}\Z$, so
$\Theta\in\frac1{9\cdot8^k\mathfrak D_{k-1}}\Z$ with $\mathfrak D_{k-1}$ from
the proof of \cref{lem:sizes}. Moreover, $2t_e\in\Z$ (the dummies $\tau_{1,1}$ and
$\tau_{2,1}$ have $t_e=9/2$),
$1/\beta_e\in\frac14\Z$, $\Lambda'=\delta N-\frac13\in\frac\delta3\Z$, and
$\kappa\in\frac1{10}\Z$. The numbers of the output are the rewards and
curvatures (among them $\beta_{e_g}=1/(q+\kappa)$ of $e_g$), the values of
$f_c$, the capacities and weights of its cut form, $C_g$, $p$, $q$, $\delta$, and
$\kappa$. Apart from $\beta_{e_g}$, each of them is a sum of $\Oh(|E|)$ products
of at most three of the numbers $\Theta$, $\Lambda'$, $\kappa$,
$\alpha_\Gamma=\delta N$, values of $g$, numbers of the parameter rule, rewards, $1/\beta_e$,
and $t_e$. So their denominators divide a common integer with
$\Oh(k+\log\muin_k+\log N)$ bits, and so do their magnitudes. Hence every number
of the output, $\beta_{e_g}$ included, has $\Oh(k+\log\muin_k+\log N)$ bits. The
values $g(E)$, $g(E\setminus\{e\})$, and $g(\{e\})$ needed for $\Theta$ and~$\Lambda$ are $2|E|+1$ evaluations of $g$. So the algorithm runs in polynomial
time.
\end{proof}

\subsection{Exact best responses and NP-completeness}\label{sec:D-npcomplete}

With the comb-gadget theorem proved, it remains to establish polynomial-size certificates and complete the proof of \cref{thm:nphard}.
Membership in NP and the hardness of computing an optimal contract use exact best responses in the bit model.
\Cref{thm:waterfilling} counts submodular minimizations, evaluations of $f$, and arithmetic operations, but not the sizes of the numbers.
For cut forms, we therefore bound the encoding length of $\xs(\alpha)$ and compute it by maximum flows on numbers of polynomial size.
The \emph{encoding length} of a rational number $u/v$, with coprime $u\in\Z$ and $v\in\Z_{>0}$, is
\[
  \enc{u/v}:=1+\lceil\log_2(|u|+1)\rceil+\lceil\log_2(v+1)\rceil.
\]
This is one bit less than the encoding length of~\cite[\S1.3]{GLS88}, which changes only constants.
For an instance $I$ of $\Ccut$, let $\enc{I}$ be the sum of the encoding lengths of the capacities, weights, rewards, and curvatures of $I$, which is at most the encoding length of $I$.
As in the proof of \cref{thm:alm}, the source and the sink of a network are $s$ and $s'$, and for a vector $y\in\Q^E$ we write $\cut_D+y$ for the function $S\mapsto\cut_D(S)+y(S)$.

\begin{lemma}\label{lem:exactbr}
Let $I=(f,m,\beta)$ be an instance of $\Ccut$ with the cut form $(D,\mathrm{cap},\omega)$, and let $\alpha\ge0$ be rational.
\begin{enumerate}[label=(\alph*)]
\item The best response $\xs(\alpha)$ is a rational vector, and each of its coordinates has encoding length at most $6\enc{\alpha}+31\enc{I}$.
\item Given a rational vector $x\in\Q^E$, one can decide in polynomial time whether $x=\xs(\alpha)$.
\item The best response $\xs(\alpha)$ can be computed in polynomial time.
\end{enumerate}
Here polynomial time means polynomial in the encoding lengths of $I$ and $\alpha$, and in (b) also of $x$.
\end{lemma}

\begin{proof}
A product or a quotient of rationals has encoding length at most the sum of their encoding lengths, and a sum at most twice the sum of their encoding lengths, as one sees by writing the terms over the product of their denominators.
Every rational $u/v$ as above satisfies $|u|<2^{\enc{u/v}}$ and $v<2^{\enc{u/v}}$.
Every value $f(S)=\cut_D(S)+\omega(S)$ is a sum of distinct capacities and weights, and $t(S)$ and $\binv(S)$ are sums of distinct quotients $m_e/\beta_e$ and $1/\beta_e$, so all three have encoding length at most $2\enc{I}$.

\textit{(a)} Let $x:=\xs(\alpha)$ and $\rho:=\rho(\alpha)=\alpha m-\beta x$, the vector of the marginals $\rho_e(\alpha)=\alpha m_e-\beta_ex_e$.
By \cref{prop:optchar}, $\rho\ge0$, and every level set $A_\theta(\alpha)=\{e\in E\colon\rho_e\ge\theta\}$ with $\theta>0$ is tight for $x$.
As in the proof of \cref{prop:optchar}, let $\varrho_1>\dots>\varrho_k$ be the distinct positive values of $\rho$, and let $A_i:= A_{\varrho_i}(\alpha)$ for $i=1,\dots,k$.
These sets are tight for $x$.
Put $A_0:=\emptyset$ and $D_i:= A_i\setminus A_{i-1}$ for $i=1,\dots,k$ (the blocks $D_i$ are subsets of $E$, not to be confused with the digraph $D$).
Then $\rho_e=\varrho_i$ for $e\in D_i$, also if $x_e=0$, and $\rho_e=0$ for $e\in E\setminus A_k$.
Summing $x_e=(\alpha m_e-\rho_e)/\beta_e$ over $D_i$ gives
\[
  f(A_i)-f(A_{i-1})=x(D_i)=\alpha t(D_i)-\varrho_i\binv(D_i),
\]
one linear equation for $\varrho_i$, with $\binv(D_i)>0$ as $D_i\ne\emptyset$.
Hence
\begin{equation}\label{eq:D-explicit}
\begin{aligned}
  \varrho_i&=\frac{\alpha t(D_i)-f(A_i)+f(A_{i-1})}{\binv(D_i)}&&\text{for }i=1,\dots,k,\\
  x_e&=(\alpha m_e-\varrho_i)/\beta_e&&\text{for }e\in D_i\text{ and }i=1,\dots,k,\\
  x_e&=\alpha t_e&&\text{for }e\in E\setminus A_k.
\end{aligned}
\end{equation}
In particular, $x$ is rational.
The numerator of $\varrho_i$ is a sum of three terms of encoding lengths at most $\enc{\alpha}+2\enc{I}$, $2\enc{I}$, and $2\enc{I}$.
So, by the rules above, $\enc{\varrho_i}\le2(\enc{\alpha}+6\enc{I})+2\enc{I}=\Omega$ for
\[
  \Omega:=2\enc{\alpha}+14\enc{I},
\]
and $\enc{x_e}\le2(\enc{\alpha}+\enc{I}+\Omega)+\enc{I}=6\enc{\alpha}+31\enc{I}$ for $e\in A_k$.
For $e\in E\setminus A_k$, we have $\enc{x_e}\le\enc{\alpha}+\enc{t_e}\le\enc{\alpha}+\enc{I}$.

\textit{(b)} For $y\in\Q^E$, let $G_y$ be the network with a source $s$, a sink $s'$, and the nodes $E$, with the arcs of $D$ and their capacities, and with an arc $s\to e$ of capacity $\max(-y_e,0)$ and an arc $e\to s'$ of capacity $\max(y_e,0)$ for every $e\in E$.
For $S\subseteq E$, the arcs that leave $\{s\}\cup S$ have total capacity
\begin{align*}
  \cut_D(S)+\sum_{e\in S}\max(y_e,0)+\sum_{e\in E\setminus S}\max(-y_e,0)
  =\cut_D(S)+y(S)+\sum_{e\in E}\max(-y_e,0),
\end{align*}
as $\max(y_e,0)-\max(-y_e,0)=y_e$.
So the minimizers of $\cut_D+y$ are the sets $S$ for which $\{s\}\cup S$ is the source side of a minimum $s$-$s'$ cut.
After multiplication by the product of their denominators, the capacities are integers of polynomial encoding length, and the algorithm of Edmonds and Karp computes a maximum flow in polynomial time \cite[Thm.~10.5]{Sch03}.
This gives the minimum of $\cut_D+y$ and its largest minimizer.
Indeed, for a maximum flow, a cut is minimum if and only if no arc of the residual network leaves its source side, so the largest minimizer is the set of nodes of $E$ from which $s'$ cannot be reached in the residual network~\cite[\S2.1--2.2]{GGT89}.

Now let $x\in\Q^E$, and let $\rho(x)$ and $A_\theta(x)$ be as in \cref{sec:bestresponse}, for the share $\alpha$.
By \cref{prop:optchar}, $x=\xs(\alpha)$ if and only if $x\ge0$, $\rho(x)\ge0$, $x(S)\le f(S)$ for every $S\subseteq E$, and the at most $|E|$ sets $A_\theta(x)$, for the positive values $\theta$ of $\rho(x)$, are tight for $x$.
Indeed, every set $A_\theta(x)$ with $\theta>0$ is one of them or empty.
The sign conditions take $2|E|$ comparisons, and the tightness takes $|E|$ evaluations of $f$, each a sum of input numbers.
The third condition says that $\cut_D+(\omega-x)$, which vanishes at $\emptyset$, has minimum $0$, and one maximum flow in $G_{\omega-x}$ decides it.

\textit{(c)} Let $\varrho_1,\dots,\varrho_k$, $A_0,\dots,A_k$, and $\Omega$ be as in the proof of (a), and put $\varrho_0:=\infty$ and $\varrho_{k+1}:=0$.
Then $A_\theta(\alpha)=A_i$ for $\varrho_{i+1}<\theta\le\varrho_i$ and $i=0,\dots,k$.
By \cref{thm:levelsets}, $A_\theta(\alpha)$ is the largest minimizer of
\[
  F_{\alpha,\theta}(S)=f(S)-\alpha t(S)+\theta\binv(S)=\cut_D(S)+(\omega-\alpha t+\theta\binv)(S),
\]
where $t$ and $\binv$ also denote the vectors $(t_e)_{e\in E}$ and $(\beta_e^{-1})_{e\in E}$.
So, for rational $\theta$, it is computed by one maximum flow in $G_{\omega-\alpha t+\theta\binv}$, as in (b).
By (a), the numerators and denominators of $\varrho_1,\dots,\varrho_k$ have absolute value less than $2^\Omega$.
So $\varrho_k>2^{-\Omega}$ and $\varrho_1<2^\Omega$, and $\varrho_i-\varrho_{i+1}>2^{-2\Omega}$ for $1\le i\le k$, as $\varrho_i-\varrho_{i+1}$ is a positive rational with a denominator less than $2^{2\Omega}$.
Let $h:=2^{-2\Omega}$.
We compute $A_1,\dots,A_k$ in turn.
Let $0\le i\le k$, and suppose that $A_i$ is known.
As $0<h<\varrho_k$, we have $A_h(\alpha)=A_k$, so if $A_h(\alpha)=A_i$, then $i=k$ and we stop.
Otherwise $i<k$, and for $j\in\{1,\dots,2^{3\Omega}\}$ we have $A_{jh}(\alpha)\supsetneq A_i$ if and only if $jh\le\varrho_{i+1}$.
As $h<\varrho_{i+1}<2^\Omega=2^{3\Omega}h$, a binary search over $j$ with at most $3\Omega+1$ maximum flows finds the largest such $j$, namely $j=\lfloor\varrho_{i+1}/h\rfloor$.
It satisfies $\varrho_{i+2}<\varrho_{i+1}-h<jh\le\varrho_{i+1}$, so $A_{jh}(\alpha)=A_{i+1}$.
Every $\theta=jh$ has encoding length at most $5\Omega+3$, so all capacities in these flows have polynomial encoding length.
After at most $|E|+1$ rounds, \eqref{eq:D-explicit} gives $\xs(\alpha)$ from $A_1,\dots,A_k$ with polynomially many further operations on numbers whose encoding lengths are bounded as in the proof of (a).
\end{proof}

We can now combine the reduction with the certificate bounds to prove NP-completeness.

\restate{\rsNphard}{thm:nphard}
\begin{proof}[Proof of \cref{thm:nphard}]
We first establish NP-hardness.
The reduction of \cref{def:reduction} runs in time polynomial in the encoding length of $(a,T)$.
Indeed, the algorithm of \cref{thm:gadget} computes $N_0$ and both combs in polynomial time, as its inputs have $\log\mu_k,\log\mu'_k=\Oh(k+\log(2+\sum_i a_i))$ and the integers $N_0$ and $N_0+T$ have bit lengths bounded in the same way (\cref{lem:sizes}, the proof of \cref{thm:gadget}, and $T\le\sum_i a_i$).
Apart from the choice of $\sigma$, the steps \ref{red:3} to \ref{red:5} take $\Oh(k)$ arithmetic operations, and $\sigma$ is a power of $2$ that is at least half the smallest right-hand side in step \ref{red:4}, so $\log(1/\sigma)$ is polynomial and $\sigma$ is found by as many comparisons.
The output lies in $\Ccut$, since its rank function is a direct sum of polymatroid rank functions whose cut forms, together with an isolated node of weight $C$ for the centering summand, form a cut form of it (\cref{lem:cutforms}), and its rewards and curvatures are positive rationals.
By \cref{prop:correct}, and as Subset Sum is NP-complete~\cite{GJ79}, deciding whether $\UP(\as)\ge\vartheta$ is NP-hard on $\Ccut$.

We next show membership in NP.
The inputs are recognized in polynomial time by \cref{lem:cutforms}(b).
As $\UP(\as)=\max_{\alpha\in[0,1]}\UP(\alpha)$, we have $\UP(\as)\ge\vartheta$ if and only if some $\alpha\in[0,1]$ has $\UP(\alpha)\ge\vartheta$.
A certificate is such a rational share $\alpha$ together with the vector $x=\xs(\alpha)$.
It is verified in polynomial time: check that $\alpha\in[0,1]$, that $x=\xs(\alpha)$ by \cref{lem:exactbr}(b), and that $(1-\alpha)\,r(x)\ge\vartheta$, in exact arithmetic.
A yes-instance $I$ has a certificate of polynomial encoding length.
Indeed, by the rules at the start of the proof of \cref{lem:exactbr}, every reward, every curvature, and every value of $f$ has encoding length at most $2\enc{I}$, so it is a quotient of two integers of absolute value less than $2^{2\enc{I}}$.
Moreover, $n\le\enc{I}$, as every reward has encoding length at least $1$.
So the hypothesis of \cref{thm:bitlength} holds with $\nu:=2\enc{I}$.
By \cref{thm:bitlength}, some optimal contract $\as$ is a rational whose numerator and denominator have bit length $\Oh(n\enc{I})$.
Then $\UP(\as)\ge\vartheta$, and by \cref{lem:exactbr}(a), each coordinate of $\xs(\as)$ has encoding length at most $6\enc{\as}+31\enc{I}=\Oh(n\enc{I})$.
So $(\as,\xs(\as))$ is a certificate of polynomial encoding length.
\end{proof}

Computing an optimal contract is NP-hard under polynomial-time Turing reductions: a polynomial-time algorithm returning a rational optimal contract $\as$ would also allow us to compute $\xs(\as)$ and $\UP(\as)$ exactly in polynomial time by \cref{lem:exactbr}(c). Comparing this value with $\vartheta$ would decide the NP-complete problem of \cref{thm:nphard}.

\section{Omitted proofs for Section~\ref{sec:fptas}}\label{app:halfrule}

\restate{\rsFptas}{thm:fptas}
\begin{proof}
Put $\varepsilon':=\varepsilon/2$, $J:=\min\{j\ge0:(1-\varepsilon')^j\le1-\bar\alpha\}$, and $s:=\lceil\log_2(4/(\varepsilon(1-\bar\alpha)))\rceil$. The scheme evaluates the dyadic shares
\[
\hat\alpha_j:=1-2^{-s}\big\lfloor2^s(1-\varepsilon')^j\big\rfloor,\qquad j=0,\ldots,J,
\]
and returns the one with largest utility. There are at most $\lceil2\varepsilon^{-1}\ln(1/(1-\bar\alpha))\rceil+1$ shares.

Let $b_j:=1-\hat\alpha_j$. Then $b_0=1$, $b_j$ is nonincreasing in $j$, and $(1-\varepsilon')^j-2^{-s}<b_j\le(1-\varepsilon')^j$. Let $\alpha\in[0,\bar\alpha]$. We show that $\UP(\hat\alpha_j)\ge(1-\varepsilon)\UP(\alpha)$ for some $j$, which proves the claim as $\bar\alpha$ is a share bound. As $b_J\le(1-\varepsilon')^J\le1-\bar\alpha\le1-\alpha$, there is a smallest $j$ with $b_j\le1-\alpha$, that is, $\hat\alpha_j\ge\alpha$. If $j=0$, then $\alpha=0=\hat\alpha_0$. Otherwise $b_{j-1}>1-\alpha$, and $(1-\varepsilon')^{j-1}>1-\bar\alpha$ by the minimality of $J$. With $2^{-s}\le\varepsilon(1-\bar\alpha)/4<\varepsilon'(1-\bar\alpha)$,
\[
b_j>(1-\varepsilon')^j-\varepsilon'(1-\bar\alpha)>(1-\varepsilon')^{j}-\varepsilon'(1-\varepsilon')^{j-1}=(1-\varepsilon)(1-\varepsilon')^{j-1}\ge(1-\varepsilon)\,b_{j-1}.
\]
By \cref{lem:monreward}, $r(\xs(\hat\alpha_j))\ge r(\xs(\alpha))$, so
\[
\UP(\hat\alpha_j)=b_j\,r(\xs(\hat\alpha_j))\ge(1-\varepsilon)\,b_{j-1}\,r(\xs(\alpha))\ge(1-\varepsilon)(1-\alpha)\,r(\xs(\alpha))=(1-\varepsilon)\UP(\alpha).
\]
Finally, $J\le\lceil\ln(1/(1-\bar\alpha))/\ln(1/(1-\varepsilon'))\rceil\le\lceil\ln(1/(1-\bar\alpha))/\varepsilon'\rceil$, and $s\le\log_2(4/(\varepsilon(1-\bar\alpha)))+1$.
\end{proof}

\restate{\rsProbe}{cor:probe}
\begin{proof}
By \cref{lem:monreward}, $0<r(\xs(a))\le r(\xs(1))$, so $a\le\bar\alpha_a<1$. For $\alpha>\bar\alpha_a$, again by \cref{lem:monreward}, $\UP(\alpha)=(1-\alpha)r(\xs(\alpha))\le(1-\alpha)r(\xs(1))<(1-\bar\alpha_a)\,r(\xs(1))=\UP(a)$. So $\sup_{\alpha\in[0,\bar\alpha_a]}\UP(\alpha)=\mathrm{OPT}$. By \cref{thm:fptas} with $\bar\alpha=\bar\alpha_a$, at most $\lceil2\varepsilon^{-1}\ln(1/(1-\bar\alpha_a))\rceil+1$ shares need to be tried, and the best responses at $a$ and at $1$ add two more.
\end{proof}

\restate{\rsFptashom}{cor:fptashom}
\begin{proof}
By \cref{thm:prule}, $\bar\alpha=1/p$ is a share bound. Substituting this bound into \cref{thm:fptas} gives the shares, their number, and their bit length. Evaluate $(1-\alpha)r(\xs(\alpha))$ at each share and return one with the largest value.
\end{proof}

\restate{\rsFptasquad}{cor:fptasquad}
\begin{proof}
Apply \cref{cor:fptashom} with $p=2$. On an explicitly described rational polytope, the unique best response minimizes the strictly convex quadratic function $\sum_e\beta_ex_e^2/2-\alpha m^\top x$ over $K$. Rational convex quadratic programming computes this response exactly in polynomial time~\cite{KTK80}, and its reward is evaluated exactly. The grid has $\Oh(1/\varepsilon)$ shares, each with $\Oh(1 + \log(1/\varepsilon))$ bits, so the total running time is polynomial in the input encoding length and in $1/\varepsilon$.

For a polymatroid, each of the $\Oh(1/\varepsilon)$ best responses takes time $\Oh(n^7\log n)\cdot\FO+\Oh(n^8\log n)$ by \cref{thm:waterfilling}. Evaluating the reward takes $\Oh(n)$ further operations per response, so the total runtime is $\Oh(n^7\log n/\varepsilon)\cdot\FO+\Oh(n^8\log n/\varepsilon)$.
\end{proof}

\subsection{Lemmas needed for the proof of \cref{thm:prule}}\label{sec:E-prule}

For $z\ge0$ we write $\|z\|_c:=c(z)^{1/p}$.

\begin{lemma}\label{lem:segment}
For all $z,w\in\R^E_{\ge0}$ and $\sigma\in[0,1]$,
\[
  c\big((1-\sigma)z+\sigma w\big)\le\big((1-\sigma)\|z\|_c+\sigma\|w\|_c\big)^p.
\]
\end{lemma}

\begin{proof}
Let $Z:=\|z\|_c$ and $W:=\|w\|_c$. For $\sigma\in\{0,1\}$ the claim is an equality, so let $\sigma\in(0,1)$. If $Z,W>0$, then $c(z/Z)=c(w/W)=1$ by homogeneity. With $H:=(1-\sigma)Z+\sigma W>0$, the point $(1-\sigma)z+\sigma w$ is $H$ times the convex combination of $z/Z$ and $w/W$ with the weights $(1-\sigma)Z/H$ and $\sigma W/H$, at which $c$ is at most $1$ by convexity. So homogeneity gives $c\big((1-\sigma)z+\sigma w\big)\le H^p$. If $Z=0$, then $c(tz)=t^pc(z)=0$ for every $t>0$, and for every $\lambda\in(0,1)$ convexity and homogeneity give
\[
  c\big((1-\sigma)z+\sigma w\big)\le(1-\lambda)\,c\Big(\frac{1-\sigma}{1-\lambda}\,z\Big)+\lambda\,c\Big(\frac\sigma\lambda\,w\Big)=\lambda^{1-p}\sigma^pc(w).
\]
Letting $\lambda\to1$ gives $c\big((1-\sigma)z+\sigma w\big)\le(\sigma W)^p$, which is the claim. The case $W=0$ is symmetric.
\end{proof}

\begin{lemma}\label{lem:foc}
Let $\alpha>0$, let $x$ maximize $\UA(\cdot\mid\alpha)$ over $K$, and let $z\in K$. Then
\[
  \alpha\,\big(r(z)-r(x)\big)\le p\,\|x\|_c^{\,p-1}\big(\|z\|_c-\|x\|_c\big).
\]
\end{lemma}

\begin{proof}
Let $X:=\|x\|_c$ and $Z:=\|z\|_c$. For $\sigma\in(0,1]$, the point $x+\sigma(z-x)$ lies in the convex set $K$. By the optimality of $x$, the linearity of $r$, and \cref{lem:segment},
\[
  \alpha\,r(x)-X^p\ \ge\ \alpha\,r(x)+\sigma\alpha\big(r(z)-r(x)\big)-\big(X+\sigma(Z-X)\big)^p.
\]
Dividing by $\sigma$ and letting $\sigma\downarrow0$ gives the claim, since the right derivative of $\sigma\mapsto(X+\sigma(Z-X))^p$ at $0$ is $pX^{p-1}(Z-X)$. This also holds for $X=0$, where the derivative is $\lim_{\sigma\downarrow0}\sigma^{p-1}Z^p=0$, as $p>1$.
\end{proof}

We are now in a position to prove \cref{thm:prule}.

\restate{\rsPrule}{thm:prule}
\begin{proof}[Proof of \cref{thm:prule}]
We first show that the reward grows at most like $\alpha^{1/(p-1)}$. Let $0<\alpha<\alpha'$, let $x$ and $y$ maximize $\UA(\cdot\mid\alpha)$ and $\UA(\cdot\mid\alpha')$ over $K$, let $X:=\|x\|_c$ and $Y:=\|y\|_c$, and let $\varrho:=(\alpha'/\alpha)^{1/(p-1)}>1$. \Cref{lem:foc} with $z=0$ and with $z=y$ for $x$, and with $z=x$ for $y$, gives
\[
  \alpha\,r(x)\ge pX^p,\qquad\alpha\,\big(r(y)-r(x)\big)\le pX^{p-1}(Y-X),\qquad\alpha'\big(r(x)-r(y)\big)\le pY^{p-1}(X-Y).
\]
In particular, $r(x)\ge0$. If $Y\le X$, the second inequality gives $r(y)\le r(x)\le\varrho\,r(x)$. If $Y>X$, the second and the third give
\[
  \frac p{\alpha'}\,Y^{p-1}(Y-X)\le r(y)-r(x)\le\frac p\alpha\,X^{p-1}(Y-X),
\]
so $X>0$ and $(Y/X)^{p-1}\le\alpha'/\alpha$, that is, $Y\le\varrho X$. Then the second and the first inequality give
\[
  r(y)\le r(x)+\frac p\alpha\,X^{p-1}(\varrho X-X)=r(x)+(\varrho-1)\,\frac{pX^p}\alpha\le\varrho\,r(x).
\]
Applied to the best responses at two shares, this shows that $R(\alpha):=r(\xs(\alpha))/\alpha^{1/(p-1)}$ is nonincreasing on $(0,\infty)$.

Next we show that $\UP$ is nonincreasing on $[1/p,1]$. For $\alpha\in(0,1]$,
\[
  \UP(\alpha)=(1-\alpha)\,r(\xs(\alpha))=h(\alpha)R(\alpha),\qquad\text{where }h(\alpha):=(1-\alpha)\,\alpha^{1/(p-1)}.
\]
The function $h$ is nonnegative on $[0,1]$, and its derivative
\[
  h'(\alpha)=\alpha^{1/(p-1)-1}\Big(\frac1{p-1}-\frac p{p-1}\,\alpha\Big)
\]
shows that $h$ is increasing on $[0,1/p]$ and strictly decreasing on $[1/p,1]$. Moreover, $R\ge0$, as $m\ge0$ and $\xs(\alpha)\ge0$. So, for $1/p\le\alpha<\alpha'\le1$,
\begin{equation}\label{eq:E-product}
  \UP(\alpha)=h(\alpha)R(\alpha)\ge h(\alpha')R(\alpha)\ge h(\alpha')R(\alpha')=\UP(\alpha'),
\end{equation}
and $\UP$ is nonincreasing on $[1/p,1]$.

Now we show that $\UP$ is strictly decreasing on $[1/p,1]$ unless $r$ vanishes on $K$. Suppose that $r(z)>0$ for some $z\in K$, and let $\alpha>0$. For $t\in(0,1]$, the point $tz$ lies in $K$ by convexity and $0\in K$, and $\UA(tz\mid\alpha)=t\alpha\,r(z)-t^pc(z)>0$ for small $t$, as $p>1$. As $c\ge0$,
\[
  \alpha\,r(\xs(\alpha))\ge\UA(\xs(\alpha)\mid\alpha)\ge\UA(tz\mid\alpha)>0.
\]
So $R>0$ on $(0,\infty)$, and the first inequality in \eqref{eq:E-product} is strict. Hence $\UP$ is strictly decreasing on $[1/p,1]$, so $\UP(\alpha)<\UP(1/p)$ for $\alpha\in(1/p,1]$, and every optimal contract has a share in $[0,1/p]$. If, in contrast, $r$ vanishes on $K$, then $\UP\equiv0$ and every share is optimal.

Finally, we show that the bound $1/p$ on the share of an optimal contract is attained. For one element with $K=[0,1]$, $m_e=1$, and $c(x)=x^p/p$, the agent maximizes $\alpha x-x^p/p$ over $[0,1]$, so $\xs(\alpha)=\alpha^{1/(p-1)}$ for $\alpha\in[0,1]$. Then $\UP=h$ on $[0,1]$, which is maximal exactly at $\alpha=1/p$.
\end{proof}

\subsection{Lower bounds}\label{sec:E-lower}

This subsection proves \cref{prop:querylb} and the bound for sets of shares that do not depend on the instance (\cref{prop:sharesets}). Both constructions have one element $e$, so an action is a number $z\in[0,f(\{e\})]$. For $\alpha<1$, the best response $\xs(\alpha)$ is the largest maximizer of $\alpha m_ez-c(z)$ over this interval, as ties are broken in favor of the principal (\cref{sec:model}), and at $\alpha=1$ the maximizer is unique in both constructions. If $c(z)=\int_0^zg(y)\,\mathrm dy$ for a nondecreasing function $g$, then $\alpha m_ez-c(z)$ is nondecreasing in $z$ as long as $g\le\alpha m_e$ to the left of $z$, and decreasing afterwards, so
\begin{equation}\label{eq:E-largest}
\xs(\alpha)=\sup\bigl\{z\in[0,f(\{e\})]:g(y)\le\alpha m_e\text{ for all }0\le y<z\bigr\}.
\end{equation}

\begin{proposition}\label{prop:sharesets}
Let $\varepsilon\in(0,1)$, and let $A\subseteq[0,1]$ contain a $(1-\varepsilon)$-optimal contract of every instance with one element $e$, $m_e=\beta_e=1$, and quadratic costs. Then $|A|\ge\lfloor1/(5\varepsilon)\rfloor$.
\end{proposition}

\begin{proof}
For $u\in(0,1/3]$, let $I_u$ be the instance with $f(\{e\})=u$. By \eqref{eq:E-largest}, its best response is $\xs(\alpha)=\min\{\alpha,u\}$, so $\UP(\alpha)=(1-\alpha)\min\{\alpha,u\}$. As $\alpha\mapsto\alpha(1-\alpha)$ is increasing on $[0,1/2]$ and $u\le1/2$, the optimal contract is $u$, and $\mathrm{OPT}=u(1-u)$. Let $\mathcal G(u)$ be the set of $(1-\varepsilon)$-optimal contracts of $I_u$.

We show that $\mathcal G(u)\subseteq[u-\frac23\varepsilon,u+\varepsilon)$. A share $\alpha\ge u$ is $(1-\varepsilon)$-optimal if and only if $(1-\alpha)u\ge(1-\varepsilon)(1-u)u$, that is, $\alpha\le u+\varepsilon(1-u)$, and $u+\varepsilon(1-u)<u+\varepsilon$. A share $\alpha\le u$ is $(1-\varepsilon)$-optimal if and only if $\alpha(1-\alpha)\ge(1-\varepsilon)u(1-u)$. The function $d\mapsto(u-d)(1-u+d)$ decreases from $u(1-u)$ to $0$ on $[0,u]$, so there is a unique $d\in[0,u]$ with $(u-d)(1-u+d)=(1-\varepsilon)u(1-u)$, and a share $\alpha\le u$ is $(1-\varepsilon)$-optimal if and only if $\alpha\ge u-d$. Expanding gives $d(1-2u+d)=\varepsilon u(1-u)$, so $d\le\varepsilon u(1-u)/(1-2u)\le\frac23\varepsilon$, as $u\mapsto u(1-u)/(1-2u)$ is increasing on $[0,1/2)$ and equals $2/3$ at $u=1/3$.

Let $N:=\lfloor1/(5\varepsilon)\rfloor$ and $u_i:=\frac53\varepsilon i$ for $i\in[N]$, so that $u_N\le1/3$. As $u_i+\varepsilon=u_{i+1}-\frac23\varepsilon$, the sets $\mathcal G(u_1),\dots,\mathcal G(u_N)$ are pairwise disjoint. The set $A$ meets each of them, so $|A|\ge N$.
\end{proof}

The instances differ only in $f$, so the bound holds even for sets of shares that depend on $m$ and $\beta$. It concerns sets of shares, not adaptive algorithms: on these instances, one best response suffices, as $\xs(1)=u$ is the optimal contract.

\begin{proposition}\label{prop:querylb}
Let $\varepsilon\in(0,1/2)$ and $\bar\alpha\in(0,1)$, and let $\mathcal A$ be a deterministic algorithm that is given $\varepsilon$ and $\bar\alpha$, learns about the costs only from best responses at shares of its choice, with ties broken in favor of the principal as in \cref{sec:model}, and returns a $(1-\varepsilon)$-optimal contract for every instance with one element, $m_e=1$, $f(\{e\})=1/(1-\bar\alpha)$, and convex costs for which $\bar\alpha$ is a share bound. Then $\mathcal A$ needs more than $\ln(1/(1-\bar\alpha))/(2\varepsilon)-3$ best-response computations on one of these instances, on which the share bound $\bar\alpha_a$ of \cref{cor:probe} satisfies $\bar\alpha_a\le\bar\alpha$ for every $a\in[\varepsilon,\bar\alpha]$.
\end{proposition}
\begin{proof}
If $\bar\alpha<\varepsilon$, then $\ln(1/(1-\bar\alpha))<\ln(1/(1-\varepsilon))\le\varepsilon/(1-\varepsilon)<2\varepsilon$, so the bound is negative and there is nothing to show. So let $\bar\alpha\ge\varepsilon$. Let $\gamma:=1+2\varepsilon$, let $l_j:=1-(1-\varepsilon)\gamma^{-j}$ for $j\ge0$, and let $K\ge0$ be the largest integer with $(1-\varepsilon)\gamma^{-K}\ge1-\bar\alpha$. Then $\varepsilon=l_0<l_1<\dots<l_K\le\bar\alpha$.

We now define a base instance $I_0$ and $K$ variants $I_1,\dots,I_K$ of it. All instances have one element $e$ with $m_e=1$ and $f(\{e\})=1/(1-\bar\alpha)$, so $r(z)=z$, and they differ only in the costs. The base instance $I_0$ has the cost $c_0(z):=\int_0^zc_0'(y)\,\mathrm dy$ with
\[
c_0'(y):=\begin{cases}\varepsilon(1-\varepsilon)\,y,&0\le y\le\frac1{1-\varepsilon},\\1-\frac1y,&\frac1{1-\varepsilon}\le y\le\frac1{1-\bar\alpha},\\\bar\alpha,&y\ge\frac1{1-\bar\alpha}.\end{cases}
\]
The pieces agree where they meet, so $c_0'$ is continuous, nonnegative, and nondecreasing, and $c_0$ is convex with $c_0(0)=0$ and $c_0\ge0$. For $k\in[K]$, the instance $I_k$ has the cost $c_k(z):=\int_0^zg_k(y)\,\mathrm dy$, where $g_k$ equals $c_0'$, except that $g_k:=l_{k-1}$ on $[\frac1{1-l_{k-1}},\frac1{1-l_k})\subseteq[\frac1{1-\varepsilon},\frac1{1-\bar\alpha}]$. There $c_0'$ increases from $l_{k-1}$ to $l_k$, so $g_k$ is still nondecreasing, and $c_k$ is convex with $c_k(0)=0$ and $c_k\ge0$.

Next we compute the best responses of these instances. By \eqref{eq:E-largest}, the best response of $I_0$ is
\[
x_0(\alpha)=\begin{cases} \frac{\alpha}{\varepsilon(1-\varepsilon)},&0\le\alpha\le\varepsilon,\\[4pt] \frac{1}{1-\alpha},&\varepsilon\le\alpha\le\bar\alpha,\\[4pt] \frac{1}{1-\bar\alpha},&\bar\alpha\le\alpha\le1.\end{cases}
\]
At $z=1/(1-l_k)$ the left and right slopes of $c_k$ are $l_{k-1}$ and $l_k$, respectively. Thus the best response of $I_k$ is the function $x_k$ that equals $x_0$ outside the \emph{window} $[l_{k-1},l_k)$ and equals $1/(1-l_k)$ on it. At $\alpha=l_{k-1}$, the maximizers of $I_k$ form the interval $[\frac1{1-l_{k-1}},\frac1{1-l_k}]$, and the tie is broken in favor of the principal. In particular, $x_k\ge x_0$ and $x_k(1)=x_0(1)=1/(1-\bar\alpha)$.

Then we show that $\bar\alpha$ is a share bound for all these instances and that every $(1-\varepsilon)$-optimal contract of $I_k$ lies in its window. As $r(z)=z$, the principal's utility in $I_0$ is $\UP^{(0)}(\alpha)=(1-\alpha)\,x_0(\alpha)$. It equals $\alpha(1-\alpha)/(\varepsilon(1-\varepsilon))\le1$ on $[0,\varepsilon]$, as $\varepsilon<1/2$, equals $1$ on $[\varepsilon,\bar\alpha]$, and equals $(1-\alpha)/(1-\bar\alpha)\le1$ on $[\bar\alpha,1]$. So $\mathrm{OPT}_0=1$ is attained on $[\varepsilon,\bar\alpha]$, and $I_0$ is a continuous counterpart of the equal-revenue instances of~\cite[Def.~5]{DFGR26}, as every share in $[\varepsilon,\bar\alpha]$ is optimal, each with a different best response. In $I_k$, the utility $\UP^{(k)}$ equals $\UP^{(0)}\le1$ outside the window, and on the window $\UP^{(k)}(\alpha)=(1-\alpha)/(1-l_k)$ decreases from $\UP^{(k)}(l_{k-1})=(1-l_{k-1})/(1-l_k)=\gamma=1+2\varepsilon$. So $\mathrm{OPT}_k=1+2\varepsilon$ is attained at $l_{k-1}\le\bar\alpha$. Hence $\bar\alpha$ is a share bound for all these instances, and, as $(1-\varepsilon)(1+2\varepsilon)=1+\varepsilon(1-2\varepsilon)>1$, no share outside the window is $(1-\varepsilon)$-optimal for $I_k$. Moreover, for $a\in[\varepsilon,\bar\alpha]$ and $k\in\{0,\dots,K\}$, the share bound of \cref{cor:probe} satisfies $\bar\alpha_a=1-(1-a)\,x_k(a)/x_k(1)\le1-(1-a)\,x_0(a)(1-\bar\alpha)=\bar\alpha$.

Now we show that $\mathcal A$ needs at least $K-1$ best-response computations on $I_0$. Let $\alpha_1,\dots,\alpha_q$ be the shares at which $\mathcal A$ computes best responses on $I_0$, and let $\hat\alpha$ be the share it returns. The windows $[l_{k-1},l_k)$, $k\in[K]$, are pairwise disjoint, so at most $q+1$ of them contain one of the shares $\alpha_1,\dots,\alpha_q,\hat\alpha$. Suppose that $q+1<K$. Then some window $[l_{k-1},l_k)$ contains none of them. As $x_k=x_0$ outside this window, $\mathcal A$ receives on $I_k$ the same best responses as on $I_0$, so, being deterministic, it computes them at the same shares and returns $\hat\alpha$ on $I_k$ too. But $\bar\alpha$ is a share bound for $I_k$, and $\hat\alpha$ is not $(1-\varepsilon)$-optimal for $I_k$, a contradiction. So $q\ge K-1$.

Finally, we bound $K$ from below. By the choice of $K$, $(1-\varepsilon)\gamma^{-K-1}<1-\bar\alpha$, so $K+1>\ln((1-\varepsilon)/(1-\bar\alpha))/\ln(1+2\varepsilon)$. Here $\ln((1-\varepsilon)/(1-\bar\alpha))\ge0$, $\ln(1+2\varepsilon)\le2\varepsilon$, and $\ln(1/(1-\varepsilon))\le\varepsilon/(1-\varepsilon)\le2\varepsilon$, as $\varepsilon<1/2$. So
\[
q\ge K-1>\frac{\ln(1/(1-\bar\alpha))-2\varepsilon}{2\varepsilon}-2=\frac{\ln(1/(1-\bar\alpha))}{2\varepsilon}-3.\qedhere
\]
\end{proof}

\subsection{Convex costs}\label{app:convex}

For convex costs in general, the best responses and an optimal contract share can be transcendental, as the following example shows. Take one element with $f(\{e\})=3$, $m_e=1$, and $c(x)=(\exp(x)-1)/8$. The agent maximizes the strictly concave function $\alpha x-(\exp(x)-1)/8$ over $[0,3]$, whose derivative is $\alpha-\exp(x)/8$. As $\ln(8\alpha)\in[0,\ln8]\subseteq[0,3]$ for $\alpha\in[1/8,1]$, this gives
\[
  \xs(\alpha)=\begin{cases}0,&\alpha<1/8,\\\ln(8\alpha),&1/8\le\alpha\le1.\end{cases}
\]
So $\UP=0$ on $[0,1/8]$ and at $1$, and $\UP(\alpha)=(1-\alpha)\ln(8\alpha)$ is strictly concave on $[1/8,1]$, with derivative zero where $(1-\alpha)/\alpha=\ln(8\alpha)$, that is, $8\alpha=\exp(1/\alpha-1)$. This solution is the optimal contract $\as\approx0.4419$. If $\as$ were algebraic, then $1/\as-1\ne0$ would be algebraic, and $\exp(1/\as-1)=8\as$ would be transcendental by the Lindemann--Weierstrass theorem, a contradiction. In the same way, $\xs(\alpha)=\ln(8\alpha)$ is transcendental for every algebraic $\alpha\in(1/8,1]$.

For the oracle analysis, let $K\subseteq\R^E_{\ge0}$ be compact with $0\in K$, and let $c\colon K\to\R_{\ge0}$ be continuous with $c(0)=0$. We use the reward bound $\rbar$ and the approximate best-response oracle of \cref{sec:convex}. Write $V(\alpha):=\max_{x\in K}\UA(x\mid\alpha)$ and let $X^\star(\alpha)$ be the set of maximizers. Then $V(0)=0$ and $V(1)=W$.

A $\delta$-optimal response bounds the reward of every best response at larger shares from below and at smaller shares from above.

\begin{lemma}\label{lem:shift}
Let $0\le\alpha''<\alpha<\alpha'$, let $\delta\ge0$, and let $\tilde{x}$ be a $\delta$-optimal response at $\alpha$. Then every $y''\in X^\star(\alpha'')$ and every $y'\in X^\star(\alpha')$ satisfy
\[
  r(y'')-\frac{\delta}{\alpha-\alpha''}\ \le\ r(\tilde{x})\ \le\ r(y')+\frac{\delta}{\alpha'-\alpha}.
\]
\end{lemma}

\begin{proof}
As in the proof of \cref{lem:monreward}, we add two inequalities. As $y'$ is a best response at $\alpha'$ and $\tilde{x}$ is $\delta$-optimal at $\alpha$,
\[
  \alpha'r(y')-c(y')\ge\alpha'r(\tilde{x})-c(\tilde{x})
  \qquad\text{and}\qquad
  \alpha r(\tilde{x})-c(\tilde{x})\ge\alpha r(y')-c(y')-\delta.
\]
Adding them gives $(\alpha'-\alpha)\big(r(y')-r(\tilde{x})\big)\ge-\delta$, which is the right-hand inequality. In the same way, $\alpha''r(y'')-c(y'')\ge\alpha''r(\tilde{x})-c(\tilde{x})$ and $\alpha r(\tilde{x})-c(\tilde{x})\ge\alpha r(y'')-c(y'')-\delta$ give $(\alpha-\alpha'')\big(r(\tilde{x})-r(y'')\big)\ge-\delta$, which is the left-hand inequality.
\end{proof}

The welfare exceeds $\mathrm{OPT}$ at most by a logarithmic factor, which gives a share bound. For finitely many actions, D\"utting, Ezra, Feldman, and Kesselheim~\cite[Obs.~4.3, Lemma~4.4]{DEFK25} bound the share of an optimal contract in the same way. They use that the welfare is at most the number of actions times the principal's optimum~\cite{DRT19}, and their agent chooses among the $2^m$ sets of $m$ actions.

\begin{lemma}\label{lem:welfare}
We have $\mathrm{OPT}\le W\le\rbar$. If $W>0$, then $\mathrm{OPT}\ge W/\big(2\ln(2\rbar/W)\big)$.
\end{lemma}

\begin{proof}
We first prove the upper bounds. Let $\alpha\in[0,1]$ and $y:=\xs(\alpha)$. Then $W\ge r(y)-c(y)=\UP(\alpha)+V(\alpha)\ge\UP(\alpha)$, as $V(\alpha)\ge0$, so $\mathrm{OPT}\le W$. And $W\le\max_{x\in K}r(x)\le\rbar$ as $c\ge0$.

Now we prove the lower bound. Let $0\le\alpha<\alpha'\le1$. The best response $\xs(\alpha')$ is feasible at $\alpha$, and $\xs(\alpha)$ at $\alpha'$, so $V(\alpha)\ge V(\alpha')-(\alpha'-\alpha)\,r(\xs(\alpha'))$ and $V(\alpha')\ge V(\alpha)+(\alpha'-\alpha)\,r(\xs(\alpha))$, that is,
\begin{equation}\label{eq:F-incr}
  (\alpha'-\alpha)\,r(\xs(\alpha))\ \le\ V(\alpha')-V(\alpha)\ \le\ (\alpha'-\alpha)\,r(\xs(\alpha')).
\end{equation}
(Summing \eqref{eq:F-incr} over fine partitions of $[0,1]$ shows $W=\int_0^1r(\xs(\alpha))\,d\alpha$, as $r(\xs(\cdot))$ is nondecreasing, but we need only the upper bound.) For $\alpha'<1$ we have $r(\xs(\alpha'))\le\mathrm{OPT}/(1-\alpha')$, as $(1-\alpha')\,r(\xs(\alpha'))=\UP(\alpha')\le\mathrm{OPT}$, and $r(\xs(1))\le\rbar$. Let $a:=1-W/(2\rbar)$, which lies in $[1/2,1)$ as $0<W\le\rbar$, let $N\ge1$, and let $\alpha_i:=ia/N$ for $i=0,\dots,N$. Summing the right-hand inequality of \eqref{eq:F-incr} over the intervals $[\alpha_{i-1},\alpha_i]$ and $[a,1]$ gives
\[
  W=V(1)-V(0)\ \le\ \mathrm{OPT}\sum_{i=1}^N\frac{\alpha_i-\alpha_{i-1}}{1-\alpha_i}+(1-a)\,\rbar .
\]
The sum is a Riemann sum of the continuous function $\alpha\mapsto1/(1-\alpha)$ on $[0,a]$, so it tends to $\int_0^a d\alpha/(1-\alpha)=\ln(2\rbar/W)$ as $N\to\infty$. Hence $W\le\mathrm{OPT}\,\ln(2\rbar/W)+W/2$, and the claim follows as $\ln(2\rbar/W)\ge\ln2>0$.
\end{proof}

So the area under the curve $\alpha\mapsto r(\xs(\alpha))$ exceeds the largest rectangle $(1-\alpha)\,r(\xs(\alpha))$ under it at most by a logarithmic factor, a continuous counterpart of the gap between welfare and linear contracts for finitely many actions, which is logarithmic in the ratio of the largest to the smallest expected reward~\cite[Thm.~5.5]{DRT19}. If $W=0$, then $\mathrm{OPT}=0$.

\Cref{alg:convex} uses an oracle that returns, for a share $\alpha\in[0,1]$ and a precision $\delta>0$, a $\delta$-optimal response at $\alpha$. It estimates the welfare, which gives a share bound $\bar\alpha$ and the precision $\delta$, and then bridges every share $\hat\alpha_j$ constructed in the proof of \cref{thm:fptas} by two shifts. The number $w$ of \cref{thm:convexfptas} is optional and serves only to recognize $W=0$.

\begin{algorithm}[ht]
\caption{Approximation scheme using approximate responses}\label{alg:convex}
\begin{algorithmic}[1]
\Require $\varepsilon\in(0,1)$, an oracle returning $\delta$-optimal responses and their reward and cost values, optionally $w\in(0,\rbar]$ with $W=0$ or $W\ge w$
\Ensure ``$W=0$'', or a share $b$ and a number $\tilde{U}\le(1-b)\,r(y)$ for all $y\in X^\star(b)$
\State $d\gets\rbar$
\Loop \Comment{estimate the welfare}
  \State $\tilde{x}\gets$ a $d$-optimal response at the share $1$, \ $\tilde{W}\gets r(\tilde{x})-c(\tilde{x})$
  \State \textbf{if} $\tilde{W}\ge2d$ \textbf{then break}
  \State \textbf{if} $w$ is given and $d<w/3$ \textbf{then return} ``$W=0$''
  \State $d\gets d/2$
\EndLoop
\State $\mathrm{LB}\gets\tilde{W}/\big(2\lceil\log_2(2\rbar/\tilde{W})\rceil\big)$, \ $\bar\alpha\gets1-\mathrm{LB}/\rbar$, \ $\delta\gets\varepsilon^2\,\mathrm{LB}/64$
\State $J\gets\min\{j\in\Z_{\ge0}:(1-\varepsilon/4)^j\le1-\bar\alpha\}$, \ $s\gets\lceil\log_2(8/(\varepsilon(1-\bar\alpha)))\rceil$
\For{$j=0,\dots,J$}
  \State $\hat\alpha_j\gets1-2^{-s}\lfloor2^s(1-\varepsilon/4)^j\rfloor$
  \State $h_j\gets\frac\varepsilon8(1-\hat\alpha_j)$, \ $\check\alpha_j\gets\hat\alpha_j+2h_j$, \ $\tilde{x}_j\gets$ a $\delta$-optimal response at the share $\hat\alpha_j+h_j$
  \State $\tilde{U}_j\gets(1-\check\alpha_j)\big(r(\tilde{x}_j)-\delta/h_j\big)$
\EndFor
\State choose $\hat\jmath$ with $\tilde{U}_{\hat\jmath}=\max_{j\le J}\tilde{U}_j$, and \Return $b:=\check\alpha_{\hat\jmath}$ and $\tilde{U}:=\tilde{U}_{\hat\jmath}$
\end{algorithmic}
\end{algorithm}

\begin{proposition}\label{prop:convexscheme}
Under the assumptions on $K$, $c$, $r$, and the response oracle at the start of \cref{app:convex}, let $\varepsilon\in(0,1)$. If $W>0$, then \cref{alg:convex}, with or without $w$, returns a share $b\in(0,1)$ and a number $\tilde{U}$ with
\[
  (1-b)\,r(y)\ \ge\ \tilde{U}\ \ge\ (1-\varepsilon)\,\mathrm{OPT}\qquad\text{for every }y\in X^\star(b).
\]
It calls the oracle at most $\lceil\log_2(3\rbar/W)\rceil+1$ times at the share $1$, with precisions $d>W/6$, and at most $\lceil8\varepsilon^{-1}\ln(3\rbar/W)\rceil+1$ times with the precision $\delta\ge\varepsilon^2W/\big(192\lceil\log_2(3\rbar/W)\rceil\big)$. If $W=0$, the loop does not end without $w$, and with $w$ the algorithm returns ``$W=0$'' after at most $\lfloor\log_2(3\rbar/w)\rfloor+2$ calls, all with precisions $d\ge w/6$.
\end{proposition}

\begin{proof}
We first analyze the welfare loop. A $d$-optimal response at the share $1$ gives $W-d\le\tilde{W}\le W$, and the calls have the precisions $d=2^{-k}\rbar$ for $k=0,1,\dots$. Let first $W>0$. We show that the loop ends after at most $\lceil\log_2(3\rbar/W)\rceil+1$ calls, all with $d>W/6$, that it does not return ``$W=0$'', and that then $\frac23W\le\tilde{W}\le W$. Whenever $d\le W/3$, we get $\tilde{W}\ge W-d\ge2d$, so the loop breaks at the latest at the least $k$ with $2^{-k}\rbar\le W/3$, and this $k$ is at most $\lceil\log_2(3\rbar/W)\rceil$. Every earlier precision is larger than $W/3$, so every precision used, including the last one, is larger than $W/6$. If $w$ is given, then $W\ge w$, so at every $d<w/3$ the test $\tilde{W}\ge2d$, which comes first, succeeds, and ``$W=0$'' is not returned. When the loop breaks, $\tilde{W}\ge2d$, so $W\le\tilde{W}+d\le\frac32\tilde{W}$. Let now $W=0$. Then $\tilde{W}\le0<2d$ in every round, so the loop does not end without $w$. With $w$, it returns ``$W=0$'' at the least $k$ with $2^{-k}\rbar<w/3$, which is $\lfloor\log_2(3\rbar/w)\rfloor+1\ge1$ as $w\le\rbar$, and $2^{-k}\rbar\ge w/6$ by the minimality of $k$. This proves the claims on the loop.

Next we show that $\mathrm{LB}<\mathrm{OPT}$, so that $\bar\alpha$ is a share bound, and we bound $\delta$ and $J$. As $0<\tilde{W}\le W\le\rbar$, we have $2\rbar/\tilde{W}\ge2$, so $\lceil\log_2(2\rbar/\tilde{W})\rceil\ge\log_2(2\rbar/\tilde{W})>\ln(2\rbar/\tilde{W})$. The function $u\mapsto u/(2\ln(2\rbar/u))$ is increasing on $(0,\rbar]$. With $\tilde{W}\le W$ and \cref{lem:welfare},
\[
  \mathrm{LB}\ <\ \frac{\tilde{W}}{2\ln(2\rbar/\tilde{W})}\ \le\ \frac{W}{2\ln(2\rbar/W)}\ \le\ \mathrm{OPT}.
\]
So $0<\bar\alpha<1$, and for $\alpha\in(\bar\alpha,1]$ we have $\UP(\alpha)\le(1-\alpha)\,\rbar\le(1-\bar\alpha)\,\rbar=\mathrm{LB}<\mathrm{OPT}$. Hence $\sup_{\alpha\in[0,\bar\alpha]}\UP(\alpha)=\mathrm{OPT}$, and $\bar\alpha$ is a share bound. As $\tilde{W}\ge\frac23W$ by the analysis of the loop, $2\rbar/\tilde{W}\le3\rbar/W$, so $\mathrm{LB}\ge W/(3\lceil\log_2(3\rbar/W)\rceil)$, which gives the bound on $\delta=\varepsilon^2\mathrm{LB}/64$. Moreover $\lceil\log_2z\rceil\le z$ for $z\ge1$, so $1/(1-\bar\alpha)=\rbar/\mathrm{LB}=(2\rbar/\tilde{W})\lceil\log_2(2\rbar/\tilde{W})\rceil\le(3\rbar/W)^2$. By \cref{thm:fptas} with the accuracy $\varepsilon/2$, the number of shares is $J+1\le\lceil4\varepsilon^{-1}\ln(1/(1-\bar\alpha))\rceil+1\le\lceil8\varepsilon^{-1}\ln(3\rbar/W)\rceil+1$.

Next we show that every $\tilde{U}_j$ is a valid bound, that is, $\check\alpha_j\in(0,1)$ and $(1-\check\alpha_j)\,r(y)\ge\tilde{U}_j$ for every $y\in X^\star(\check\alpha_j)$. With $\varepsilon/2$ in place of $\varepsilon$, the grid in the proof of \cref{thm:fptas} uses $(1-\varepsilon/4)^j$ and $2^{-s}\le\varepsilon(1-\bar\alpha)/8$. If $J\ge1$, the minimality of $J$ gives $(1-\varepsilon/4)^{J-1}>1-\bar\alpha$, so $(1-\varepsilon/4)^J>\frac34(1-\bar\alpha)>2^{-s}$, and if $J=0$, then $(1-\varepsilon/4)^J=1>2^{-s}$. So $1-\hat\alpha_j=2^{-s}\lfloor2^s(1-\varepsilon/4)^j\rfloor>(1-\varepsilon/4)^j-2^{-s}\ge(1-\varepsilon/4)^J-2^{-s}>0$ for every $j\le J$. Hence $h_j>0$ and $1-\check\alpha_j=(1-\varepsilon/4)(1-\hat\alpha_j)\in(0,1)$. Let $y\in X^\star(\check\alpha_j)$. The right-hand inequality of \cref{lem:shift} with $\alpha=\hat\alpha_j+h_j$ and $\alpha'=\check\alpha_j$ gives $r(y)\ge r(\tilde{x}_j)-\delta/h_j$, so $(1-\check\alpha_j)\,r(y)\ge\tilde{U}_j$. For $j=\hat\jmath$, this is the left-hand inequality of the claim.

Finally, we show that some $\tilde{U}_j$ is large, namely $\tilde{U}_j\ge(1-\varepsilon)\,\mathrm{OPT}$ for some $j$. The proof of \cref{thm:fptas} uses only \cref{lem:monreward}, which holds for all maximizers, so the theorem applies to the best responses $\xs(\alpha)$ and the share bound $\bar\alpha$ shown above. It gives a $j\le J$ with $\UP(\hat\alpha_j)\ge(1-\varepsilon/2)\,\mathrm{OPT}$. The left-hand inequality of \cref{lem:shift} with $\alpha''=\hat\alpha_j$, $\alpha=\hat\alpha_j+h_j$, and $y''=\xs(\hat\alpha_j)$ gives $r(\tilde{x}_j)\ge r(\xs(\hat\alpha_j))-\delta/h_j$. As $1-\check\alpha_j=(1-\varepsilon/4)(1-\hat\alpha_j)$, $h_j=\frac\varepsilon8(1-\hat\alpha_j)$, and $\delta=\varepsilon^2\mathrm{LB}/64$,
\begin{align*}
  \tilde{U}_j&\ \ge\ (1-\check\alpha_j)\Big(r(\xs(\hat\alpha_j))-\frac{2\delta}{h_j}\Big)
  =\Big(1-\frac\varepsilon4\Big)\Big(\UP(\hat\alpha_j)-\frac{\varepsilon\,\mathrm{LB}}4\Big)\\
  &\ \ge\ \Big(1-\frac\varepsilon4\Big)\Big(1-\frac\varepsilon2\Big)\mathrm{OPT}-\frac\varepsilon4\,\mathrm{OPT}
  \ \ge\ (1-\varepsilon)\,\mathrm{OPT},
\end{align*}
using $\mathrm{LB}<\mathrm{OPT}$. As $\tilde{U}\ge\tilde{U}_j$, this proves the right-hand inequality of the claim.
\end{proof}

We are now in a position to prove \cref{prop:approxresponses}.

\restate{\rsApproxresponses}{prop:approxresponses}
\begin{proof}
Apply \cref{prop:convexscheme} with the supplied $w$. If $W>0$, then $W\ge w$, so its call bound is $\Oh(\varepsilon^{-1}\log(3\rbar/w))$, and the returned contract is $(1-\varepsilon)$-optimal. If $W=0$, it detects this with $\Oh(\log(3\rbar/w))$ calls. The required oracle precisions and the guarantee without a supplied $w$ are given in \cref{prop:convexscheme}.
\end{proof}

\subsubsection{Computing \texorpdfstring{$\delta$}{delta}-optimal responses}\label{sec:F-ellipsoid}

The ellipsoid method computes a $\delta$-optimal response in a number of iterations polynomial in $n$ and $\log((\rbar+G\,f(E))/\delta)$. We use the exact real-arithmetic model of \cref{sec:convex}, counting submodular minimizations, exact calls of the oracle for $c$, and arithmetic operations and comparisons, including square roots. The oracle is queried at feasible real ellipsoid centers.

\begin{lemma}\label{lem:ellipsoid}
Let $\alpha\in[0,1]$ and $0<\delta<\rbar+G\,f(E)$. Then the central-cut ellipsoid method computes a $\delta$-optimal response at $\alpha$ with at most $\big\lceil2n^2\ln\big(2n^{3/2}(\rbar+G\,f(E))/\delta\big)\big\rceil$ iterations, each of which makes at most one submodular minimization, at most one call of the oracle for $c$, and $\Oh(n^2)$ further arithmetic operations. For $\delta\ge\rbar+G\,f(E)$, the response $0$ is $\delta$-optimal.
\end{lemma}

\begin{proof}
We first reduce to the case that $f(\{e\})>0$ for all $e$. Let $E':=\{e\in E:f(\{e\})>0\}$. For $e\notin E'$, every $x\in\Pp{f}$ has $x_e=0$, and $f(S)=f(S\cap E')$ for all $S\subseteq E$, since $f(S\cap E')\le f(S)\le f(S\cap E')+\sum_{e\in S\setminus E'}f(\{e\})$ by monotonicity and submodularity. So $\Pp{f}$ is $\Pp{f|_{E'}}$, padded with zeros, and we may assume that $f(\{e\})>0$ for all $e\in E$. If $E'=\emptyset$, then $\Pp{f}=\{0\}$ and there is nothing to compute.

Next we place $\Pp{f}$ between two cubes. In the coordinates $y_e:=x_e/f(\{e\})$, the image $K$ of $\Pp{f}$ satisfies $[0, 1/n]^E\subseteq K\subseteq[0,1]^E$. Indeed, if $0\le x_e\le f(\{e\})/n$ for all $e$ and $\emptyset\ne S\subseteq E$, then $x(S)\le |S| \max_{e\in S}f(\{e\}) /n \le f(S)$ by monotonicity, and every $x\in\Pp{f}$ has $x_e\le f(\{e\})$. So $K$ contains a ball of radius $1/(2n)$ and lies in a ball of radius $\sqrt n/2$, and the ratio of the radii is $n^{3/2}$.

A separation oracle needs one submodular minimization. At a center $y$ in normalized coordinates, set $z_e:=f(\{e\})y_e$. Either $z_e<0$ for some $e$, which gives the cut $x_e\ge0$, or one submodular minimization of $T\mapsto f(T)-z(T)$ returns a set $T$ with $z(T)>f(T)\ge x(T)$ for all $x\in\Pp{f}$, or it shows $z\in\Pp{f}$. In normalized coordinates the violated inequality is $y_e\ge0$ or $\sum_{e\in T} f(\{e\})y_e\le f(T)$. Its normal gives the central cut through the current center.

Next we bound the objective. We minimize $-\UA(\cdot\mid\alpha)=c-\alpha\,r$, which is convex on $\Pp{f}$ with the subgradient $g(x)-\alpha m$ at $x$. For $x\in\Pp{f}$, the subgradient inequality with $y=0$ gives $0=c(0)\ge c(x)-g(x)^\top x$, so $0\le c(x)\le G\,x(E)\le G\,f(E)$, and $0\le\alpha\,r(x)\le\rbar$. Hence $|\UA(\cdot\mid\alpha)|\le M:=\rbar+G\,f(E)$ on $\Pp{f}$.

Finally, we apply the ellipsoid method. If a convex body in $\R^n$ lies in a ball of radius $R_{\mathrm{out}}$ and contains a ball of radius $R_{\mathrm{in}}$, and a continuous convex function has absolute values at most $M$ on it, then after $t\ge2n^2\ln(R_{\mathrm{out}}/R_{\mathrm{in}})$ iterations of the ellipsoid method, which uses a separation oracle for the body and subgradients of the function at the centers in the body, some center lies in the body, and the best such center $x_t$ is within $(2MR_{\mathrm{out}}/R_{\mathrm{in}})\exp(-t/(2n^2))$ of the minimum~\cite[Thm.~2.4]{Bub15}. In dimension one, the method is bisection, which halves the length of the interval in each step and so satisfies the same bound. We apply this to $K$ and to $-\UA(\cdot\mid\alpha)$ in the coordinates $y$, where the subgradients are scaled by $f(\{e\})$, with $R_{\mathrm{out}}/R_{\mathrm{in}}=n^{3/2}$. For $t:=\lceil2n^2\ln(2n^{3/2}M/\delta)\rceil$, which is positive and exceeds $2n^2\ln(n^{3/2})$ since $\delta<M$, this gives $\UA(x_t\mid\alpha)\ge V(\alpha)-\delta$, so $x_t$ is a $\delta$-optimal response. The values of $c$ at the centers in $\Pp{f}$ select $x_t$, and each iteration updates the ellipsoid with $\Oh(n^2)$ operations. If $\delta\ge M$, then $V(\alpha)-\UA(0\mid\alpha)=V(\alpha)\le\rbar\le\delta$.
\end{proof}

The lemma uses exact separating cuts, subgradients, and objective values. In particular, its proof does not establish a finite-precision implementation: approximate oracle answers would require error bounds for the cuts, the welfare tests, and the utility certificates in \cref{alg:convex}.

\renewcommand{\thissection}{\cref{sec:convex}}%
\restate{\rsConvexfptas}{thm:convexfptas}
\begin{proof}
Run \cref{alg:convex} with $w$ and with \cref{lem:ellipsoid} as the oracle. By \cref{prop:convexscheme}, it returns ``$W=0$'' if and only if $W=0$, and then $\mathrm{OPT}=0$ by \cref{lem:welfare}, so every share is optimal. Otherwise $W\ge w$, and the share $b$ has the claimed guarantee. As $u\mapsto u/\lceil\log_2(3\rbar/u)\rceil$ is nondecreasing on $(0,\rbar]$, the bounds of \cref{prop:convexscheme} hold with $w$ in place of $W$. So, in both cases, $\Oh(\varepsilon^{-1}\log(3\rbar/w))$ responses are computed, each with a precision at least $\delta_0:=\varepsilon^2w/(192\lceil\log_2(3\rbar/w)\rceil)$, as $w/6\ge\delta_0$. With $M:=\rbar+G\,f(E)\ge w$ and $\lceil\log_2(3\rbar/w)\rceil\le3M/w$, we get $\ln(M/\delta_0)\le\ln(576/\varepsilon^2)+2\ln(M/w)$, so each response takes $\Oh(n^2\log(2nM/(\varepsilon w)))$ iterations by \cref{lem:ellipsoid}. The remaining work, the $n+1$ evaluations of $f$ for the coordinate bounds $f(\{e\})$ used by the ellipsoid method and for $f(E)$, the values of $r$ at the responses (those of $c$ come with them), the number $\lceil\log_2(2\rbar/\tilde{W})\rceil$, and the $J+1$ shares constructed in the proof of \cref{thm:fptas} with $s=\Oh(\log(1/(\varepsilon(1-\bar\alpha))))$ bits each, takes a number of operations polynomial in $n$, $1/\varepsilon$, and $\log(3\rbar/w)$, as $\ln(1/(1-\bar\alpha))\le2\ln(3\rbar/W)\le2\ln(3\rbar/w)$.
\end{proof}

\section*{AI methodology}
The authors used the large language model Claude (Anthropic) to assist with the research and the writing of this paper. Additionally, the authors used ChatGPT (OpenAI) for further copy editing and proofreading. The authors posed the research questions and directed the work. Independently of Claude, they worked on linear and quadratic costs. They designed the algorithms for both, the greedy algorithm with its candidate shares and water-filling with the tracing of its pieces (\cref{sec:linear,sec:bestresponse}), programmed visualizations of them, and wrote much of the text on them, which this paper draws from their earlier write-up. Claude proposed and worked out constructions and proofs, among them the instances with exponentially many breakpoints and polynomial bit length (\cref{thm:polybits}), the bounds for polymatroids with polynomially many breakpoints (\cref{rem:subclasses}), the comb gadget with its parameter rule and the reduction (\cref{sec:hardness} and \cref{app:hard}), the $1/p$ rule (\cref{thm:prule}), the scheme for convex costs (\cref{sec:convex}), and the lower bounds of \cref{prop:querylb,prop:sharesets}. It checked statements and proofs with exact rational computations, among them the inequalities of the parameter rule in \cref{app:hard} and the constructions of the lower bounds, and it formalized the results of this paper in the Lean proof assistant. It also drafted the remaining text from the authors' notes and earlier drafts, and revised the whole text on their instructions. The tool thus materially affected \cref{sec:bestresponse,sec:hardness,sec:fptas}, all appendices, and the presentation throughout. The authors verified the correctness and originality of all content, including the references.

\end{document}